\documentclass{aa}  

\usepackage{graphicx}
\usepackage{txfonts}
\usepackage{stfloats}
\usepackage[colorlinks,citecolor=blue]{hyperref}
\usepackage{orcidlink}

\begin{document} 

\title{Synthetic pulsar lightcurves from global kinetic simulations and comparison with the {\em Fermi}-LAT catalog}

\titlerunning{Synthetic pulsar lightcurves and comparison with {\em Fermi} pulsars}

\author{Beno\^it Cerutti\inst{1}\orcidlink{0000-0001-6295-596X} \and Enzo Figueiredo\inst{1}\orcidlink{0009-0006-7103-1965} \and Guillaume Dubus\inst{1}\orcidlink{0000-0002-5130-2514}}

\institute{Univ. Grenoble Alpes, CNRS, IPAG, 38000 Grenoble, France\\
           \email{benoit.cerutti@univ-grenoble-alpes.fr}
           }

\date{Received \today; accepted \today}

 
\abstract
{Rotation-powered pulsars represent the main class of identified gamma-ray sources in the Galaxy. The wealth of observational data collected by the AGILE and \emph{Fermi} gamma-ray space telescopes in the GeV range, and by ground-based Cherenkov telescopes in the TeV band provide invaluable insights into how relativistic plasmas dissipate and accelerate particles.}
{Decoding the information contained in the gamma-ray pulses profile is an important step to understand how pulsars work. In this study, we aim at putting an ab initio plasma model of pulsar magnetospheres to the test, in light of the most recent gamma-ray observations in the GeV and TeV bands.}
{To this end, we present of a new series of global particle-in-cell simulations of an inclined pulsar magnetosphere. High-quality synthetic pulse profiles in the synchrotron and inverse Compton channels are reconstructed to study in greater details their morphology and their energy dependence. We also perform a fit of observed lightcurves with the model, using the third \emph{Fermi}-LAT gamma-ray pulsar catalog.}
{Reconnection in the wind current sheet powers synchrotron and inverse Compton emission. The modeled pulse profiles reproduce some of the salient features of observed gamma-ray pulsars, including the mysterious Vela-like lightcurves, such as: the generic double-peaked structure, the presence of a bridge or third peak in between the main pulses, the pulse narrowing with increasing energy. The bolometric synchrotron radiative efficiency is strictly limited by the reconnection rate.}
{Our global kinetic simulations are able to match observed pulse profiles. Such direct comparisons will help drive and focus future simulation developments.}

\keywords{acceleration of particles -- magnetic reconnection -- radiation mechanisms: non-thermal -- methods: numerical -- pulsars: general -- stars: winds, outflows}

\maketitle


\section{Introduction}

Rotation-powered young and millisecond pulsars represent the dominant class of identified high-energy gamma-ray sources in the Galaxy \citep{2019A&A...627A..13B, 2020ApJS..247...33A}. They are firmly established as such via the detection of short pulses of light modulated with the pulsar spin period. The remarkable coherence and stability of the gamma-ray pulse profile \citep{2022ApJ...934...30K} indicates that the emitting zone must be compact and close to the star where the energy densities are highest, naturally pointing towards a magnetospheric origin. How and where pulses are formed in the magnetosphere is a matter of interpretation. In this regard, the gamma-ray emission plays a special role because most of the power radiated away by pulsars is in the gamma-ray band. Current data estimate that about $1\%$ to $100\%$ of the pulsar spindown power, in other words of the full energy reservoir of the system, is channeled above $100$\,MeV gamma rays alone \citep{2010ApJS..187..460A, 2013ApJS..208...17A, 2023ApJ...958..191S}. In contrast, the radio emission represents a negligible fraction of the spindown power ($\lesssim 10^{-6}$, \citealt{2004hpa..book.....L, 2022ARA&A..60..495P}). All things considered, understanding how the high-energy emission is produced is crucial to map the main dissipative zones in the magnetosphere.

This information is encoded in the gamma-ray pulse profiles. The majority of observed gamma-ray pulsars present two well-separated pulses per spin period. They are in general not aligned in phase and significantly wider than their radio counterparts (if any, and set apart notable exceptions like the Crab pulsar where the radio and the gamma-ray pulses are nearly aligned, \citealt{2010ApJ...708.1254A}). Based solely on statistical arguments, observations show that the gamma-ray beam also appears wider than in radio, that is to say that more pulsars are observed in gamma rays than in radio at a fixed sensitivity. These clues suggest that the gamma-ray and radio emitting regions are probably distinct. The polar cap of the star is not a favorable location for the gamma-ray emission, because it does not generically produce a double-peaked lightcurve unless the magnetic axis is nearly perpendicular to the spin axis of the star. In addition, the polar cap is opaque via magnetic conversion to gamma rays above a few GeV. The discoveries of pulsed emission above $\gtrsim 1\rm{GeV}$, and up to the TeV range in the Crab and Vela pulsars \citep{2016A&A...585A.133A, 2023NatAs...7.1341H} irrevocably push the gamma-ray emitting region far away from the stellar surface regardless of the radiation mechanism, and thus, favoring the outskirts of the magnetosphere.

In recent years, the current sheet forming beyond the light-cylinder surface \citep{1969Natur.223..277M, 1971CoASP...3...80M, 1990ApJ...349..538C, 1999A&A...349.1017B, 1999ApJ...511..351C, 2006ApJ...648L..51S} has become one of the main suspect for the location of the gamma-ray emission, and more generally of the incoherent non-thermal pulsed emission (i.e., from optical to gamma rays, \citealt{1971ApJ...163L..17P, 1983ApJ...274..369P, 1996A&A...311..172L}, as opposed to the coherent radio emission). This scenario offers multiple attractive features, to cite a few: (i) it naturally explains the generic double-peaked structure and the phase-shift with the putative polar-cap emitting radio pulse \citep{2002A&A...388L..29K, 2010ApJ...715.1282B, 2010MNRAS.404..767C, 2011MNRAS.412.1870P, 2013A&A...550A.101A, 2014ApJ...793...97K, 2023ApJ...954..204K}, (ii) it is optically thin to magnetic conversion process, (iii) it is an obvious location for particle acceleration via magnetic reconnection \citep{1996A&A...311..172L, 2001ApJ...547..437L, 2003ApJ...591..366K, 2012MNRAS.424.2023P, 2014ApJ...780....3U}, and (iv) it can explain abrupt swings of the polarization angle \citep{2005ApJ...627L..37P, 2009MNRAS.397..103S, 2010MNRAS.404..767C, 2013MNRAS.434.2636P, 2016MNRAS.463L..89C, 2017ApJ...840...73H}. This scenario is further supported by global particle-in-cell (PIC) plasma simulations of pulsar magnetospheres \citep{2014ApJ...785L..33P, 2014ApJ...795L..22C, 2015MNRAS.448..606C, 2015MNRAS.449.2759B, 2015ApJ...801L..19P, 2018ApJ...858...81B, 2022ApJ...939...42H, 2024A&A...690A.229C, 2024A&A...690A.170S}. In the limit where the plasma supply is abundant in the magnetosphere, simulations consistently show that reconnection in the current sheet consumes $\sim 10\%$ of the pulsar spindown power within a few light-cylinder radii. The amount of dissipation is solely governed by the universal relativistic reconnection rate \citep{2020A&A...642A.204C, 2023ApJ...943..105H}. The energy dissipated is then efficiently channeled into non-thermal particle acceleration and bright synchrotron radiation. Synthetic pulse profiles reconstructed from the global PIC models present a robust double-peaked pattern in qualitative agreement with the current sheet scenario \citep{2016MNRAS.457.2401C, 2018ApJ...855...94P, 2018ApJ...857...44K, 2023ApJ...954..204K}. A pulse of light is received each time the current sheet crosses the observer line of sight, which occurs for most viewing angles twice per spin period.

In this work, we aim at putting further the global PIC model to the test, in light of the most recent gamma-ray observations in the GeV and TeV bands. To this end, we produce synthetic observables of high enough quality so that they can be directly compared to observations, and in particular to the third \emph{Fermi}-LAT catalog \citep{2023ApJ...958..191S}. Here, we focus our attention on the pulse profiles and their energy dependence in both the synchrotron channel, and the inverse Compton channel considering a uniform target photon bath and the anisotropic emission from the neutron star surface. Thereby, our objectives are twofold: (i) provide a new set of reference simulations and synthetic observables (Sects.~\ref{sect::setup}-\ref{sect::results}), and (ii) perform lightcurve fitting of the \emph{Fermi}-LAT gamma-ray pulsar catalog to explore the validity and the implications of the PIC model at the scale of a population (Sect.~\ref{sect::fitting}). We summarize our findings in Sect.~\ref{sect::summary}.

\section{PIC simulation setup}\label{sect::setup}

\subsection{Initial configuration}

This new series of global models of inclined pulsar magnetospheres is performed with the {\tt Zeltron} PIC code \citep{2013ApJ...770..147C, 2016MNRAS.457.2401C}. The numerical setup is similar to the one presented in \citet{2016MNRAS.457.2401C} that we repeat here for completeness. The most important difference is how secondary pairs are injected in the simulation as explained below.

We use a spherical coordinate system $(r,\theta,\phi)$ where grid cells are equally spaced in $\log r$, $\theta$ and $\phi$. The grid is composed of $1024\times256\times512$ cells along the $r$-, $\theta$-, and $\phi$- directions respectively. The inner radial boundary is fixed at the neutron star surface, $r_{\rm min}=r_{\star}$. The outer radial boundary is fixed at $r_{\rm max}=28 r_{\star}$ but a buffer zone encompassing the whole domain absorbs all particles and fields beyond $r_{\rm abs}=25 r_{\star}$ \citep{2015MNRAS.448..606C}, which effectively plays the role of an open boundary. Axial symmetry is imposed to the fields at both $\theta$-boundaries \citep{1983ITNS...30.4592H} and periodic boundary conditions are applied along the $\phi$-direction.

The initial magnetic field is a pure dipole inclined at an angle $\chi$ from the rotation axis, the latter is aligned with the grid axis,
\begin{equation}
\mathbf{B}=\frac{3\left(\mathbf{r}\cdot\boldsymbol{\mu}\right)\mathbf{r}}{r^5}-\frac{\boldsymbol{\mu}}{r^3},
\end{equation}
where $\mu=B_{\star}r^3_{\star}$ is the magnetic moment, and $B_{\star}$ is the surface polar magnetic field. The dipolar field is frozen-into the neutron star surface and spins at the neutron star angular velocity, $\Omega$. The light-cylinder radius, where the co-rotation velocity equals the speed of light, is set at $R_{\rm LC}\equiv c/\Omega=5 r_{\star}$. The rotation of the field lines induces an ideal electric field, $\mathbf{E}=-\left(\boldsymbol{\Omega}\times\mathbf{r}\right)\times\mathbf{B}/c$, that is enforced on the inner boundary at each time step. General relativistic effects are neglected in this work (see however, \citealt{2015ApJ...815L..19P, 2018ApJ...855...94P, 2020ApJ...889...69C, TORRES2024102261}).

The magnetosphere is initially set in vacuum. Plasma is gradually injected into the simulation box via two channels. The first source directly originates from the star surface where a fraction of the surface charge density is delivered at each time step. This density is given by the mismatch between the radial electric field right above the star surface and the ideal co-rotation value according to the jump condition as in \citet{2015MNRAS.448..606C}. To avoid over-injection, the maximum injected density is limited by the Goldreich-Julian number density, $n_{\rm GJ}=\left|\boldsymbol{\Omega}\cdot\mathbf{B}\right|/2\pi e c$, where $e$ is electron charge \citep{1969ApJ...157..869G}. This population is modeled in this work with one particle per cell per species injected at every time step within the first row of cells above the surface. The purpose of this source is to simulate the formation of the primary beam of charges ripped off the stellar crust by the surface electric field \citep{1971ApJ...164..529S, 1975ApJ...196...51R}.

Once injected, these particles are accelerated to high energies as they undergo a large fraction of the vacuum potential drop of the star. Thanks to the strong surface magnetic field and field line curvature, they trigger intense pair creation \citep{1971ApJ...164..529S, 2013MNRAS.429...20T, 2015ApJ...810..144T}. This mechanism represents the second and dominant channel of plasma supply in active magnetospheres which is the main focus of this work. Pair creation is modeled as in \citet{2014ApJ...795L..22C, 2015ApJ...801L..19P}: a pair is produced if the parent particle Lorentz factor, $\gamma$, reaches a fraction of the polar-cap potential drop, $\gamma_{\rm pc}=e\mu\Omega^2/m_{\rm e}c^4$, where $m_{\rm e}$ is the electron mass. Here, this threshold is fixed at $\gamma_{\rm th}=0.05\gamma_{\rm pc}$ and is constant throughout the whole domain. Secondary pairs are boosted along the parent particle direction of motion with a Lorentz factor $\gamma_{\rm s}=0.1 \gamma_{\rm th}$. The energy of the secondary pair is removed from the primary particle energy. This simplified procedure is effective at filling the magnetosphere with abundant pairs, and it differs from \citet{2016MNRAS.457.2401C} where a high-multiplicity plasma is injected from the surface without pair creation elsewhere in the magnetosphere. Ions have a dynamically negligible role in the magnetosphere because they are significantly outnumbered by the pairs. In addition, ions do not participate to pair production and to the radiative output due to their heavier mass. Therefore, ions are not included in the simulations presented here (see, however, \citealt{2014ApJ...795L..22C, 2020A&A...635A.138G, 2024A&A...690A.170S}).

The magnetic field strength is limited by the numerical resolution. The fiducial plasma skindepth scale at the surface of the star, $d_{\rm e}=(m_{\rm e}c^2/4\pi n_{\rm GJ} e^2)^{1/2}$, is resolved by two cells for a particle with $\gamma=1$. Note that this is a conservative estimate since most pairs will have a Lorentz factor $\gamma\gtrsim \gamma_{\rm s}$. In our simulations, we have $\gamma_{\rm pc}=2.6\times 10^3$ so that $\gamma_{\rm th}=1.3\times 10^2$ and $\gamma_{\rm s}=13$. For secondary pairs, the plasma skindepth scale is increased by $d^{\rm s}_{\rm e}=\sqrt{\gamma_{\rm s}}d_{\rm e}$, and hence is resolved by about $7$ cells. This series of new simulations is complemented by a large split-monopole 3D PIC simulation with $\chi=60^{\circ}$ already presented in \citet{2020A&A...642A.204C}. In the latter simulation, the box size extends up to $50$ light-cylinder radii. Our purpose is to reanalyse that simulation to explore the evolution of the emitted radiation at large radii in the wind region.

\subsection{Synchrotron-curvature emission and feedback}

The field strength, the curvature of field lines and the ultra-relativistic nature of the plasma ($\gamma\gg 1$) make curvature and synchrotron cooling efficient and dynamically important in pulsar magnetospheres. The radiative feedback on the particle motion is captured in simulations by including the Landau-Lifshitz radiation-reaction force to the particle equation of motion, $\mathbf{g}$, that depends only on local quantities of the electromagnetic field \citep{1971ctf..book.....L},
\begin{equation}
\mathbf{g}\approx\frac{2}{3}r^2_{\rm e}\left[\left(\mathbf{E}+\boldsymbol{\beta}\times\mathbf{B}\right)\times\mathbf{B}+\left(\boldsymbol{\beta}\cdot\mathbf{E}\right)\mathbf{E}\right]-\frac{2}{3}r^2_{\rm e}\gamma^2\tilde{B}^2_{\perp}\boldsymbol{\beta},
\label{eq::frad}
\end{equation}
where $r_{\rm e}=e^2/m_{\rm e}c^2$ is the classical radius of the electron, $\beta=v/c$ is the particle 3-velocity divided by the speed of light, and
\begin{equation}
\tilde{B}_{\perp}=\sqrt{\left(\mathbf{E}+\boldsymbol{\beta}\times\mathbf{B}\right)^2-\left(\boldsymbol{\beta}\cdot\mathbf{E}\right)^2},
\end{equation}
is the effective magnetic field strength perpendicular to the particle momentum measured in the lab (simulation) frame \citep{2016MNRAS.457.2401C}. Assuming that the emission is perfectly optically thin, the radiation power spectrum emitted by each simulation particle is given by the classical synchrotron-curvature formula (e.g., \citealt{1970RvMP...42..237B})
\begin{equation}
\frac{d\mathcal{P}_{\rm syn}}{d\nu}=\frac{\sqrt{3}e^3 \tilde{B}_{\perp}}{m_{\rm e}c^2}\left(\frac{\nu}{\nu_{\rm c}}\right)\int_{\nu/\nu_{\rm c}}^{+\infty}K_{5/3}(x)dx,
\label{eq_sync}
\end{equation}
where $K_{5/3}$ is the modified Bessel function of $5/3$ order, $\nu$ is the radiation frequency, and
\begin{equation}
\nu_{\rm c}=\frac{3e\tilde{B}_{\perp}\gamma^2}{4\pi m_{\rm e}c}.
\end{equation}
Given the extreme relativistic boosting effect for $\gamma\gg 1$, we assume that photons are produced along the emitting particle's momentum. The frequency-integrated radiated power per lepton is
\begin{equation}
\mathcal{P}_{\rm syn}=\frac{2}{3}r^2_{\rm e}c\gamma^2 \tilde{B}^2_{\perp}.
\label{eq::psyn}
\end{equation}

This classical approach is valid if $\gamma\tilde{B}_{\perp}/B_{\rm QED}\ll~1$, where $B_{\rm QED}=m^2_{\rm e} c^3/\hbar e$ is the critical magnetic field, which is a good approximation in gamma-ray pulsars, except perhaps close to the star surface where this ratio can be close to unity. Because of the unrealistic field strength accessible in simulations, radiative cooling is rescaled by a large constant factor to achieve the relevant radiation-reaction-limited regime in the magnetosphere (e.g., \citealt{2014ApJ...780....3U}). This regime corresponds to the strong cooling limit in which the acceleration rate is balanced by radiative losses. It sets another energy scale in the problem, $\gamma_{\rm rad}$, given by \citep{2012ApJ...746..148C}
\begin{equation}
\gamma_{\rm rad}=\sqrt{\frac{3eE}{2r^2_{\rm e}\tilde{B}^2_{\perp}}},
\label{eq::grad}
\end{equation}
such that $\gamma_{\rm rad}\ll\gamma_{\rm pc}$ in active pulsars. In the simulations presented here, the classical radius of the electron in Eq.~(\ref{eq::grad}) is amplified by a factor $f_{\rm rad}=5\times 10^8$, giving $\gamma_{\rm rad}\approx 4$ at the star surface, and $\gamma^{\rm LC}_{\rm rad}\approx 45 \lesssim \gamma_{\rm th} \ll \gamma_{\rm pc}$ at the light cylinder (assuming $E=\tilde{B}_{\perp}$), and hence preserving a relevant hierarchy of scales. It was shown by \citet{2024A&A...690A.170S} that this rescaling procedure does not affect the results for a fixed ratio between the above energy scales. The shortest synchrotron cooling time is resolved by $4$ simulation time steps. A modified Boris pusher is used in the simulation to incorporate the full radiation-reaction force along with the Lorentz force in the equation of motion \citep{2010NJPh...12l3005T, 2013ApJ...770..147C}. The total number of iteration per pulsar spin period is then fixed to $P/\Delta t=2\pi/\Omega\Delta t\approx 3.3\times 10^5$.

\subsection{Inverse Compton emission}

In contrast to synchrotron and curvature radiation, inverse Compton cooling and its feedback onto the dynamic are negligible in isolated systems. The cooling timescale of TeV electrons onto CMB photons or those emitted by the surrounding nebula is way longer than the pulsar spin period. If, however, the background photon field is concentrated on magnetospheric scales, the IR and optical fluxes in the Crab and Vela pulsars can lead to non-negligible losses \citep{2023NatAs...7.1341H}. In both cases though, synchrotron radiation remains the main cooling channel, and thus the radiation-reaction force induced by Compton scattering is neglected in this work. The computation of the emitted spectrum is done under the same assumptions as above: the medium is optically thin and the upscattered photons are beamed along the emitting particle direction of motion. The background radiation field is fixed and photons produced in the simulation do not contribute it. 

While the curvature and synchotron emissivities are fully determined by the simulation native quantities, the geometry and the energy density of the background radiation field are not well constrained and must be prescribed. In this work, we explore two simple geometries: (i) the simulation is immersed into a uniform isotropic photon bath, which could represent CMB photons or diffuse emission emitted at larger scales such as the nebula surrounding the magnetosphere; and (ii) a radial point-like source located at the neutron star center and decaying as $1/r^2$ meant to simulate thermal photons or polar-cap hotspots present on the star surface. The point-like approximation is valid at light-cylinder scales and beyond in the wind current sheet (since $R_{\rm LC}\gg r_{\star}$). For simplicity, we assume that the target radiation field has a single photon energy, $\epsilon_0$ (normalized to $m_{\rm e}c^2$), and we explore both the Thomson and Klein-Nishina regimes.

In the laboratory frame, the inverse Compton spectrum emitted by a relativistic electron scattering off a soft photon with a pitch angle $\theta_0$ between the electron and incoming photon momenta, is given by \citep{1981Ap&SS..79..321A, 2008A&A...477..691D}
\begin{equation}
\frac{dN_{\rm ic}}{dtd\epsilon_1}=\frac{2\pi r^2_{\rm e}c}{\gamma^2\epsilon_0}\left[1+2q\left(q-1\right)+\frac{1}{2}\frac{\left(\Gamma_{\rm ic} q\right)^2}{1+\Gamma_{\rm ic} q}\right],
\label{eq::aic}
\end{equation}
where
\begin{equation}
q\equiv\frac{\epsilon_1}{\Gamma_{\rm ic}\left(\gamma-\epsilon_1\right)},
\label{eq::q}
\end{equation}
\begin{equation}
\Gamma_{\rm ic}\equiv2\epsilon^{\prime}_0=2\gamma\left(1-\cos\theta_0\right)\epsilon_0,
\end{equation}
where $\epsilon_1$ is the upscattered photon energy (normalized to $m_{\rm e}c^2$), and $\epsilon^{\prime}_0$ is the (normalized) soft photon energy in the rest frame of the electron. Kinematic constraints yield
\begin{equation}
0\leq\epsilon_1\leq \epsilon_+=\gamma\frac{\Gamma_{\rm ic}}{1+\Gamma_{\rm ic}}.
\end{equation}
These formula are valid in the ultra-relativistic limit (head-on approximation, $\gamma\gg 1$) for both the Thomson ($\Gamma_{\rm ic}\ll 1$) and the Klein-Nishina ($\Gamma_{\rm ic}\gg 1$) regimes. Integrating over all frequencies and assuming that $\epsilon_1\gg\epsilon_0$, the radiated power per lepton is given by \citep{1965PhRv..137.1306J,2000APh....12..335B}
\begin{equation}
\mathcal{P}_{\rm ic}=\int_{0}^{\epsilon_+}\epsilon_1\frac{dN_{\rm ic}}{dtd\epsilon_1}d\epsilon_1=\frac{2\pi r^2_{\rm e}c}{\epsilon_0}\mathcal{F}\left(\epsilon^{\prime}_0\right),
\label{eq::Pic}
\end{equation}
where
\begin{eqnarray}
\mathcal{F}\left(x\right)=\frac{-10x^4+93x^2+51x\left(1+x^2\right)+9}{3x\left(1+2x\right)^3} \\ \nonumber
+\frac{\left(x-3\right)\left(x+1\right)}{2x^2}\ln\left(1+2x\right).\hspace{0.5cm}
\end{eqnarray}
Asymptotic limits give
\begin{equation}
\mathcal{F}(x)\approx \frac{4}{3}x^2,~x\ll 1,
\label{eq::PicTH}
\end{equation}
valid in the Thomson regime, and
\begin{equation}
\mathcal{F}(x)\approx \frac{1}{2}\left(\ln 2x -\frac{5}{6}\right),~x\gg1,
\label{eq::PicKN}
\end{equation}
that is relevant to approximate the deep Klein-Nishina regime.

For an isotropic photon field, the emitted spectrum is obtained by averaging Eq.~(\ref{eq::aic}) over all pitch angles,
\begin{equation}
\frac{dN_{\rm iso}}{dtd\epsilon_1}=\frac{1}{2}\int_{-1}^{1}\frac{dN_{\rm ic}}{dtd\epsilon_1}d\left(\cos\theta_0\right).
\end{equation}
Using the variable $q$ defined in Eq.~(\ref{eq::q}) instead of $\cos\theta_0$ offers a straightforward way to perform the integration and to recover the well-known \citet{1968PhRv..167.1159J} formula,
\begin{eqnarray}
\frac{dN_{\rm iso}}{dtd\epsilon_1}=\frac{2\pi r^2_{\rm e}c}{\gamma^2\epsilon_0}\Bigg[2q_j\ln q_j+\left(1+2q_j\right)\left(1-q_j\right) \\ \nonumber
+\frac{1}{2}\frac{\left(\Gamma_j q_j\right)^2}{1+\Gamma_j q_j}\left(1-q_j\right)\Bigg],\hspace{0.7cm}
\label{eq:iso}
\end{eqnarray}
where
\begin{equation}
q_j\equiv\frac{\epsilon_1}{\Gamma_j\left(\gamma-\epsilon_1\right)},
\end{equation}
\begin{equation}
\Gamma_j\equiv4\gamma\epsilon_0,
\end{equation}
and
\begin{equation}
0\leq\epsilon_1\leq\gamma\frac{\Gamma_{j}}{1+\Gamma_{j}}.
\end{equation}
The frequency-integrated power does not admit a simple analytical expression in the isotropic case. In practice, we use the anisotropic expression in Eq.~(\ref{eq::Pic}) and draw $\cos\theta_0$ as a random number uniformly distributed between $-1$ and $1$ for each electron in the simulation.

\section{Synthetic pulse profiles}\label{sect::results}

We briefly outline the salient features of the magnetosphere (Sect.~\ref{sect::struct}), before proceeding to a detailed analysis of the synthetic synchrotron and inverse Compton emission patterns (Sects.~\ref{sect::skymaps}-\ref{sect::width}), which are the main focus of this part.

\subsection{Magnetospheric structure}\label{sect::struct}

\begin{figure*}
\centering
\includegraphics[width=9.5cm]{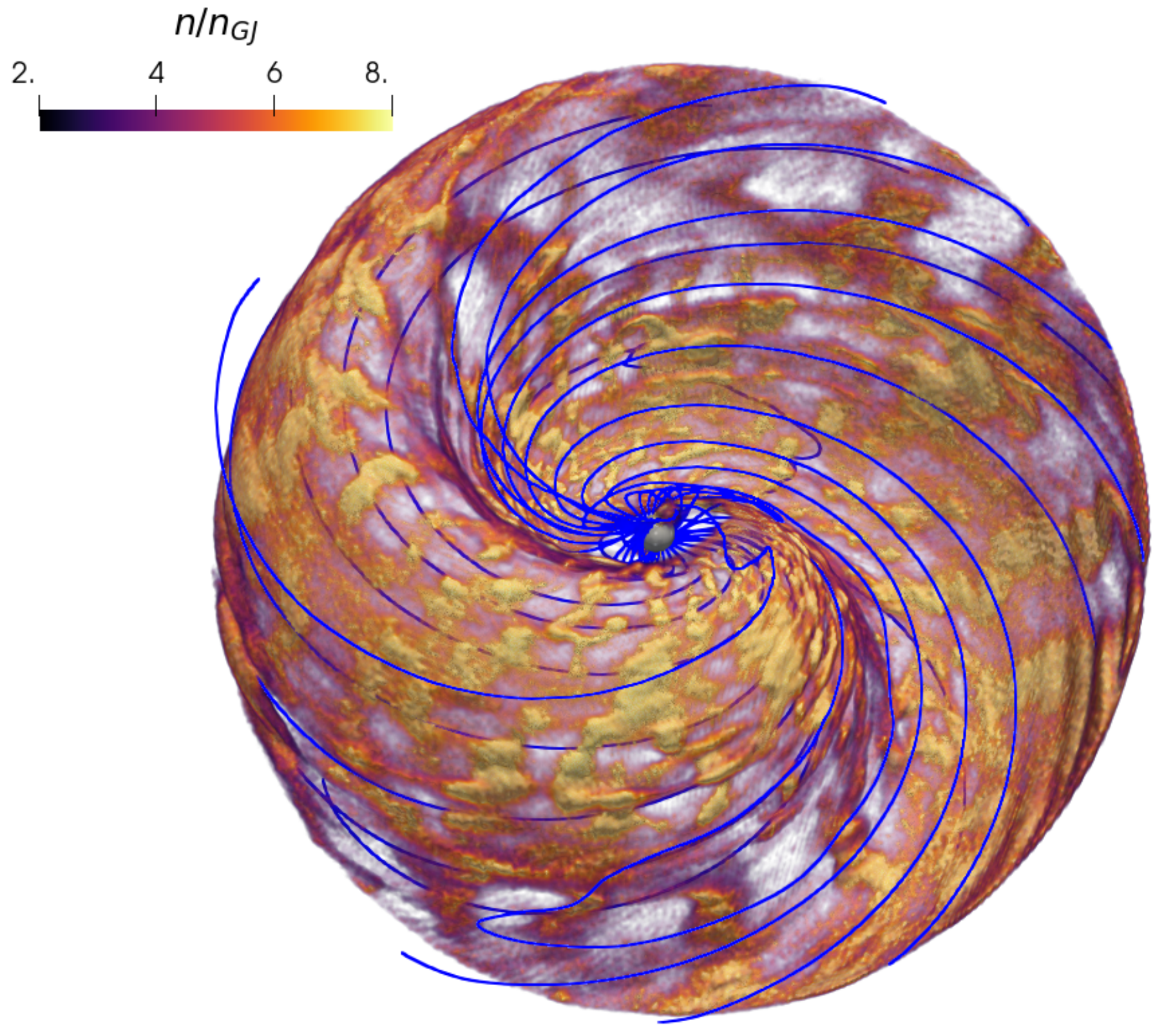}
\includegraphics[width=5.5cm]{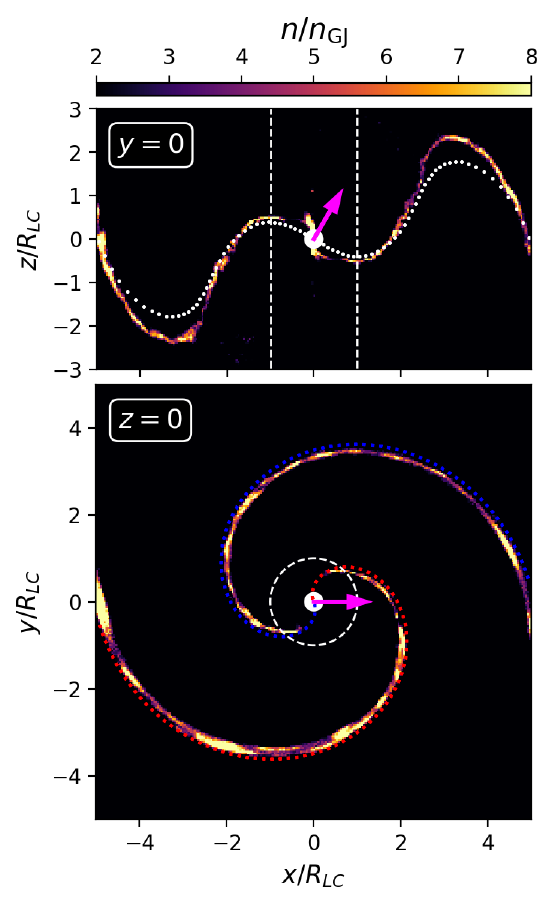}
\caption{Global plasma and magnetic structures forming in a pulsar magnetosphere of magnetic obliquity $\chi=30^{\circ}$. Left panel: Three-dimensional rendering of the plasma density (color-coded) normalized by the surface Goldreich-Julian density and multiplied by $(r/r_{\star})^2$ to compensate for the effect of expansion. Closed and open field lines encompassing the equatorial current layer are shown by blue lines. The star is the gray sphere in the center. Right panels: Poloidal (top, $y=0$) and toroidal (bottom, $z=0$ or $\theta=90^{\circ}$) slice of the plasma density. The light cylinder is shown with the white dashed line. The dotted lines represent the split-monopole solution, and the magenta arrow shows the orientation of the magnetic moment of the star at this time.}
\label{fig::fig1}
\end{figure*}

\begin{figure}
\centering
\includegraphics[width=\hsize]{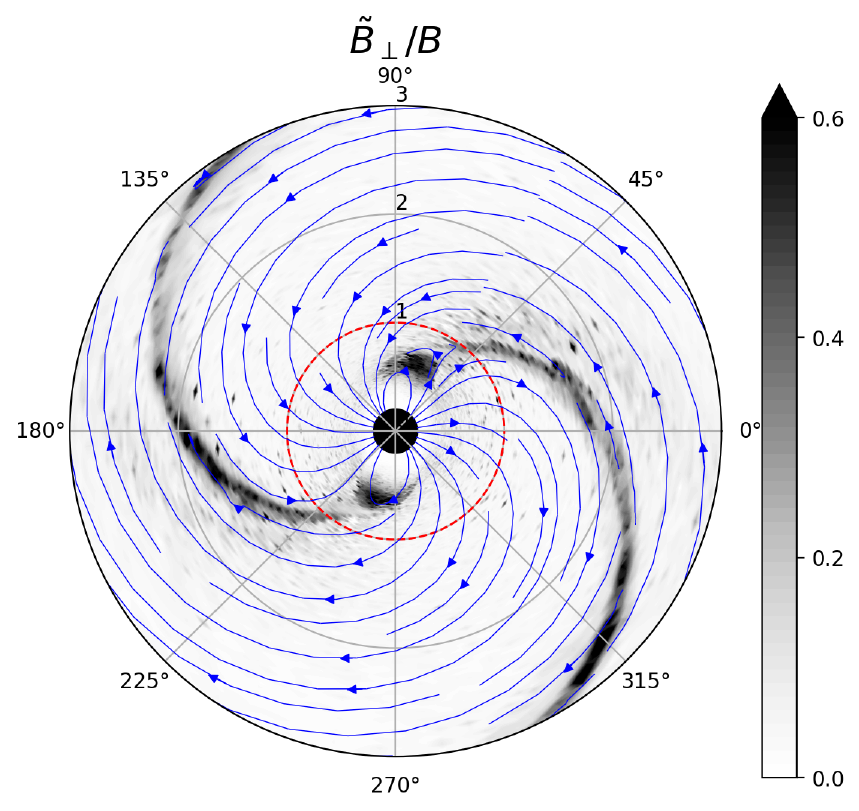}
\caption{Ratio of the effective perpendicular field, $\tilde{B}_{\perp}$, to the total field strength, $B$, in the pulsar midplane ($\theta=90^{\circ}$) for the $\chi=60^{\circ}$ simulation. Blue streamlines represent the magnetic field lines in the plane, and the red dashed circle shows the location of the light cylinder.}
\label{fig::fig_bperp}
\end{figure}

A nearly steady state is established after a few star spin periods. Figure~\ref{fig::fig1} shows a snapshot of a mature magnetospheric solution with a magnetic obliquity $\chi=30^{\circ}$, reached after about two periods once the initial transient has left the domain. The plasma concentrates in the wind current layer that forms beyond the light cylinder where magnetic reconnection, particle acceleration and pair creation take place. The shape of the layer is well captured by the geometry of the magnetic null line predicted by the inclined split-monopole model \citep{1999A&A...349.1017B}, whose location fulfills the condition
\begin{equation}
\cos\theta\cos\chi+\sin\theta\sin\chi\cos\left[\phi-\Omega\left(t-\frac{r}{c}\right)\right]=0.
\label{eq::monopole}
\end{equation}
It is worth noting that this solution is a very good approximation even close to the light cylinder, where the dipolar field is still significant. However, we observe a discrepancy in the poloidal plane ($y=0$, Fig.~\ref{fig::fig1}) where there is an offset in the maximum amplitude of the meanders of the layer. This is especially visible at low inclinations, possibly caused by kink modes growing in the layer (see also in \citealt{2020A&A...642A.204C}), or the effect of a finite magnetization in the wind zone. Overall, the portion of the pulsar wind containing the current layer, or striped region, is well delimited by the expected values: $\pi/2-\chi<\theta<\pi/2+\chi$. In the $r\phi$ plane, the current layer splits into two perfect Archimedean spirals of wavelengths $2\pi R_{\rm LC}$ moving in solid rotation with the star. In the equatorial plane, both arms are diametrically opposed from each other (see the $z=0$ plane, Fig.~\ref{fig::fig1}). The current layer is highly clumpy due to its fragmentation under the tearing instability \citep{2017A&A...607A.134C, 2020A&A...642A.204C, 2023ApJ...943..105H}, which in turn mediates fast reconnection. Underdense regions correspond to magnetic null lines where the field reconnects, while dense regions are flux ropes collecting energetic pairs that have been processed by reconnection.

\subsection{Emission pattern and geometric origin of pulses}\label{sect::skymaps}

\begin{figure*}
\centering
\includegraphics[width=18cm]{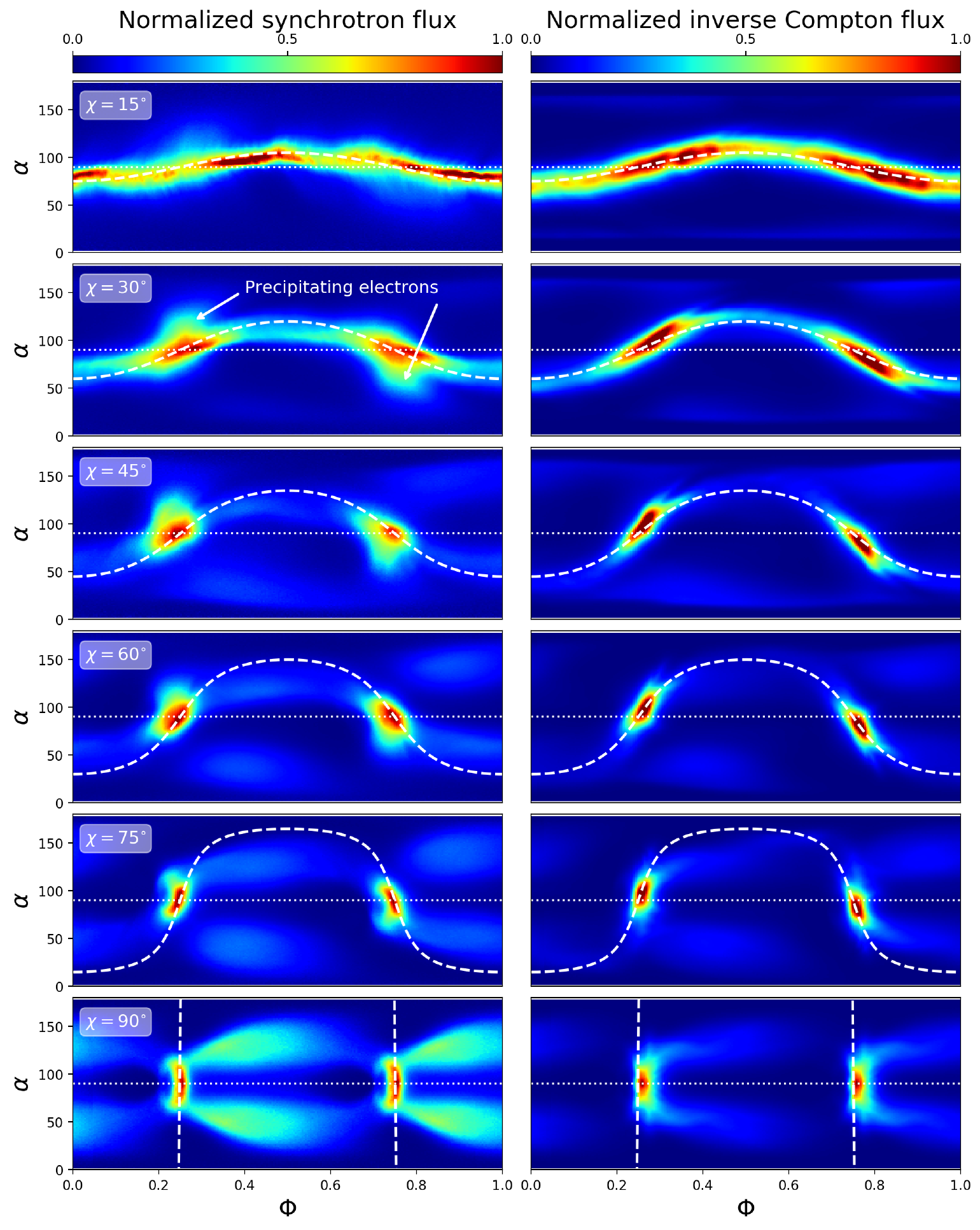}
\caption{Synthetic synchrotron (left panels) and inverse Compton (isotropic radiation field, right panels) fluxes as a function of the observer's viewing angle, $\alpha$, and pulsar phase, $\Phi$. Each map is normalized by its maximum value and includes the emission from both electrons and positrons. The white dashed line represents the location of the current sheet in the split monopole geometry.}
\label{fig::fig2}
\end{figure*}

\begin{figure}
\centering
\includegraphics[width=\hsize]{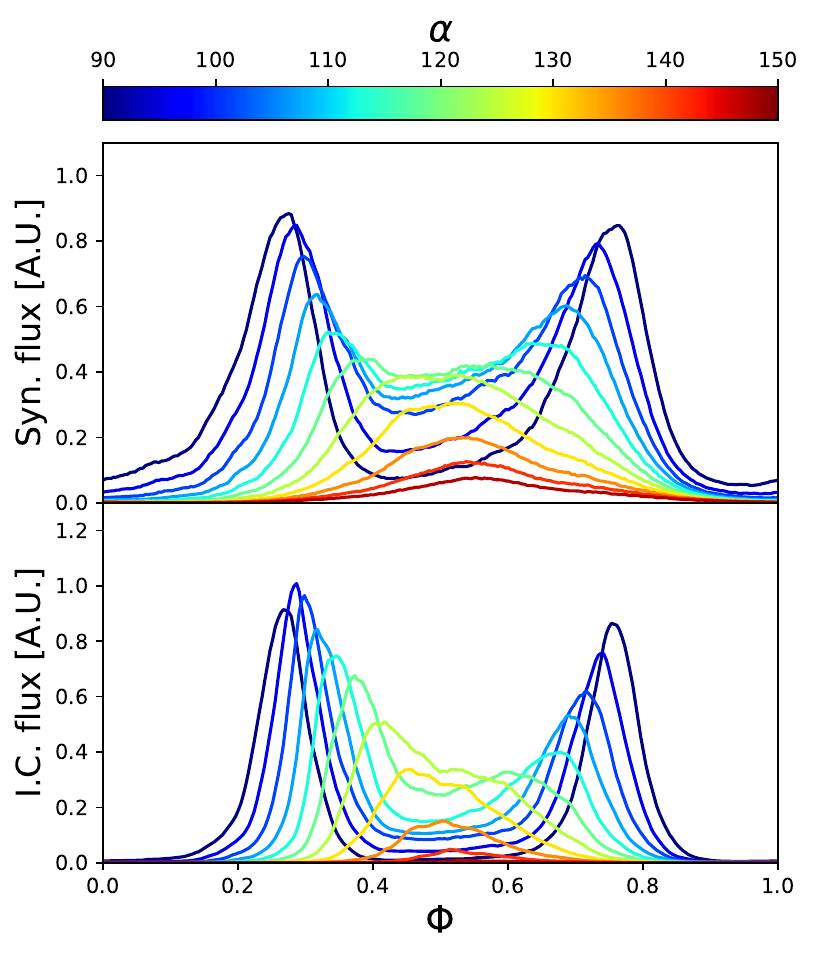}
\caption{Synthetic phase-resolved synchrotron (top) and inverse Compton (isotropic radiation field, bottom) fluxes (in arbitrary units) for $\chi=30^{\circ}$ color-coded by the observer's viewing angle (emission from positrons only).}
\label{fig::fig3}
\end{figure}

\begin{figure}
\centering
\includegraphics[width=\hsize]{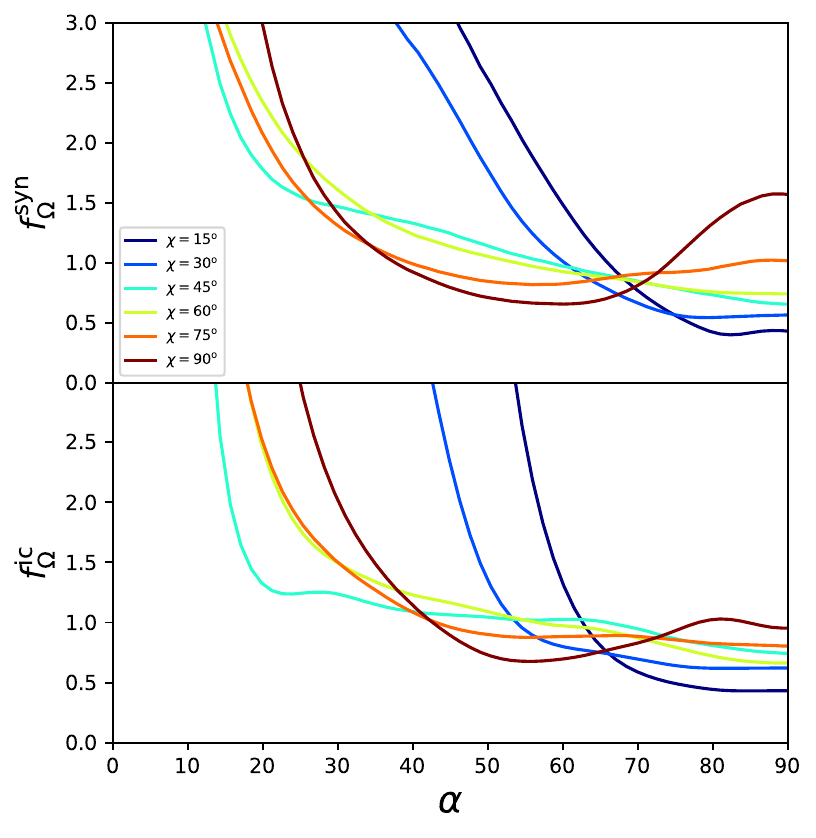}
\caption{Beam correction factor, $f_{\Omega}$, inferred from the synchrotron (top) and inverse Compton (isotropic radiation field, bottom) skymaps for all the magnetic obliquities simulated in this work.}
\label{fig::fig_omega}
\end{figure}

\begin{figure}
\centering
\includegraphics[width=\hsize]{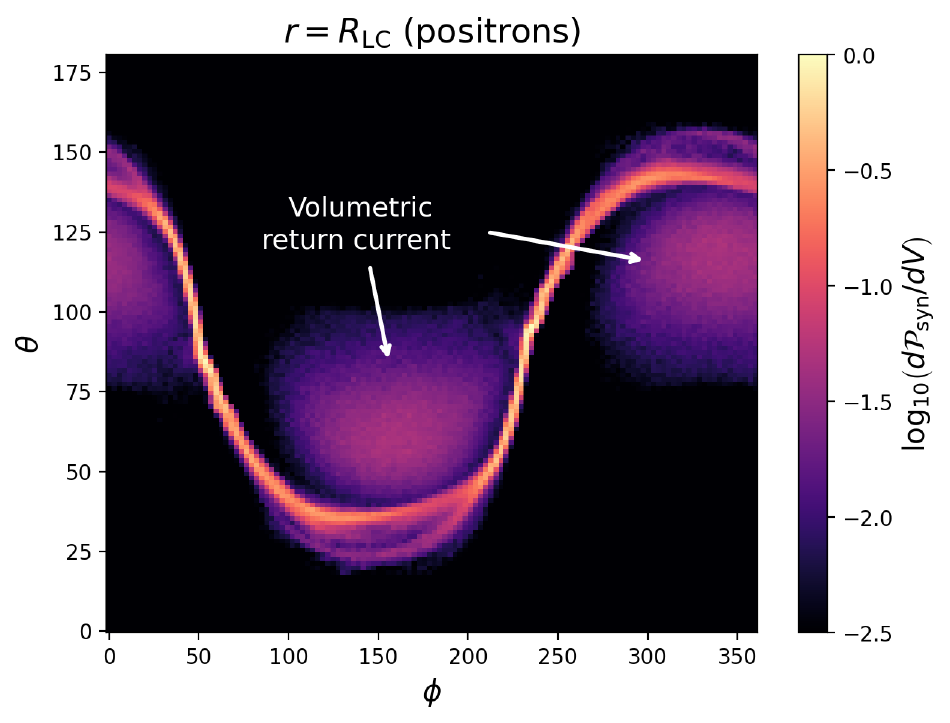}
\caption{Spatial distribution of the synchrotron power per unit of volume, $d\mathcal{P}_{\rm syn}/dV$, emitted by the positrons only at the light cylinder for $\chi=60^{\rm o}$. This figure highlights the role of the volumetric return current in explaining the Vela-like interpulse.}
\label{fig::fig_interpulse}
\end{figure}

\begin{figure*}
\centering
\includegraphics[width=5.5cm]{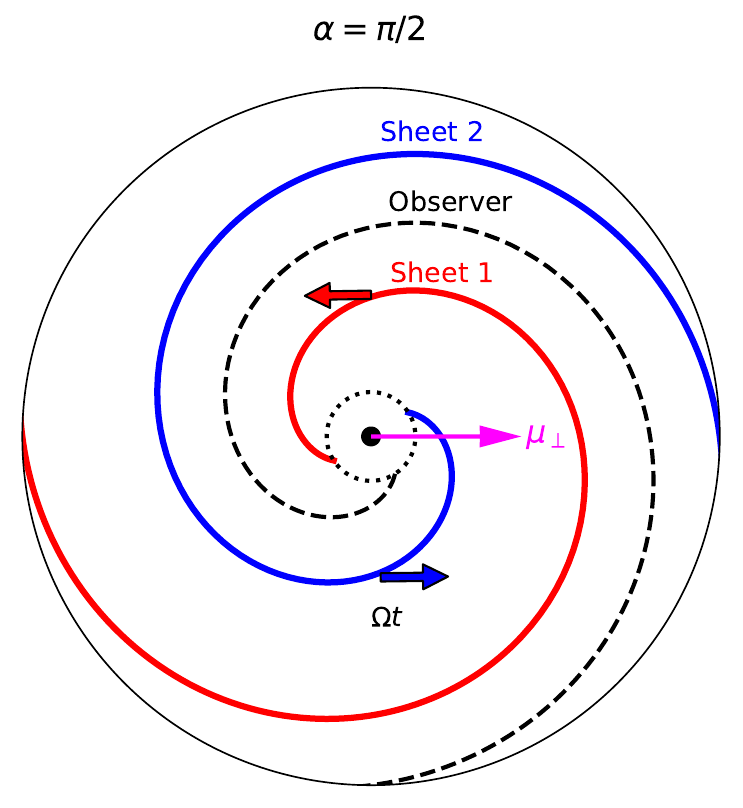}
\includegraphics[width=5.5cm]{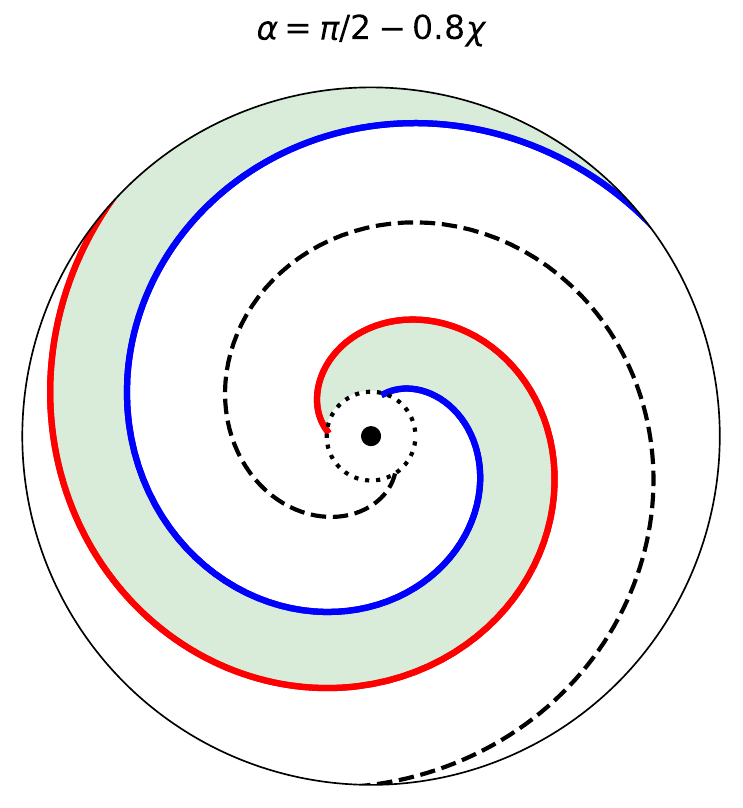}
\includegraphics[width=5.5cm]{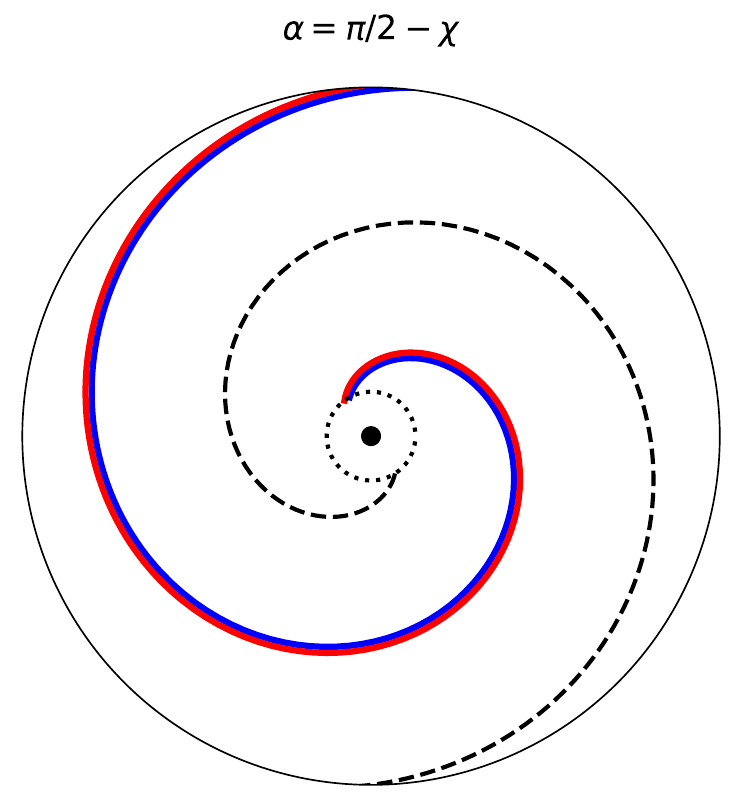}
\includegraphics[width=5.5cm]{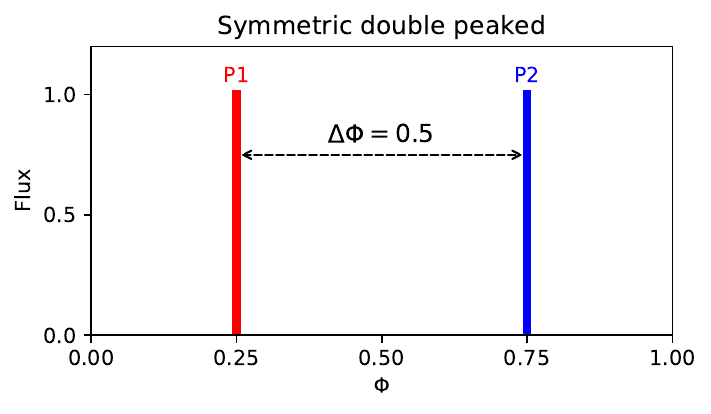}
\includegraphics[width=5.5cm]{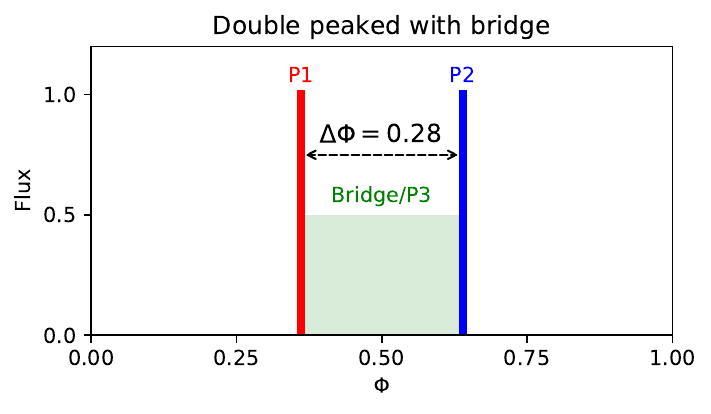}
\includegraphics[width=5.5cm]{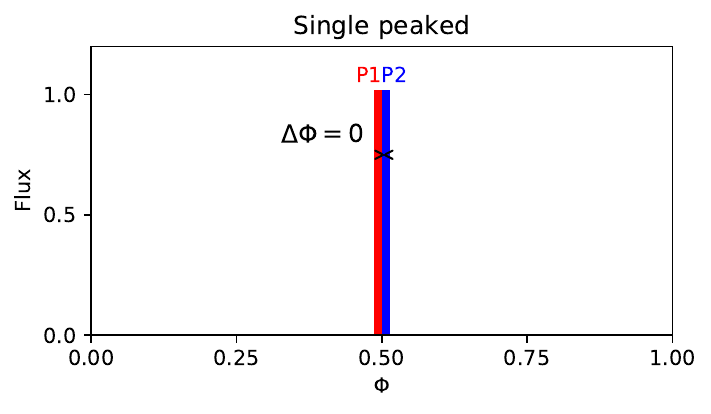}
\caption{This schematic representation illustrates how pulses form relying on simple geometric considerations based on the split-monopole configuration. Here $\chi=60^{\circ}$ for illustrative purposes, but this picture applies for all inclinations (except the aligned and orthogonal rotators). In this diagram, the pulse P1 (P2) is emitted once the sheet 1 (respectively 2) rotating at the angular velocity of the star, $\Omega$, overlaps with the fixed black dashed line, an iso-phase contour set by the observer. Three typical configurations are presented: (i) an equatorial view ($\alpha=\pi/2$) leads to two symmetric pulses separated by $\Delta\Phi=0.5$, (ii) an off-equator view within the striped wind region ($\alpha=\pi/2-0.8\chi$) where the pulses are asymmetric and separated by $\Delta\Phi=0.28$ and contains significant emission in between (a bridge or third peak, green region), and (iii) the degenerate single pulse case ($\Delta\Phi=0$) where the observer's line of sight grazes the outer boundary of the stripes ($\alpha=\pi/2-\chi$).}
\label{fig::fig4}
\end{figure*}

In agreement with previous studies, dissipation mostly occurs beyond the light cylinder where reconnection kicks in and powers efficient particle acceleration and non-thermal radiation \citep{2016MNRAS.457.2401C, 2018ApJ...855...94P}. Taking into account time delays due to the propagation of photons towards a distant observer, we compute the emission pattern as a function of the viewing angle $\alpha$ defined with respect to the star rotation axis, and the normalized pulsar phase $\Phi$ whose origin is fixed by the plane containing $\boldsymbol{\Omega}$ and $\boldsymbol{\mu}$ (see \citealt{2016MNRAS.457.2401C} for more details). Particles energized by reconnection primarily emit synchrotron radiation near the light cylinder where the field is strongest (see next section). Synchrotron largely dominates over curvature radiation because, although the particle (guiding center) trajectories are bent near the light cylinder, their radius of curvature (of order $R_{\rm c}\gtrsim R_{\rm LC}$) is much larger than their Larmor radius (of the order of the current layer thickness). Fig.~\ref{fig::fig_bperp} shows the ratio of the effective perpendicular magnetic field to the total field strength. From this figure, we can see that $\tilde{B}_{\perp}\lesssim B$ within the current layer (see also \citealt{2023ApJ...959..122C} in their local Harris sheet study), indicating that synchrotron losses are dominant there. Indeed, the ratio of the Larmor radius to the light-cylinder scale is $\xi=\gamma_{\rm th}m_{\rm e}c^2/e\tilde{B}_{\perp}R_{\rm c}\lesssim 3\times 10^{-4}$ in the simulation, and therefore the ratio of curvature to synchrotron radiation power scale as $P_{\rm curv}/P_{\rm syn}=\xi^2\sim 10^{-7}$. In contrast, in the wind zone, particles stream with negligible synchrotron losses because they are drifting at the local $\mathbf{E}\times\mathbf{B}$-drift velocity, thus $\tilde{B}_{\perp}\approx 0$ (see Fig.~\ref{fig::fig_bperp}). High-energy particles accelerated in the current layer can also upscatter low-energy photons via inverse Compton scattering.

Figure~\ref{fig::fig2} shows the emission maps, or ``skymaps'' in the following, for all the inclinations simulated in this work (except the aligned case) in the synchrotron channel (left panels), and in the inverse Compton channel for an isotropic radiation field (right panels). These maps include the contribution from both electrons and positrons. Synchrotron fluxes are frequency-integrated above the fiducial critical frequency, $\nu_0=3eB_{\star}/4\pi m_{\rm e}c$. Similarly, to isolate the high-energy inverse Compton component, we only consider particles with $\gamma>10$ (i.e., including all high-energy pairs and their secondaries with $\gamma\gtrsim \gamma_{\rm s}$), and integrate the upscattered photon spectrum $\epsilon_1>30$. These emission maps were reconstructed over multiple timesteps to wash out the effect of plasma irregularities and rapid intermittencies inherent to a reconnecting current layer \citep{2022A&A...661A.130A}, and to collect a large enough sample of photons to limit the shot noise. This averaging is also relevant in comparing our results to observations since measured gamma-ray pulse profiles are reconstructed over a very large number of spin periods due to the poor photon statistics, and therefore they should be considered as mean profiles. It explains the qualitative differences with the results presented in \citet{2016MNRAS.457.2401C} that were not of good enough quality to perform the quantitative comparison with observations that we propose below in Sect.~\ref{sect::fitting}. A synthetic lightcurve can be generated for any observer by cutting through these emission maps at a fixed value of the viewing angle $\alpha$. For illustrative purposes, Fig.~\ref{fig::fig3} presents a collection of synthetic lightcurves for synchrotron and inverse Compton radiation generated from the $\chi=30^{\circ}$ emission maps (considering only the positrons, see the justification later below).

The synchrotron and inverse Compton emission maps are very similar and aligned in phase, confirming that both components are produced by the same particle population. Interestingly, the inverse Compton features are sharper, meaning that pulses are systematically narrower than their synchrotron counterpart. The origin of this important feature is further investigated and explained in Sect.~\ref{sect::width}. We also confirm the strong concentration of the emission within the equatorial plane of the pulsar regardless of the magnetic inclination (see \citealt{2014ApJ...793...97K} in their ``FIDO'' model, and \citealt{2016MNRAS.457.2401C, 2018ApJ...855...94P, 2018ApJ...857...44K} for PIC studies), for both the inverse Compton and the synchrotron channels. This effect is due to the higher concentration of the Poynting flux in the equatorial regions and the nearly independent dissipation rate with latitude within the striped wind region \citep{2020A&A...642A.204C}. Fig.~\ref{fig::fig_omega} shows the beam correction factor defined as \citep{2009ApJ...695.1289W, 2009ApJ...707..800V},
\begin{equation}
f_{\Omega}\left(\chi,\alpha\right)=\frac{\iint F\left(\chi, \alpha^{\prime},\Phi\right) \sin\alpha^{\prime}d\alpha^{\prime}d\Phi}{2\int F\left(\chi, \alpha,\Phi\right) d\Phi},
\label{eq::fomega}
\end{equation}
where $F$ is the photon flux received by a given observer. This quantity corrects for the anisotropy of the emission to infer the true luminosity. It is of the order $f_{\Omega}\sim 1$ in the equatorial regions where most of the emission comes from, and it quickly rises for any observer looking at the polar regions where little emission is predicted (see also \citealt{2023ApJ...954..204K}). The emission pattern presents a bright sinusoidal-like component (or ``caustic'') at low obliquities which breaks up into two bright hotspots at higher obliquities ($\chi\gtrsim 30^{\rm o}$). The skymaps present a central symmetry with respect to the center of the maps, located here in the equatorial plane at the pulsar phase $\Phi=0.5$. Slight asymmetries between the left and the right caustics can be seen in some solutions like in the $\chi=45^{\rm o}$ solution; these are due to statistical fluctuations and we anticipate that they would vanish for a longer integration time. A pulse of light is received by a distant observer each time the current layer passes through the line of sight, which happens in general twice per pulsar period. The central symmetry in the emission maps is due to the periodic change in the direction of the current density direction carried by the layer (mostly oriented towards $\pm\theta$ directions) between two consecutive crossings, leading to a preferential emission in the northern/southern hemispheres in each hotspot.

Overall, the emission pattern closely follows the geometry of the current layer projected onto the sky. The latter is rather well approximated by the split-monopole solution (Eq.~\ref{eq::monopole}) given by
\begin{eqnarray}
\alpha\left(\Phi,\chi\right)&=&\arctan{\left(\frac{1}{\tan\chi\cos2\pi\Phi}\right)},~0<\alpha<\pi/2,\\
\alpha\left(\Phi,\chi\right)&=&\pi+\arctan{\left(\frac{1}{\tan\chi\cos2\pi\Phi}\right)},~\pi/2<\alpha<\pi.
\end{eqnarray}
It is shown by the white dashed line in Fig.~\ref{fig::fig2}. We recover the typical double-peaked structure reminiscent of gamma-ray pulsars. For $\alpha=90^{\circ}$, both peaks are perfectly symmetric and separated in phase by $\Delta\Phi\approx 0.5$, regardless of magnetic obliquity. The peaks become unequal due to the central symmetry highlighted above, and their separation in phase decreases for an observer looking away from the equatorial plane. The latter behavior is also consistent with the split-monopole solution that predicts \citep{2011MNRAS.412.1870P}
\begin{equation}
\Delta\Phi\left(\alpha,\chi\right)=\frac{1}{\pi}\arccos\left|\cot\alpha\cot\chi\right|.
\end{equation}
As expected, pulsations disappear outside the domain of the striped wind region ($\alpha\lesssim\pi/2-\chi$ or $\alpha\gtrsim\pi/2+\chi$).

However, what the split-monopole model does not predict is the presence of a significant interpulse, that sometimes even appears as a third peak particularly visible in the synchrotron lightcurves (see Figs.~\ref{fig::fig3}-\ref{fig::fig5}). This is also a common feature observed in gamma-ray pulsars. This interpulse should not be confused with the bridge emission that arises when the observer intercepts the outer boundaries of the striped wind region ($\pi/2-\chi\lesssim\alpha\lesssim\pi/2+\chi$) that is as visible at low obliquities, but that disappears at high obliquities (Fig.~\ref{fig::fig2}, \citealt{2016MNRAS.457.2401C, 2018ApJ...855...94P}; this feature is also observed in many pulsars. The presence of a third peak is a clearly recognizable, yet mysterious, property of the Vela gamma-ray pulse profile \citep{1988A&A...204..117G, 2009ApJ...691.1618P, 2010ApJ...713..154A}. We checked that this third component is emitted in the wind zone by cutting off all the emission coming from inside the light cylinder. It is mostly prominent at high-inclinations ($\chi\gtrsim 30^{\circ}$) in the synchrotron channel and seems to originate from the wind region contained within the meanders of the layer for $\chi=45^{\circ}-75^{\circ}$, but it is unclear how particles are accelerated and heated (i.e., $\tilde{B}_{\perp}\neq 0$) in this sector. It is possible that this particular region is under the strong influence of the current layer that surrounds it. However, the $\chi=90^{\circ}$ solution shows that the interpulse emission is distributed at intermediate latitudes far from the boundaries of the layer that extends all the way to the poles, approximately at the same location as for $\chi=60^{\circ}$ and $75^{\circ}$. Another possibility is that high-energy particles accelerated in the sheet are escaping in the upstream medium, as recently shown in local 3D reconnection studies \citep{2021ApJ...922..261Z}, but the extension in phase of that component suggest that high-energy particles are distributed all over the wind region, which is difficult to reconcile with the latter scenario. In addition, this effect would probably be limited to ions only due to strong radiative losses experienced by the leptons \citep{2023ApJ...959..122C}.

What appears to be the most convincing scenario is that the Vela-like interpulse is produced within the volumetric return current. Fig.~\ref{fig::fig_interpulse} shows the spatial distribution of the synchrotron power emitted near the light cylinder by positrons (this feature is mostly produced by positrons). Outside the current sheet, the two hotspots coincide with regions of strong return current connecting the polar cap to the current sheet. This special region appears at large obliquities and is known to accelerate and create pairs because the amount of charges extracted at the polar cap are insuffisant to sustain the current required by the magnetosphere \citep{2013MNRAS.429...20T, 2015ApJ...801L..19P}. Interestingly, this scenario also explains the origin of the high-viewing angle components in the skymaps ($\Phi\sim 0.4,~0.9$, $\alpha\gtrsim 50^{\rm o},~140^{\rm o}$). These regions are consistent with polar-cap field lines carrying super-Goldreich-Julian current densities, and they are mostly sustained by electrons. In the orthogonal rotator, these components become the volumetric return current by symmetry. It is unclear though how particle acceleration along these field lines remains active this far from the star surface. We leave this question to a future study.

What is clear however, is that this component is emitted in a special part of the inter-layer wind. In contrast to the (unstriped) polar regions, the wind between two consecutive layers is not an homogeneous medium, there is a strong plasma density and magnetic field strength gradient. The wind region leading the current layer is highly depleted in plasma and magnetic field strength, while the wind region trailing the layer contains denser, more energetic plasma, and stronger fields (see \citealt{2017A&A...607A.134C, 2020A&A...642A.204C}). All in all, these clues suggest that this mysterious third peak may be a by-product of the wind dynamic, plasma reorganization induced by the presence of the reconnecting layers, and the overall magnetospheric current density.

Figure~\ref{fig::fig4} graphically summarizes how pulses form in the simulations. As pointed out above, the shape of the current layer is remarkably well fitted by the split monopole solution, even close to the light cylinder. Therefore, mapping the emission onto this surface is sufficient to render the main features of the lightcurves. A fixed observer probes a surface of constant latitude $\theta=\alpha$, then the equation of each layer is given by
\begin{equation}
\phi(r)=\pm\arccos\left(-\cot{\alpha}\cot{\chi}\right)+\Omega t-r/R_{\rm LC},
\end{equation}
which corresponds to an Archimedean spiral (for sheet 1 and 2 in Fig.~\ref{fig::fig4}). The spiral pattern is moving in solid rotation with the star --but not the plasma-- in virtue of Ferraro's isorotation law. Thus, the emitting zone of photons arriving at the same phase at the observer has an Archimedean spiral shape as well (dashed line in Fig.~\ref{fig::fig4}). A simple way to see this is to consider a photon emitted radially at a time $t$ towards the observer located along the $\phi=\phi_0$ direction. After a time $\delta t$, the photon will propagate by a distance $\delta r=c\delta t$ while the emitting zone will have shifted by $\delta\phi=\phi-\phi_0=\Omega\delta t$. At all radii where new photons can be created, the location of all the emitting zones probed by the observer at a given time draws a spiral of equation $\phi=\phi_0-r/R_{\rm LC}$. Contrary to the sheets, this pattern is fixed in time since the observer is assumed static in the simulation frame. In this geometric construction, a pulse of light will be observed when the sheet overlaps with the observer's spiral. This happens twice per spin period within the striped wind region. At $\alpha=\pi/2\pm\chi$, the layer degenerates into a single sheet leading to a single pulse. In this framework, interpulse emission can be produced in between the layers.

\subsection{Asymmetries and energy dependence}\label{sect::energy}

\begin{figure}
\centering
\includegraphics[width=\hsize]{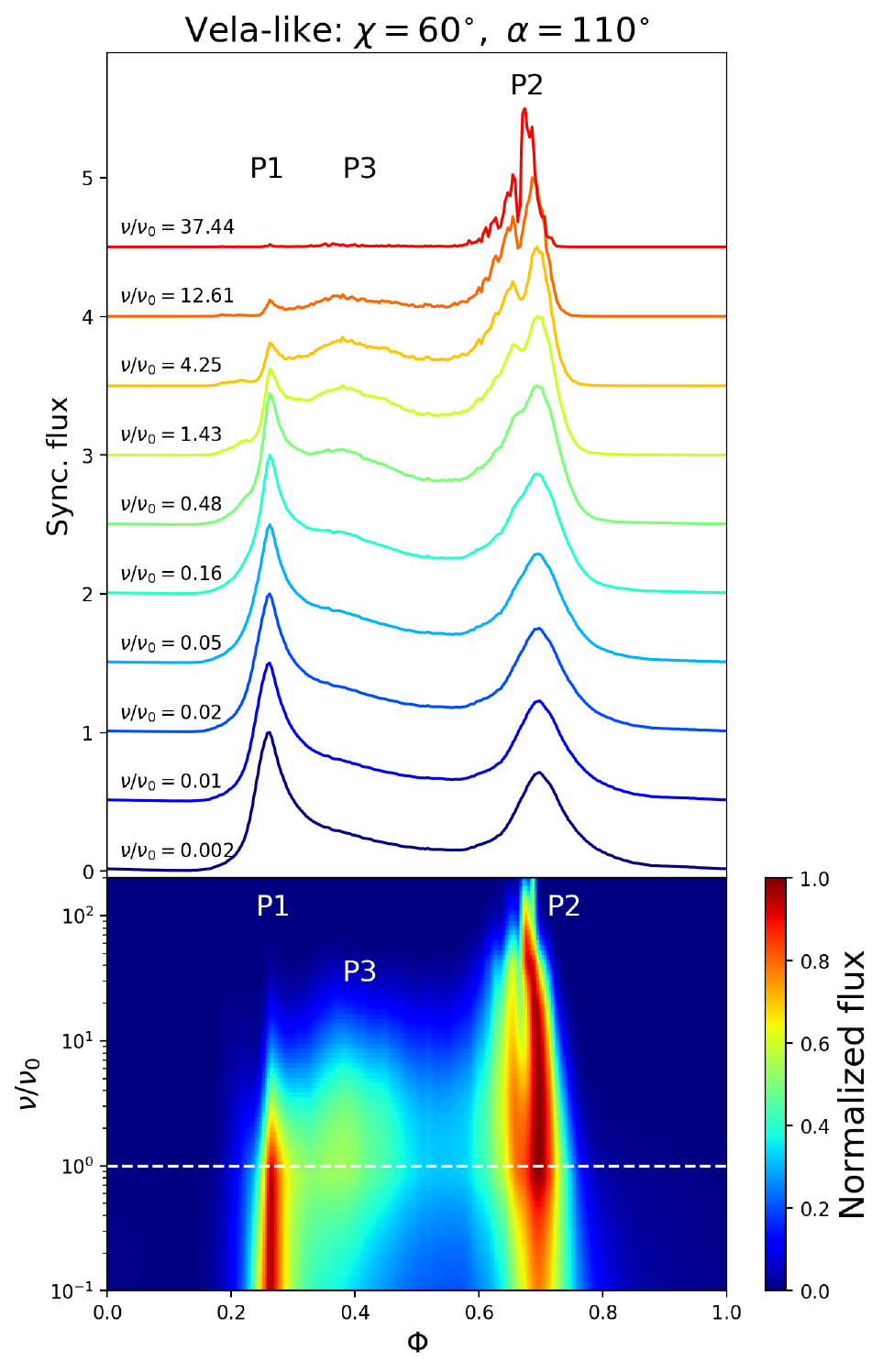}
\caption{Energy dependence of the synthetic synchrotron pulse profile with the photon frequency, $\nu$ normalized to $\nu_0$, for a Vela-like solution ($\chi=60^{\circ}$, $\alpha=110^{\circ}$). Top panel: Lightcurves equally spaced in $\log\nu/\nu_0$ from the lowest frequencies (bottom) to the highest frequencies (top). The bottom panel shows the full phase-frequency emission map. In both panels, lightcurves are normalized to the maximum flux for each frequency band. This figure only shows the contribution from positrons.}
\label{fig::fig5}
\end{figure}

Low-inclination solutions present additional emission features visible in the synchrotron maps. They appear as two humps, one above (in the $\Phi<0.5$ sector), and one below (in the $\Phi>0.5$ sector) the main emission pattern from the layer (Fig.~\ref{fig::fig2}). This feature was already reported in \citet{2016MNRAS.457.2401C}. It corresponds to high-energy electrons accelerated near the base of the current layer precipitating towards the star. This asymmetry between both species is explained by the overall polarisation of the magnetosphere: the plasma is negatively charged along magnetic polar regions and is positively charged in the equatorial regions (and the other way around, if $\boldsymbol{\Omega}\cdot\boldsymbol{\mu}<0$). Although this global polarization effect should remain at realistic scales, the relative contribution from precipitating electrons to the total emission synchrotron output may be overestimated due to the modest plasma multiplicities achieved in the simulations, $\kappa\equiv n/n_{\rm GJ}\gtrsim 5$ (Fig.~\ref{fig::fig1}), as opposed to observed gamma-ray pulsars where $\kappa \gtrsim 10^2-10^6$. A higher plasma multiplicity should lead to a smaller fraction of precipitating particles.

Although our scale separation is modest, we observe a significant dependence of the pulse profile with the photon energy. Fig.~\ref{fig::fig5} represents a synthetic lightcurve reminiscent of the Vela gamma-ray pulse profile, that is composed of two main pulses (P1 and P2 separation by $\Delta\Phi\approx 0.43$) and a weaker broad peak in between (P3) obtained from the $\chi=60^{\circ}$ solution for a viewing angle $\alpha=110^{\circ}$. At low frequencies ($\nu/\nu_0\lesssim 0.1$), P1 slightly dominates over P2 in height. Both peaks are connected by a smooth, asymmetric bridge. P3 appears for $\nu/\nu_0\gtrsim 0.1$ close to P1 and drifts away towards P2 at higher frequencies. The amplitude of P1 relative to P2 dramatically decreases for $\nu/\nu_0\gtrsim 1$ (see also \citealt{2015ApJ...804...84B, 2022ApJ...925..184B}). P1 and then P3 nearly completely vanish at the highest frequencies where only P2 remains. These features are in very good agreement with {\em Fermi}-LAT observations of Vela \citep{2010ApJ...713..154A} and more generally of gamma-ray pulsars \citep{2023ApJ...958..191S}. This clean cut property can be drawn in the simulations only if considering the positronic emission from the magnetosphere. Adding the contribution from electrons makes this conclusion less robust because of the precipitating particles, that we argue may be overproduced in the simulation.

\subsection{Radiated power and efficiencies}\label{sect::efficiencies}

\begin{figure}
\centering
\includegraphics[width=\hsize]{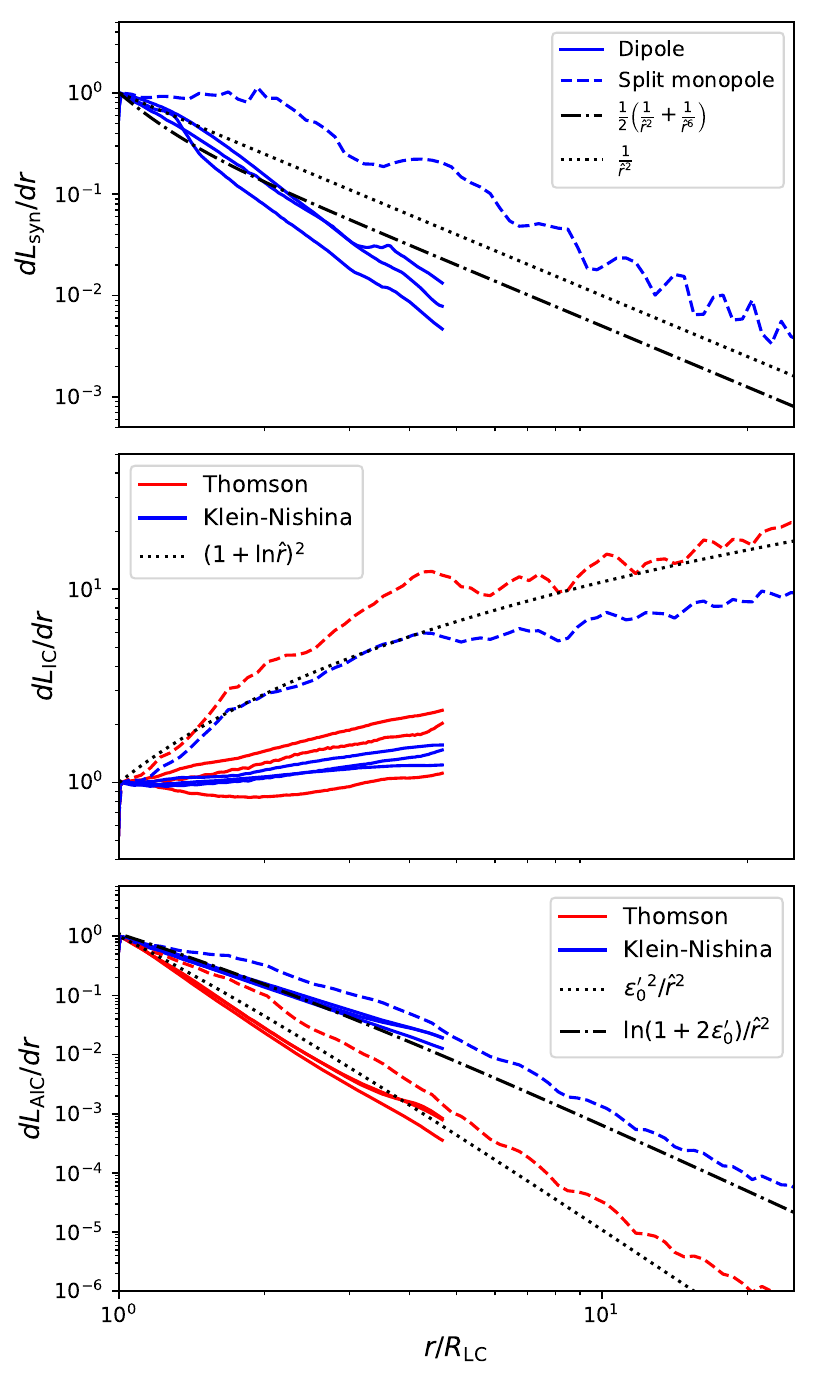}
\caption{Radial distribution of the frequency-integrated radiated power in the synchrotron channel (top panel), in the inverse Compton channel for both an isotropic (middle panel) and an anisotropic (from the star, bottom panel) target photon field in the Thomson (red lines) and deep Klein-Nishina (blue lines) regimes. Simulations with an initial dipolar field are shown by solid lines ($\chi=30^{\circ}$, $45^{\circ}$, $60^{\circ}$), while the split-monopole simulation is shown by dashed lines. Simple analytical expressions are also overplotted for comparison (black dashed-dotted and dotted lines). All quantities are normalized by their value at the light cylinder.}
\label{fig::fig6}
\end{figure}

\begin{figure}
\centering
\includegraphics[width=\hsize]{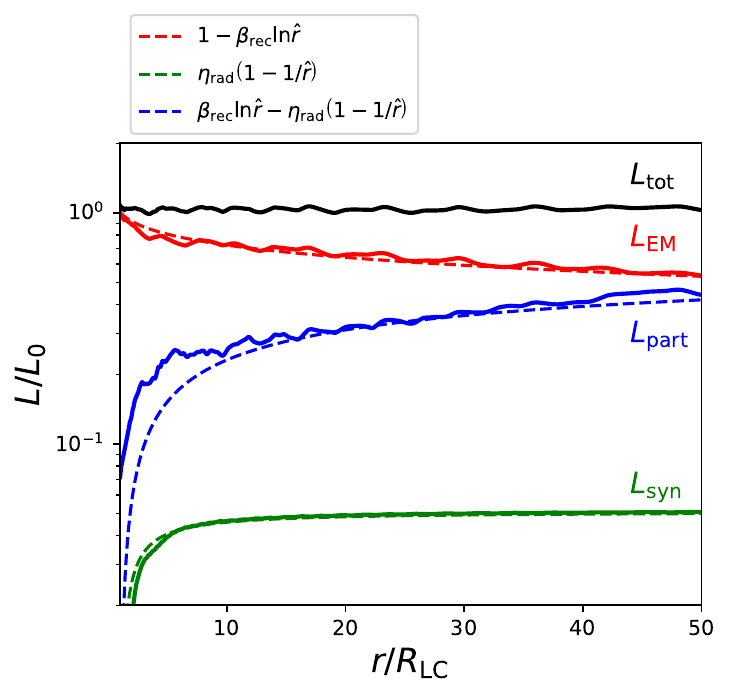}
\caption{Radial evolution of all energy channels in the split-monopole solution, including the synchrotron luminosity, normalized to the pulsar spindown power ($L_0$). The inferred reconnection rate is $\beta_{\rm rec}\approx 0.12$ and the radiative efficiency is $\eta_{\rm rad}\approx 0.05$. The inverse Compton power is not shown since radiative energy losses through this channel are neglected in this work.}
\label{fig::fig7}
\end{figure}

Fig.~\ref{fig::fig6} presents how is the radiated power distributed in the wind region at all radii probed by the simulations. The large-scale split-monopole simulation is shown for comparison and highlights trends at large radii. As expected, the synchrotron power rapidly falls off with radius because of the decrease of the magnetic field strength with radius, since $\mathcal{P}_{\rm syn}\propto \tilde{B}^2_{\perp}$ (Eq.~\ref{eq::psyn}). The radial evolution of the effective perpendicular field felt by the particles within the sheets is well modeled by the superposition of the initial vacuum poloidal field, $B_{\rm pol}$, and a purely toroidal component, $B_{\phi}$, where both components are equal at the light cylinder, such that
\begin{equation}
\tilde{B}^2_{\perp}\sim B^2_{\rm pol} +B^2_{\phi}\sim \frac{B^2_{\rm LC}}{2}\left(\frac{1}{\hat{r}^{2a}}+\frac{1}{\hat{r}^2}\right),
\label{eq::bperp}
\end{equation}
where $\hat{r}=r/R_{\rm LC}$, $a=3$ for the dipolar simulations, and $a=2$ for the split-monopole simulation, and $B_{\rm LC}=B_{\star}\left(r_{\star}/R_{\rm LC}\right)^a$ is an estimate of the field strength at the light cylinder.

The synchrotron power also depends on the square of the mean particle Lorentz factor in the current layer, but it increases at best as $\ln \hat{r}$ \citep{2020A&A...642A.204C}. This weak dependence with radius is well visible in the inverse Compton power for an isotropic radiation bath. In this case, the emitted power depends only on the particle Lorentz factor, as $\mathcal{P}_{\rm ic}\propto \gamma^2$ in the Thomson regime and $\mathcal{P}_{\rm ic}\propto \ln\gamma$ in the Klein-Nishina regime. In both regimes, the inverse Compton power slowly increases with radius. In contrast, if the target photon field originates from the star, the Compton power steeply drops with radius. This is, for one part, due to the decrease of the target photon number density with radius as $1/\hat{r}^2$ and, for the other part, due to nearly rear-end collisions between the stellar photons and the particles that are mostly moving radially outward in the wind. The latter effect is most pronounced in the Thomson regime. This is explained by the dependence of the (anisotropic) inverse Compton power with $\epsilon^{\prime}_0=\gamma\left(1-\cos\theta_0\right)\epsilon_0$. Assuming that the particle direction of motion is given by the $\mathbf{V}_{\rm D}=c\mathbf{E}\times\mathbf{B}/B^2$ drift velocity of the monopole wind and the purely radial stellar photons, then the pitch angle is given by (e.g., \citealt{2017SSRv..207..111C})
\begin{equation}
\cos\theta_0=\mathbf{V_{\rm D}}\cdot\mathbf{e_r}/V_{\rm D}=\left(1+\frac{1}{\hat{R}^2}\right)^{-1/2}\approx 1-\frac{1}{2\hat{R}^2},
\end{equation}
where $\hat{R}=\hat{r}\sin\theta\sim \hat{r}\gg 1$ is the cylindrical radius, and hence
\begin{equation}
\epsilon^{\prime}_0\propto \frac{1+\ln\hat{r}}{\hat{r}^2},
\end{equation}
where we assume once again that $\gamma$ increases with $\ln\hat{r}$. In the Thomson regime, $\mathcal{P_{\rm ic}}\propto {\epsilon^{\prime}}^2_0$ (Eq.~\ref{eq::PicTH}), while $\mathcal{P_{\rm ic}}\propto \ln\left(2\epsilon^{\prime}_0\right)$ (Eq.~\ref{eq::PicKN}) when Klein-Nishina corrections are important. Thus, anisotropic effects are reduced in the Klein-Nishina regime and result in a slower decay of the radiated power than in the Thomson regime. Note that there is no obvious dependence with the magnetic inclination angle.

We can deduce from the above scaling laws an approximate expression for the synchrotron radiative efficiency and the radial evolution of the total (frequency-integrated) radiated power, and compare it with the other energy channels. Fig.~\ref{fig::fig7} shows the radial evolution of the outward radial Poynting ($L_{\rm EM}$) and pairs kinetic energy ($L_{\rm part}$) fluxes passing through a spherical shell of radius $r$ as reported in \citet{2020A&A...642A.204C}. In this work, we compute the contribution of the radiated synchrotron power, $L_{\rm syn}$, and neglect the power lost via inverse Compton scattering. Assuming that the radiated power mostly depends on the evolution of $\tilde{B}^2_{\perp}$, using Eq.~(\ref{eq::bperp}), we can derive a crude analytical expression for $L_{\rm syn}$ normalized to the pulsar spindown power, $L_0$,
\begin{equation}
\frac{L_{\rm syn}}{L_0}\approx \eta_{\rm rad}\frac{\int_1^{\hat{r}}\tilde{B}^2_{\perp}d\hat{r}}{\int_1^{+\infty}\tilde{B}^2_{\perp}d\hat{r}},
\end{equation}
\begin{equation}
\frac{L_{\rm syn}}{L_0}\approx\eta_{\rm rad}\left(1-\frac{1}{\hat{r}}\right),
\label{eq::lsyn}
\end{equation}
for a pure toroidal field, or
\begin{equation}
\frac{L_{\rm syn}}{L_0}\approx \eta_{\rm rad}\left(1-\frac{5}{6\hat{r}}-\frac{1}{6\hat{r}^5}\right),
\label{eq::lsyndipole}
\end{equation}
for a toroidal and dipolar field components. The radiative efficiency, $\eta_{\rm rad}$, is defined as the ratio of the asymptotic synchrotron power to the pulsar spindown power, $\eta_{\rm rad}=L^{\infty}_{\rm syn}/L_0$. The spindown power is measured here as the Poynting flux at the light cylinder, and thus assumes that there is negligible dissipation within the light cylinder. For the split-monopole simulation, dissipation depends mildly on the magnetic inclination and the radiative efficiency is $\eta_{\rm rad}\approx 5\%$ (Fig.~\ref{fig::fig7}). In contrast, the dipole simulations present a significant variation of the radiative efficiency that is not translated in Eq.~(\ref{eq::lsyndipole}), it varies from about $\sim 6\%$ in the aligned case down to $\lesssim 1\%$ for the orthogonal rotator (see also, \citealt{2016MNRAS.457.2401C}).

Following \citet{2020A&A...642A.204C}, the Poynting flux evolution with radius is approximately given by
\begin{equation}
\frac{L_{\rm EM}}{L_0}=1-\beta_{\rm rec}\ln\hat{r},
\end{equation}
where $\beta_{\rm rec}$ is the dimensionless reconnection rate that quantifies how efficient reconnection is at dissipating the Poynting flux. By virtue of energy conservation, the amount of power dissipated and channeled to particle kinetic and synchrotron power is then
\begin{equation}
\frac{L_{\rm part}+L_{\rm syn}}{L_0}=\beta_{\rm rec}\ln\hat{r}.
\label{eq::lpartlsyn}
\end{equation}
In the extreme case where all the dissipated power goes directly into radiation, that is to say if $L_{\rm part}=0$, then using Eq.~(\ref{eq::lsyn}), we see that $\eta_{\rm rad}\leq \beta_{\rm rec}$, meaning that the maximum bolometric radiative efficiency is bounded by the reconnection rate. Here, $\eta_{\rm rad}\approx 0.05<\beta_{\rm rec}\approx 0.12$ in the split-monopole simulation. However, gamma-ray pulsars are most likely in this extreme cooling regime, as also indicated by recent full-scale hybrid simulations of millisecond pulsars \citep{2024A&A...690A.170S}, such that $\eta_{\rm rad}=\beta_{\rm rec}$. Then, assuming a purely toroidal field geometry yields
\begin{equation}
\frac{L_{\rm syn}}{L_0}=\beta_{\rm rec}\left(1-\frac{1}{\hat{r}}\right),
\label{eq::lsyn2}
\end{equation}
and thus using Eq.~(\ref{eq::lpartlsyn}), the particle kinetic energy flux evolution can be inferred as
\begin{equation}
\frac{L_{\rm part}}{L_0}=\beta_{\rm rec}\left(\ln\hat{r}+\frac{1}{\hat{r}}-1\right).
\end{equation}

This important result can be recovered with the following argument for an aligned rotator. In the most efficient synchrotron cooling regime, the bulk of particles accelerated at the base of the current layer at the light cylinder will be in the radiation-reaction-limited regime, which implies that the power radiated away per electron, $\mathcal{P}_{\rm syn}$, balances the acceleration rate estimated as
\begin{equation}
\mathcal{P}_{\rm syn}\sim e c E_{\rm rec},
\end{equation}
where $E_{\rm rec}=\beta_{\rm rec}B_{\rm LC}$ is the reconnection electric field accelerating particles in the layer. Assuming that the emitting particles are confined within a flat layer of inner radius $R_{\rm LC}$ and outer radius $2R_{\rm LC}$ of thickness $\delta$, the total number of particles is
\begin{equation}
N_{\rm part}\sim \kappa n_{\rm GJ}\left(4\pi R^2_{\rm LC}-\pi R^2_{\rm LC}\right)\delta,
\end{equation}
where $\delta$ is the layer thickness. According to Eq.~(\ref{eq::lsyn2}), $50\%$ of the synchrotron power is radiated away within a few light-cylinder radii, suggesting that the synchrotron emission is highly concentrated close to the base of the layer as assumed here. Following \citet{2017A&A...607A.134C}, the layer thickness at the light cylinder is given by $\delta\sim R_{\rm LC}/\kappa$. With $n_{\rm GJ}=\Omega B_{\rm LC}/2\pi e c$, and ignoring factors of order unity yields
\begin{equation}
N_{\rm part}\sim \frac{\Omega B_{\rm LC} R^3_{\rm LC}}{ec}.
\end{equation}
Then, the total synchrotron power can be estimated as $L_{\rm syn}\sim N_{\rm part}\mathcal{P}^{\rm LC}_{\rm syn}$. Noticing that $B_{\rm LC}\sim B_{\star}r^3_{\star}/R^3_{\rm LC}=\mu/R^3_{\rm LC}$, we obtain
\begin{equation}
L_{\rm syn}\sim \beta_{\rm rec} L_0,
\end{equation}
where $L_0=\mu^2\Omega^4/c^3$ is the force-free spindown of the aligned rotator. A similar argument is also exposed in \citet{2014ApJ...780....3U}.

The fact that all pulsars should have the same bolometric radiative efficiency may appear in contradiction with the {\em Fermi}-LAT data that show a clear decrease of the gamma-ray efficiency above $100$~MeV with increasing spindown power, ranging from $\sim 10-100$\% for millisecond pulsars down $\sim 1$\% in powerful Crab-like pulsars. However, the third {\em Fermi}-LAT catalog also shows that high-spindown pulsars present significantly softer spectra than millisecond pulsars \citep{2023ApJ...958..191S}, suggesting that a sizeable fraction of the radiative output is below the {\em Fermi}-LAT threshold and thus missed in the estimation of the bolometric luminosity. This trend may be explained by a higher yield of pair production in high spindown power pulsars. Secondary pairs then redistributes the radiative power at lower frequencies over a broader spectral range. In this regard, a telescope operating in the MeV range would bring valuable constraints for the models. Unfortunately, given the limited scale separation achieved in the simulations and the simplified pair production model adopted in this work, we cannot elaborate more on this scenario at this stage.

\subsection{Pulse width}\label{sect::width}

\begin{figure}
\centering
\includegraphics[width=\hsize]{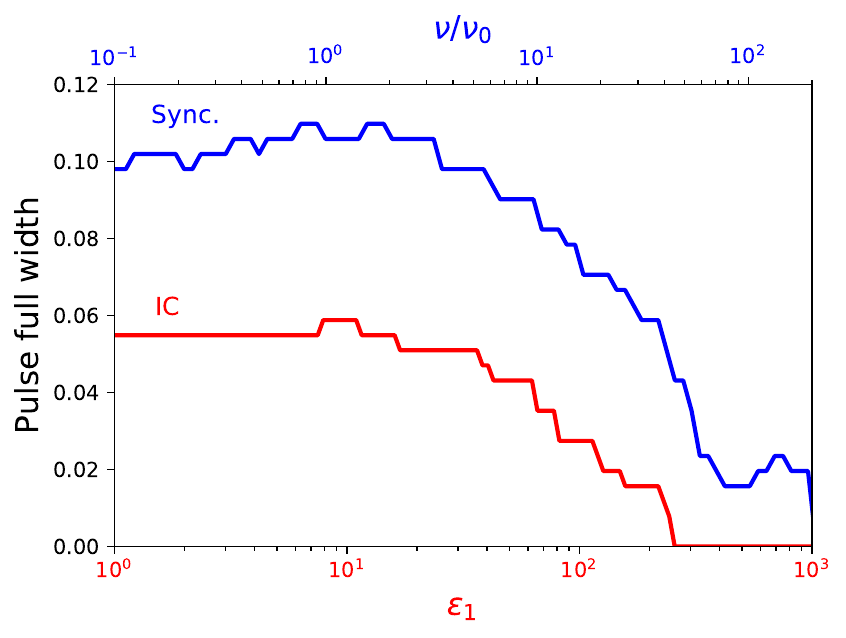}
\caption{Narrowing of the pulse full width at half maximum with the energy of the emitted photons in the inverse Compton (isotropic radiation field, red line) and synchrotron channels (blue line) for the dipole simulation with $\chi=60^{\circ}$ and $\alpha=110^{\circ}$.}
\label{fig::fig55}
\end{figure}

\begin{figure}
\centering
\includegraphics[width=\hsize]{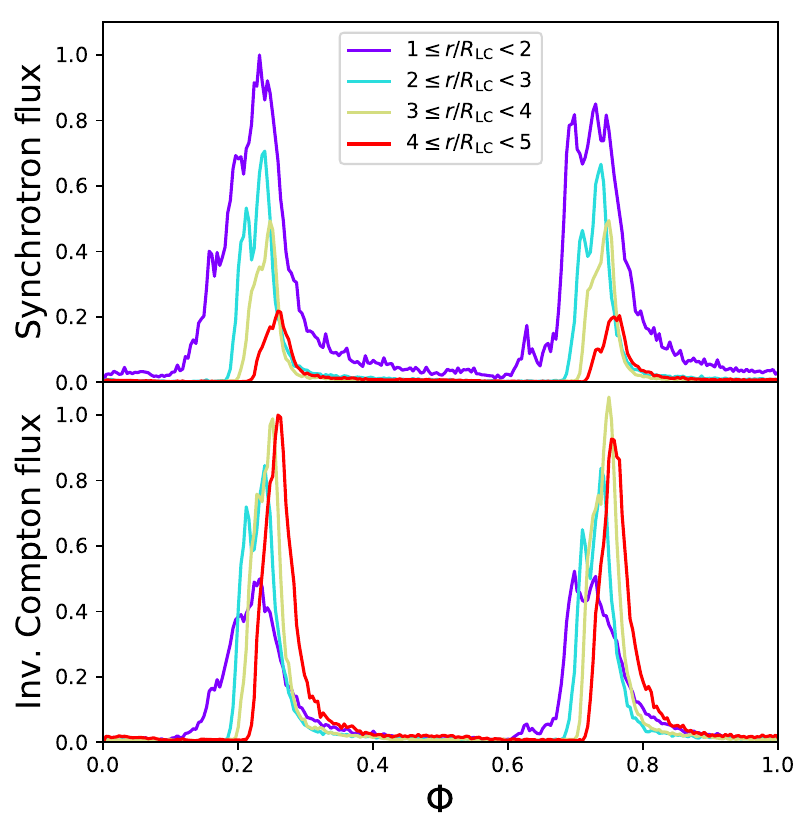}
\includegraphics[width=\hsize]{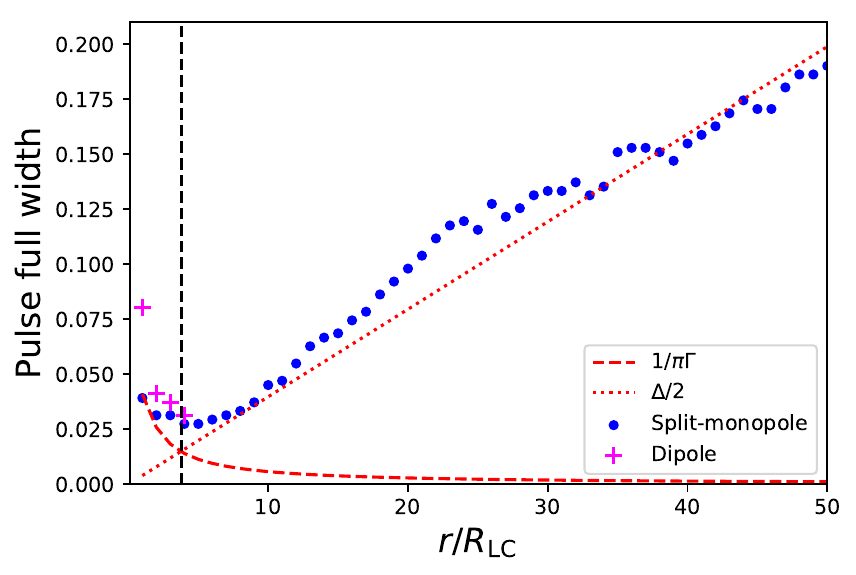}
\caption{Evolution of the observed lightcurve as a function of the radius of emission. Top panel: Synthetic synchrotron and inverse Compton (isotropic radiation field) lightcurves as a function of the emitting radius, from the light-cylinder radius up to $5_{\rm LC}$ for the dipole simulation ($\chi=60^{\circ}$, $\alpha=90^{\circ}$). Bottom panel: Full width at half maximum of the inverse Compton pulse width as a function of radius for the dipole (magenta crosses) and split-monopole (blue dots) simulations. The red dotted and dashed lines shows the two asymptotic behaviors discussed here. The black vertical dashed line marks the location of the critical radius $r_{\rm m}$ where the emitted pulses are the narrowest.}
\label{fig::fig89}
\end{figure}

Another important feature of gamma-ray pulsars is the narrowing of the pulse width with increasing photon energy \citep{2010ApJ...713..154A, 2010ApJ...708.1254A, 2010ApJ...720..272A}. This property is also reproduced by the simulations. Fig.~\ref{fig::fig55} shows the evolution of the full width at half maximum for a Vela-like configuration with the photon energy for the synchrotron and the inverse Compton (for an isotropic radiation bath) emission channels. Below $\nu_0$, the synchrotron pulse width represents about $\Delta\Phi\approx 0.1$ in phase, and drops down to $\Delta\Phi\approx0.02$ at the high-end of the synchrotron spectrum ($\nu/\nu_0\gtrsim 10$). The inverse Compton pulse is about twice thinner than the synchrotron pulse at low energy, and decreases in a similar manner as the synchrotron pulse width with increasing energy ($\epsilon_1\gtrsim 10$). Since the emission originates from high-energy particles accelerated in the wind current layer, this result indicates that the highest-energy particles are concentrated deep into the layer. This is reminiscent of the kinetic beaming effect reported in local reconnection studies \citep{2012ApJ...754L..33C, 2020MNRAS.498..799M}, where the angular distribution of the particles decreases with energy. This effect is operating in the strong synchrotron cooling regime because particles do radiate high-energy photons before they isotropize in plasmoids (in the wind comoving frame). This regime is relevant near the base of the current layer.

The acceleration of the wind bulk Lorentz factor with radius also leads to the sharpening of the pulses due to relativistic beaming effects \citep{2002A&A...388L..29K, 2011MNRAS.412.1870P}. Fig.~\ref{fig::fig89} shows the emitted pulse profile as a function of the radius of emission in the wind within a $\Delta r=R_{\rm LC}$ thick radial bin. The synchrotron and inverse Compton peaks become thinner with radius. In a given radial bin, the pulses have a similar width in both channels. However, the observed synchrotron lightcurve presents broader peaks because most of the synchrotron emission is produced near the light cylinder where the wind is slowest. In contrast, the observed inverse Compton lightcurve for a uniform photon field has narrower peaks because the majority of the flux is produced at larger radii where the bulk Lorentz factor is higher. This is consistent with the increase of the inverse Compton power with radius reported in Sect.~\ref{sect::efficiencies} and in Fig.~\ref{fig::fig6}. This effect explains the narrower pulses in the inverse Compton channel compared to the synchrotron lightcurve (Fig.~\ref{fig::fig55}). Conversely, if the target photon field originates from the star, the inverse Compton pulse widths are comparable to the synchrotron ones, hence confirming the important role of the photon field geometry in shaping the pulse profile.

The above conclusion is valid as long as the geometric thickness of the layer is very small compared to the separation between two successive current layers in the wind, meaning $\Delta\equiv\delta/\pi R_{\rm LC}\ll 1$. In this regime, the pulse width is shaped by relativistic beaming effects. Following \citet{2016JPlPh..82e6302P}, the pulse width emitted by an infinitely thin Archimedean spiral shape current sheet is given by
\begin{equation}
\delta\Phi\sim\frac{1}{\pi\Gamma},
\label{eq::doppler}
\end{equation}
where $\Gamma$ is the bulk Lorentz factor of the emitting particles. In the split-monopole solution, the latter evolves as
\begin{equation}
\Gamma_{\rm M}=\sqrt{1+\hat{R}^2},
\end{equation}
inside the fast magnetosonic point and saturates beyond this point to $\Gamma_{\infty}\sim \mu^{1/3}_{\rm M}=(B^2/4\pi n m_{\rm e} c^2 )^{1/3}$ \citep{1994PASJ...46..123T, 1998MNRAS.299..341B, 2020A&A...642A.204C}. The law given in Eq.~(\ref{eq::doppler}) explains well the synthetic pulse width and its radial evolution near the light cylinder ($r\lesssim 5 R_{\rm LC}$, see Fig.~\ref{fig::fig89}) if the particles in the layer begins with a mildly relativistic boost,
\begin{equation}
\Gamma=\Gamma_{\rm LC}\Gamma_{\rm M},
\end{equation}
where $\Gamma_{\rm LC}\approx 5.5$ best reproduce the reported simulation data. At larger radii, as the wind expands, the layer thickness increases with radius, then the pulse width due to the finite thickness of the layer is \citep{1990ApJ...349..538C, 2017A&A...607A.134C}
\begin{equation}
\delta\Phi\sim\frac{\Delta}{2}=\frac{1}{2\pi\Gamma_{\rm LC}\kappa}\hat{r},
\label{eq::expansion}
\end{equation}
meaning that the pulse width widens linearly with radius. In the simulations, we have $\Delta/2\approx 0.004\hat{r}$ (Fig.~\ref{fig::fig89}). Thus, there is a radius, $r_{\rm m}$, where the pulse width is minimum and beyond which the effect of expansion dominates over relativistic beaming. Matching Eq.~(\ref{eq::doppler}) with Eq.~(\ref{eq::expansion}) gives this critical radius (for $\theta=\pi/2$)
\begin{equation}
r_{\rm m}=\sqrt{\frac{\sqrt{1+16\kappa^2}-1}{2}}\approx\sqrt{2\kappa},~\rm{for}~\kappa\gg1,
\end{equation}
if $r_{\rm m}$ is inside the fast magnetosonic point, else
\begin{equation}
r_{\rm m}=2\kappa\left(\frac{\Gamma_{\rm LC}}{\Gamma_{\infty}}\right).
\end{equation}
Only the split-monopole simulation is large enough to observe the effect of the layer expansion on the pulse width. In this simulation, the bulk Lorentz factor saturates $\hat{r}\gtrsim 10$, thus $r_{\rm m}$ is best estimated with Eq.~(\ref{eq::expansion}). With $\kappa\sim 7$ (Fig.~\ref{fig::fig1}) gives $r_{\rm m}\approx 3.7$ (Fig.~\ref{fig::fig89}). For more realistic multiplicities ($\kappa\sim 10^2-10^6$), $r_{\rm m}\sim 10-10^3$. Pulsations fully disappear if $\Delta=1$ at
\begin{equation}
r_{\rm c}=\pi\Gamma_{\rm LC}\kappa,
\end{equation}
so that $r_{\rm c}\sim 10^2-10^6$ in active pulsars.

\subsection{Implications for the pulsed TeV emission}\label{sect::tev}

This work constrains the location of the pulsed TeV emission discovered in the Crab and Vela pulsars within the inner parts of the pulsar wind. The target photon field must therefore remain compact. A promising candidate for emitting this low-energy radiation are secondary pairs produced in the vicinity of the current sheet --like a sheath surrounding the layer-- near the light cylinder and shining synchrotron radiation \citep{1996A&A...311..172L, 2019ApJ...877...53H}. This photon field would appear isotropic seen from the high-energy pairs accelerated deep inside the reconnection layer, which could then be upscattered into the TeV range. In contrast, any diffuse target photon field on much larger scales, from the surrounding nebula, the CMB, or from a nearby main-sequence companion star, would smear out any pulsation by adding up a dominant contribution from large radii.

The detection of 20~TeV pulsed emission from the Vela pulsar implies that pairs must be accelerated up to at least $\gamma_{\rm max}\gtrsim 4\times 10^7$ \citep{2023NatAs...7.1341H}. This is a challenging constraint for the models because this energy scale is significantly above the radiation-reaction-limited particle energy estimated at the light cylinder, $\gamma^{\rm LC}_{\rm rad}\approx 4\times 10^5\ll \gamma_{\rm max}$ (assuming $E=\tilde{B}_{\perp}=B_{\rm LC}\approx 5\times 10^4$G, see Eq.~\ref{eq::grad}). One possible explanation for this discrepancy involves X-point acceleration, where radiative losses are substantially reduced, meaning that particle acceleration could continue nearly unimpeded up to the upstream magnetization limit, $\gamma_{\rm max}\gtrsim \sigma_{\rm LC}$ \citep{2012ApJ...746..148C, 2016ApJ...833..155K, 2023ApJ...959..122C}. This would in turn imply that the magnetization must be much higher than usually assumed in Vela. 

Another effect that could alleviate this issue is particle reacceleration at larger radii, as reported in Sect.~\ref{sect::width}, because reconnection pursues magnetic dissipation while radiative losses diminish thanks to the weakening of the magnetic field strength with distance. Note that the spatial offset between the peak of the synchrotron and the inverse Compton emission would not lead to any significant changes in pulse phase, thanks to the caustic effect highlighted in Sect.~\ref{sect::skymaps}. 
However, it remains uncertain whether this effect alone can account for the required hundred times increase in particle energy, from $\gamma_{\rm rad} \approx 4\times 10^5$ to $\approx 4 \times 10^7$, on reasonably compact scales (i.e., $\hat{r}\lesssim r_{\rm m}$). Despite these challenges, a combination of both mechanisms may render this scenario viable, but further investigation is necessary to explore these questions in greater detail.

\section{Comparison with the third {\em Fermi} pulsar catalog}\label{sect::fitting}

Here, we compare the synthetic lightcurves predicted from the synchrotron emission with the observed phase-folded lightcurves of the 294 pulsars listed in the third {\em Fermi}-LAT pulsar catalog (3PC, \citealt{2023ApJ...958..191S}). We keep only the positron emission since the emission from the precipitating electrons is likely to be strongly overestimated (Sect. \ref{sect::energy}). Skymaps are frequency-integrated above $\nu_0$ defined in Sect.~\ref{sect::skymaps}, which represents the typical synchrotron frequency emitted by pairs energized at the light cylinder. We use this frequency as a proxy for the \emph{Fermi}-LAT low-energy threshold.

\subsection{Model fitting procedure}

The synthetic skymaps were rebinned by a factor of 2 in phase and inclination $\alpha$ to increase their signal-to-noise ratio (S/N). We also assumed that both hemispheres have the same emission properties and, hence, enforced the skymap symmetry by averaging the values at ($\Phi$,$\alpha$) and ($\Phi+180\degr$,$180\degr-\alpha$), further increasing the S/N of the skymaps and reducing the scope of the parameter range to explore. 

\begin{figure*}
\centering
\includegraphics[width=6cm]{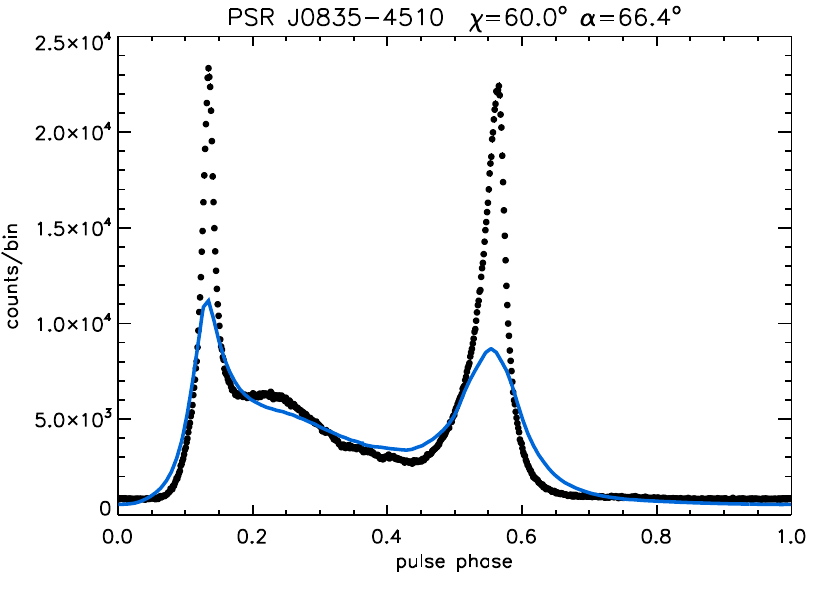}
\includegraphics[width=6cm]{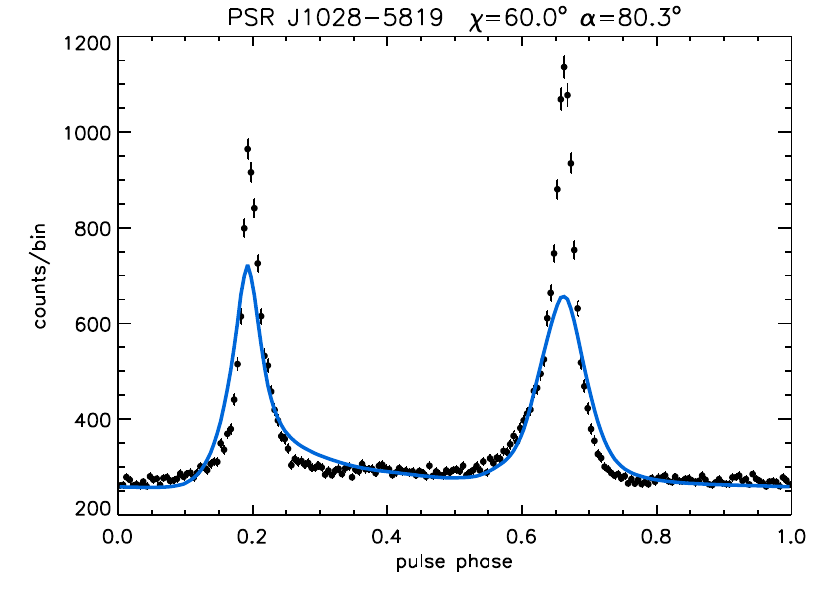}
\includegraphics[width=6cm]{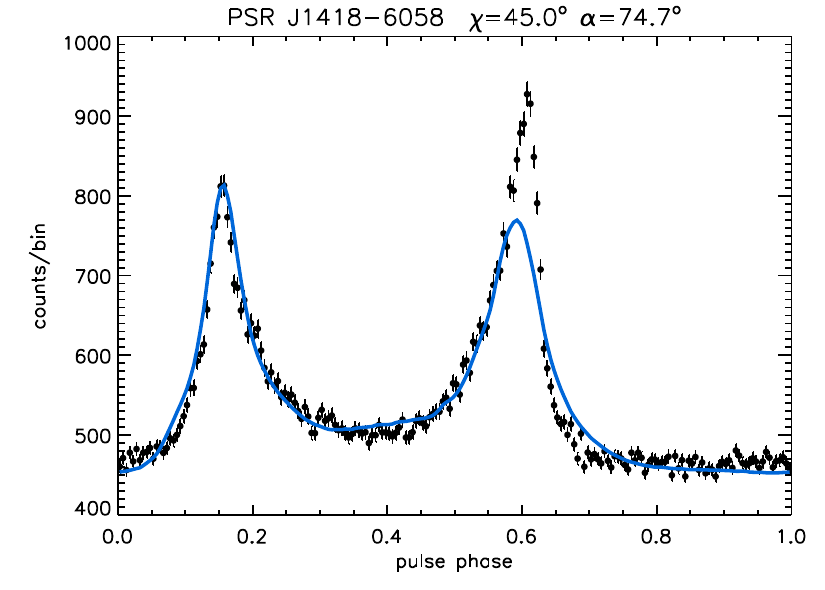}
\includegraphics[width=6cm]{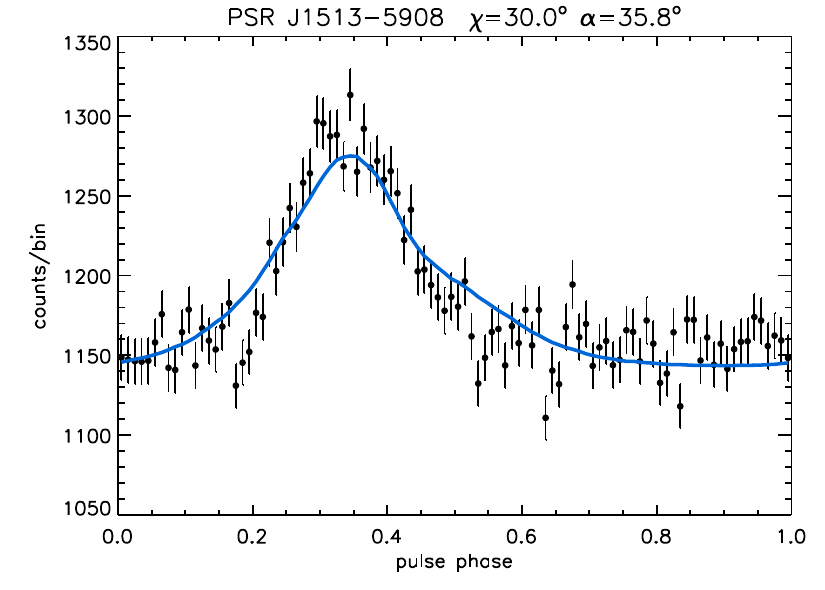}
\includegraphics[width=6cm]{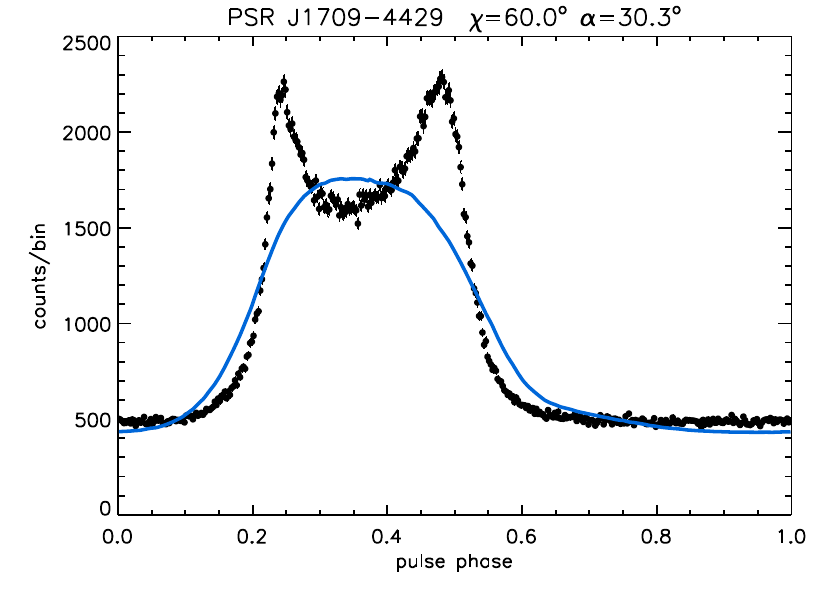}
\includegraphics[width=6cm]{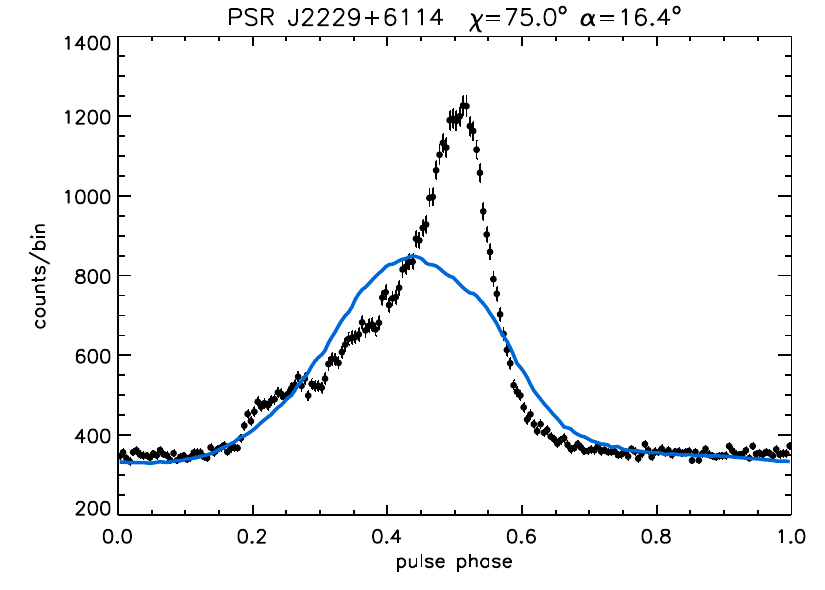}
\caption{Best fit (full line) based on the model skymaps to 6 pulse profiles (data points) taken from the {\em Fermi}-LAT 3rd pulsar catalog. The top row illustrates double-peaked pulse profile fits, including the Vela pulsar (top left). The bottom row illustrates single-peaked fits.}
\label{fig::fits}
\end{figure*}

The observed folded lightcurve consists of $n$ phase measurements $m_i$ with an associated error $\sigma_i$. The best-fit synthetic lightcurve was found by minimizing the chi-square,
\begin{equation}
{C}^2=\sum_{i=1}^{n} \frac{\left(m_i-K S_i-\mathcal{B}\right)^2}{\sigma^2_i},
\end{equation}
where $K$ is a constant scaling factor, and $\mathcal{B}$ is a constant background. Some pulsars present a strong DC component that may be of magnetospheric origin, but the model presented here does not predict any significant contribution to the background emission, we are thus considering $\mathcal{B}$ independently from the physical model. $S_i$ is the synthetic lightcurve at phase $i$ for a given obliquity $\chi$, inclination $\alpha$, dephased by $\Phi_0$ to adjust the data. The origin of the phase in the 3PC is usually set by the main radio pulse (for radio-loud pulsars), while it is set by the magnetic pole in the model. While these two definitions may be compatible if the radio pulse originates from the polar cap of the star (assuming a dipole), they should differ otherwise. Given the large diversity of the radio phenomenology (radio quiet/loud, number of pulses, phase-(mis)alignment with the gamma rays), we chose to treat the origin of phases as a free parameter in the fitting procedure. $C^2$ can be analytically minimised for the nuisance parameters $K$ and $\mathcal{B}$. However, a full search must be performed over the parameter space in $\Phi_0$, $\chi$ and $\alpha$ to find the minimum. We found that fitting with equal-sized phase bins gave acceptable results so we did not pursue more elaborate fitting routines aiming to take better into account narrow, high-amplitude pulses such as bin weighting \citep{2010ApJ...714..810R} or fixed count binning \citep{2015A&A...575A...3P}. Although this allows a first confrontation of the model to the observations, an important caveat is that the fit is usually not good statistically given the quality of the {\em Fermi}-LAT lightcurves and the limitations of the model simulation \citep[e.g.][]{2010ApJ...714..810R,2014ApJS..213....6J,2015A&A...575A...3P}.

\subsection{Best fit lightcurves}

Figure \ref{fig::fits} shows examples of best fits to the {\em Fermi}-LAT pulse profiles integrated above 100 MeV (see the full atlas in Appendix~\ref{appendix-A}). The obliquity $\chi$ and inclination $\alpha$ of the best fit are shown on top of each panel. The top row shows double-peaked pulse profiles, obtained when the line-of-sight is well within the opening angle of the current sheet. The best fit for the Vela pulsar (PSR J0835-4510) is  consistent with the values used to illustrate the energy dependence of the pulse profile (Fig.~\ref{fig::fig5}) since $\alpha\approx 70\degr$ is equivalent to $\alpha\approx 110\degr = 180\degr-70\degr$ with the map symmetry. Overall, our fit to the Vela pulse profile compares well with, e.g., the elaborate slot gap, outer gap, and two-pole caustic model fits in \citet{2016ApJ...832..107B}. Our values of $\alpha$ and $\chi$ are also consistent with their values. The bridge emission and the peak positions are well reproduced by the model. However, the observed peaks are narrower and their amplitude is greater than predicted. This is also seen in the fits for the two other pulsars on the top row of Fig. \ref{fig::fits}, PSR J1028-5819 and  PSR J1418-6058, both of which are seen at high inclination (leading to two pulses). This shortcoming is likely due to the limited range in radius and to the simulation scaling, which impact the bulk Lorentz factor and energy of the particles (Sect. \ref{sect::width}). The choice of energy range for integrating the emission can also play a role since pulses are narrower at higher energies (Fig.~\ref{fig::fig5}).

The bottom row illustrates some of the challenging variety seen in pulse profiles. In those cases, the best fit ended up being a single-pulse profile. The low-amplitude pulse of PSR J1513-5908 is consistent with a solution in which the observer line of sight grazes the emission region ($\alpha\approx \chi$). PSR J1709-4429 illustrates a difficulty of the model when there are two peaks close in phase. The expectation is that the two peaks seen at $\alpha=90\degr$ get closer in phase as the inclination $\alpha$ approaches $\chi$. However, the skymaps in Fig.~\ref{fig::fig2} show that this does not happen because the amplitude of the peaks drops quickly away from $\alpha\approx 90\degr$ for large obliquities and because the peaks are broad and smeared into a single pulse for low obliquities. Thus, the model tends to fit a single pulse. Last, PSR J2229+6114 illustrates a slow rise and fast decay profile seen in several pulsars that is difficult to understand. Indeed, the current sheet has a sharp density jump when it moves into the line-of-sight, with the density gradually falling off as it rotates away, i.e. a fast rise and slow decay would be more natural. The fits for PSR J1709-4429 and PSR J2229+6114 are both localised in a low-emission area of the skymap. The low radiative efficiency is unlikely to be compatible with the observed flux even though both are young pulsars with high $\dot{E}$ (3.5$\times 10^{36}$ and 2.2$\times 10^{37}\rm\,erg\,s^{-1}$, respectively). Notwithstanding the statistically poor fit, we therefore regard the solutions for PSR J1709-4429 and PSR J2229+6114 as unphysical. 

\subsection{Angle distributions}

Fitting all the 3PC pulsar profiles, we found that the global distribution in $\alpha$ or $\chi$ differs significantly from a randomly oriented distribution of angles. We verified through a Monte Carlo simulation that this is not due to a bias in the population selection or in the fitting procedure. To do this, simulated lightcurves were constructed from the synthetic skymaps, with a random orientation for $\alpha$ and $\chi$, and scaled so that their baseline flux and amplitude correspond to one of the 294 observed pulsars.  We then randomly picked an observed flux for each phase by assuming Poisson statistics at the predicted count rate for each phase, before adding a random phase shift. Simulating 100 observations of the 294 3PC pulsars clearly showed that our fits to the 3PC population have an excess of solutions with low inclination $\alpha\approx 20\degr$ and high obliquity $\chi\approx 60\degr$. We attribute these excesses to unphysical fits such as those described above, which bias the distribution.

To get a measure of the biases introduced by the model on the distribution of angles, we investigated fitting using model skymaps with different energy selections or using only electron or all particle emission. Interestingly, the variance of the best fit $C^2$ is lowest when using only emission from the positrons, all other combinations yielded poorer results.  Although the fits are relatively consistent regardless of the skymap when an acceptable solution exists (e.g. the pulsars on the top row of Fig.~\ref{fig::fits}), no robust trends emerge from the global distributions. Restricting to the best fits or to sub-populations of pulsars (such as radio loud vs radio quiet or millisecond pulsars vs young pulsars) did not yield more significant results. Directly fitting the model to the observed pulsar population is not yet accurate enough to provide useful results on the distribution of their inclination and obliquity. Previous works, which also systematically fitted pulsar profiles to various models, encountered the same difficulty as they showed that not a single model fares well on all types of pulsar profiles \citep{2010ApJ...714..810R,2012ApJ...744...34V,2014ApJS..213....6J,2015A&A...575A...3P} 

\subsection{Number of peaks}

To circumvent this limitation in directly fitting data, \citet{2016A&A...588A.137P} compared the ability of various pulsar models to reproduce major morphological characteristics of the pulse profile such as the number of peaks. Similarly, we estimated from the skymaps that $\ga 62\%$ of the observed pulse profiles should be double-peaked, assuming that $\alpha$ and $\chi$ are randomly-oriented and that all pulsars are equally detectable. The 3PC catalog lists 74\% pulsars with 2 or more peaks. Double peaks are associated with stronger emission (Fig.\,\ref{fig::fig2}), so the detection bias is consistent with observing a higher fraction than expected. \citet{2016A&A...588A.137P} pointed out a higher fraction of single-pulsed profiles in radio loud pulsars ($\approx 30\%$) compared to radio quiet pulsars ($\approx 10\%$) based on the sample young and middle-aged pulsars in the second \emph{Fermi}-LAT pulsar catalog. Radio quiet pulsars are more likely to be seen edge-on since radio emission is associated with the polar caps, hence more likely to produce double-peaked pulse profiles. Alternatively, a more aligned population ($\chi \rightarrow 0\degr$) would be less likely to produce double-peaked profiles. In the 3PC catalog there are 93 pulsars with a spindown timescale $\la 250\rm\,kyr$ and with a listed number of peaks : 51 are detected in radio and 42 are radio-quiet. Of those, 20/51=39\% of the radio-loud and 11/42=26\% of the radio-quiet have a single-pulse profile, which a Z-test shows is not a significant difference. In general, we did not find significant differences ($>2\sigma$) in the number of peaks when comparing sub-populations of pulsars in the 3PC catalog. 

\section{Summary}\label{sect::summary}

To summarize, the salient features reported in the modeling of synthetic pulsar lightcurves are:
\begin{itemize}
\item Particle acceleration powered by reconnection in the wind current layer leads to synchrotron and inverse Compton radiation.
\item The synchrotron and inverse Compton emission patterns are similar, but not identical, and in phase. They present two bright caustics leading, in general, to two main pulses per spin period.
\item The separation in phase of both pulses is explained by the geometry of the reconnection layer forming in the pulsar wind, and is well approximated by the split-monopole solution.
\item The current layer is not uniformly bright at all latitudes, it is mostly shining in the equatorial plane. As a direct consequence, pulsars with nearly symmetric pulses separated by $0.5$ in phase should present the highest apparent luminosities.
\item An interpulse emission, or third peak, generically appears away from the equatorial regions at intermediate latitudes, and it is most visible in the synchrotron lightcurves. This feature most convincingly corresponds to the volumetric return current densities regions, as well as reflect the strong asymmetric nature of the wind between two consecutive layers. Additional features at high viewing angles are also consistent with super-Goldreich-Julian current densities regions. More work dedicated to fully characterize the origin of these extra components is required.
\item The emission pattern has a central symmetry with respect to the center of the maps, located here in the equatorial plane at phase $\Phi=0.5$. This symmetry is related to the direction of the electric current carried by the layer, which leads to asymmetric pulse profiles seen by an observer looking away from the equatorial plane.
\item The asymmetry between both pulses increases with the photon energy. In particular, Vela-like lightcurves with dominant P2 at high energies are reproduced.
\item The width of the lightcurve peaks is set by two competing effects: (i) the acceleration of the expanding wind implies relativistic beaming that narrows the pulse width, and (ii) the geometric expansion of the layer which increases the pulse width. The emitted pulses are thinnest at a radius fully determined by the plasma multiplicity in the current layer, $r_{\rm m}=\sqrt{2\kappa}$.
\item The inverse Compton pulses are about half narrower than their synchrotron counterpart, provided that the target photon field is uniform within $r_{\rm m}$. Conversely, if the photon field is uniform beyond $r_{\rm c}=\pi\Gamma_{\rm LC}\kappa$, pulsations in the inverse Compton channel vanish.
\item The synchrotron and inverse Compton lightcurve peaks sharpen with increasing photon energy. The mean particle energy increases with radius so that more distant parts of the wind contribute to the high-energy emission where beaming is more important and synchrotron cooling is weaker. This may also be due to kinetic beaming effects in the reconnection layer (the higher the particle energy is, the deeper and the more focused its trajectory is in the layer).
\item There is no notable differences between the synchrotron and inverse Compton lightcurves if the target photon field originates from the neutron star surface. Anisotropic effects and the dilution of the target photon field diminish the radiated power. Anisotropic effects are strongest in the Thomson regime, leading to a steeper decay of the radiated power with the distance to the star.
\item The reconnection rate sets an upper limit to the bolometric synchrotron radiative efficiency, $\eta^{\rm max}_{\rm rad}\lesssim \beta_{\rm rec}\sim 0.1$.
\end{itemize}

This study paves the way for comparisons with observations of synthetic pulse profiles constructed from ab initio PIC simulations. The synthetic pulse profiles reproduce some of the salient features of observed gamma-ray pulsars, including the generic double-peaked structure, the presence of a bridge or third peak in between the main pulses, the pulse narrowing with increasing energy. Hence, fitting the pulse profiles of the {\em Fermi}-LAT 3rd pulsar catalog shows promising results despite the inherent limitations of a `one-size-fits-all' approach with a single (millisecond) pulsar model sampling a limited scale range and where the only parameters are the inclination and obliquity. Using full-scale hybrid simulations \citep{2024A&A...690A.170S} may be the way forward to match observations more closely.

\begin{acknowledgements}
We thank Alexander Philippov, Alexander Chernoglazov, Christo Venter, and the referee Constantinos Kalapotharakos for useful comments regarding this work. This project has received funding from the European Research Council (ERC) under the European Union’s Horizon 2020 research and innovation program (Grant Agreement No. 863412). Computing resources were provided by TGCC under the allocation A0130407669 made by GENCI and PRACE (call 17).
\end{acknowledgements}

\bibliographystyle{aa}
\bibliography{pulsar_lcfitting}

\begin{appendix}

\section{Atlas of the 3PC best fit lightcurves}\label{appendix-A}

\begin{figure*}[hb]
\centering
\includegraphics[width=4.5cm]{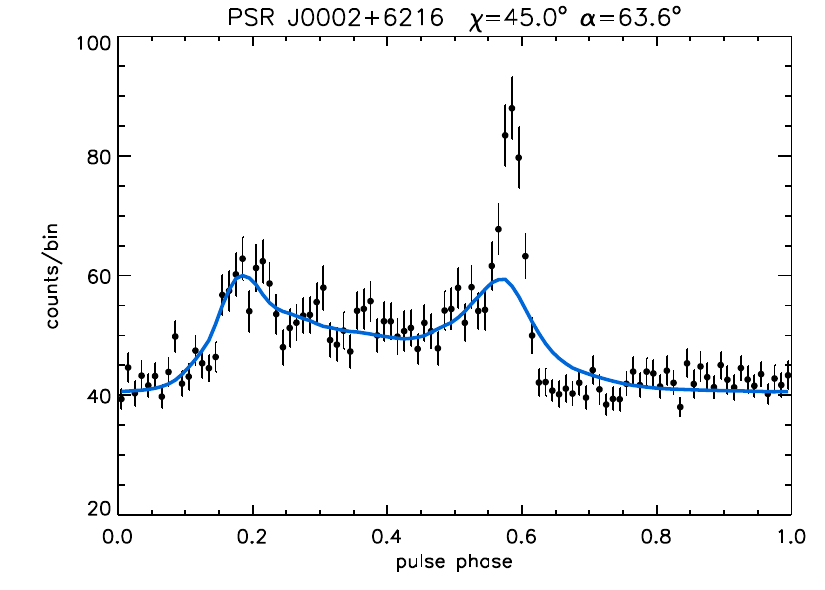}
\includegraphics[width=4.5cm]{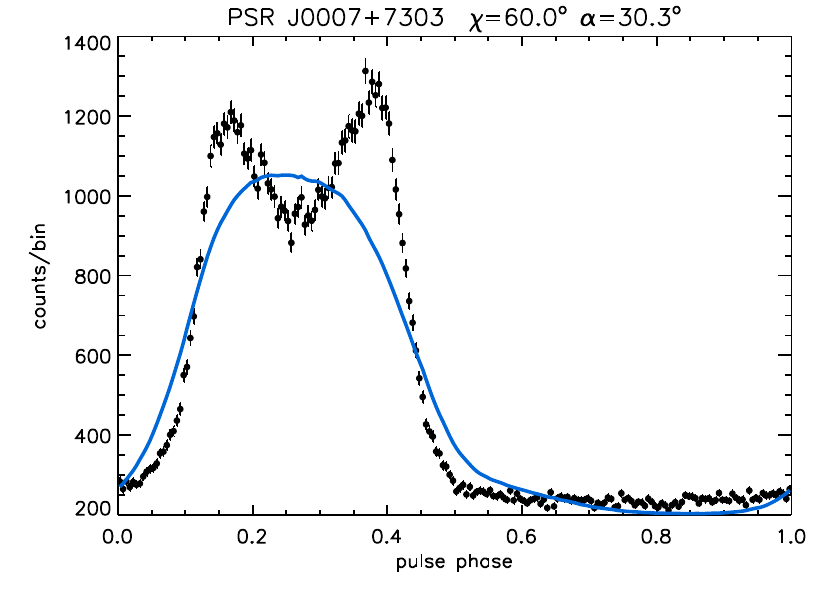}
\includegraphics[width=4.5cm]{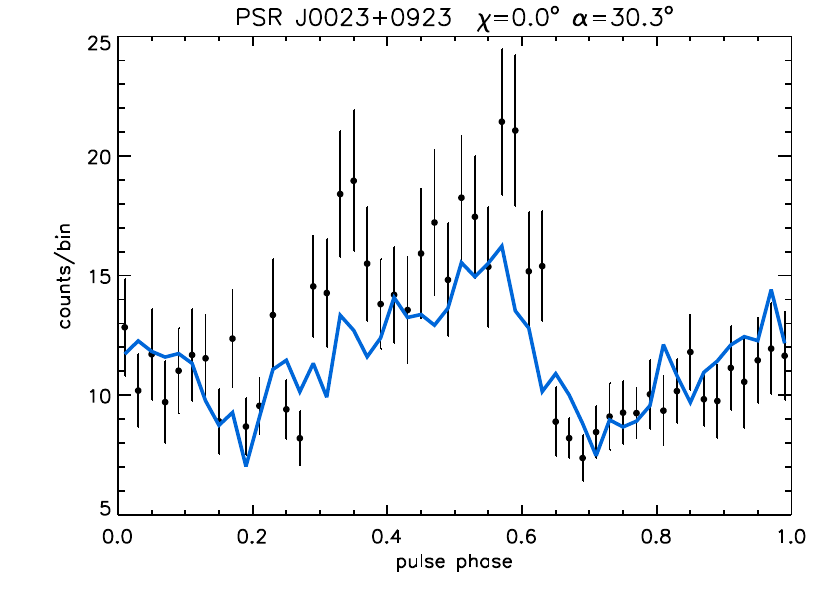}
\includegraphics[width=4.5cm]{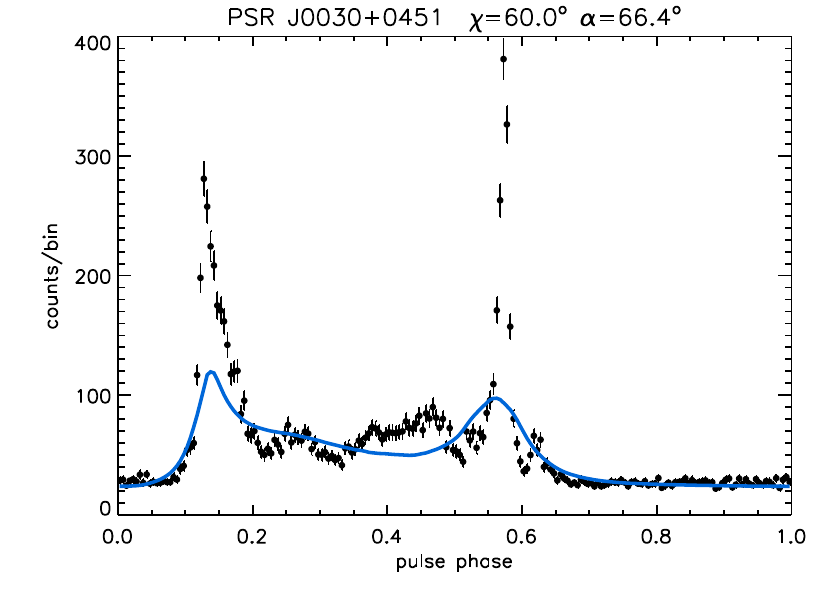}
\includegraphics[width=4.5cm]{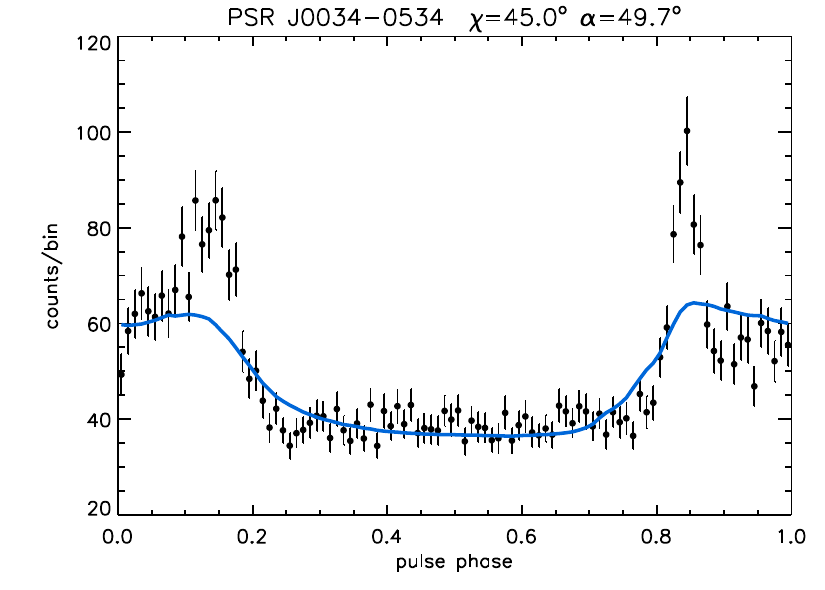}
\includegraphics[width=4.5cm]{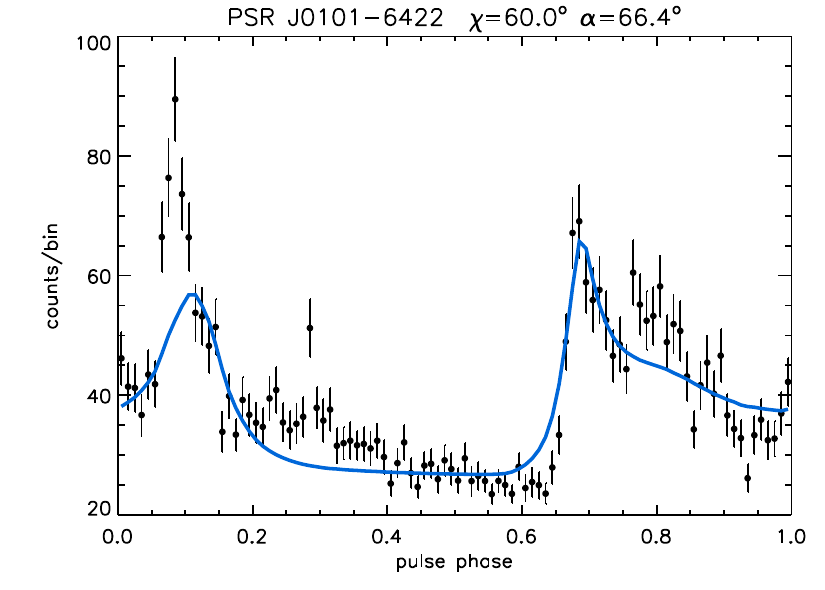}
\includegraphics[width=4.5cm]{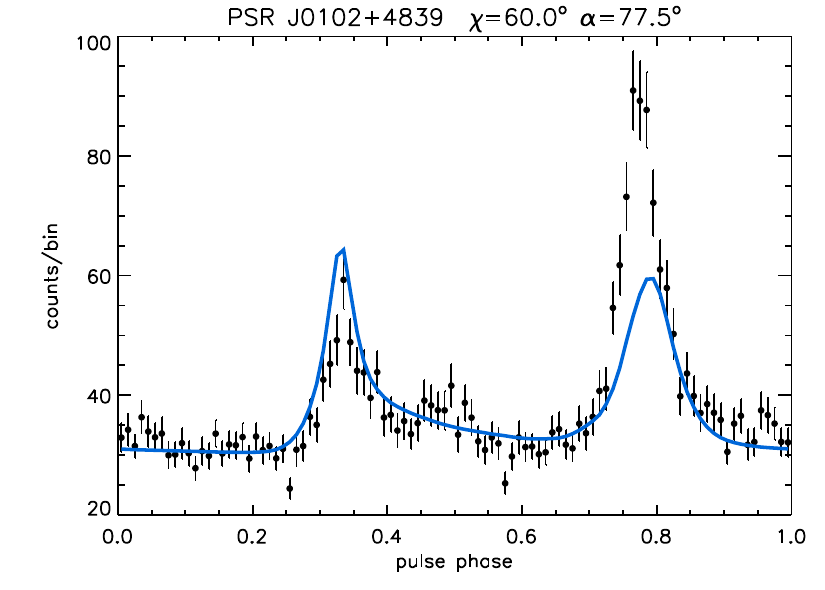}
\includegraphics[width=4.5cm]{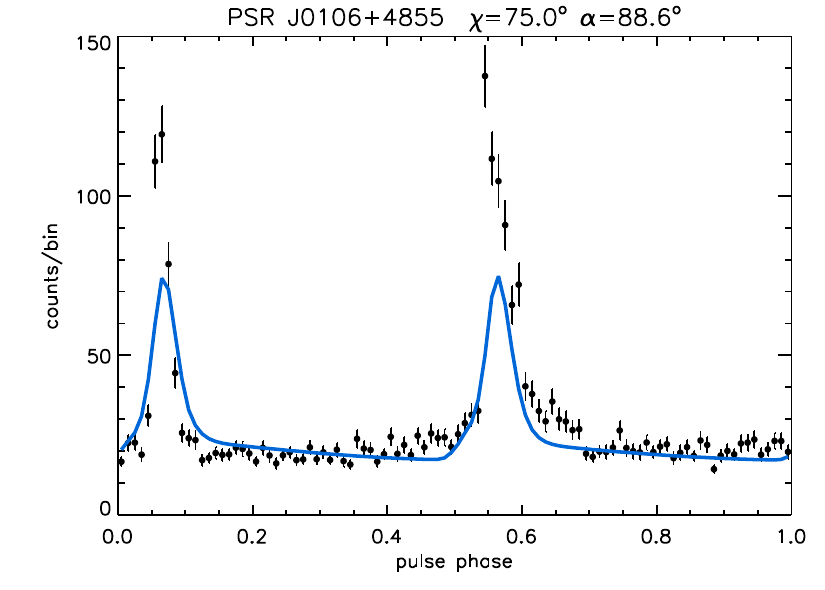}
\includegraphics[width=4.5cm]{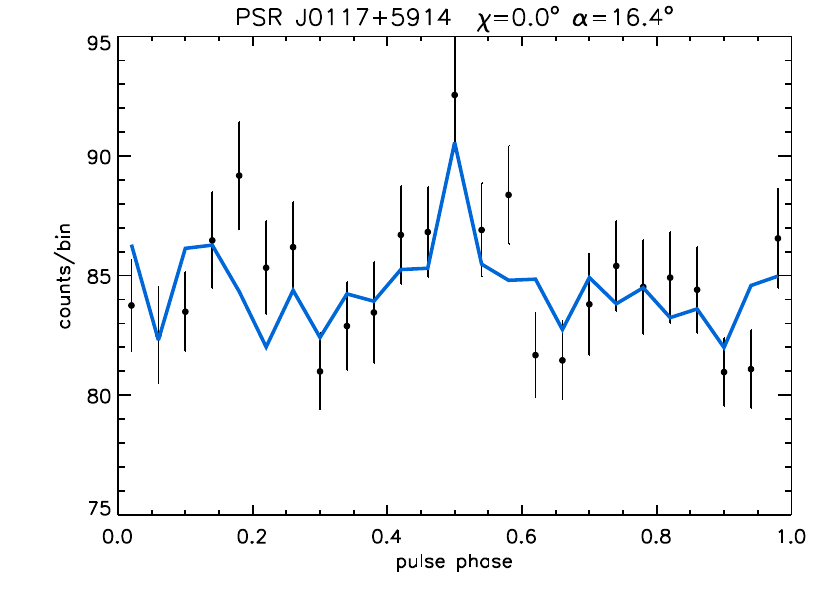}
\includegraphics[width=4.5cm]{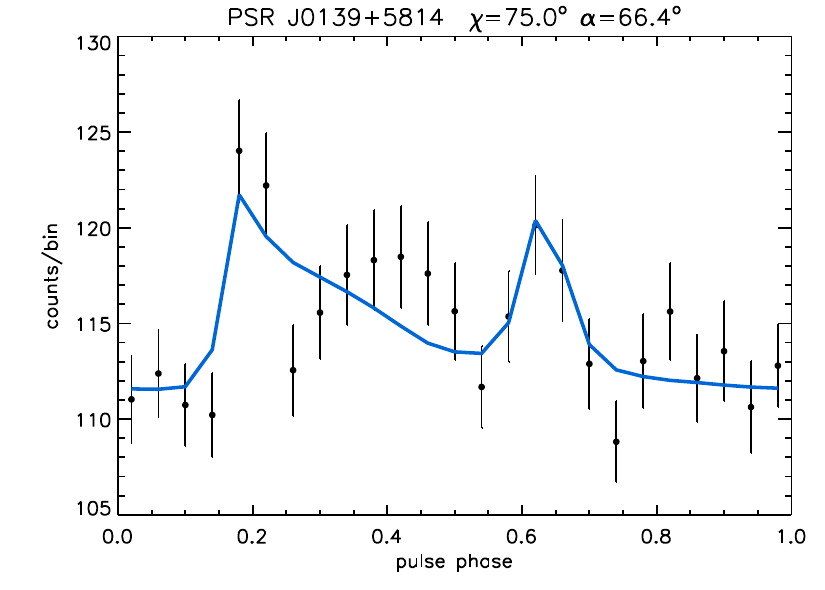}
\includegraphics[width=4.5cm]{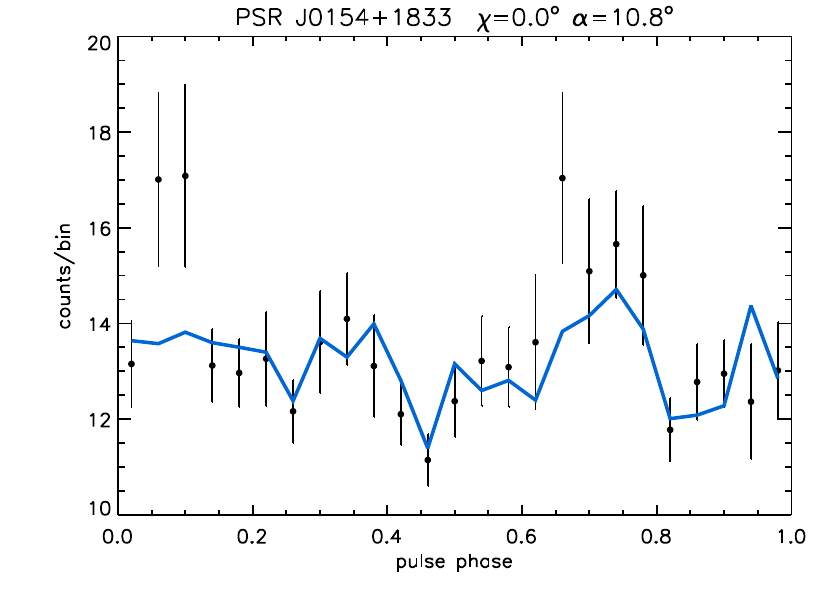}
\includegraphics[width=4.5cm]{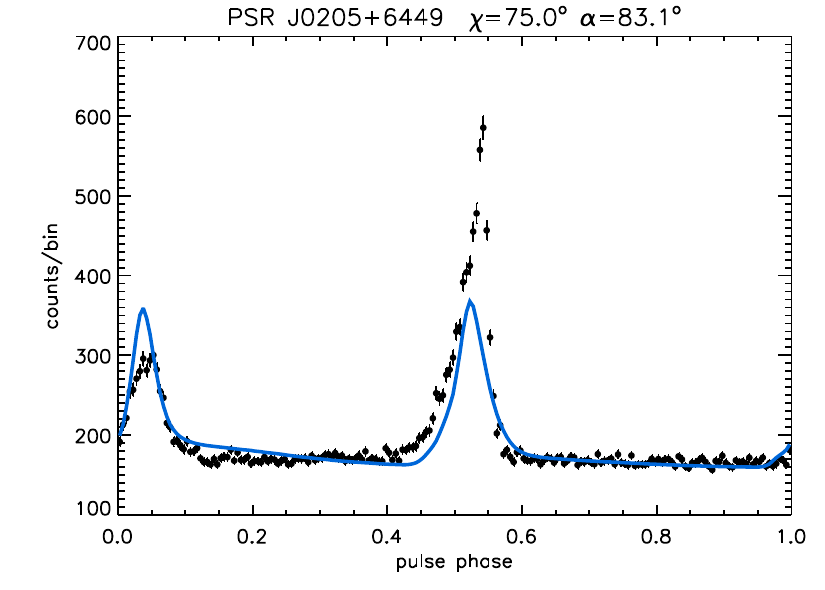}
\includegraphics[width=4.5cm]{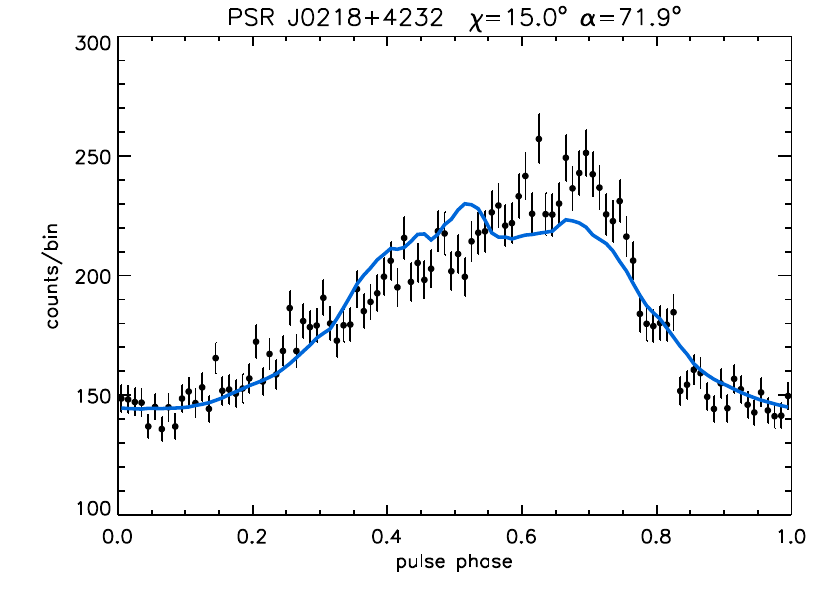}
\includegraphics[width=4.5cm]{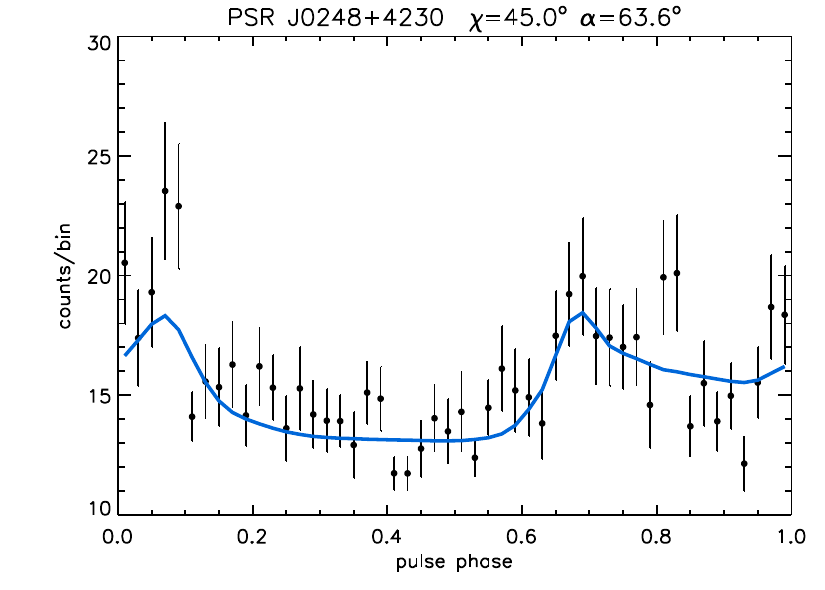}
\includegraphics[width=4.5cm]{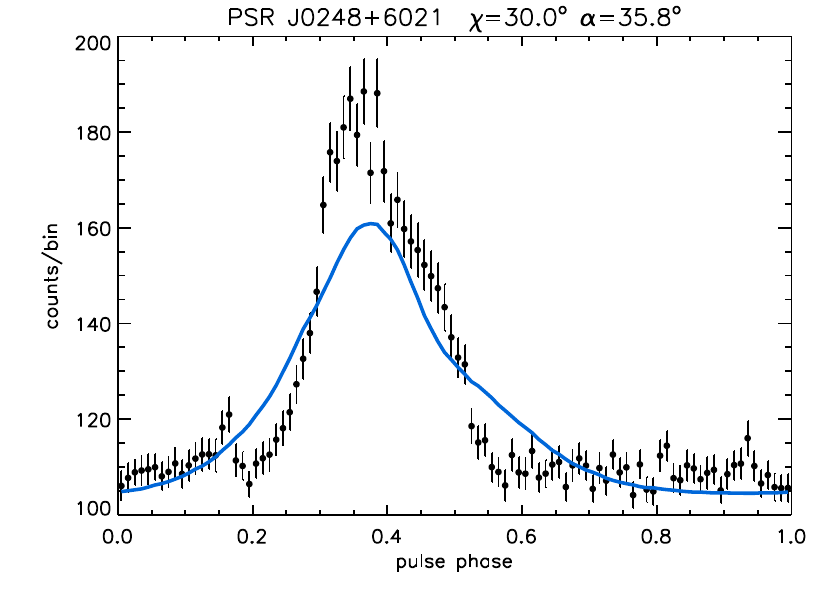}
\includegraphics[width=4.5cm]{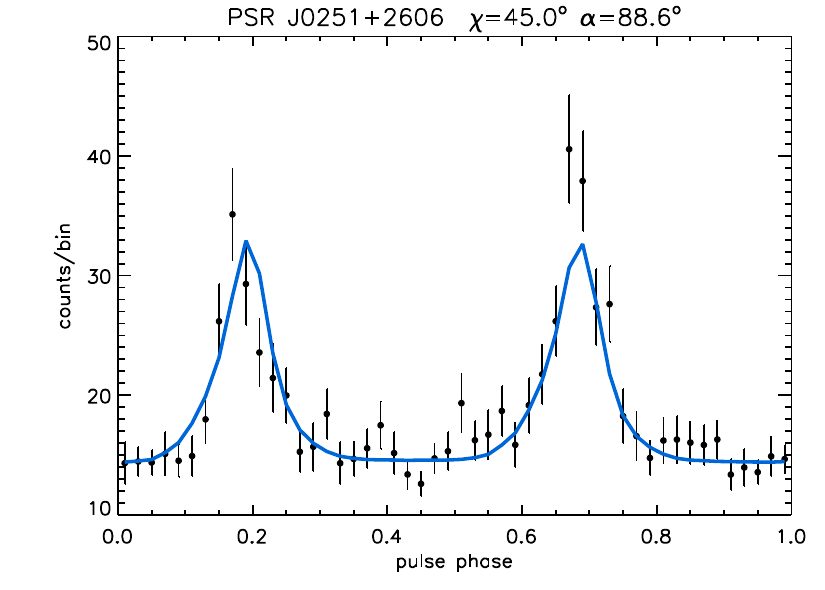}
\includegraphics[width=4.5cm]{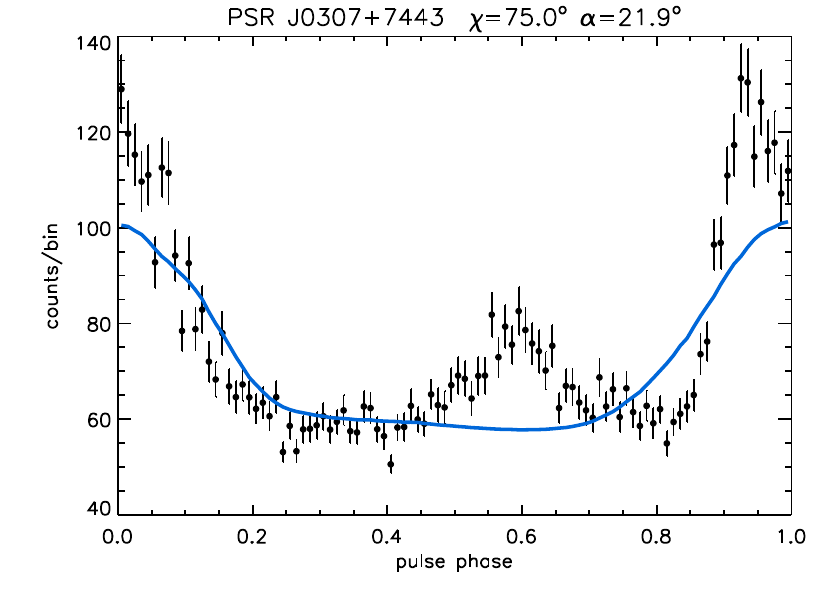}
\includegraphics[width=4.5cm]{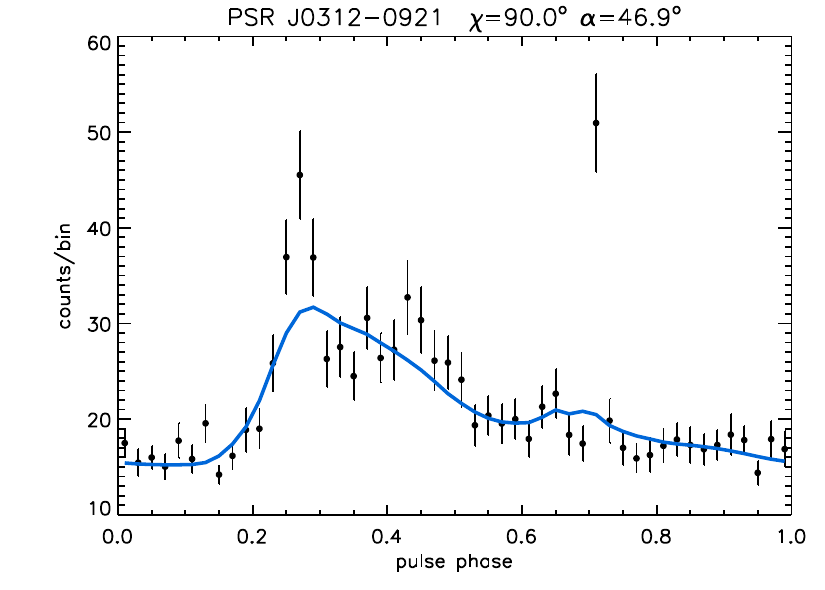}
\includegraphics[width=4.5cm]{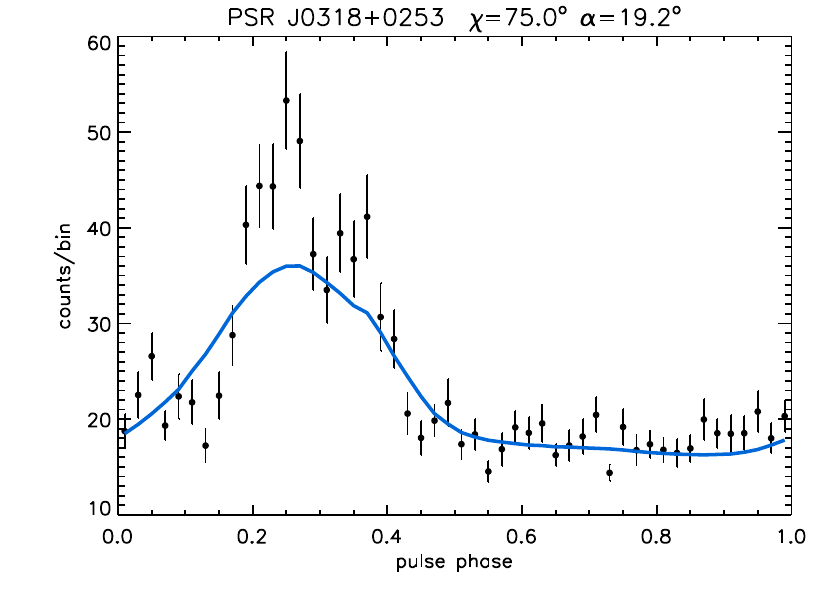}
\includegraphics[width=4.5cm]{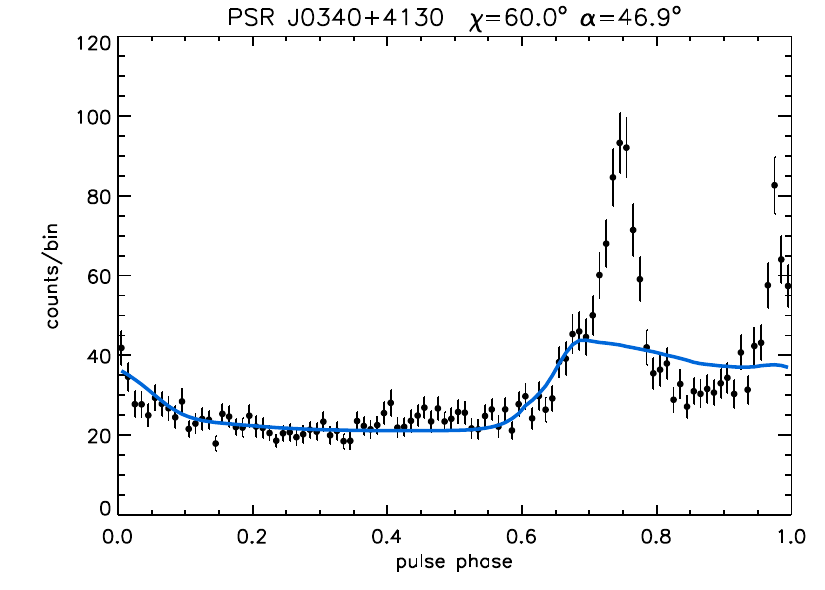}
\includegraphics[width=4.5cm]{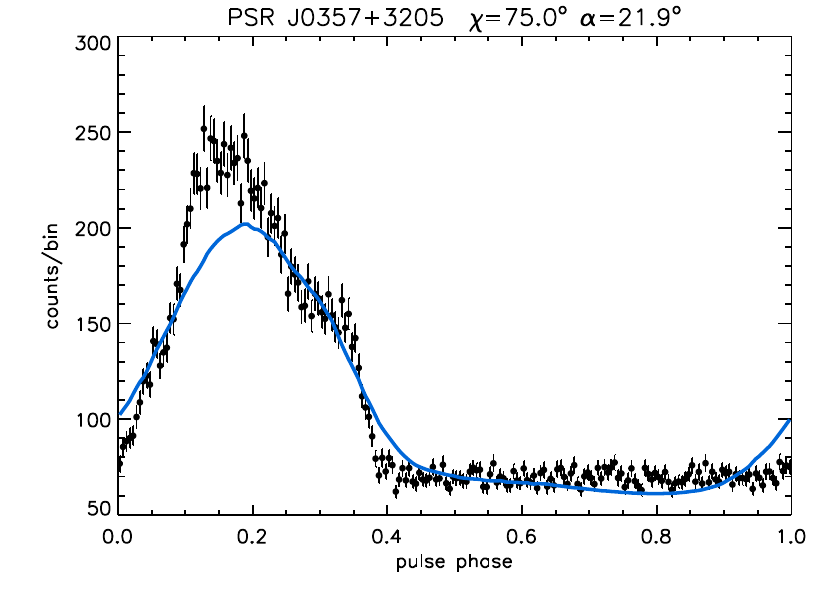}
\includegraphics[width=4.5cm]{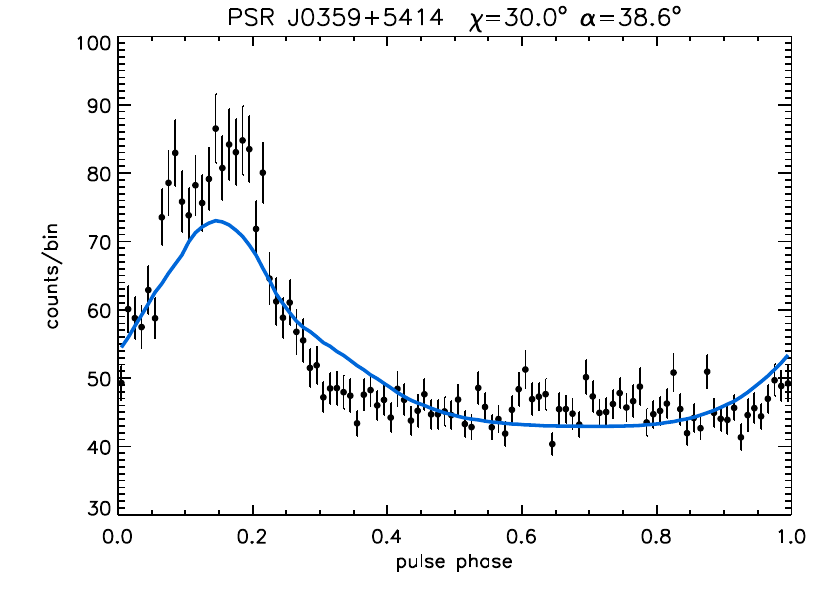}
\includegraphics[width=4.5cm]{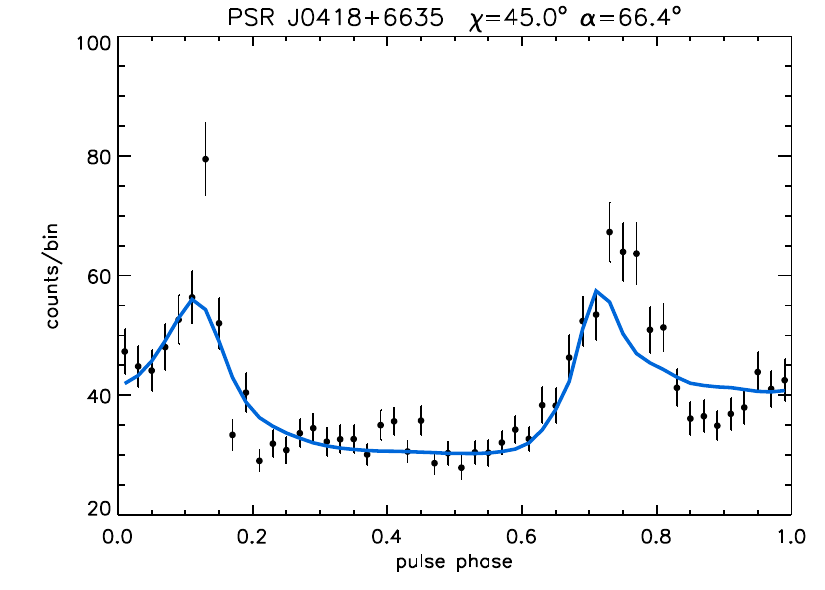}
\includegraphics[width=4.5cm]{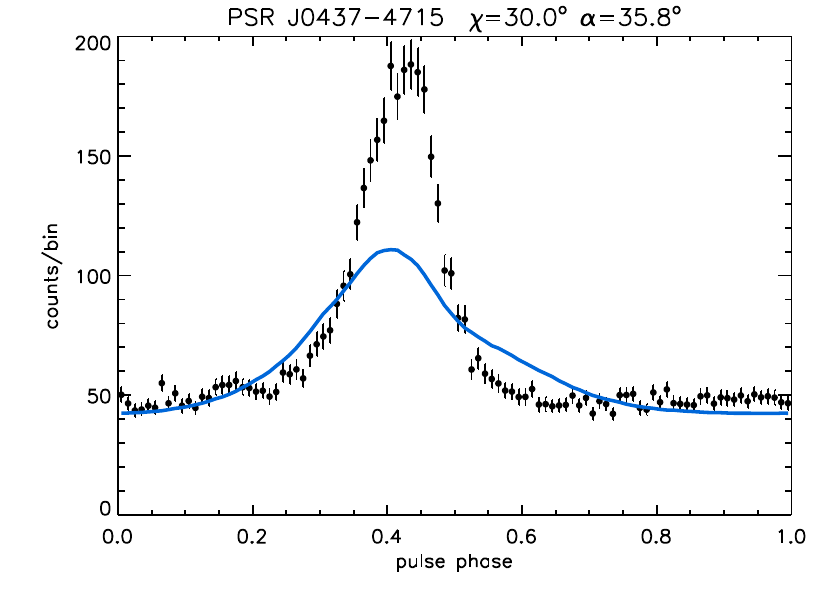}
\caption{Best fit lightcurves (full line) based on the model skymaps of all the 3rd {\em Fermi}-LAT pulsar catalog pulse profiles (data points). This figure continues on the next pages.\vspace{1.8cm}}
\label{fig::fitsall}
\end{figure*}

\begin{figure*}
\addtocounter{figure}{-1}
\centering
\includegraphics[width=4.5cm]{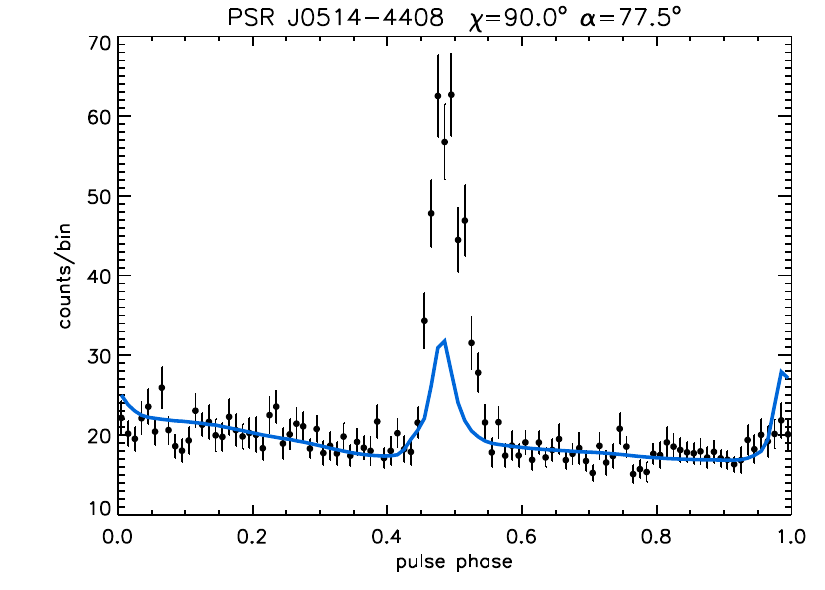}
\includegraphics[width=4.5cm]{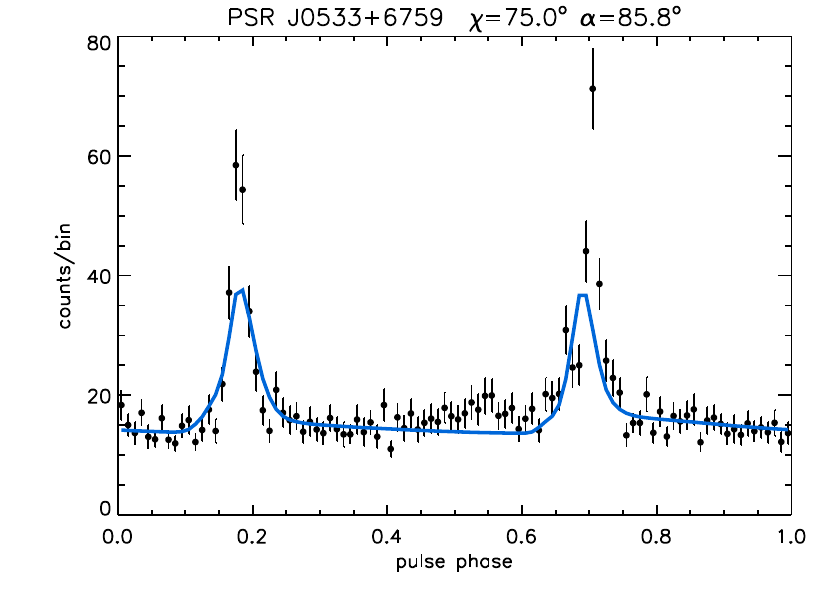}
\includegraphics[width=4.5cm]{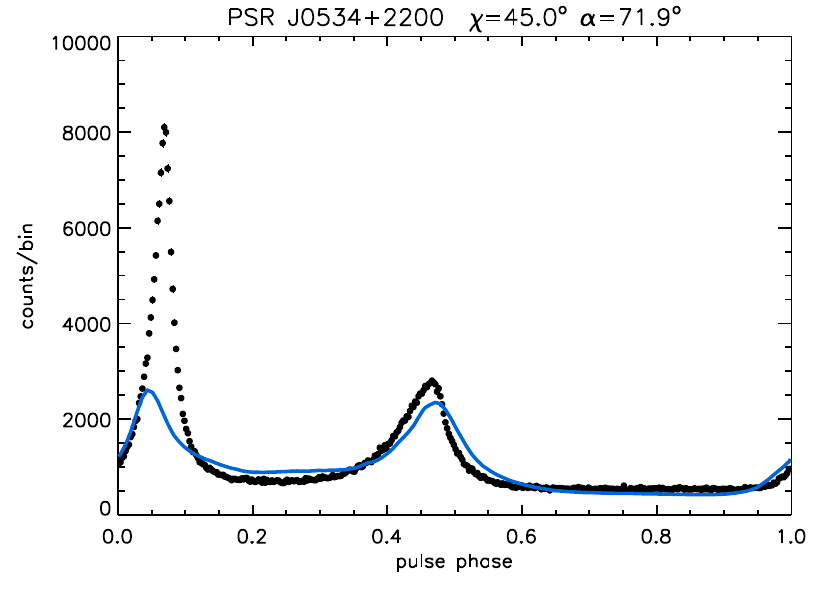}
\includegraphics[width=4.5cm]{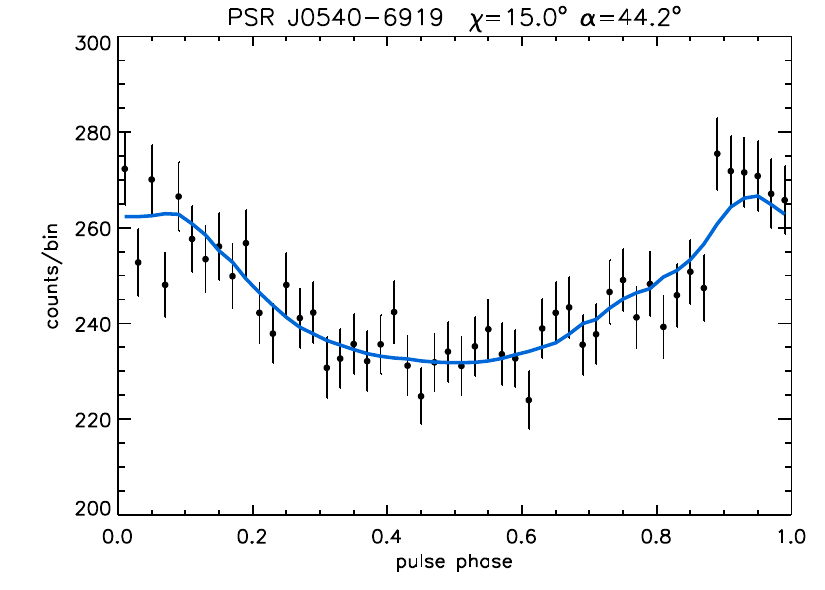}
\includegraphics[width=4.5cm]{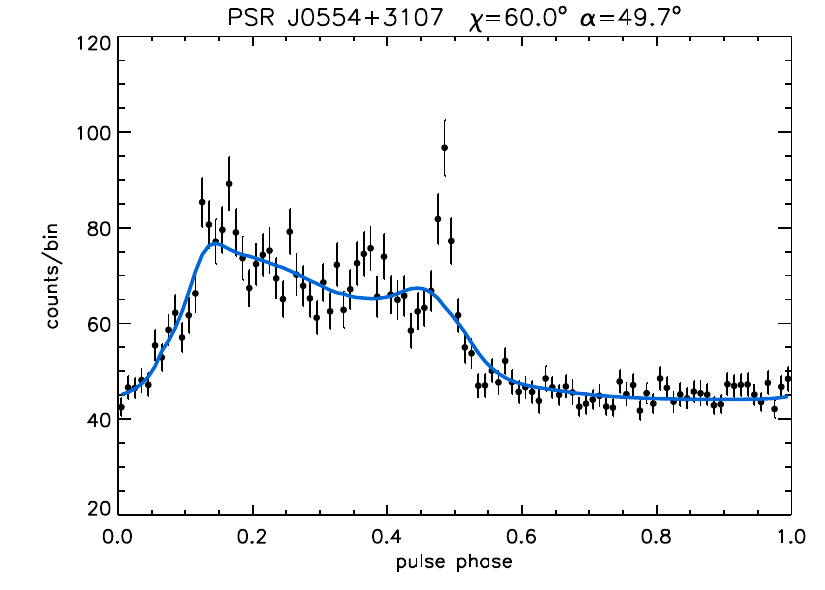}
\includegraphics[width=4.5cm]{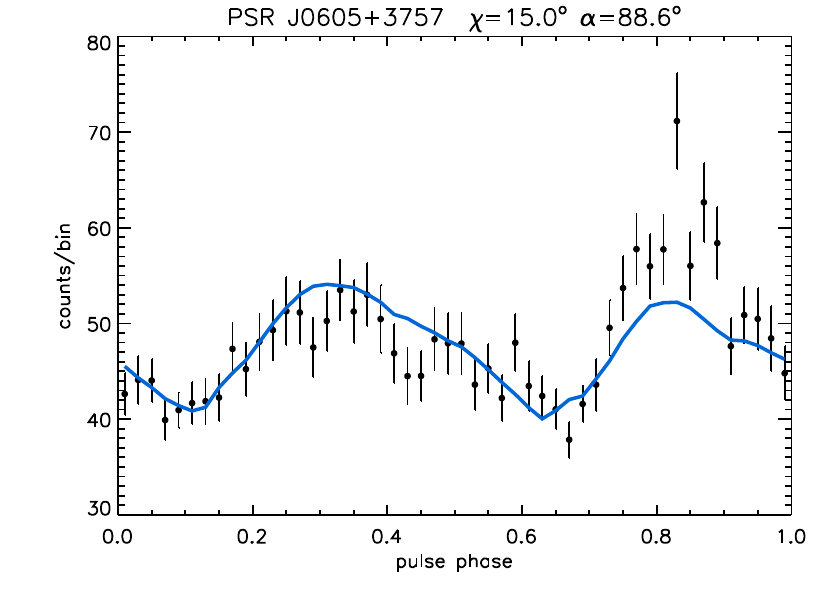}
\includegraphics[width=4.5cm]{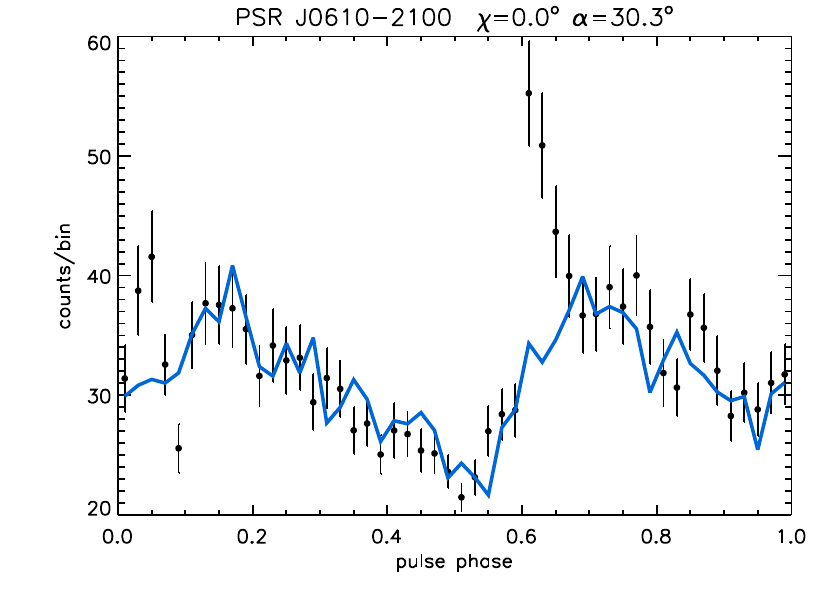}
\includegraphics[width=4.5cm]{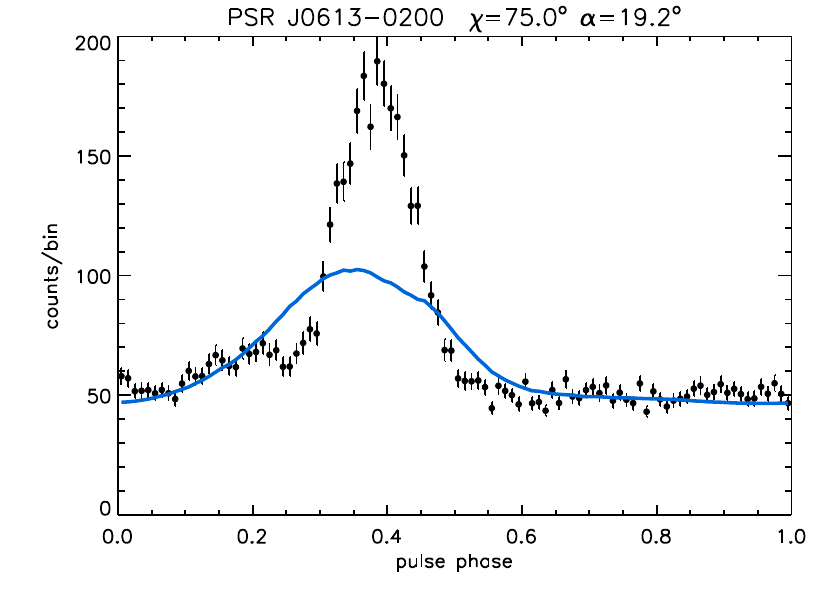}
\includegraphics[width=4.5cm]{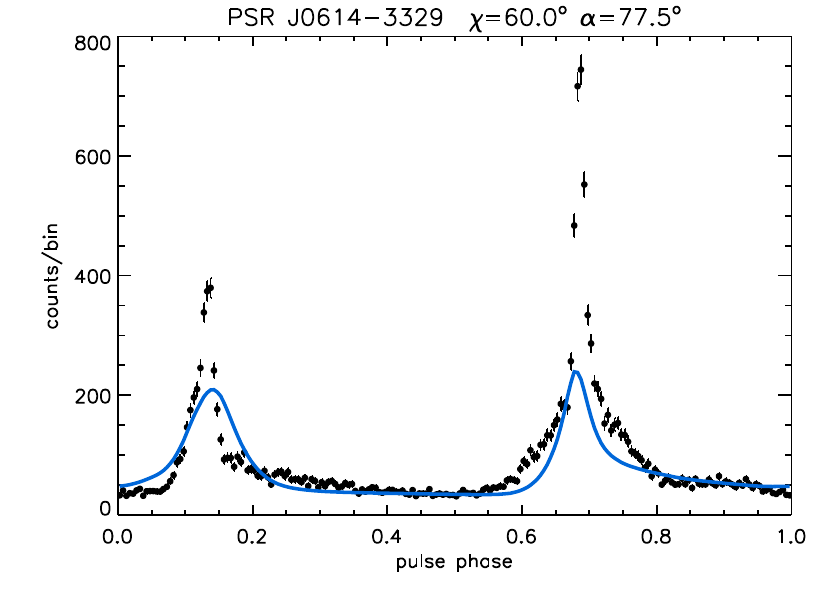}
\includegraphics[width=4.5cm]{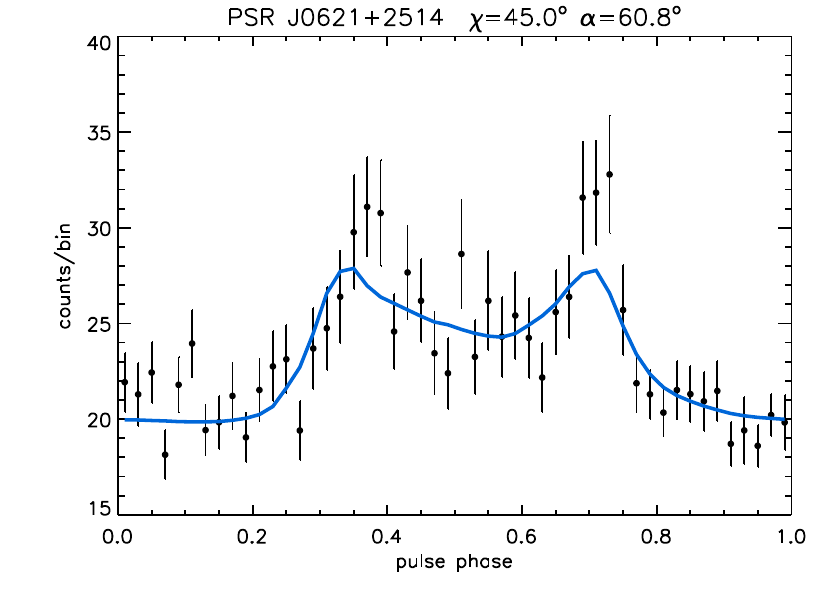}
\includegraphics[width=4.5cm]{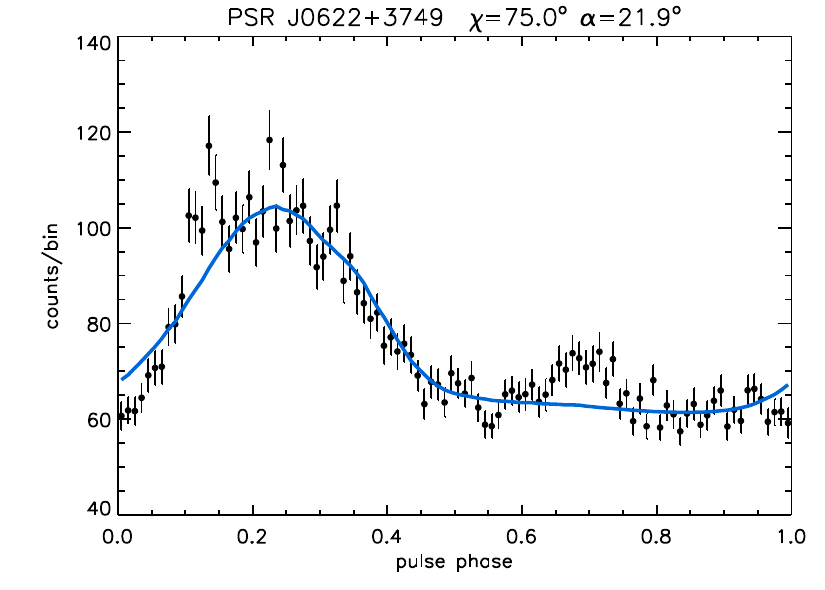}
\includegraphics[width=4.5cm]{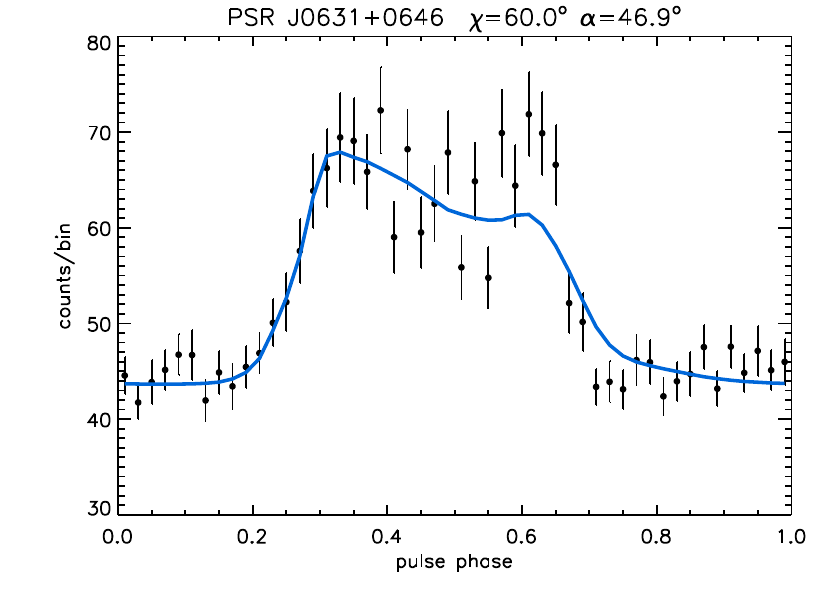}
\includegraphics[width=4.5cm]{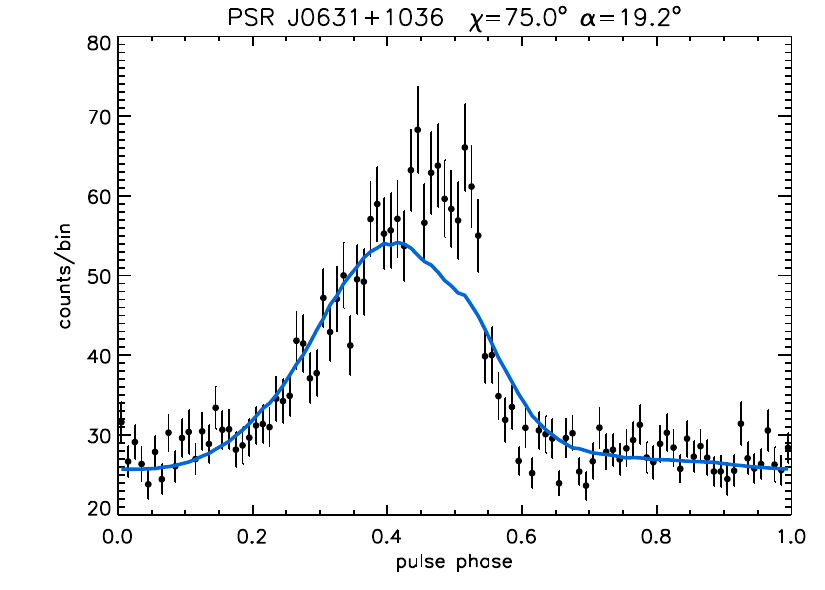}
\includegraphics[width=4.5cm]{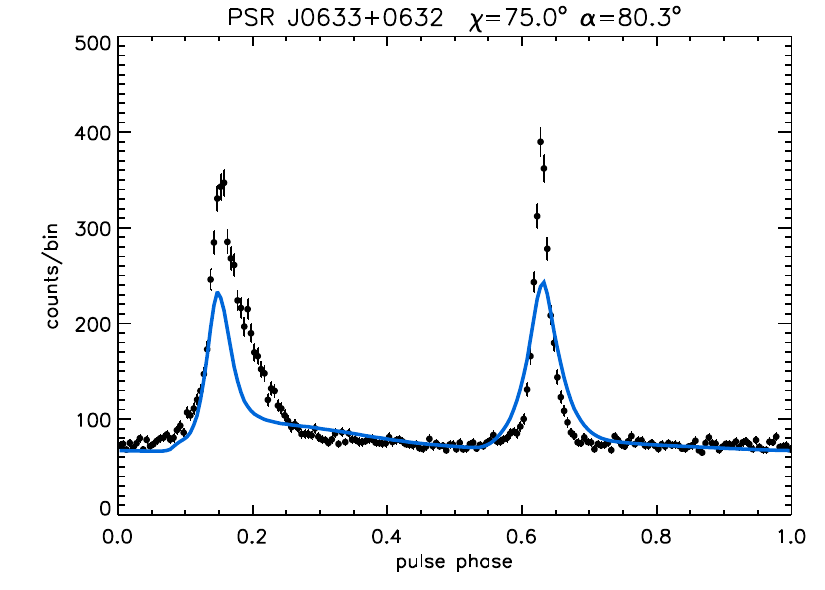}
\includegraphics[width=4.5cm]{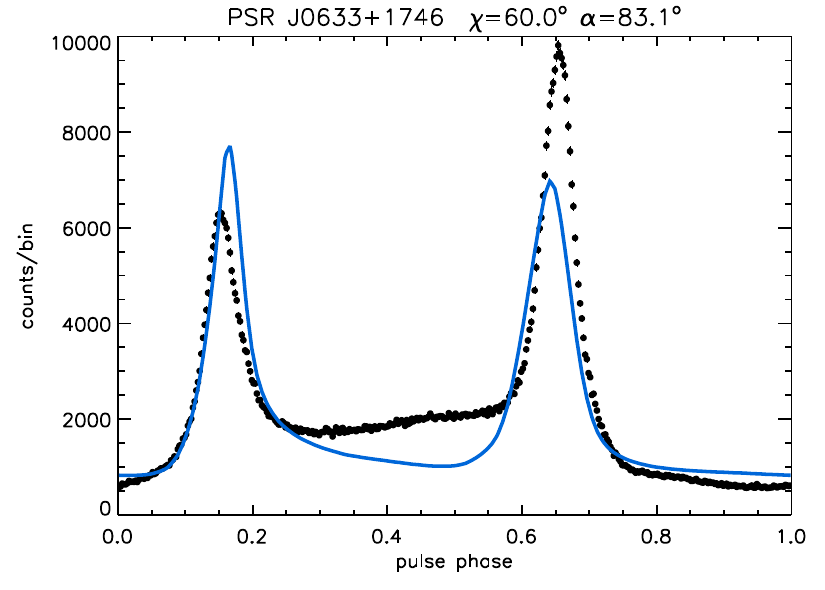}
\includegraphics[width=4.5cm]{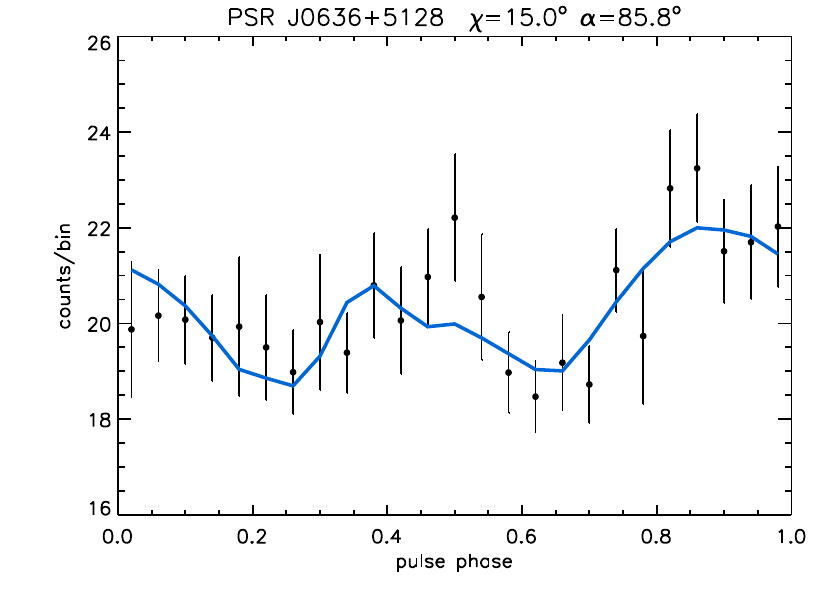}
\includegraphics[width=4.5cm]{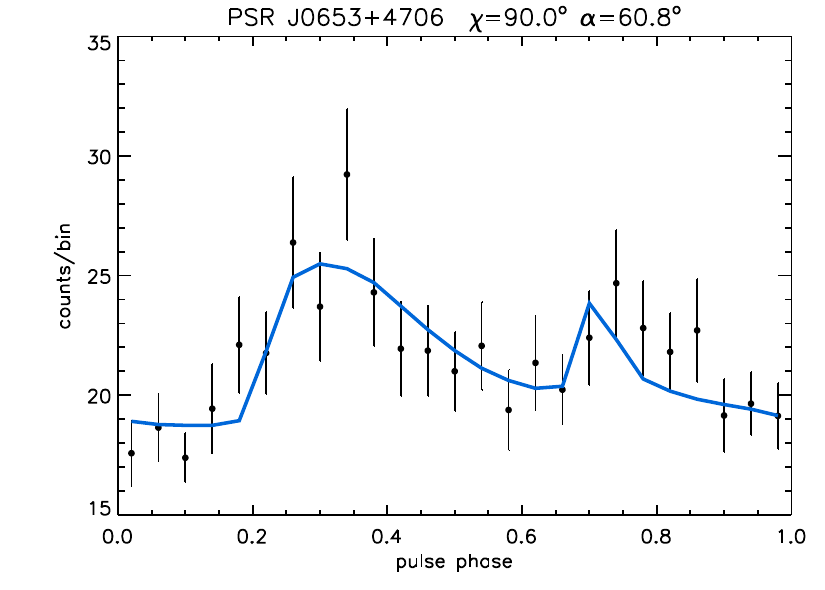}
\includegraphics[width=4.5cm]{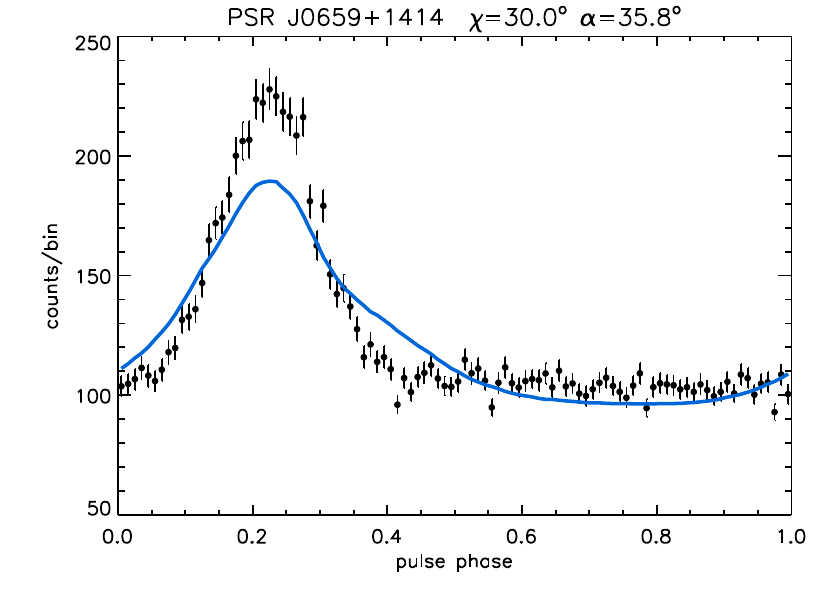}
\includegraphics[width=4.5cm]{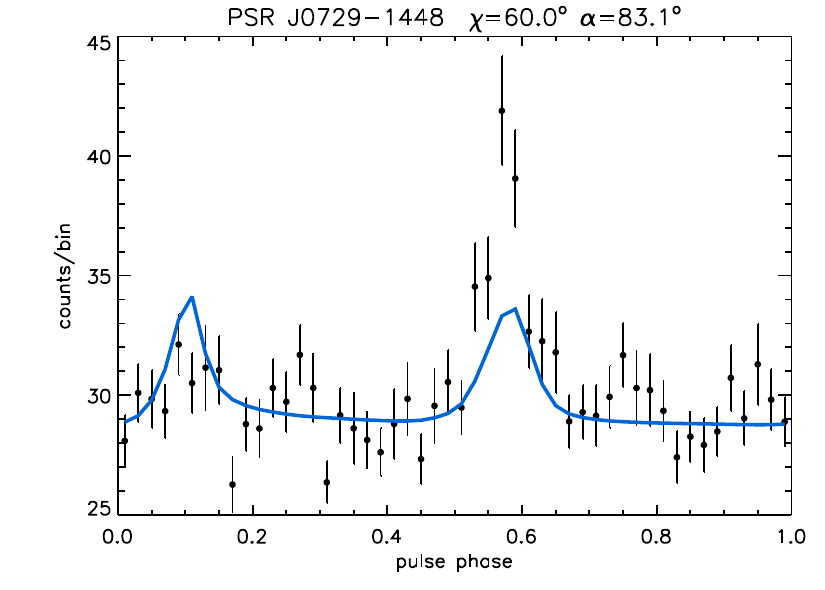}
\includegraphics[width=4.5cm]{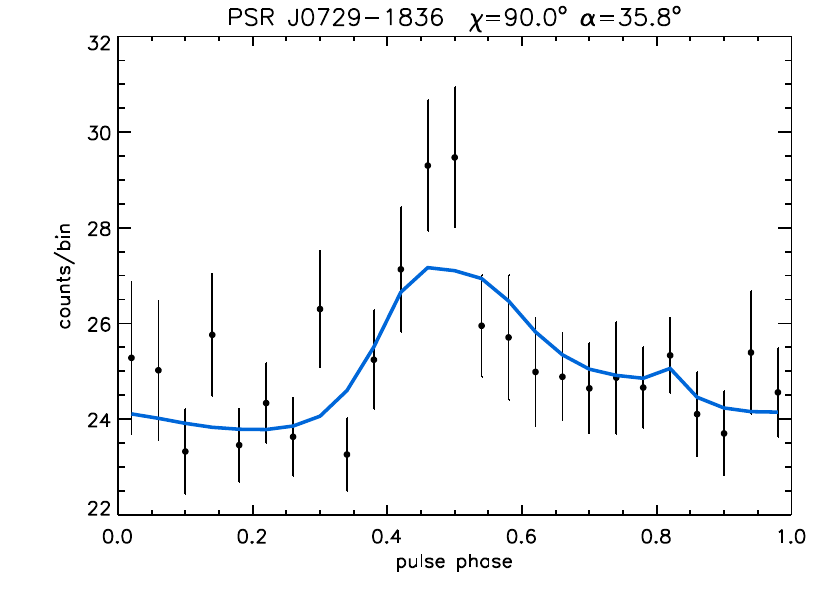}
\includegraphics[width=4.5cm]{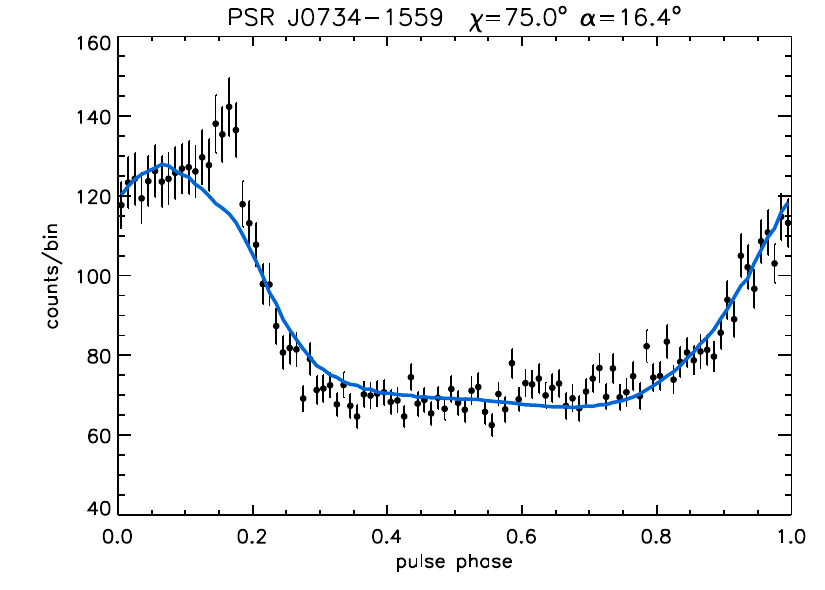}
\includegraphics[width=4.5cm]{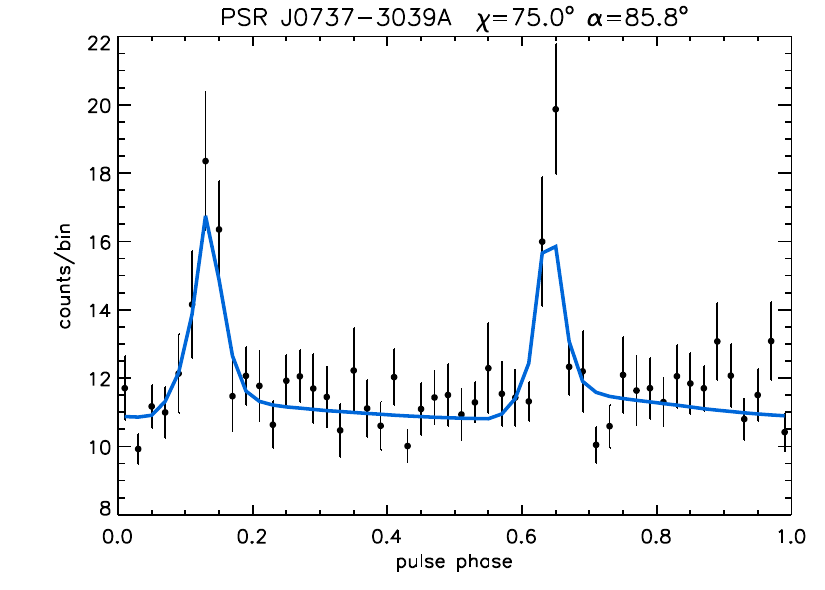}
\includegraphics[width=4.5cm]{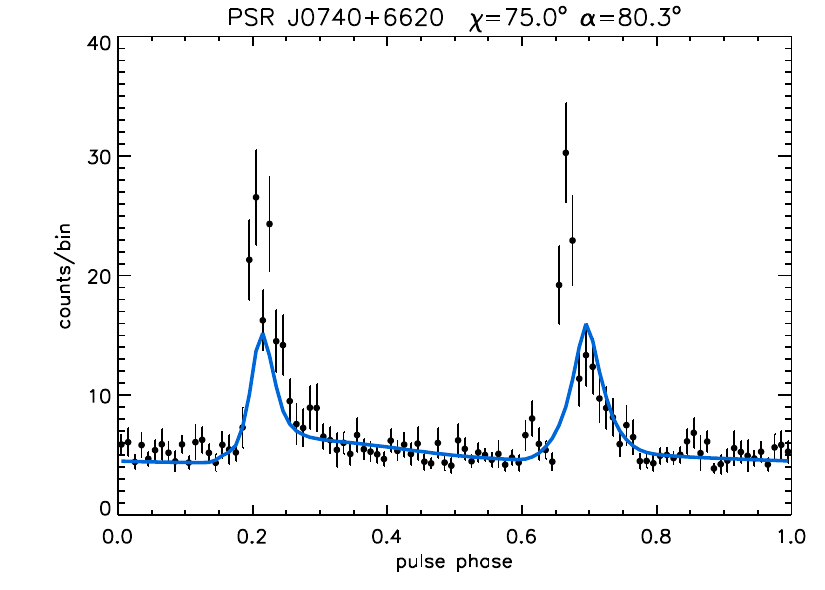}
\includegraphics[width=4.5cm]{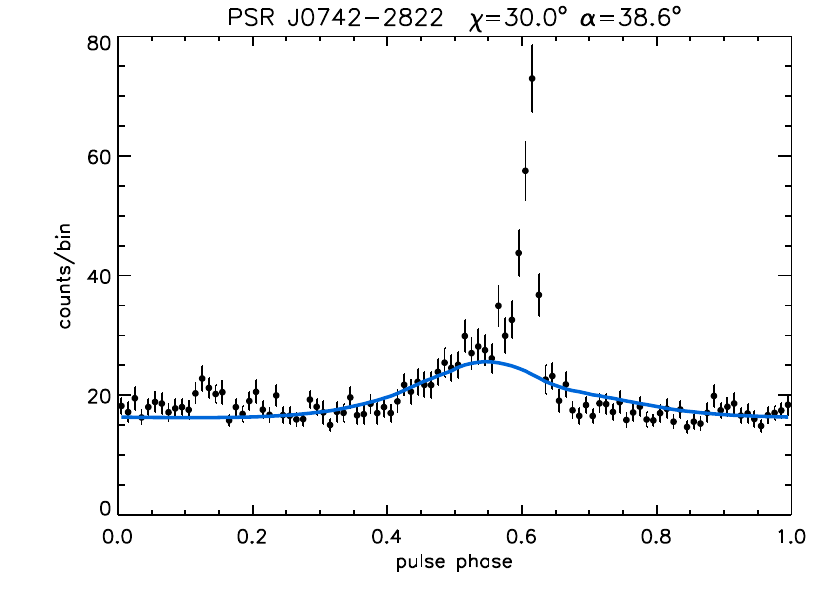}
\includegraphics[width=4.5cm]{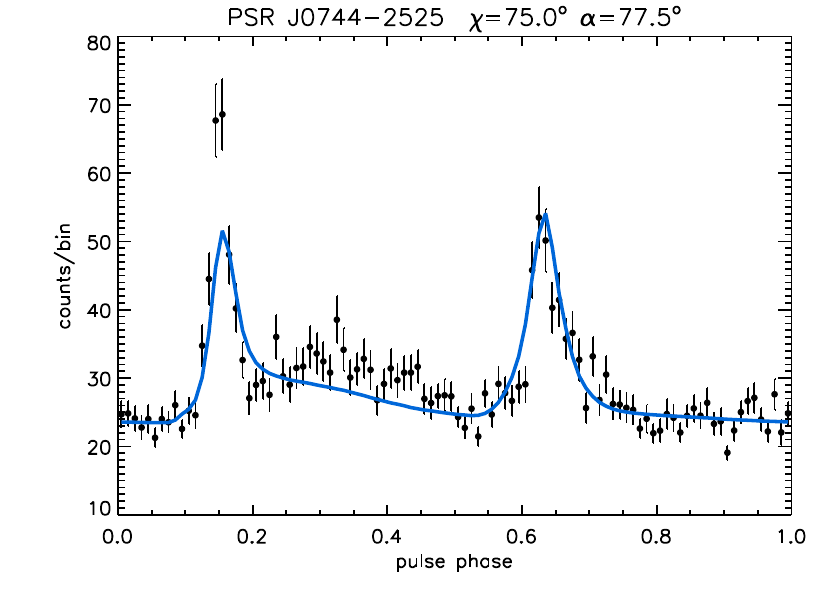}
\includegraphics[width=4.5cm]{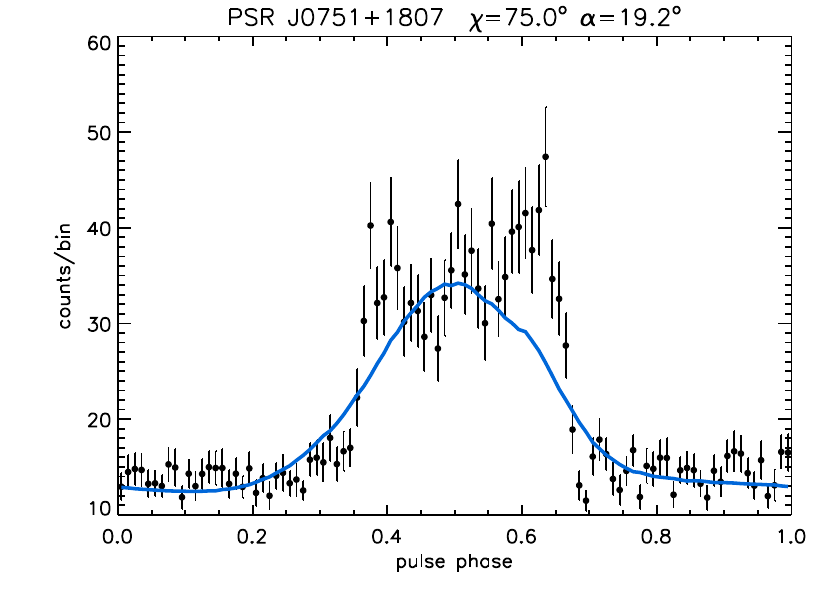}
\includegraphics[width=4.5cm]{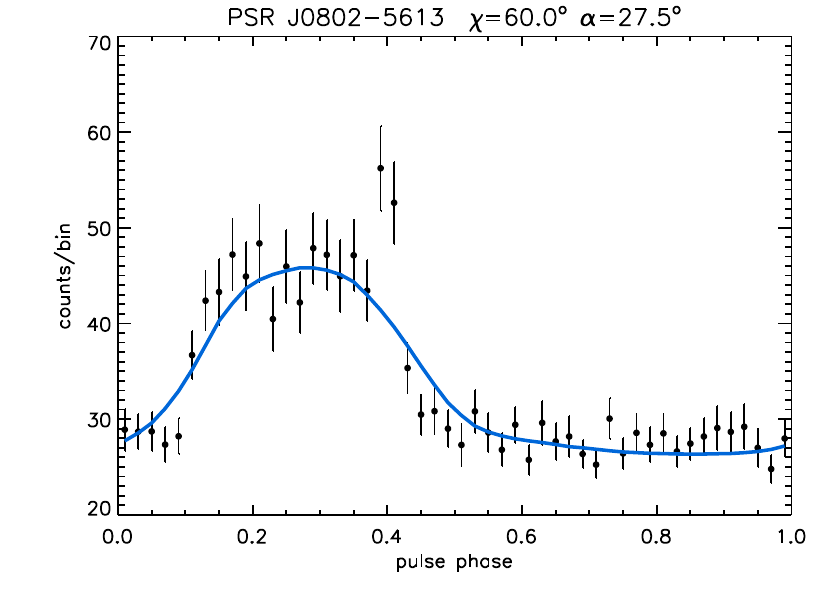}
\includegraphics[width=4.5cm]{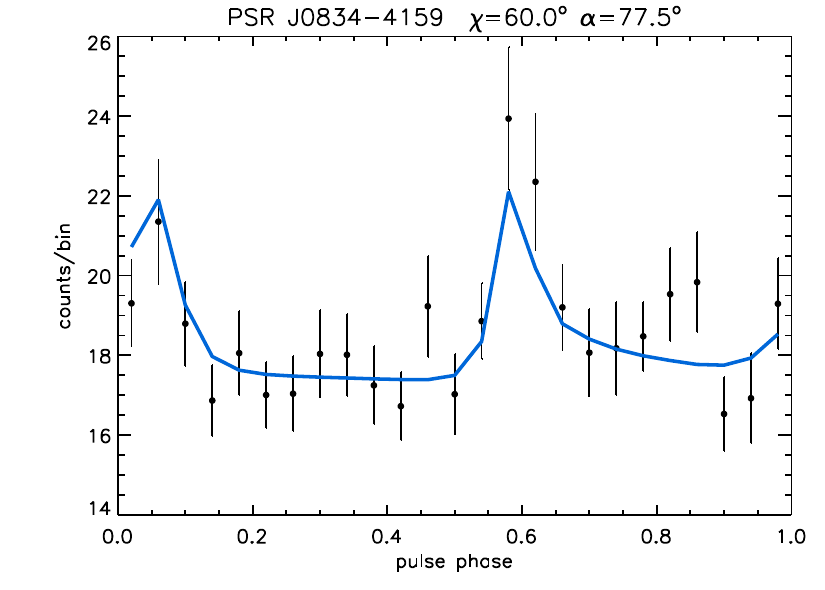}
\caption{continued.}
\end{figure*}

\begin{figure*}
\addtocounter{figure}{-1}
\centering
\includegraphics[width=4.5cm]{J0835-4510.pdf}
\includegraphics[width=4.5cm]{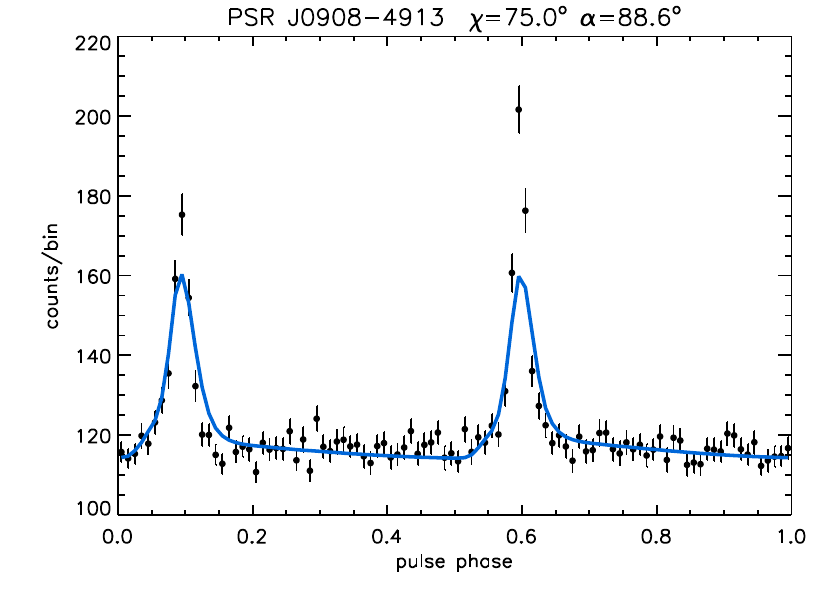}
\includegraphics[width=4.5cm]{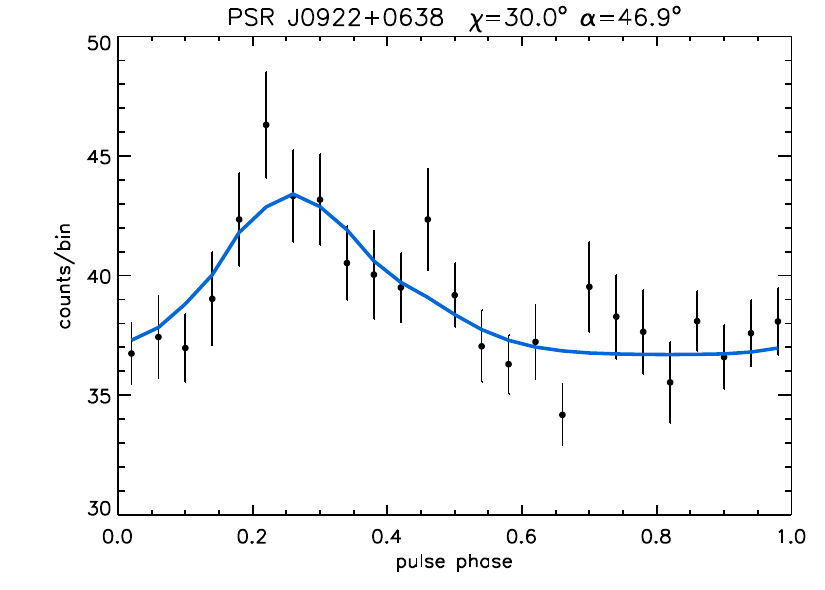}
\includegraphics[width=4.5cm]{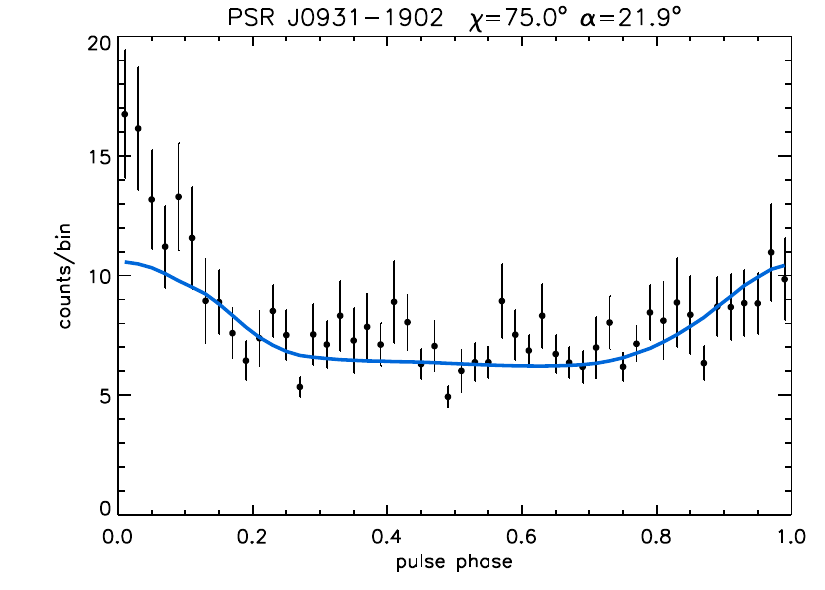}
\includegraphics[width=4.5cm]{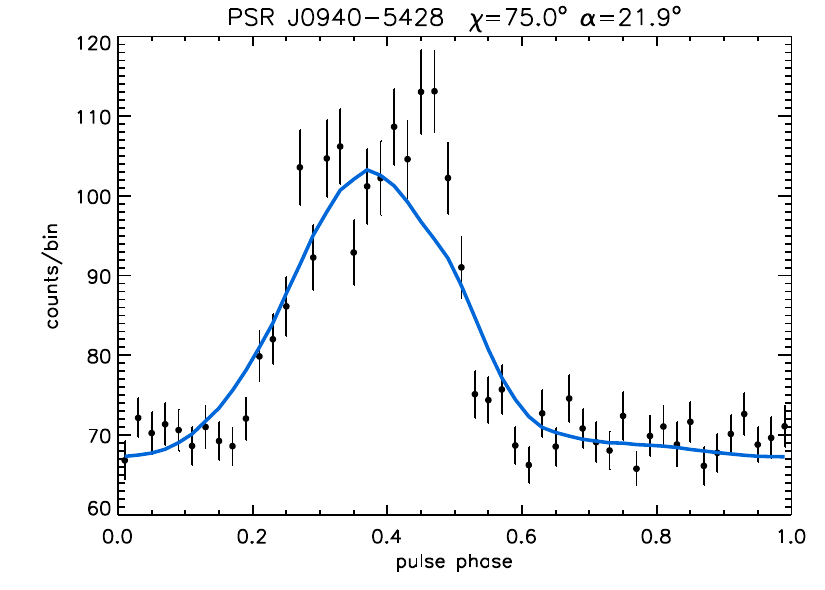}
\includegraphics[width=4.5cm]{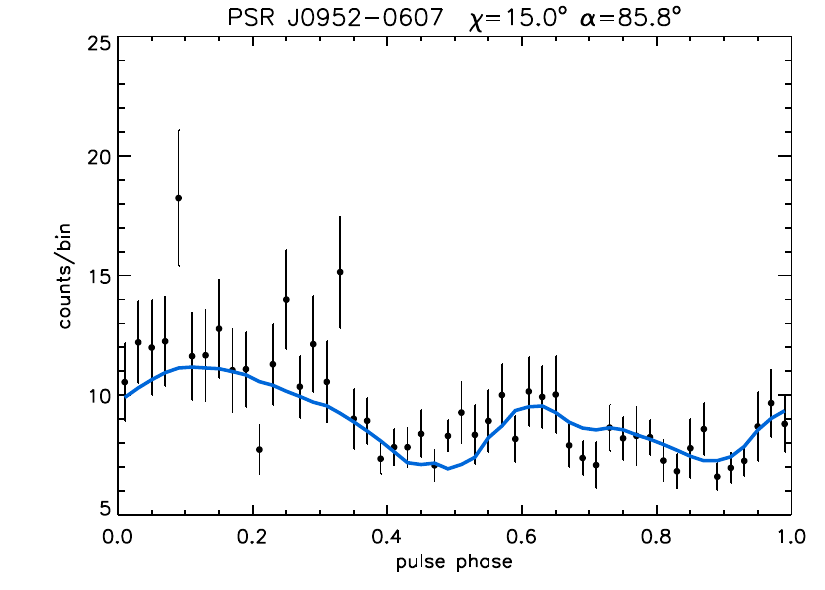}
\includegraphics[width=4.5cm]{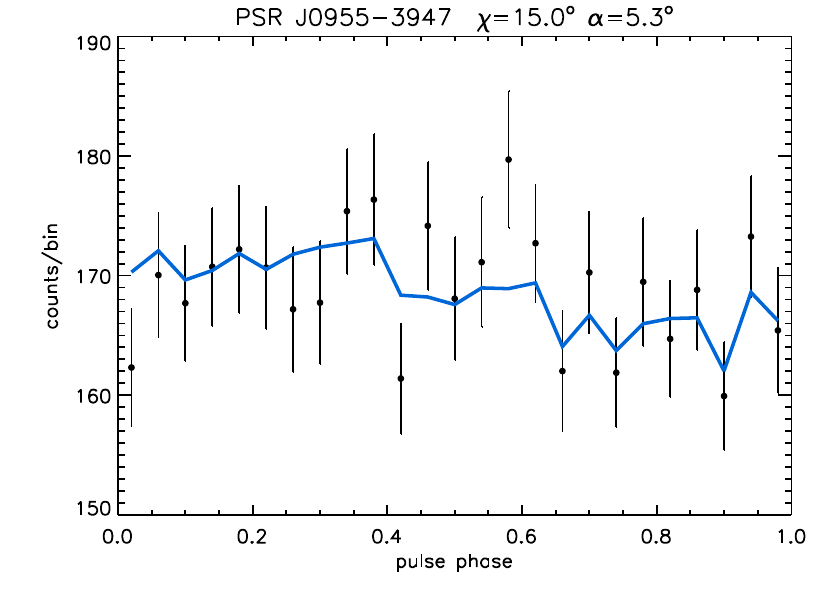}
\includegraphics[width=4.5cm]{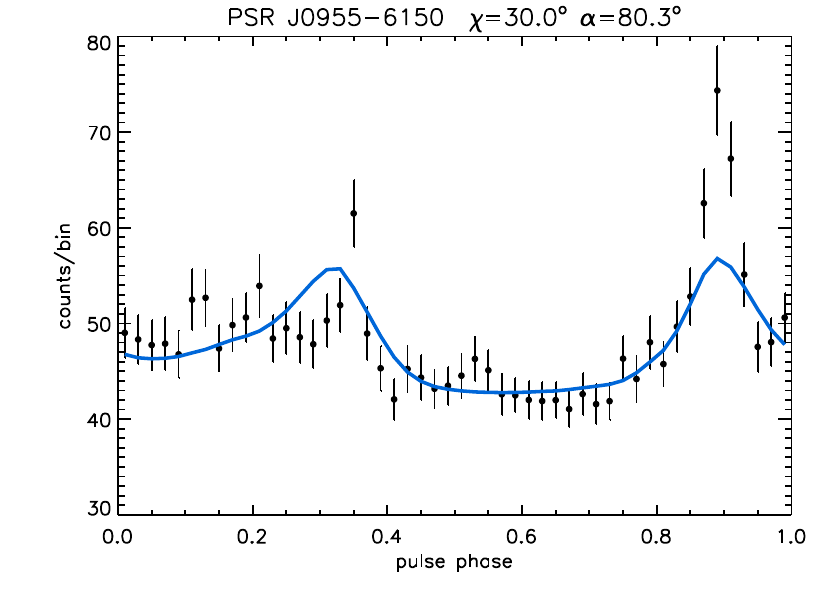}
\includegraphics[width=4.5cm]{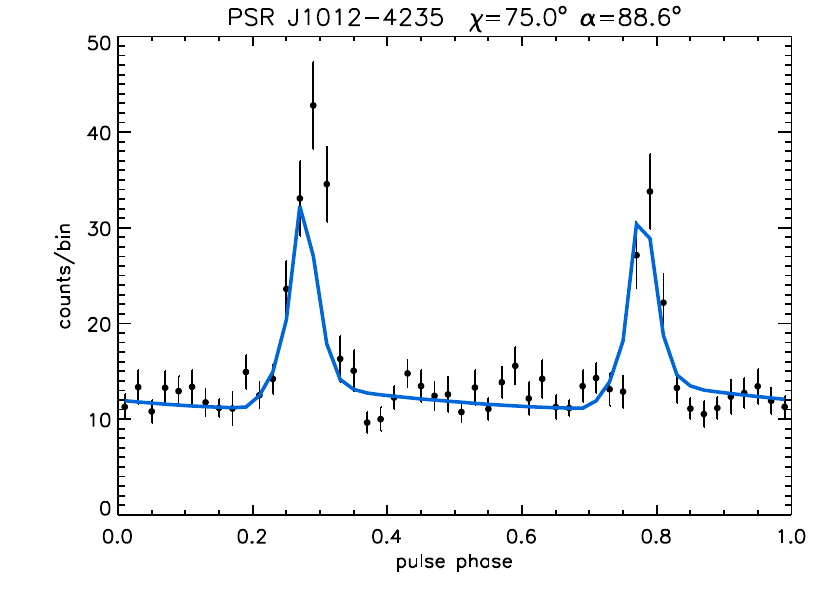}
\includegraphics[width=4.5cm]{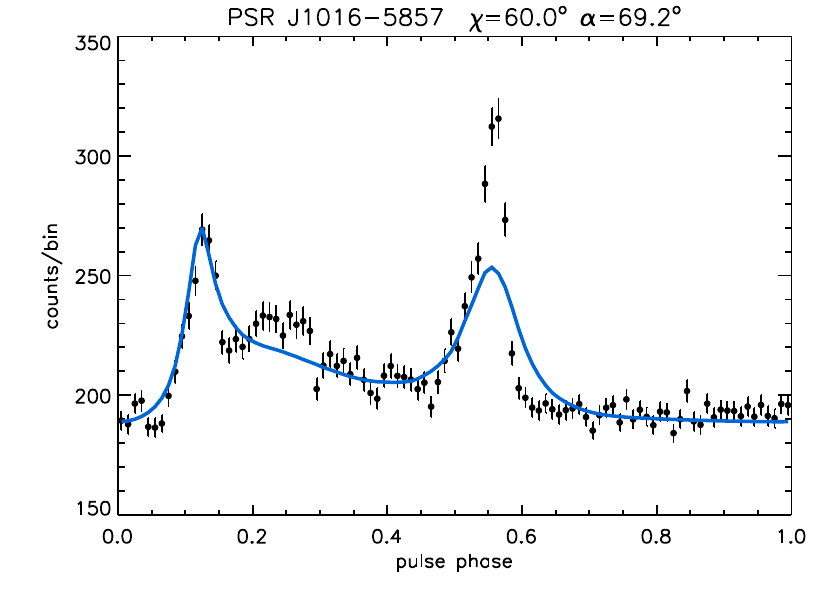}
\includegraphics[width=4.5cm]{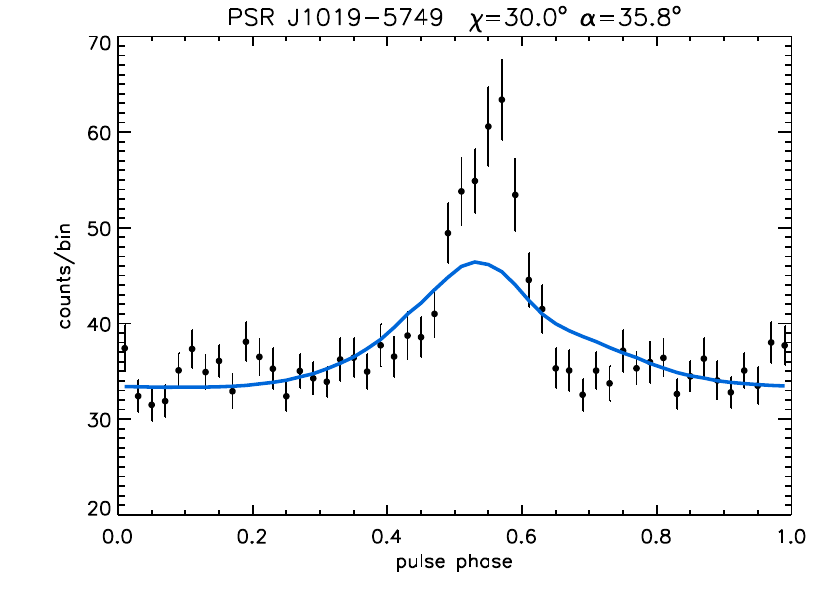}
\includegraphics[width=4.5cm]{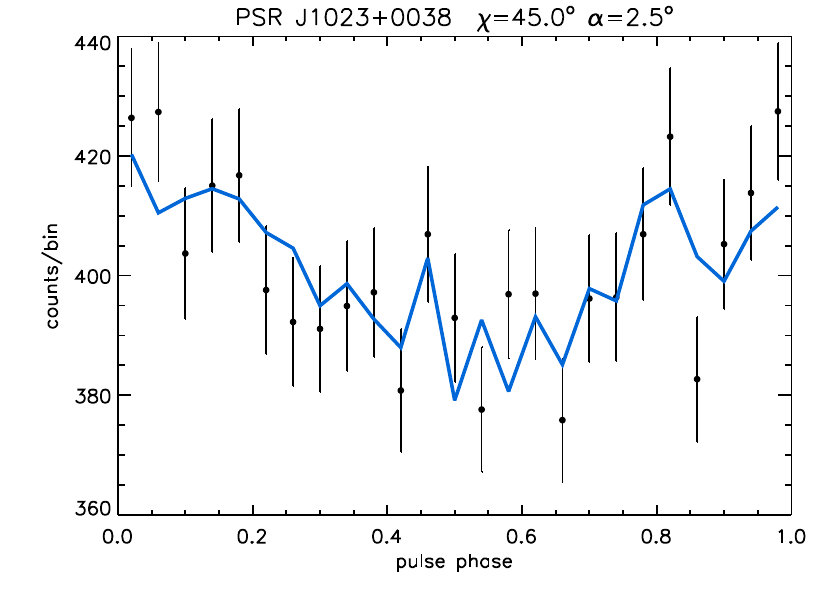}
\includegraphics[width=4.5cm]{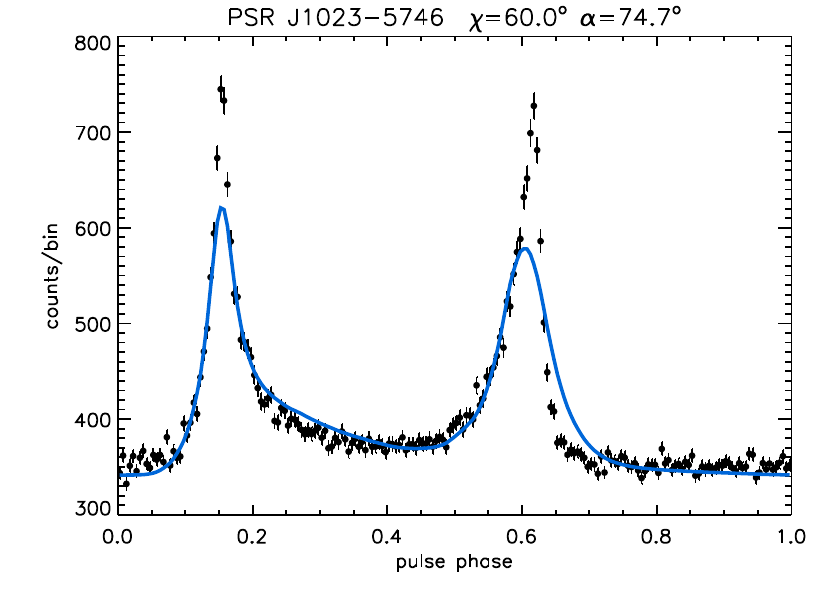}
\includegraphics[width=4.5cm]{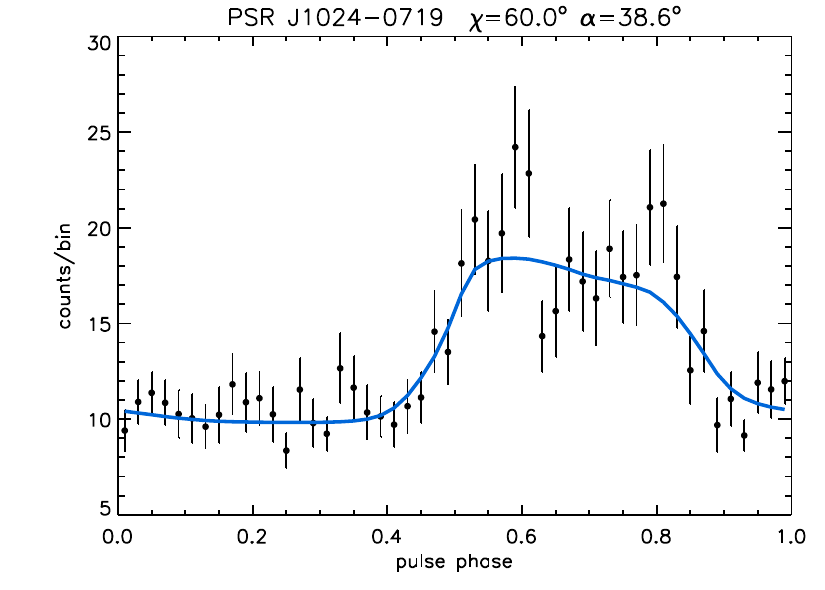}
\includegraphics[width=4.5cm]{J1028-5819.pdf}
\includegraphics[width=4.5cm]{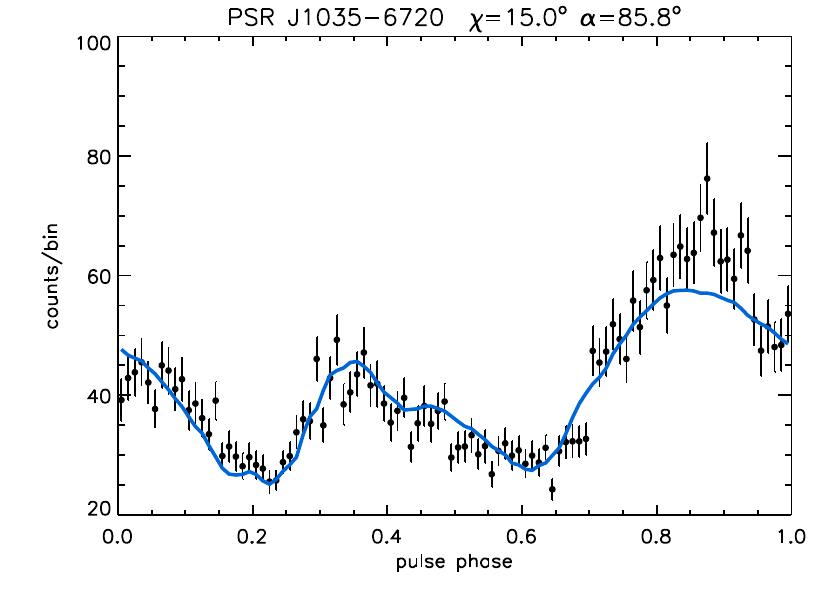}
\includegraphics[width=4.5cm]{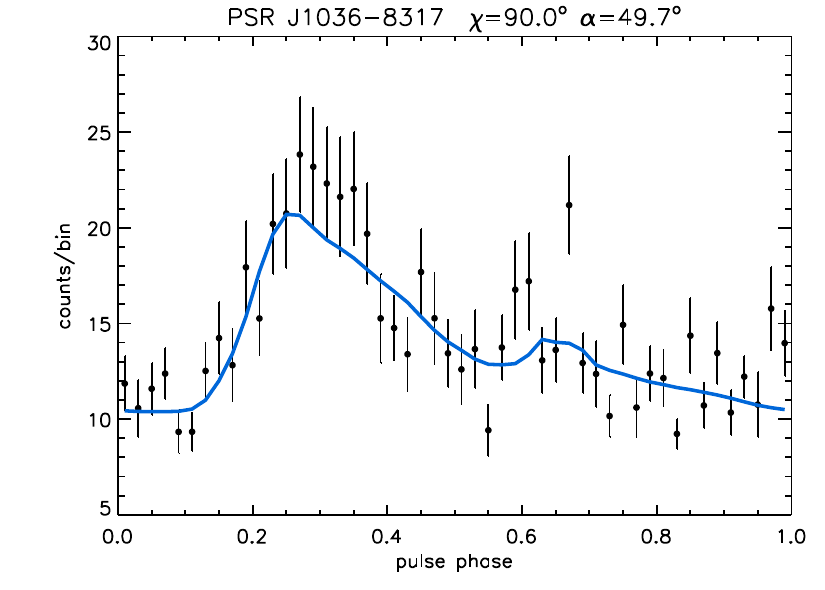}
\includegraphics[width=4.5cm]{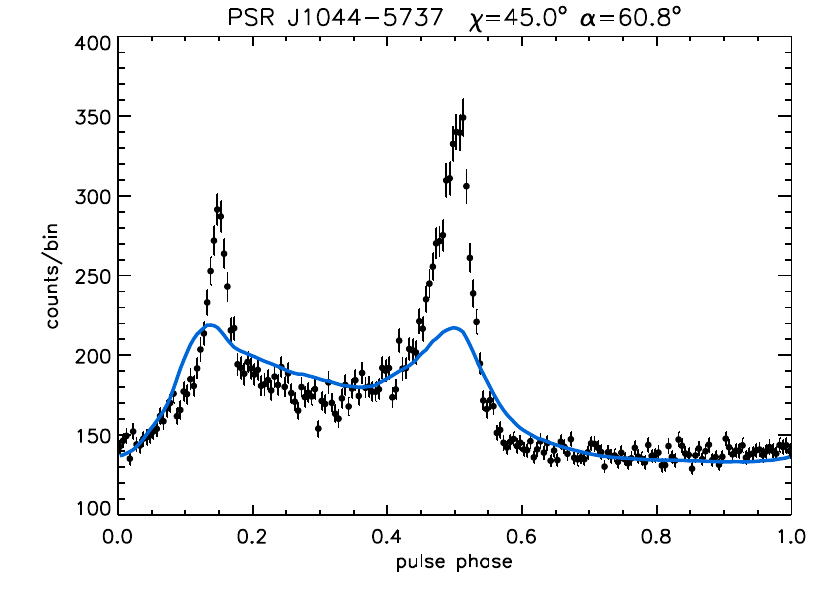}
\includegraphics[width=4.5cm]{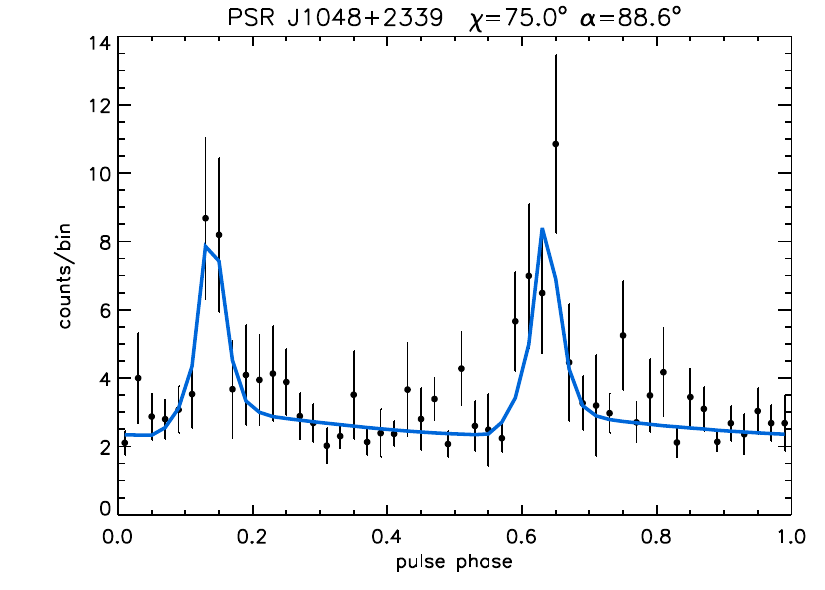}
\includegraphics[width=4.5cm]{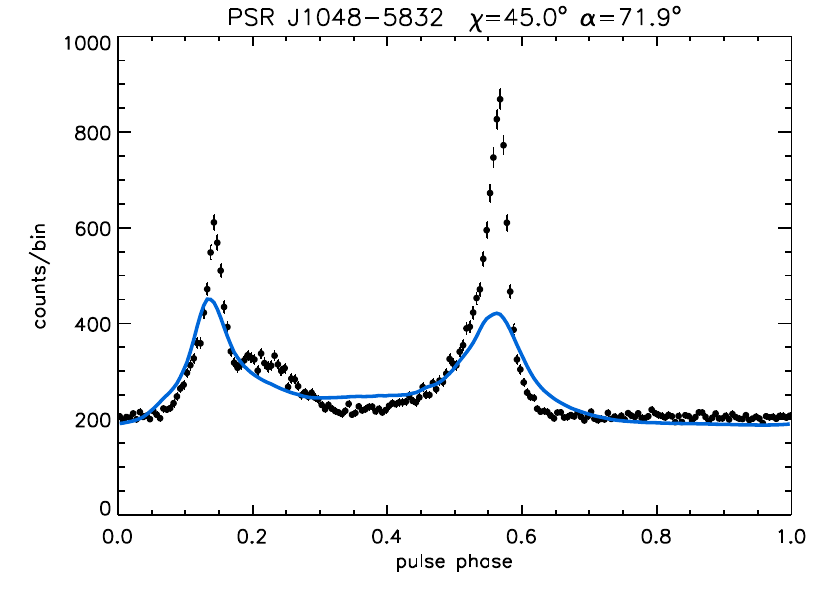}
\includegraphics[width=4.5cm]{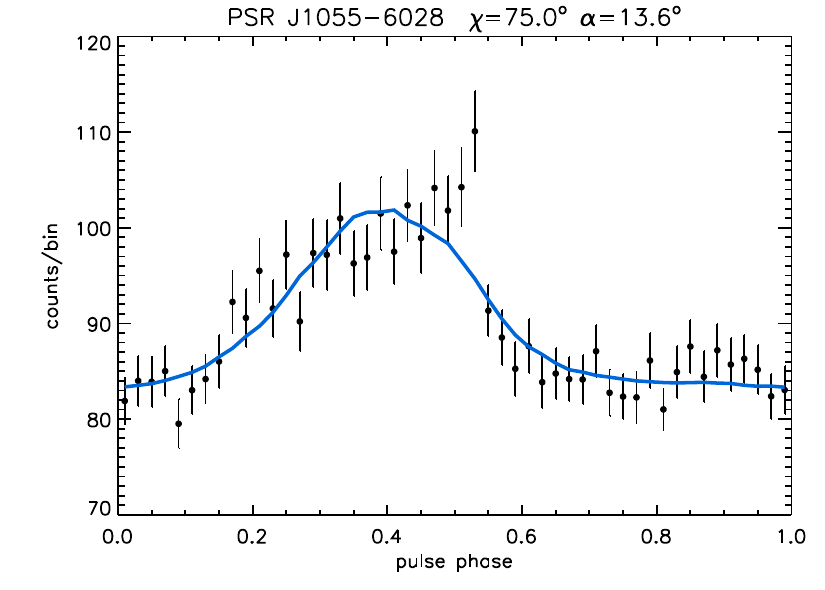}
\includegraphics[width=4.5cm]{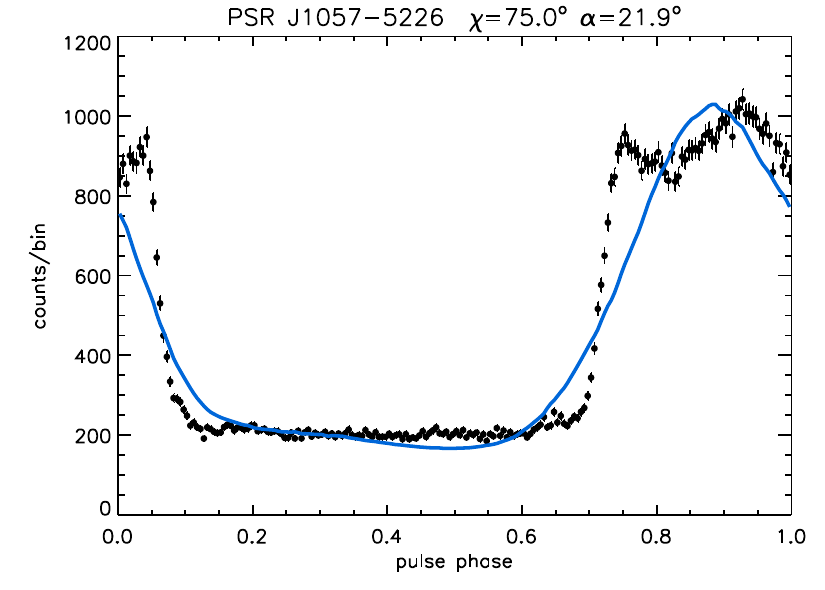}
\includegraphics[width=4.5cm]{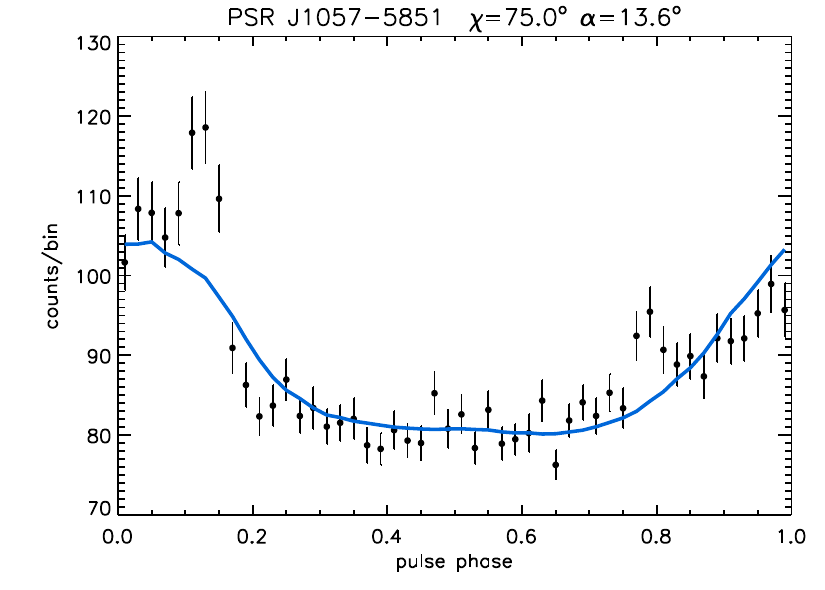}
\includegraphics[width=4.5cm]{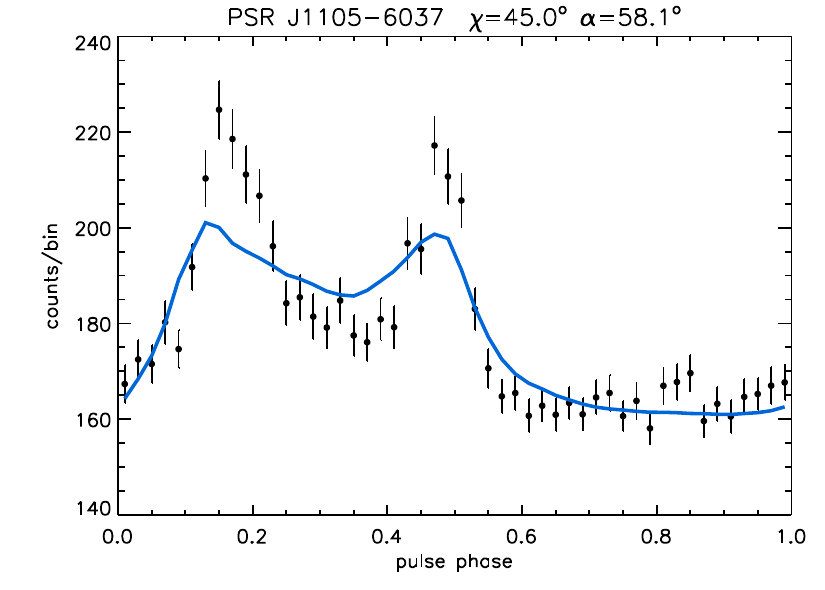}
\includegraphics[width=4.5cm]{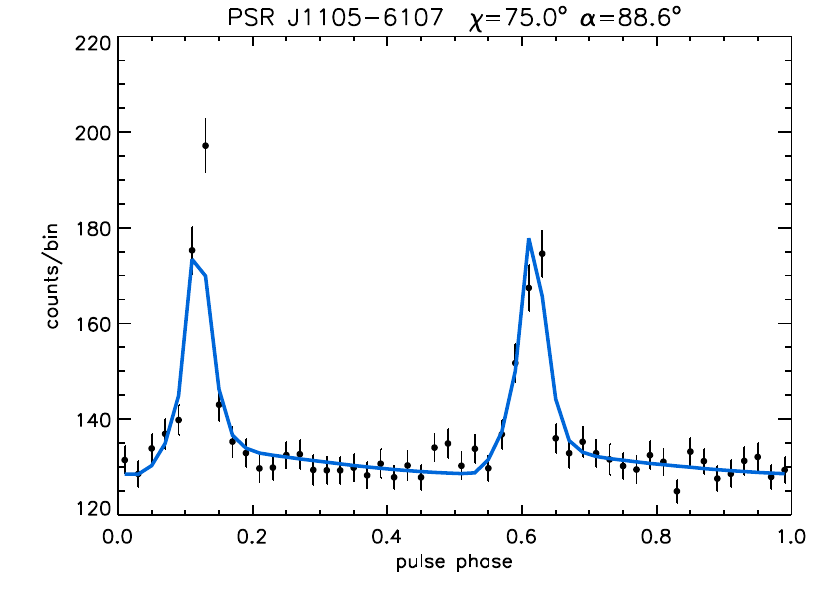}
\includegraphics[width=4.5cm]{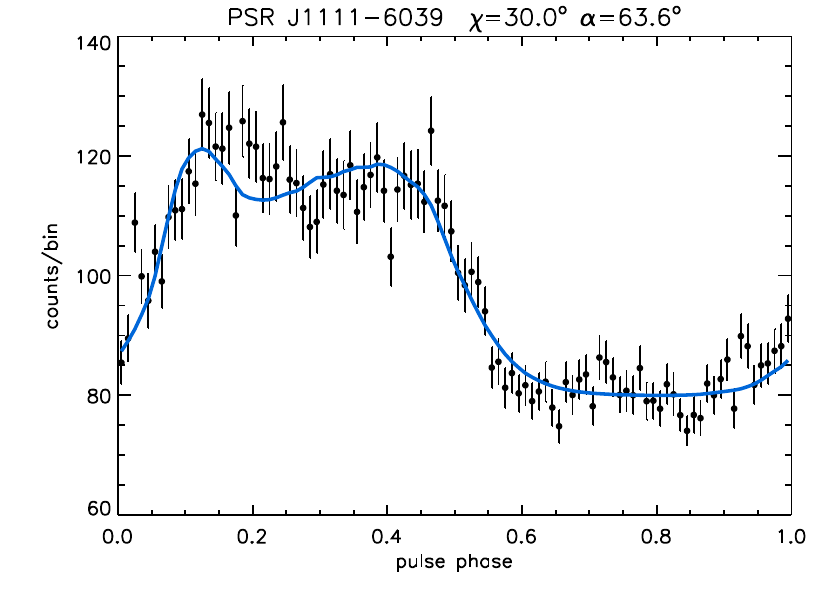}
\includegraphics[width=4.5cm]{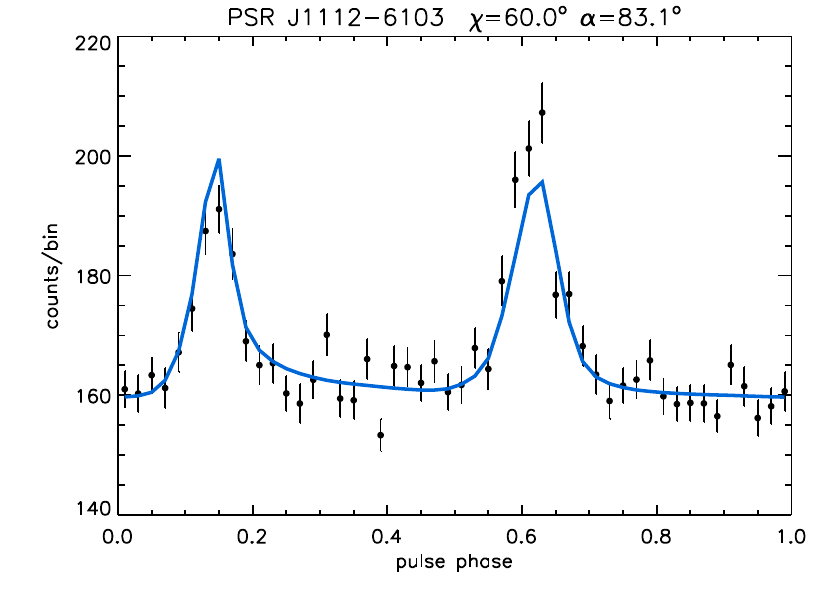}
\includegraphics[width=4.5cm]{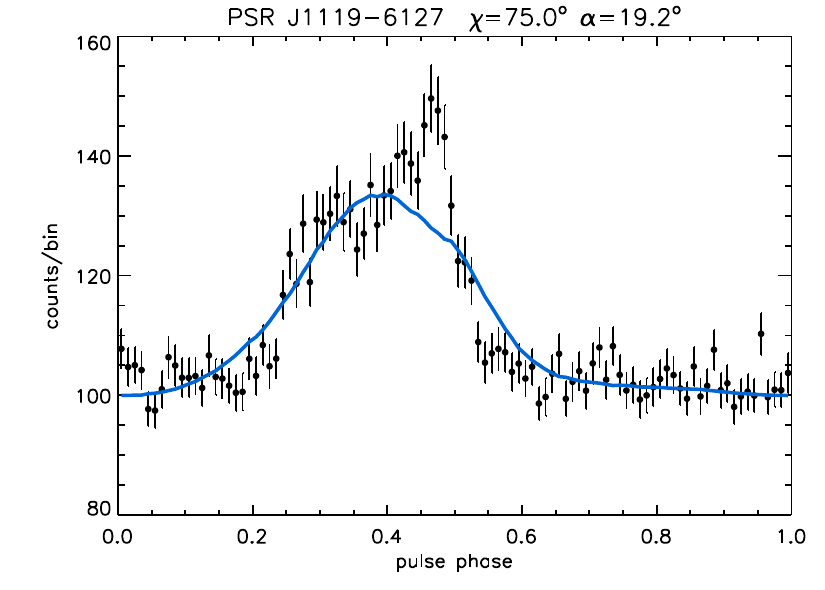}
\caption{continued.}
\end{figure*}

\begin{figure*}
\addtocounter{figure}{-1}
\centering
\includegraphics[width=4.5cm]{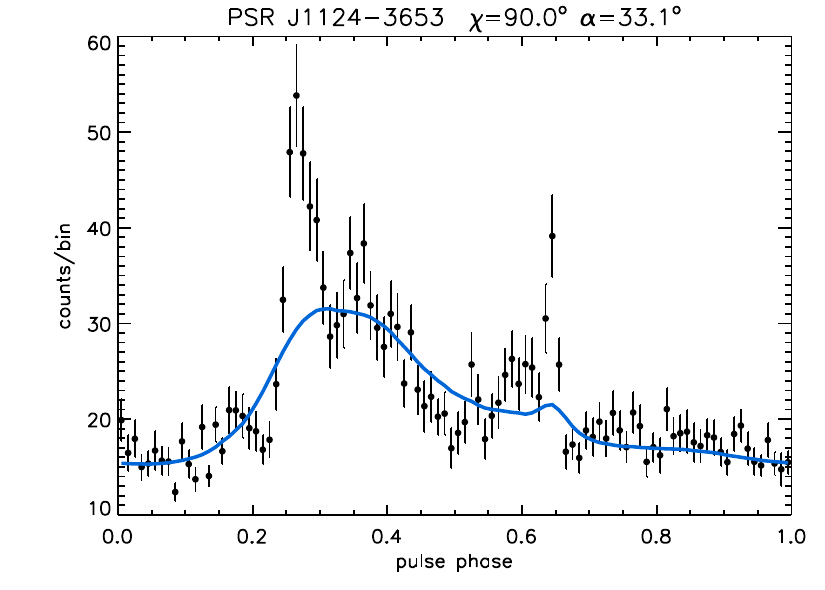}
\includegraphics[width=4.5cm]{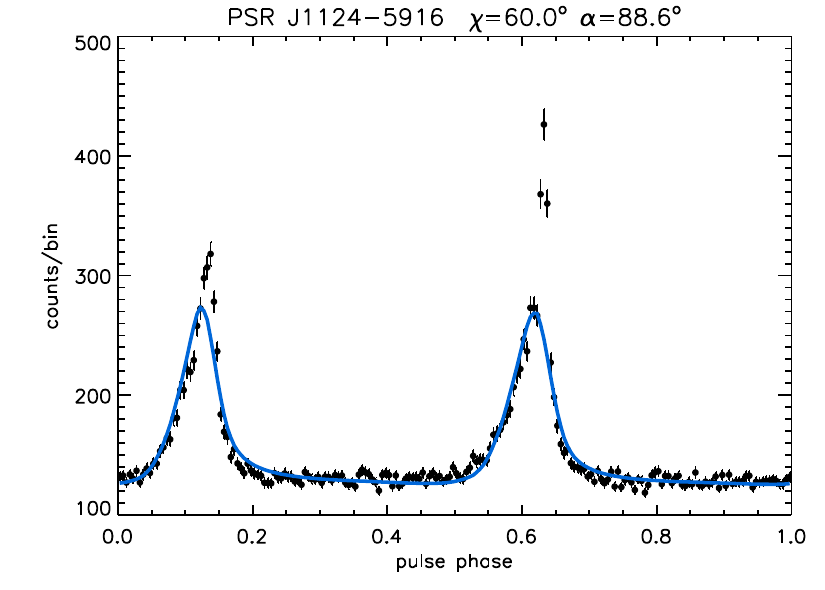}
\includegraphics[width=4.5cm]{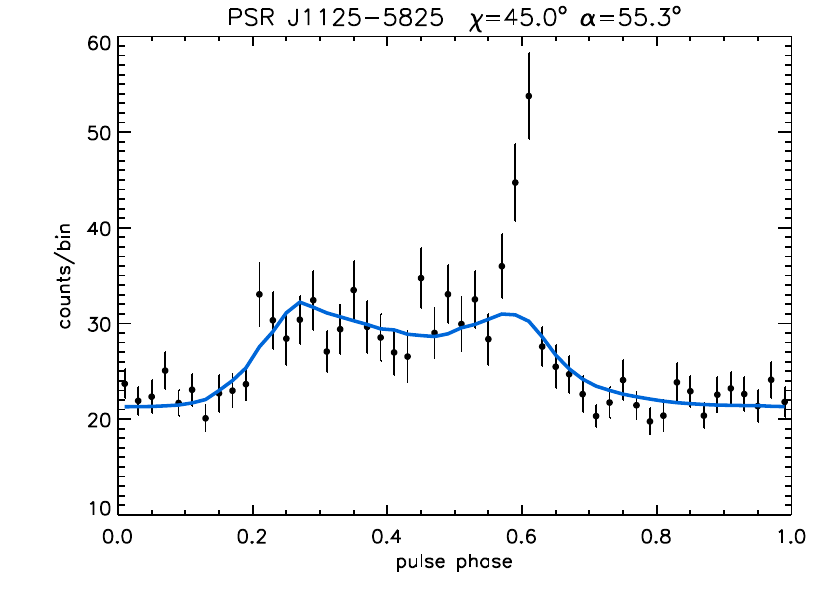}
\includegraphics[width=4.5cm]{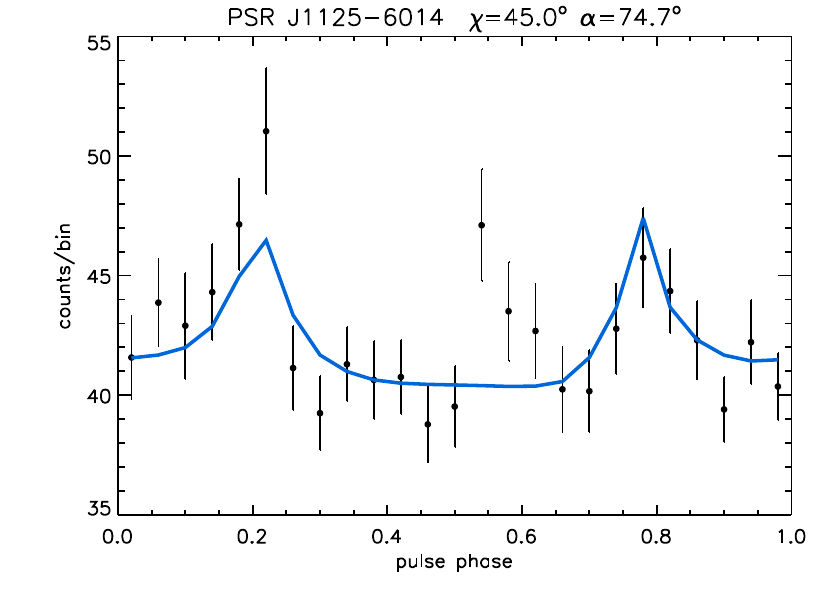}
\includegraphics[width=4.5cm]{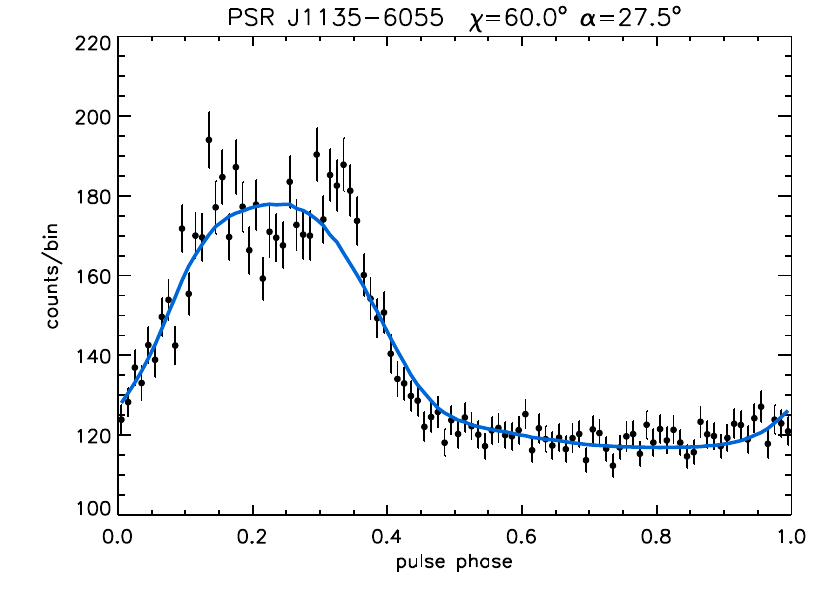}
\includegraphics[width=4.5cm]{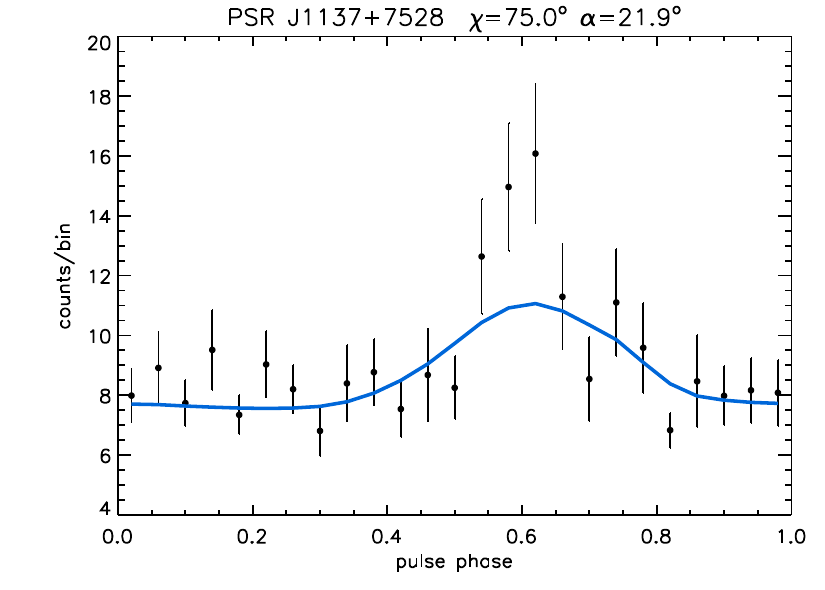}
\includegraphics[width=4.5cm]{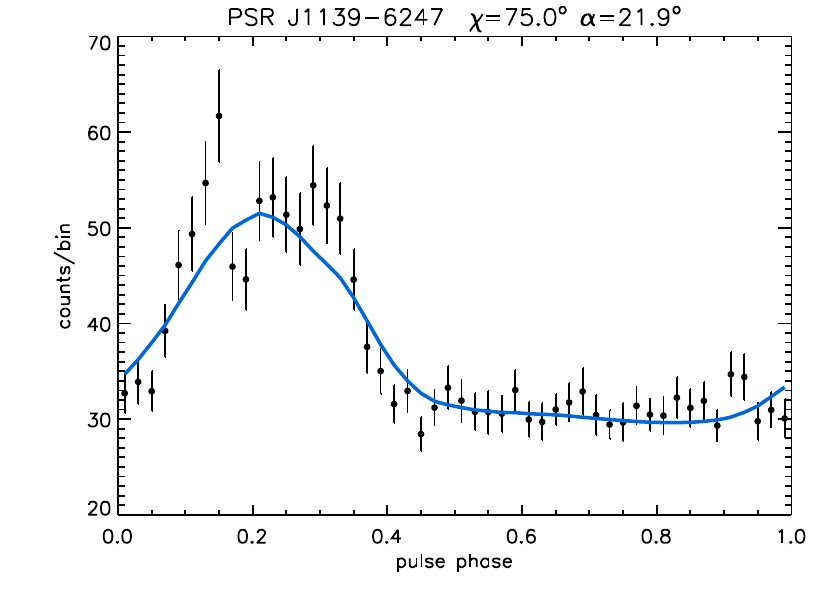}
\includegraphics[width=4.5cm]{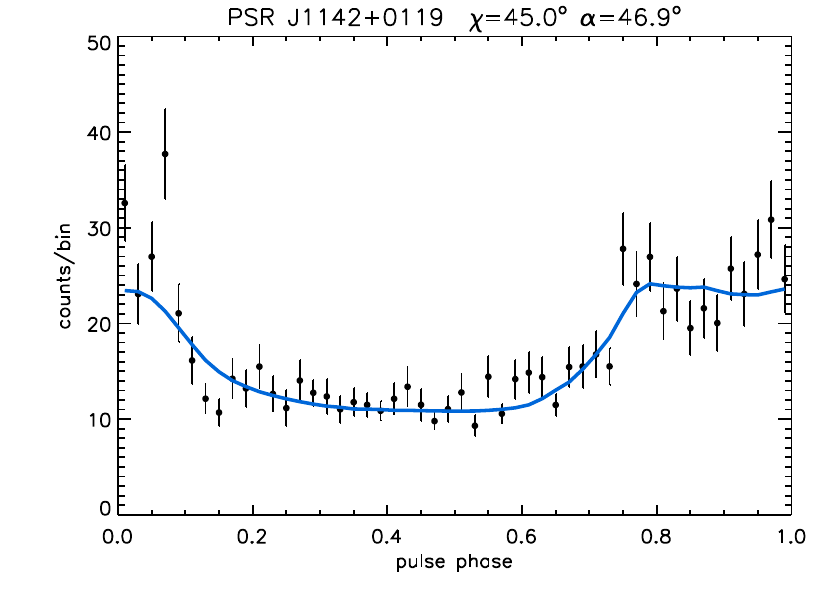}
\includegraphics[width=4.5cm]{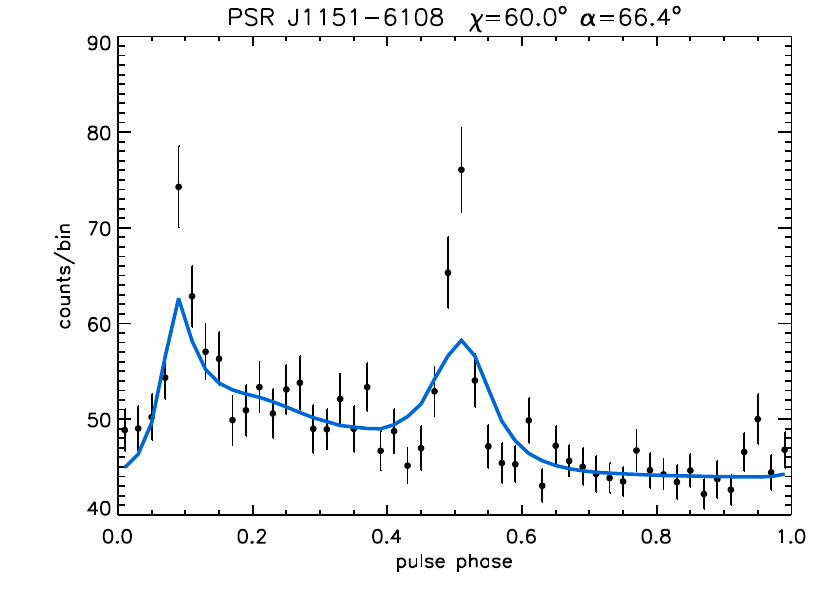}
\includegraphics[width=4.5cm]{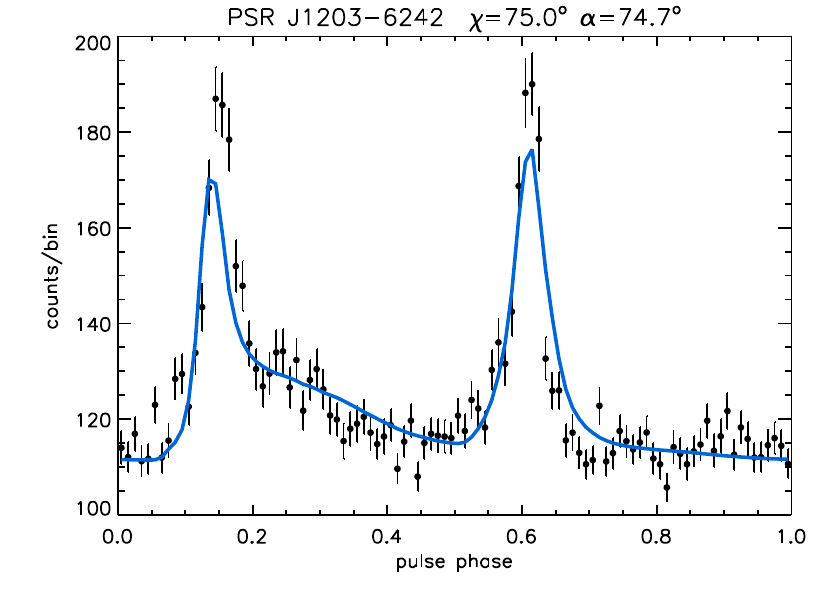}
\includegraphics[width=4.5cm]{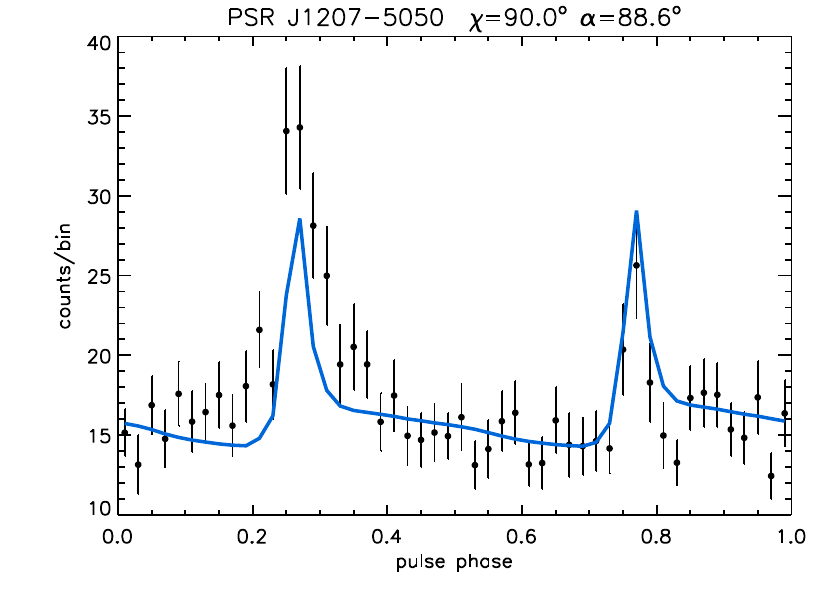}
\includegraphics[width=4.5cm]{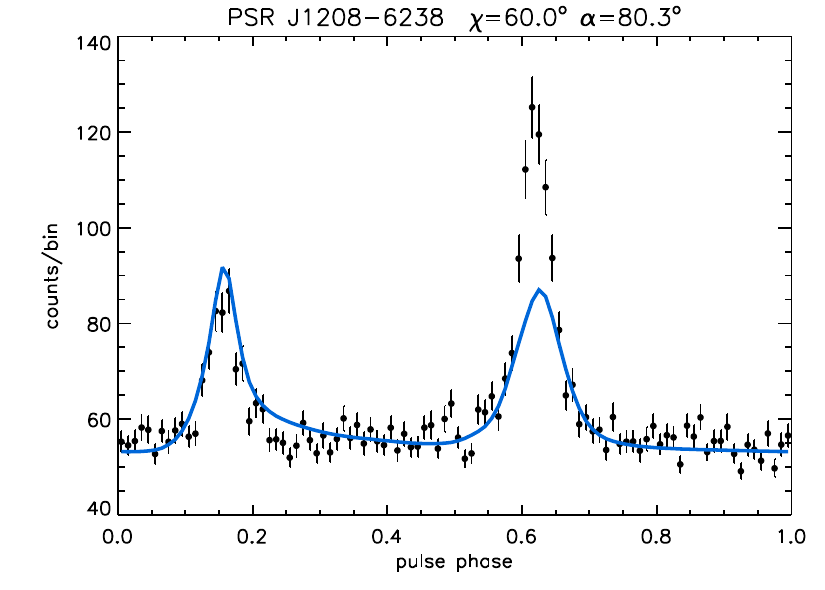}
\includegraphics[width=4.5cm]{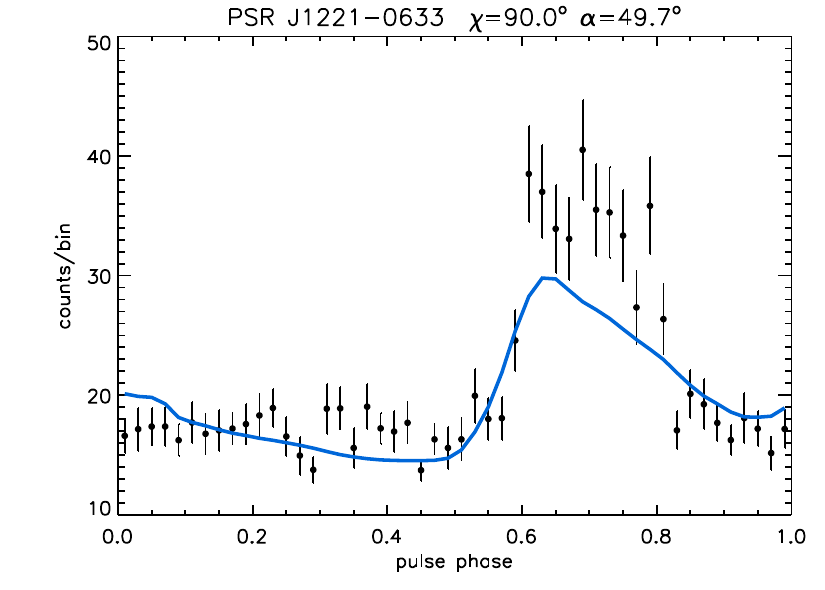}
\includegraphics[width=4.5cm]{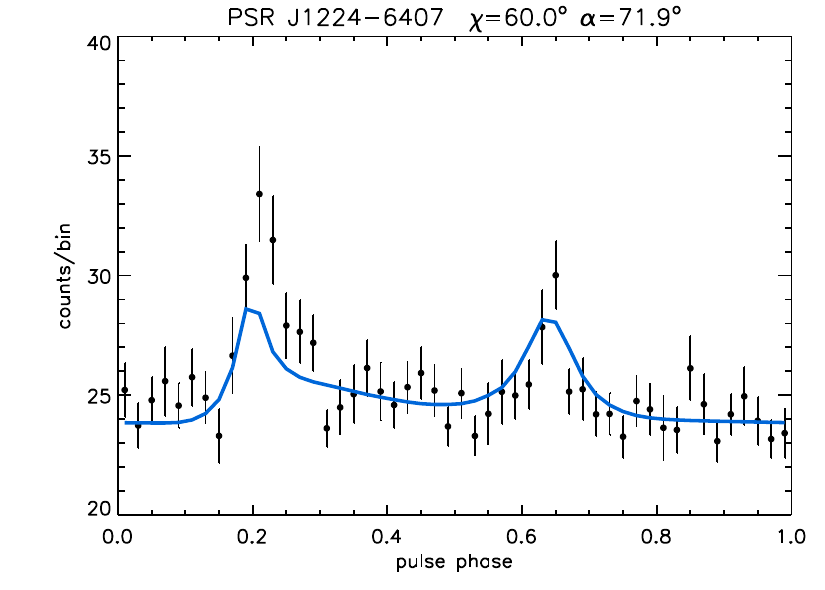}
\includegraphics[width=4.5cm]{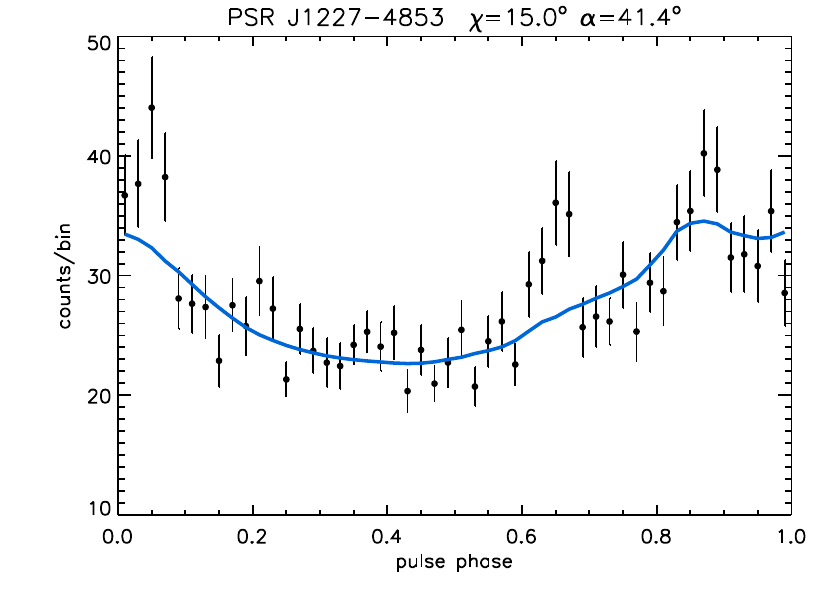}
\includegraphics[width=4.5cm]{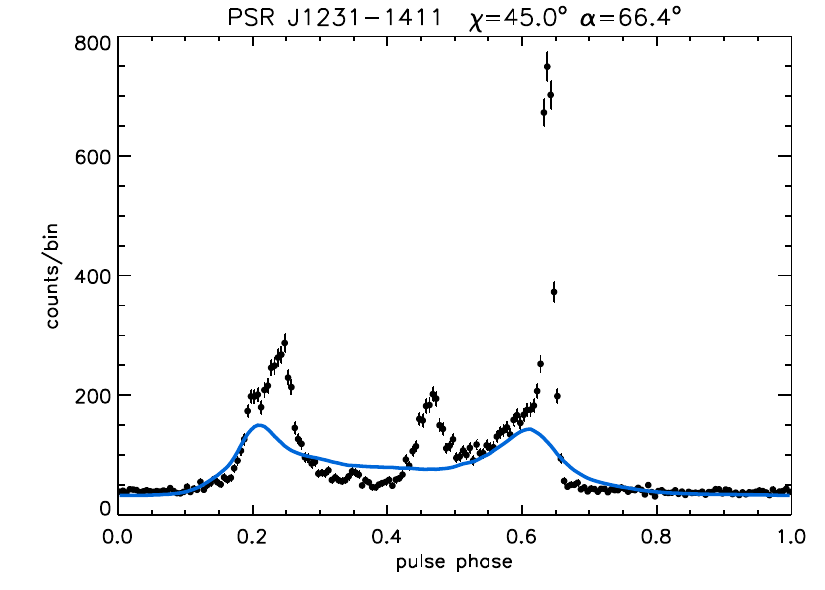}
\includegraphics[width=4.5cm]{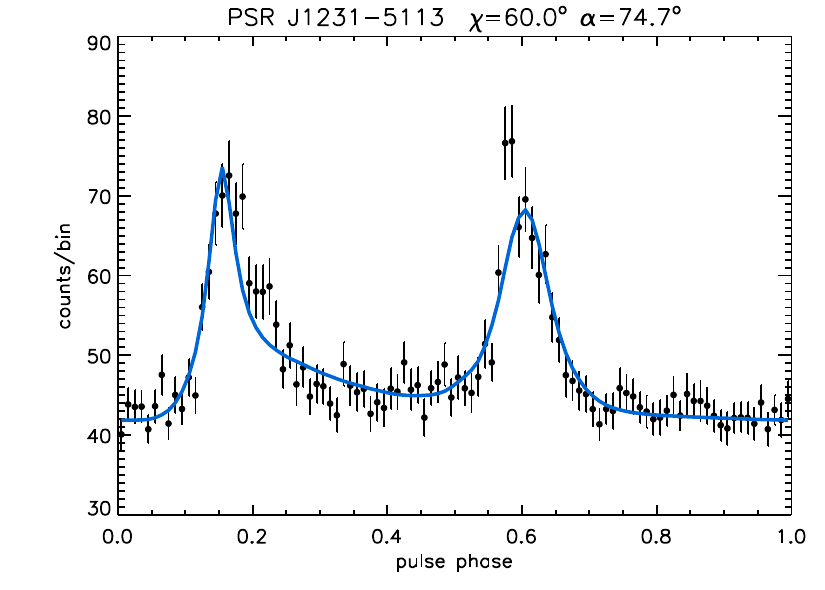}
\includegraphics[width=4.5cm]{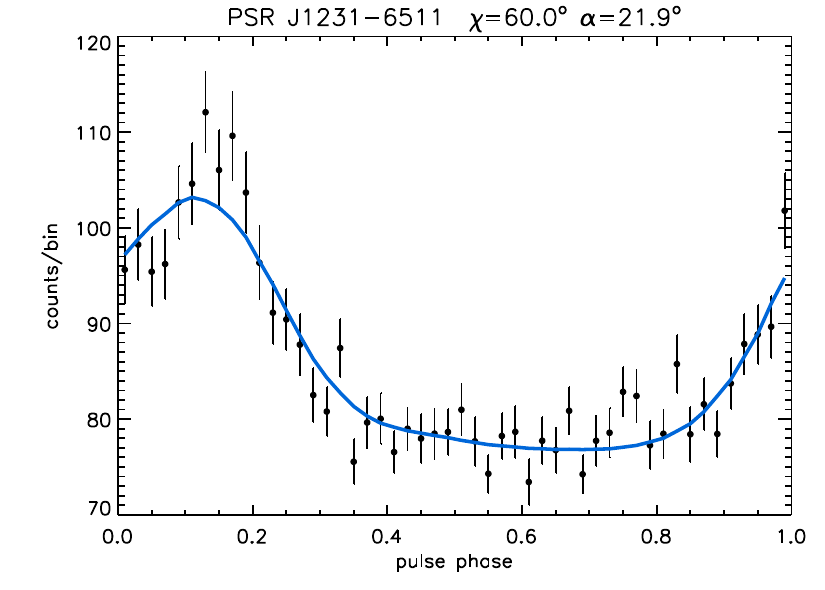}
\includegraphics[width=4.5cm]{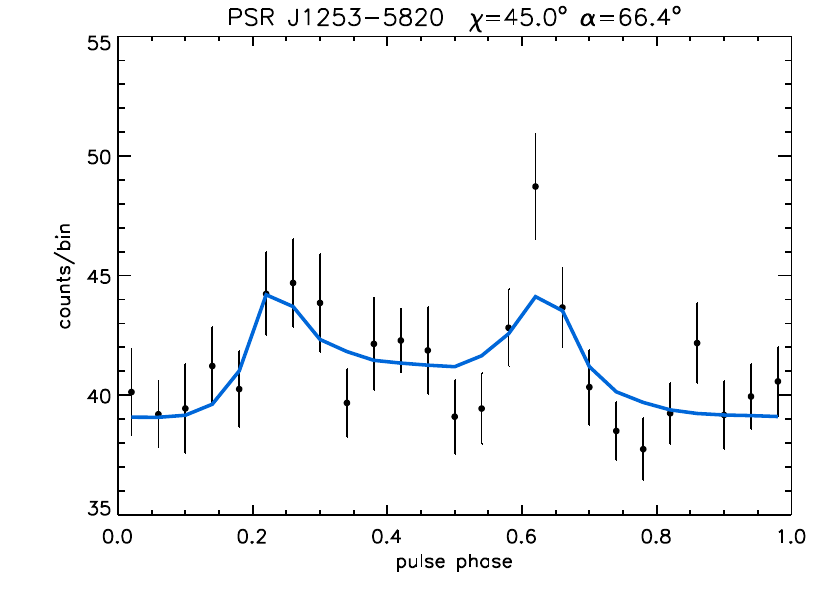}
\includegraphics[width=4.5cm]{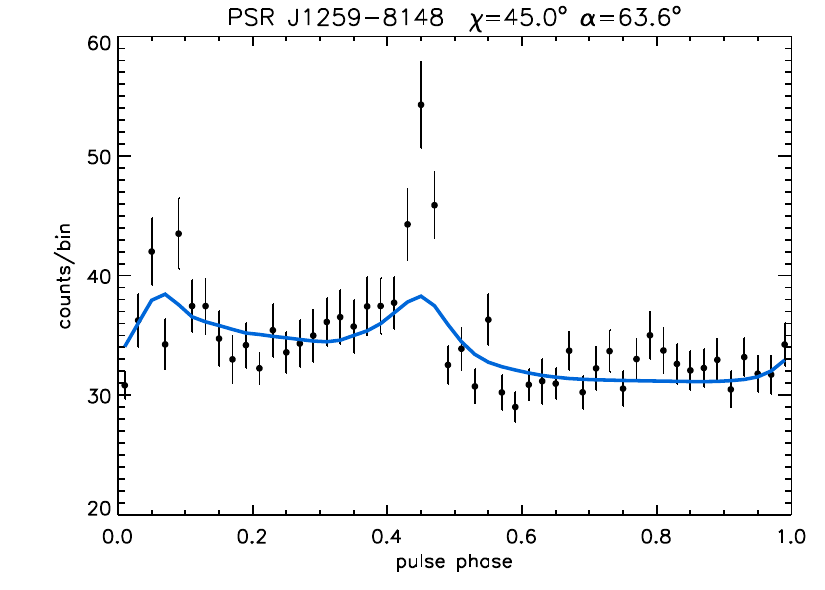}
\includegraphics[width=4.5cm]{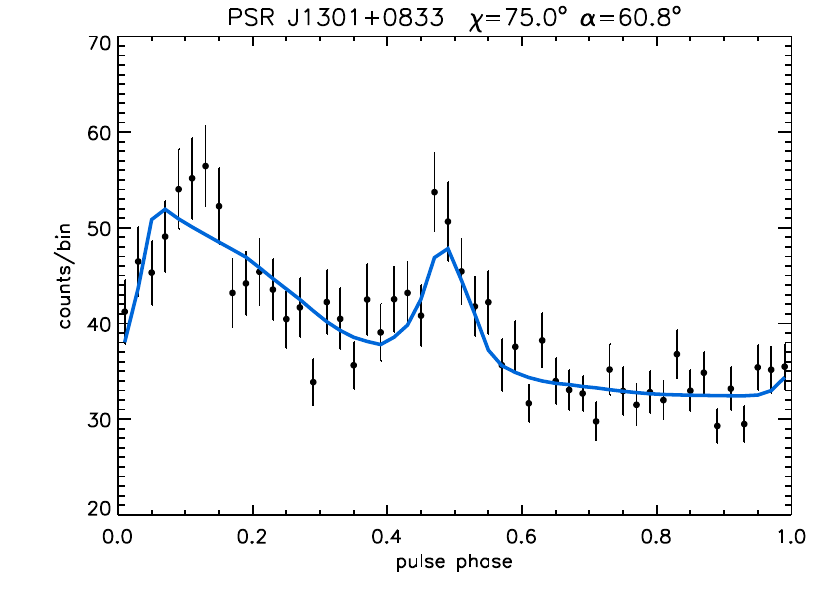}
\includegraphics[width=4.5cm]{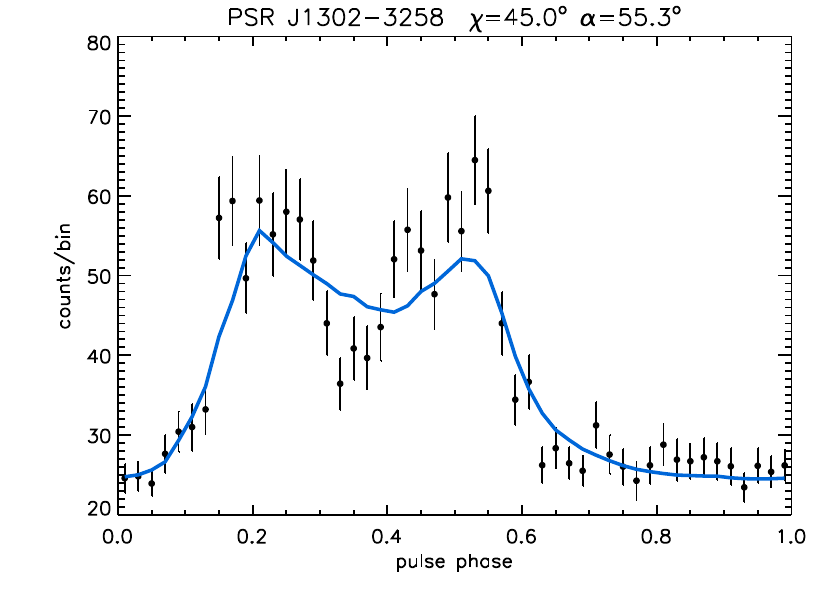}
\includegraphics[width=4.5cm]{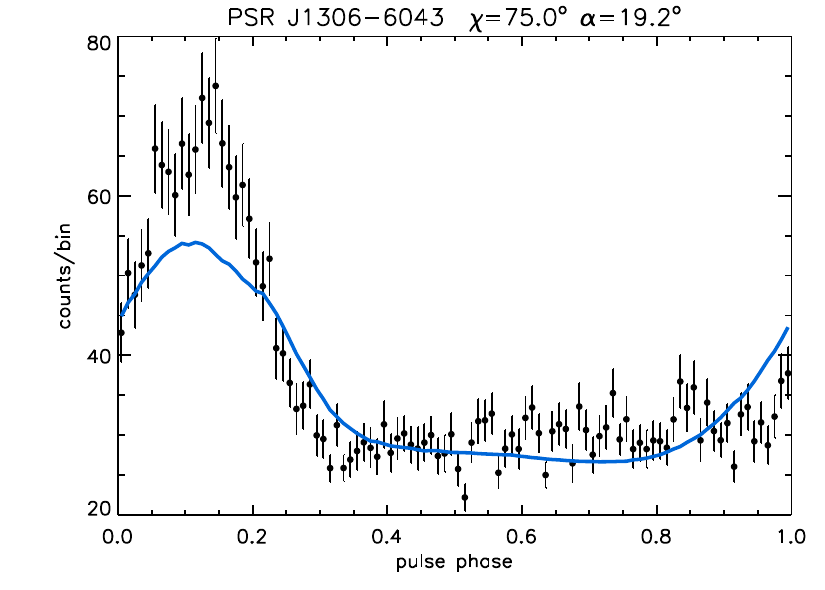}
\includegraphics[width=4.5cm]{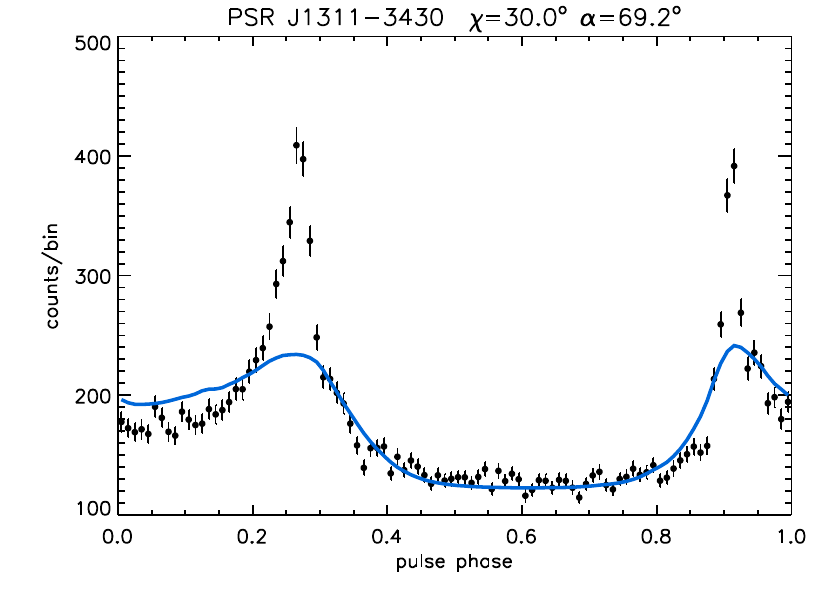}
\includegraphics[width=4.5cm]{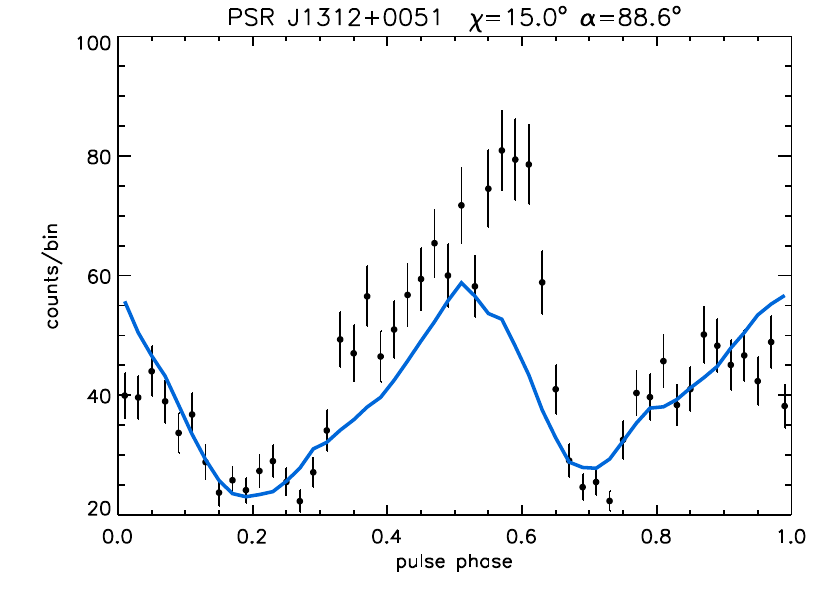}
\includegraphics[width=4.5cm]{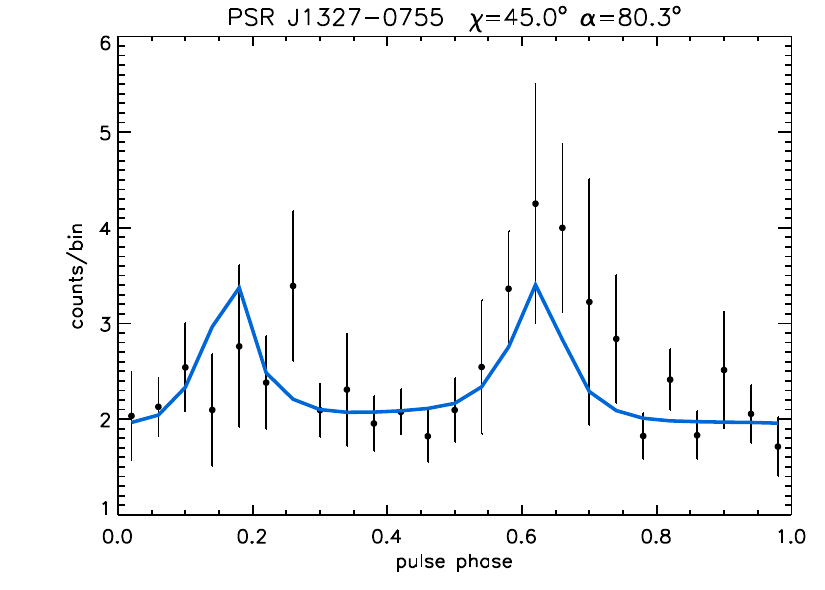}
\includegraphics[width=4.5cm]{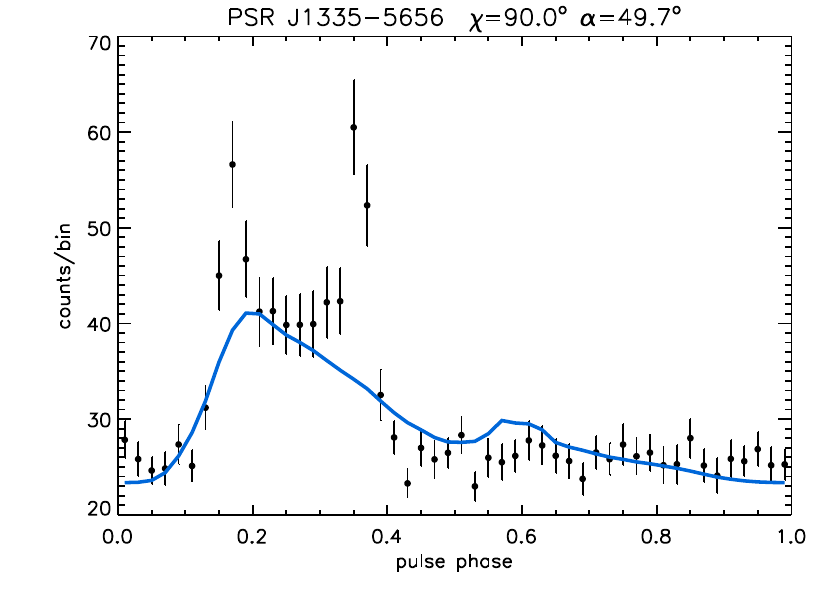}
\includegraphics[width=4.5cm]{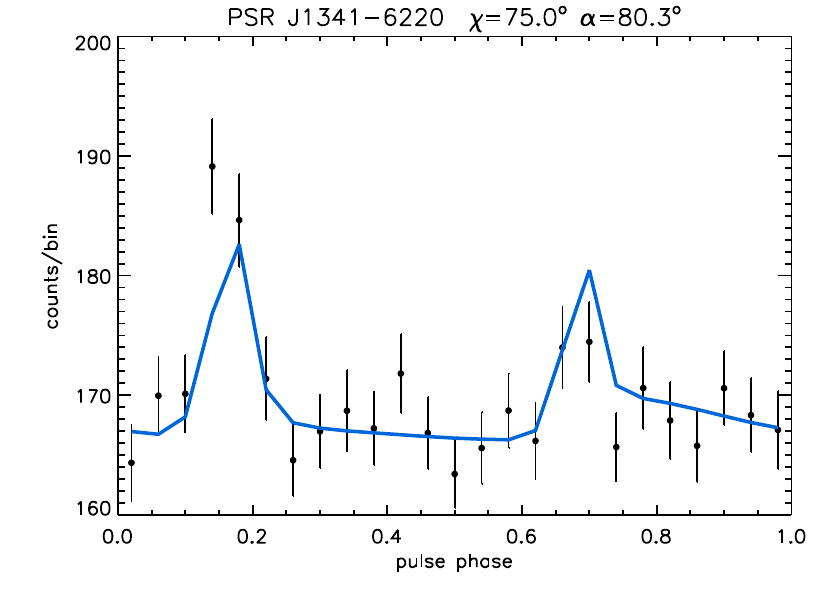}
\caption{continued.}
\end{figure*}

\begin{figure*}
\addtocounter{figure}{-1}
\centering
\includegraphics[width=4.5cm]{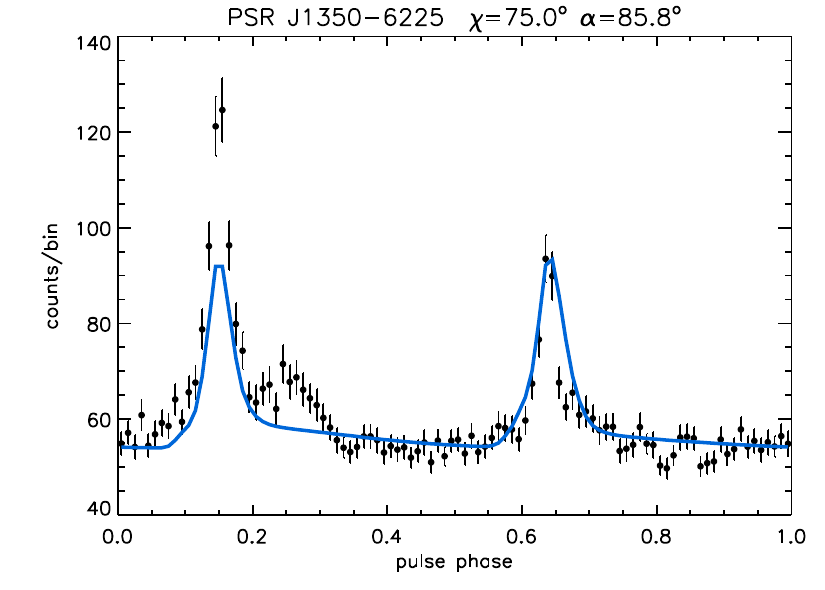}
\includegraphics[width=4.5cm]{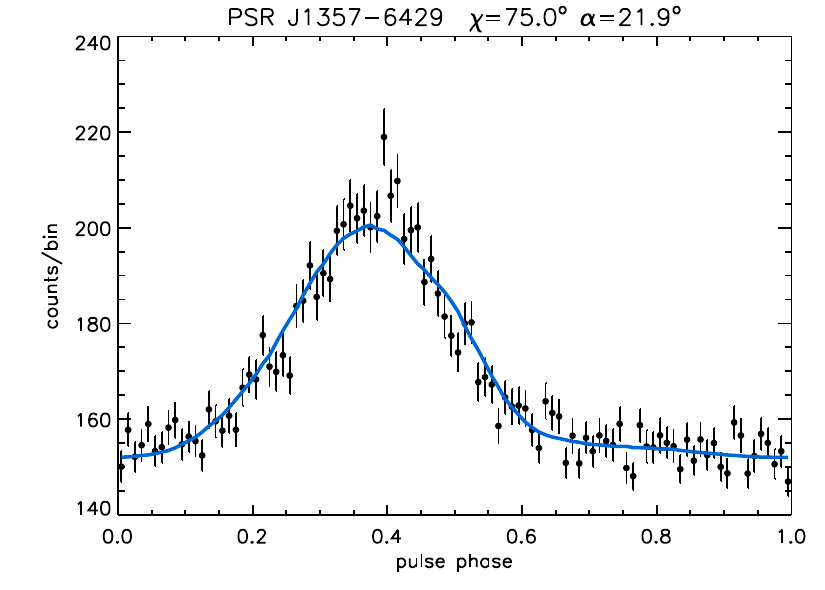}
\includegraphics[width=4.5cm]{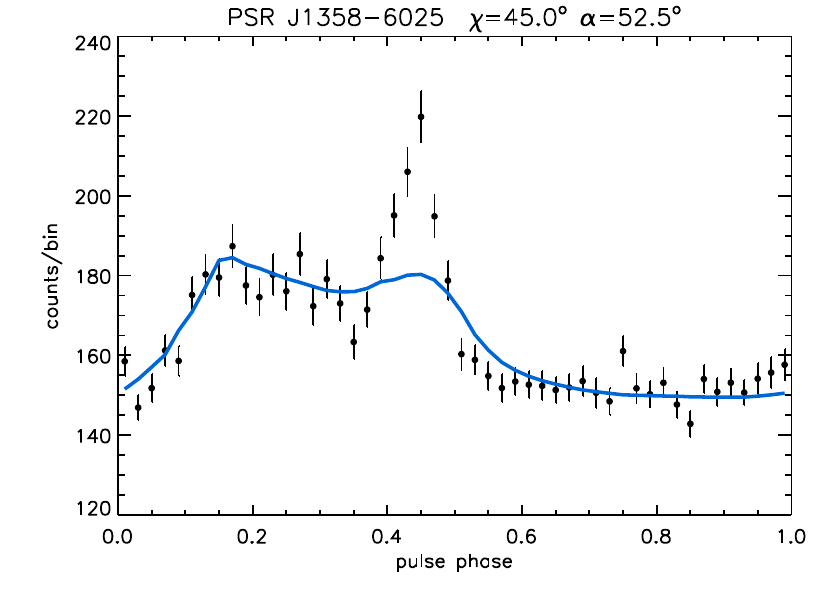}
\includegraphics[width=4.5cm]{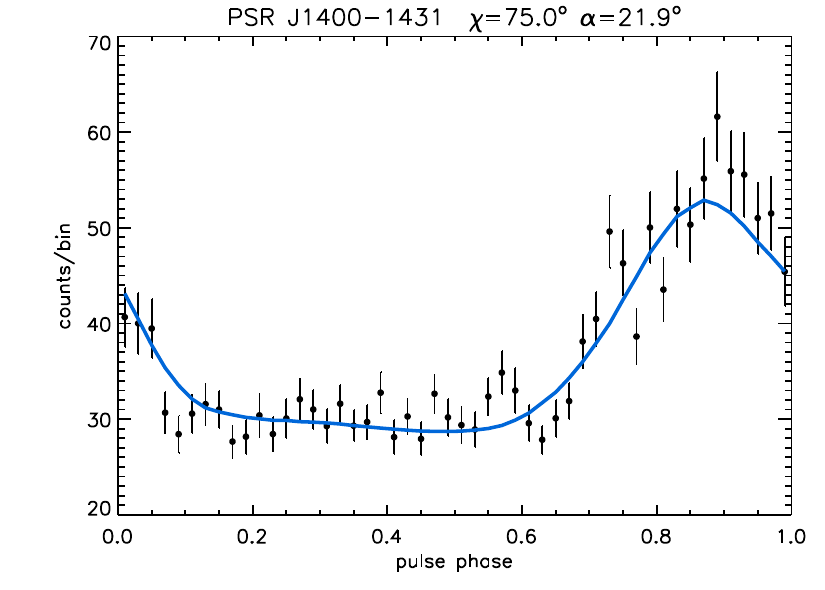}
\includegraphics[width=4.5cm]{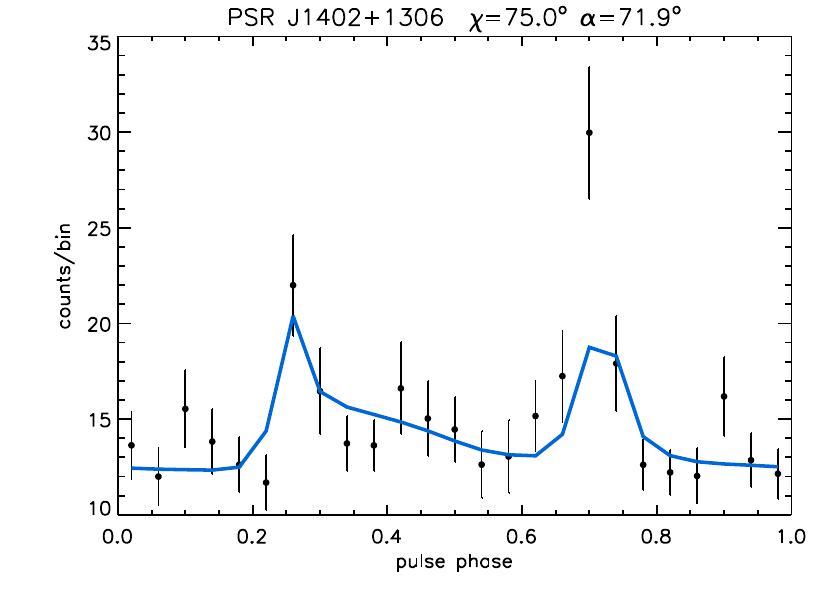}
\includegraphics[width=4.5cm]{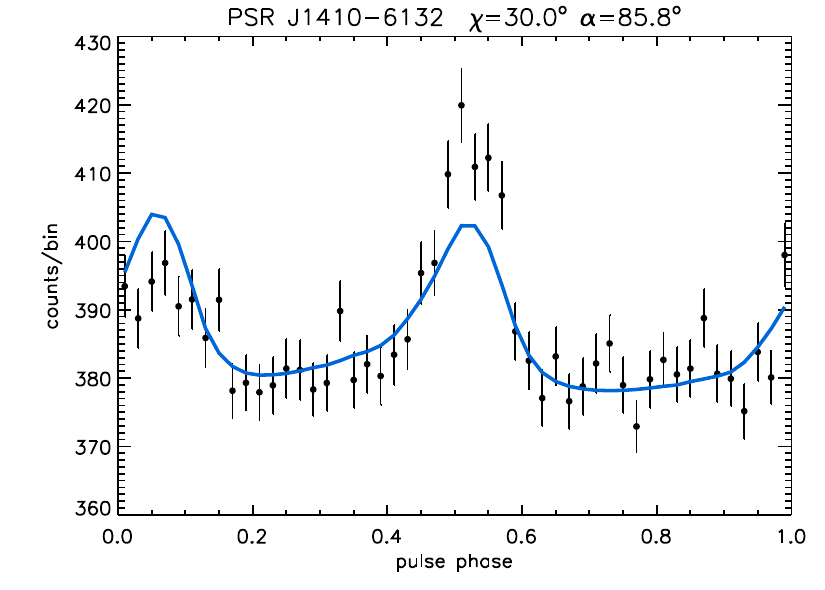}
\includegraphics[width=4.5cm]{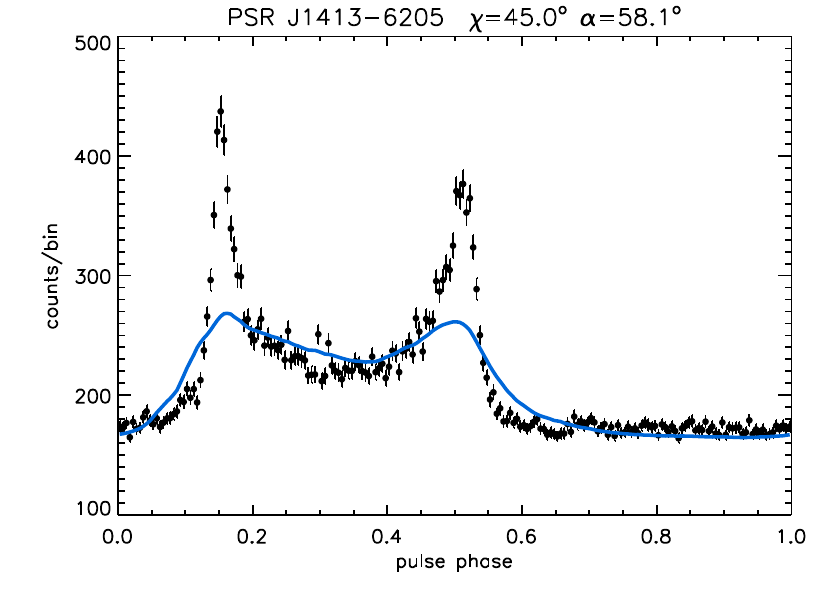}
\includegraphics[width=4.5cm]{J1418-6058.pdf}
\includegraphics[width=4.5cm]{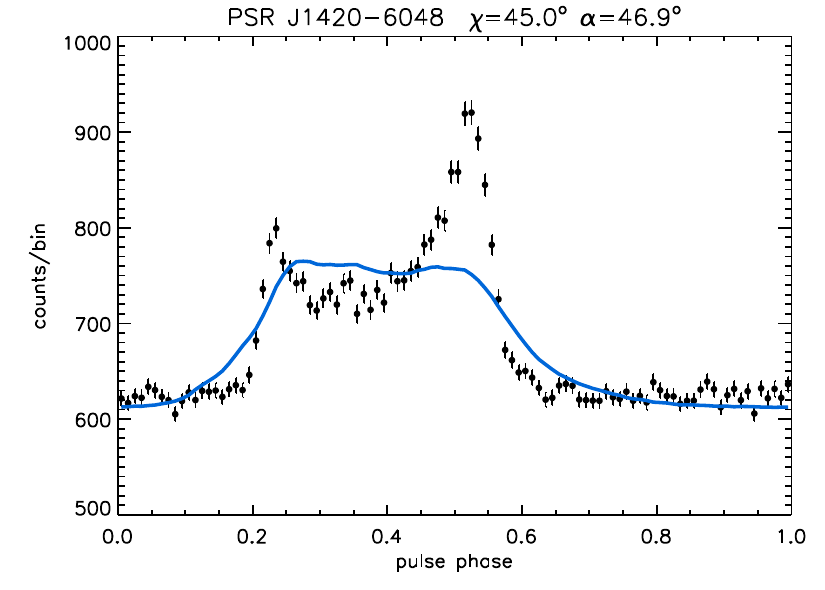}
\includegraphics[width=4.5cm]{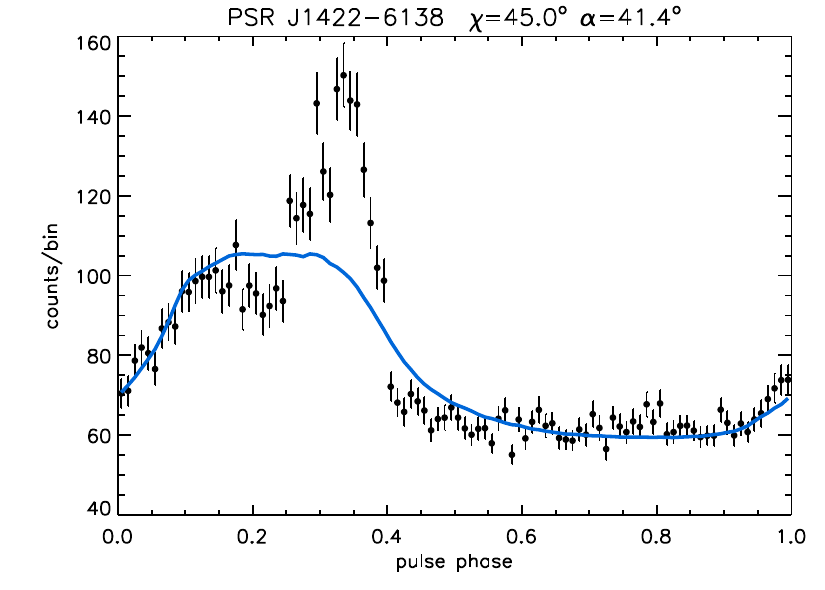}
\includegraphics[width=4.5cm]{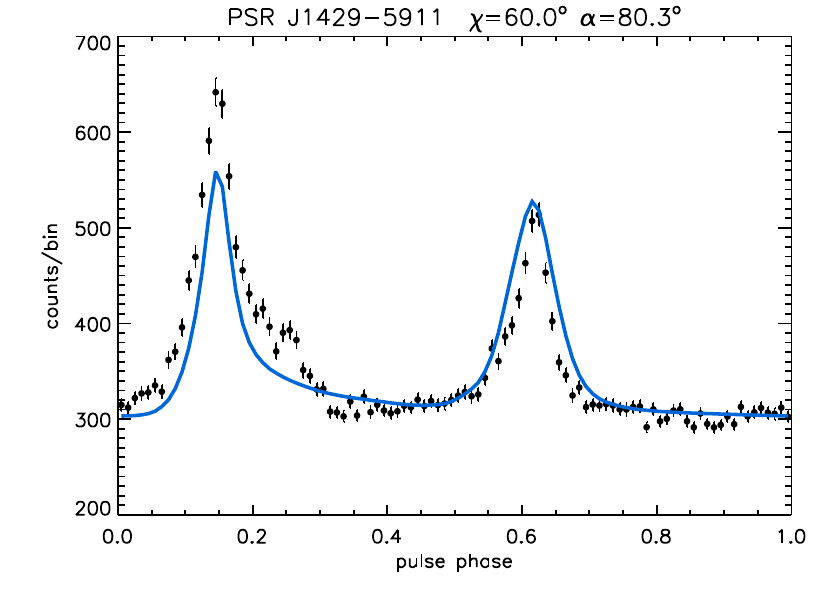}
\includegraphics[width=4.5cm]{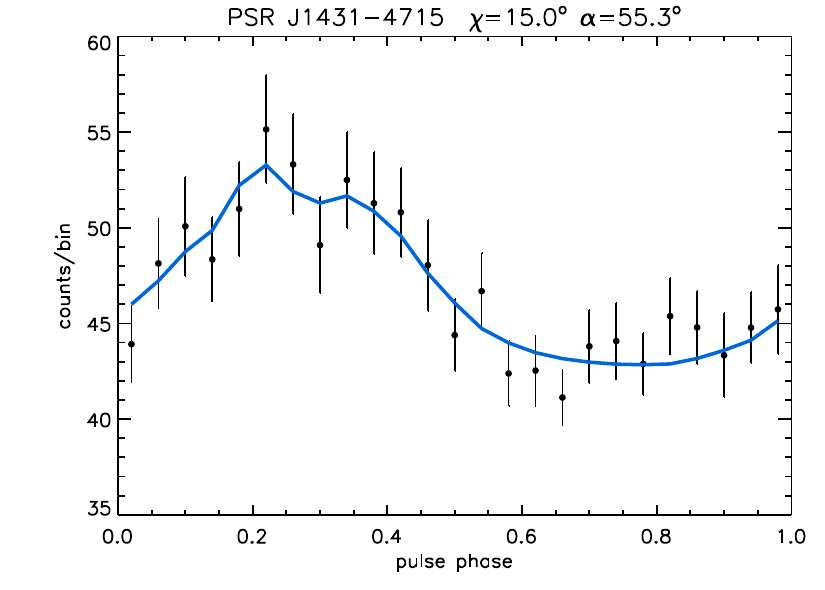}
\includegraphics[width=4.5cm]{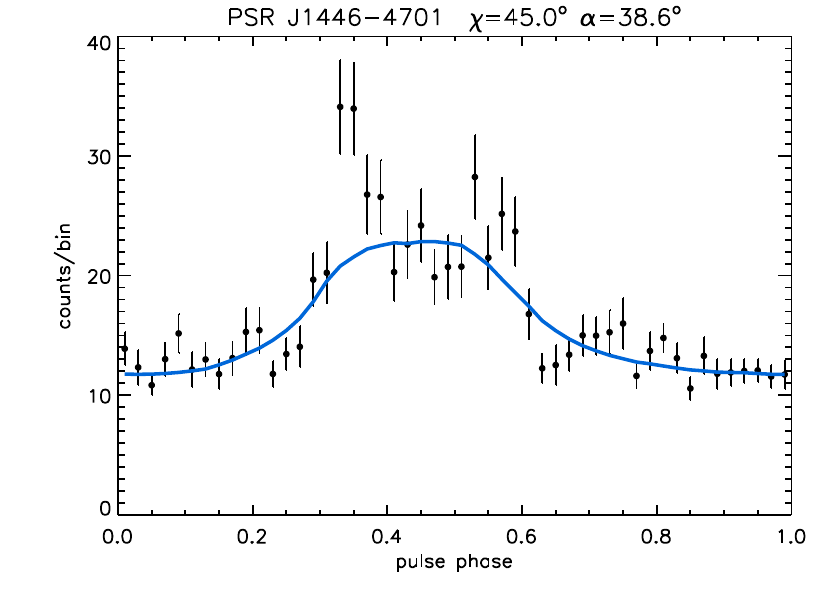}
\includegraphics[width=4.5cm]{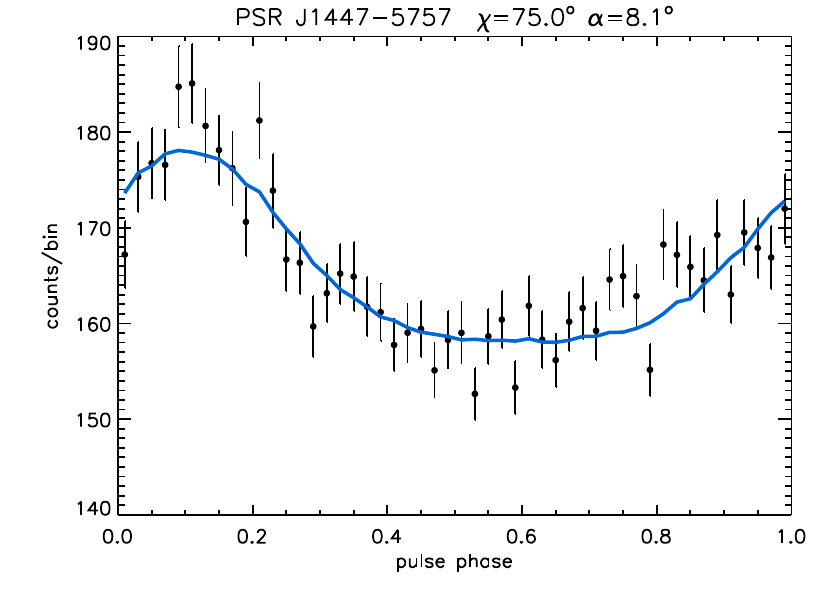}
\includegraphics[width=4.5cm]{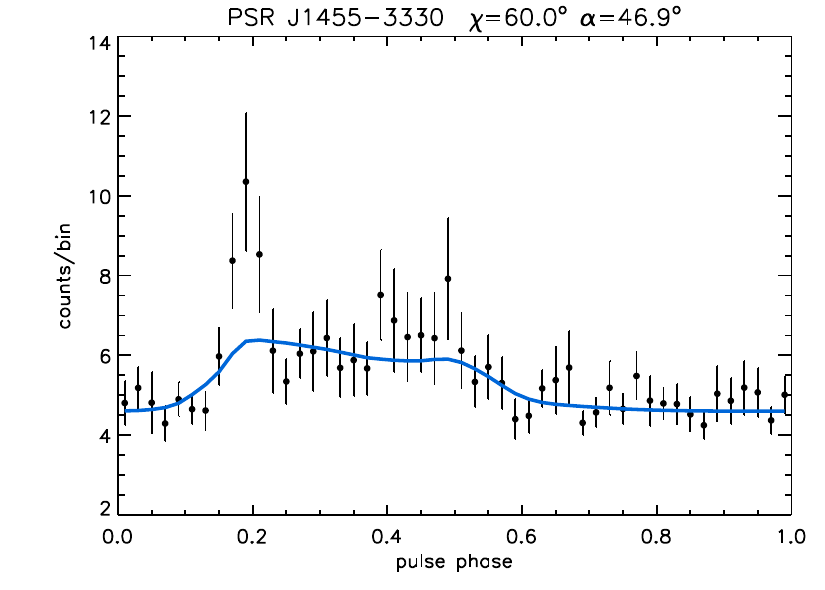}
\includegraphics[width=4.5cm]{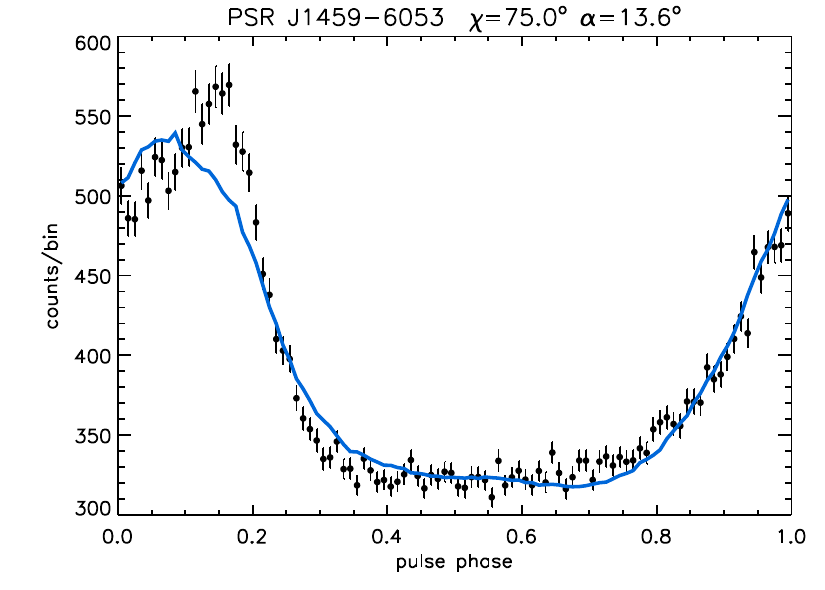}
\includegraphics[width=4.5cm]{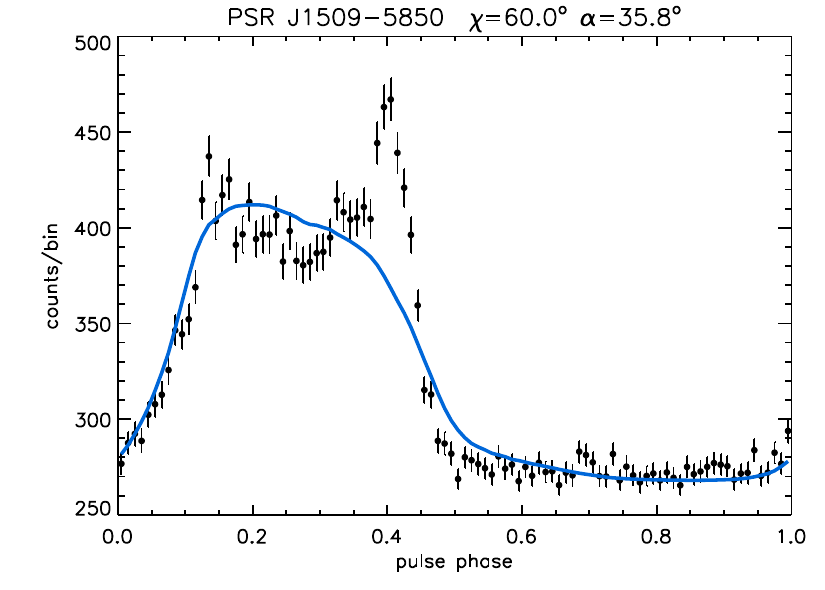}
\includegraphics[width=4.5cm]{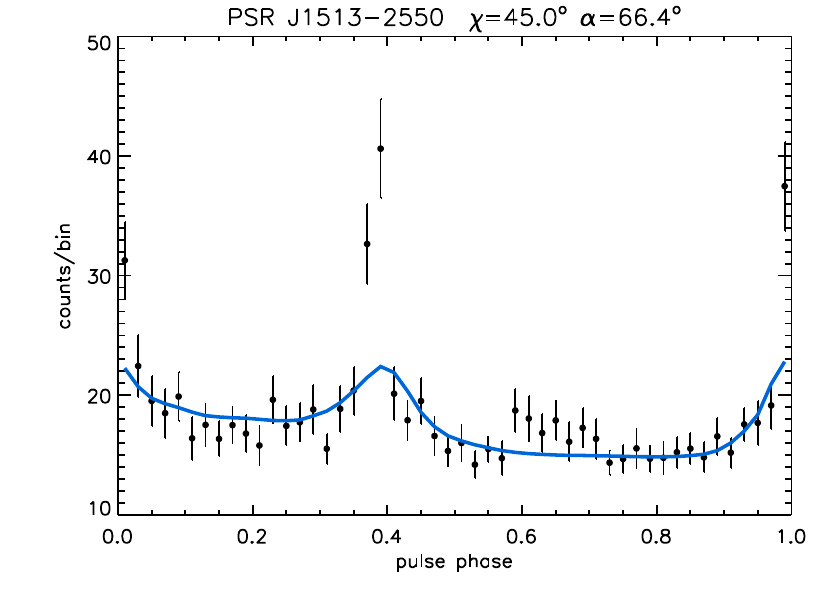}
\includegraphics[width=4.5cm]{J1513-5908.pdf}
\includegraphics[width=4.5cm]{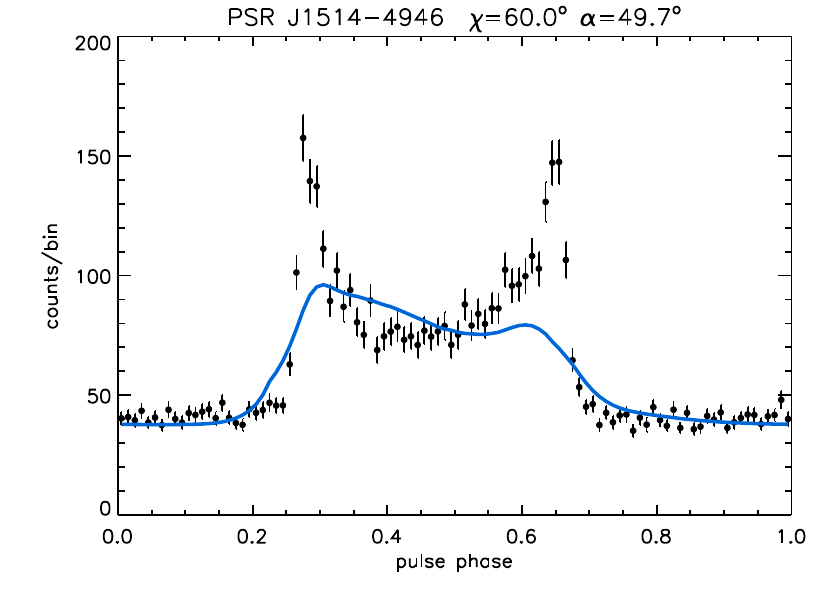}
\includegraphics[width=4.5cm]{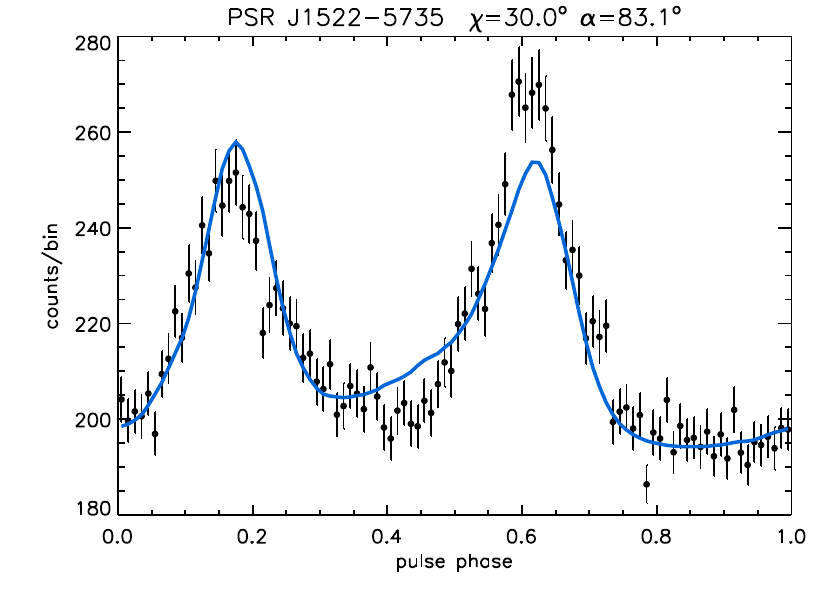}
\includegraphics[width=4.5cm]{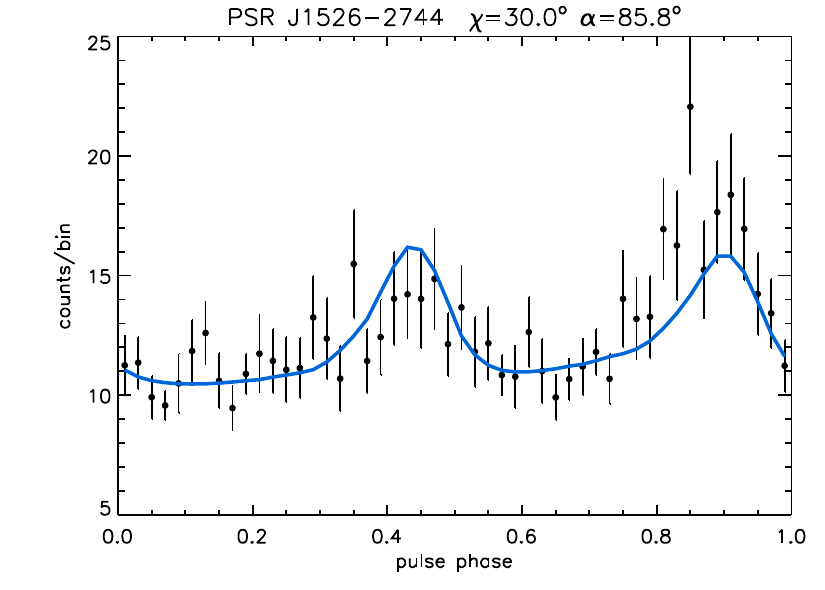}
\includegraphics[width=4.5cm]{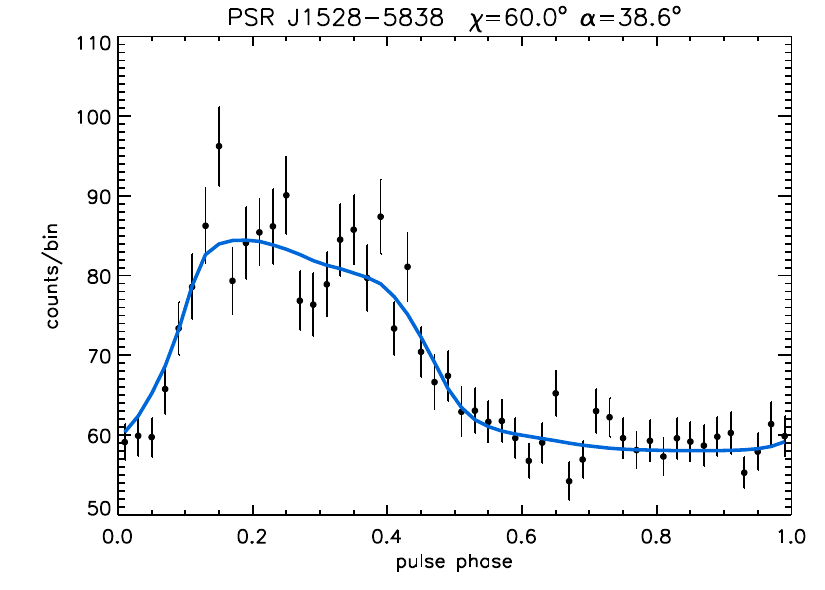}
\includegraphics[width=4.5cm]{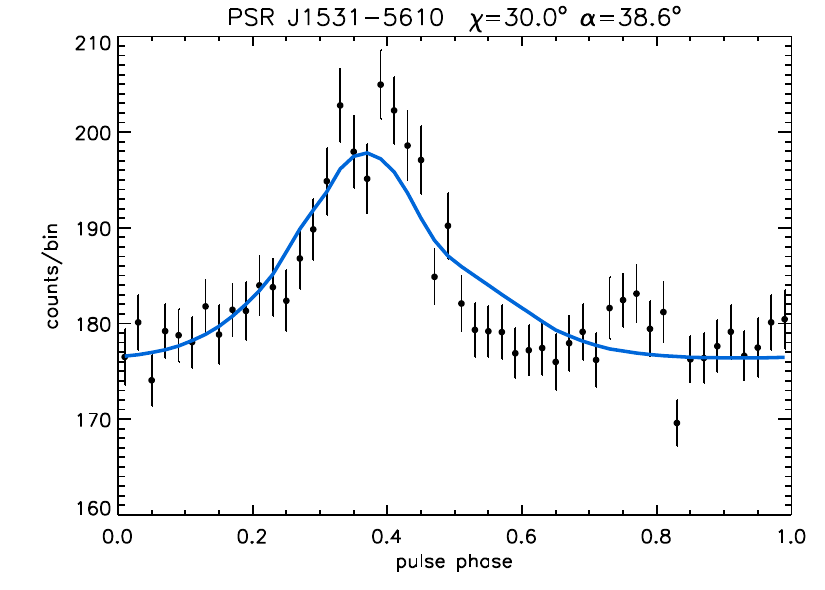}
\includegraphics[width=4.5cm]{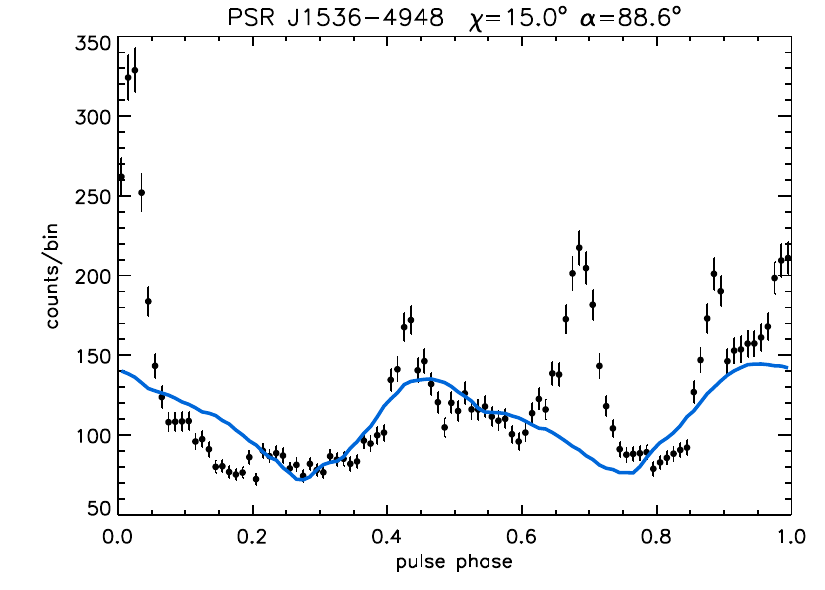}
\includegraphics[width=4.5cm]{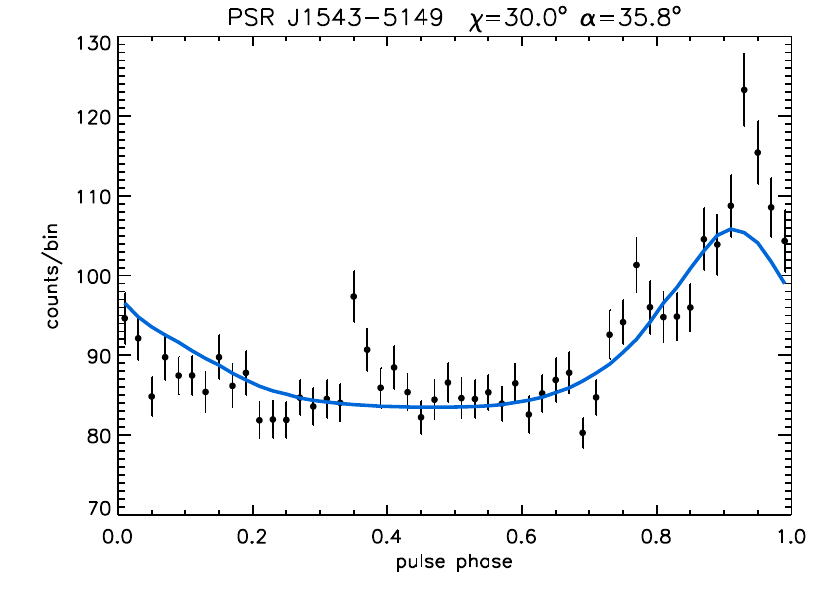}
\includegraphics[width=4.5cm]{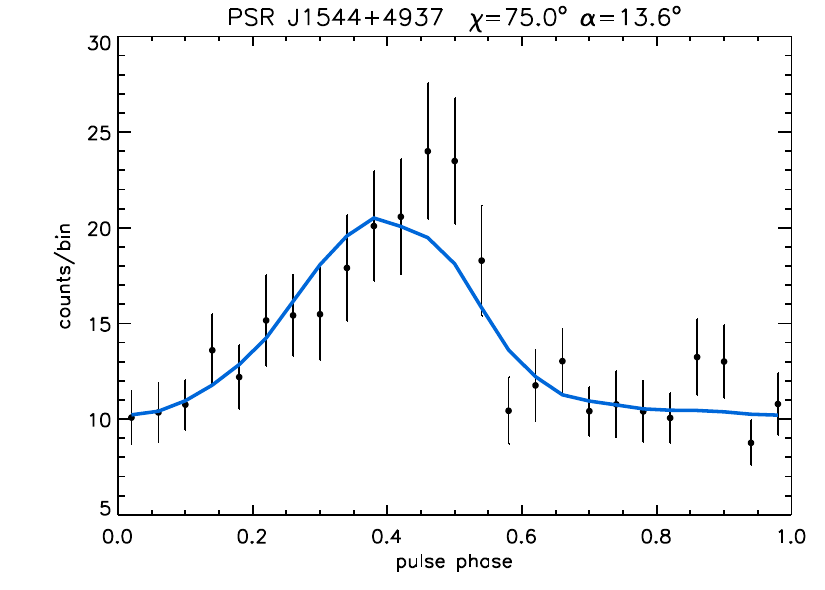}
\includegraphics[width=4.5cm]{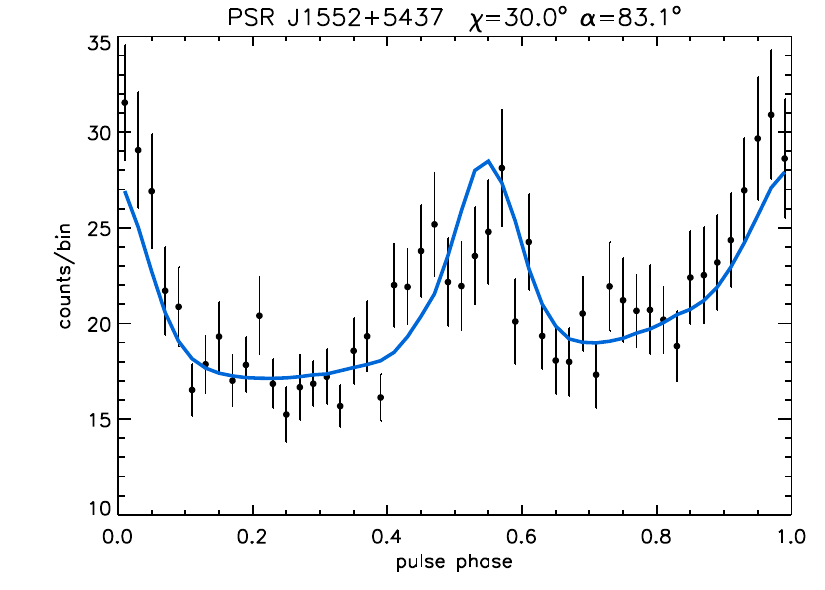}
\caption{continued.}
\end{figure*}

\begin{figure*}
\addtocounter{figure}{-1}
\centering
\includegraphics[width=4.5cm]{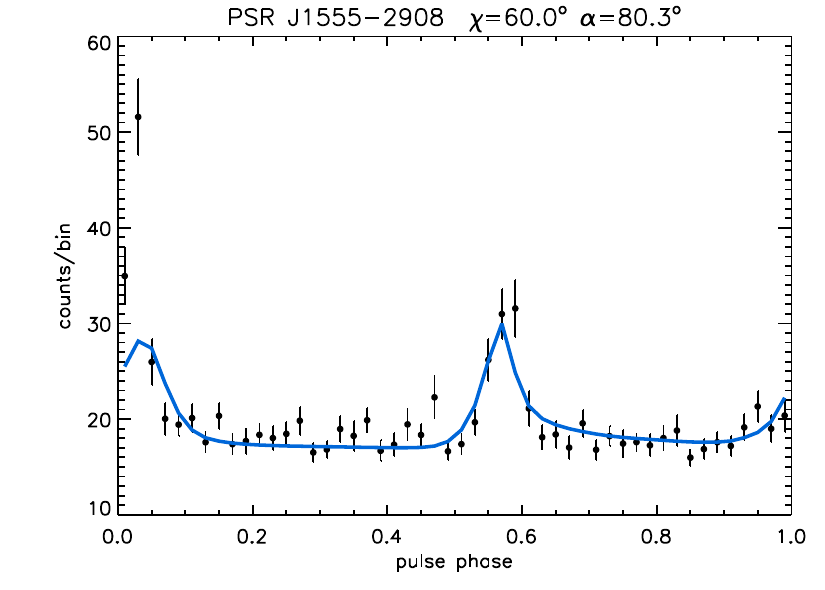}
\includegraphics[width=4.5cm]{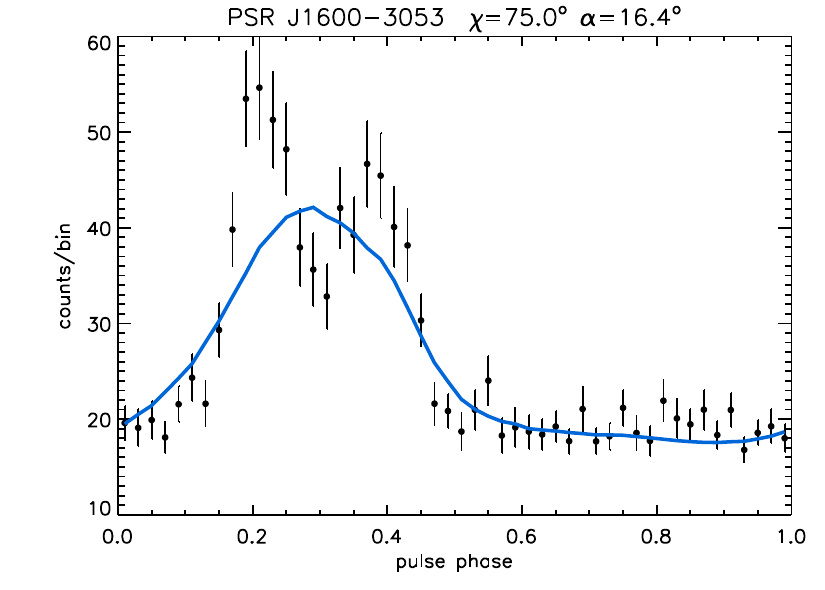}
\includegraphics[width=4.5cm]{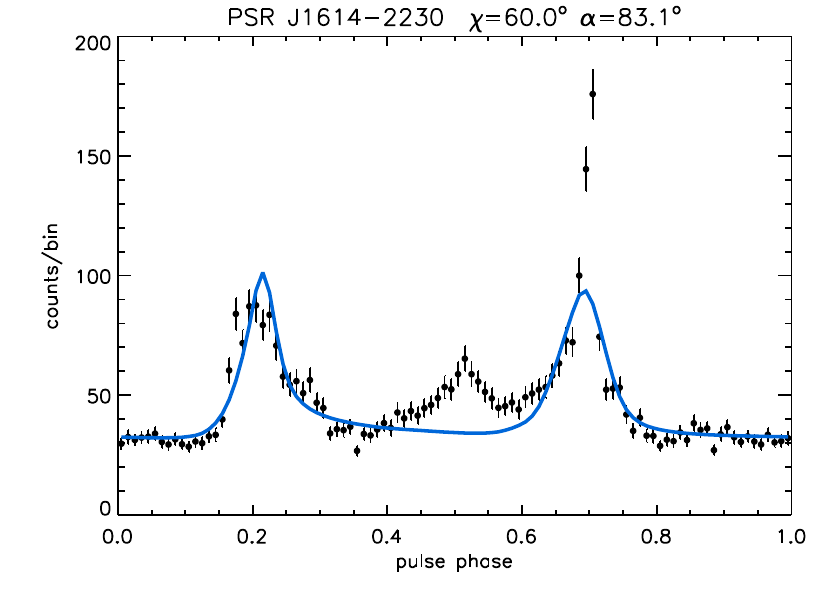}
\includegraphics[width=4.5cm]{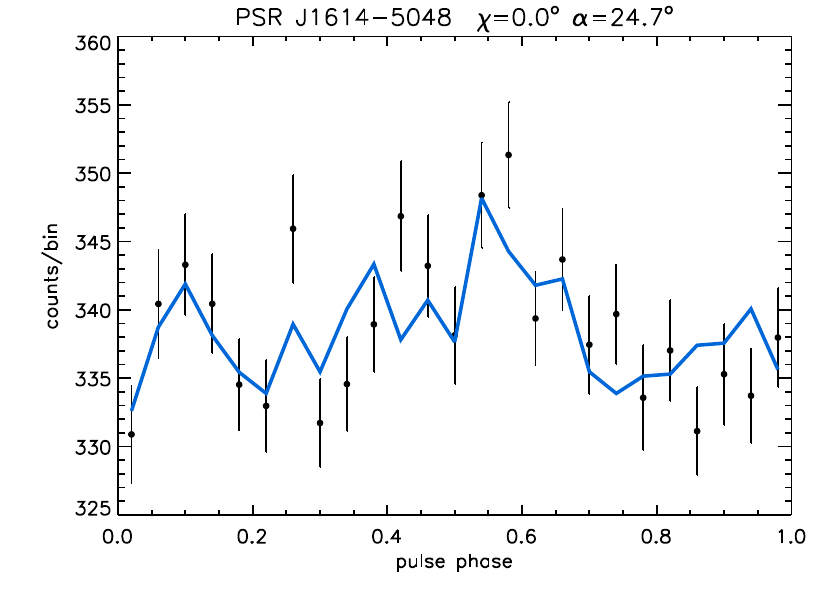}
\includegraphics[width=4.5cm]{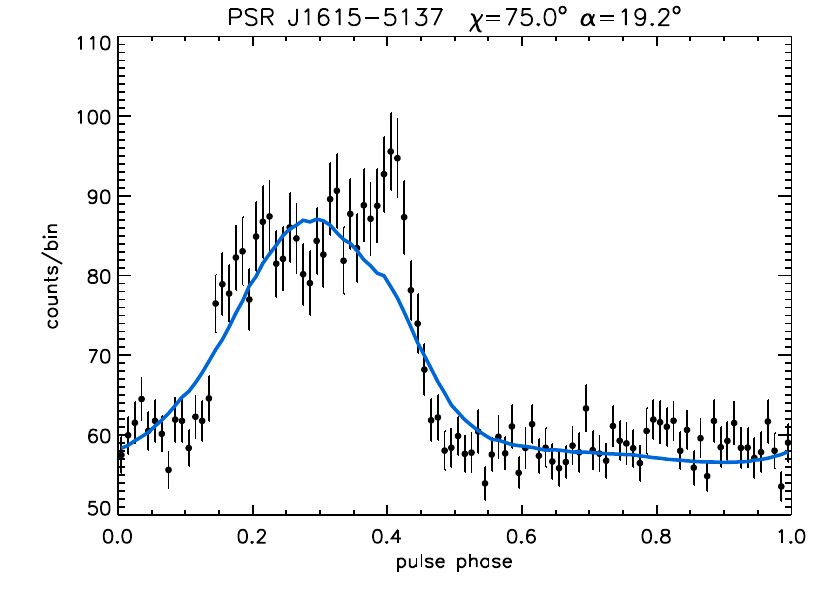}
\includegraphics[width=4.5cm]{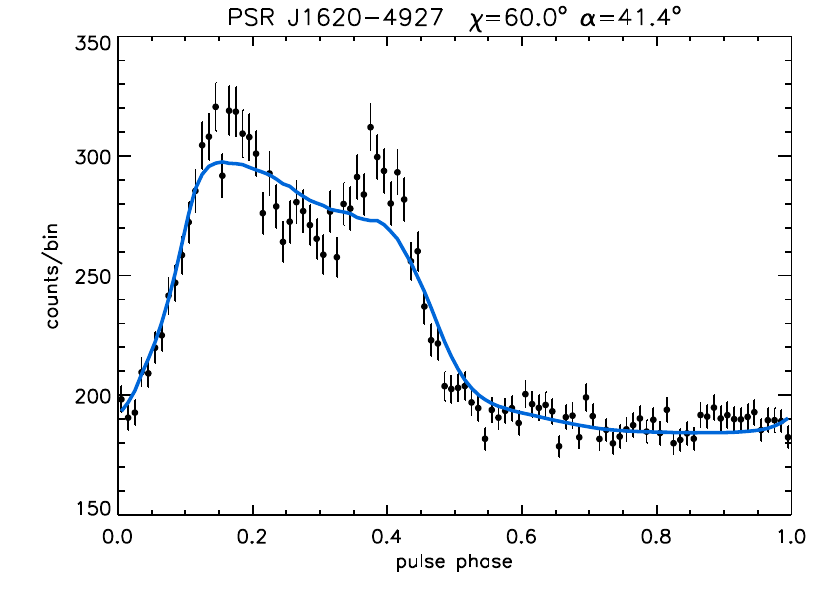}
\includegraphics[width=4.5cm]{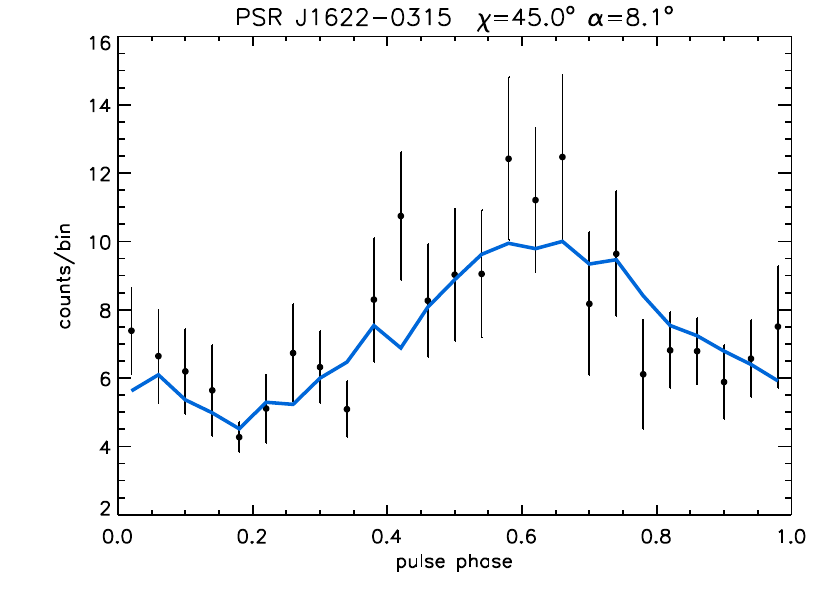}
\includegraphics[width=4.5cm]{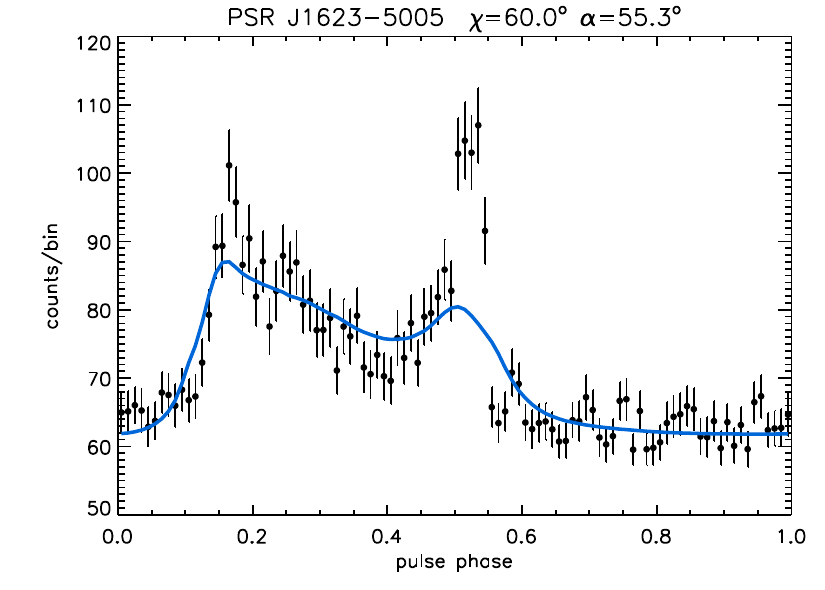}
\includegraphics[width=4.5cm]{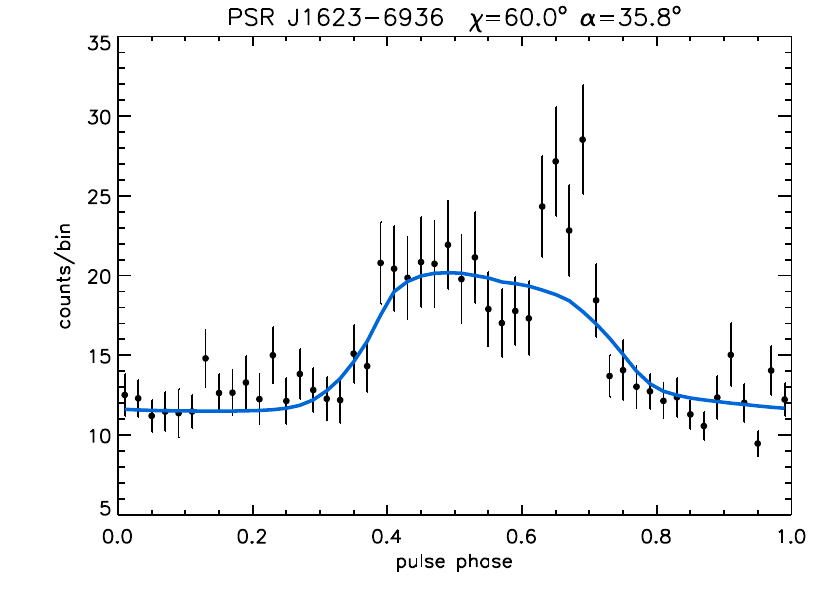}
\includegraphics[width=4.5cm]{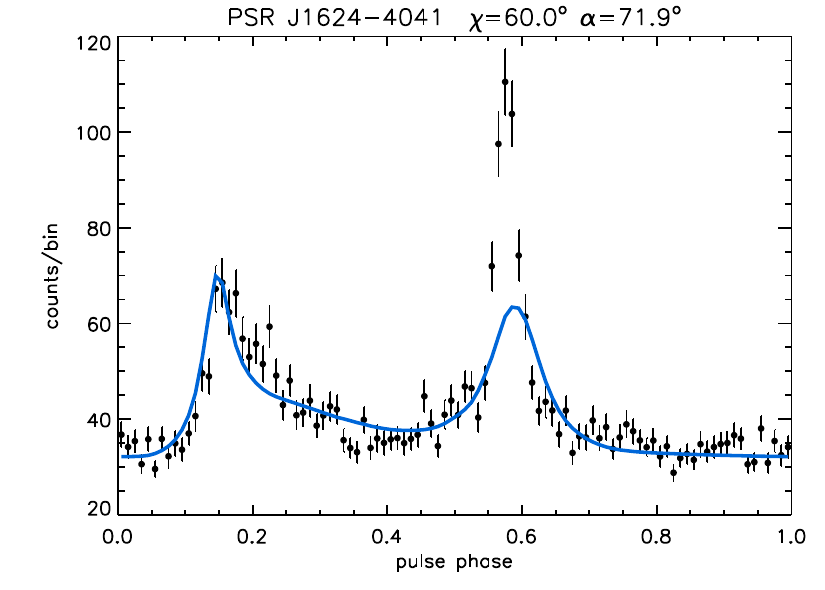}
\includegraphics[width=4.5cm]{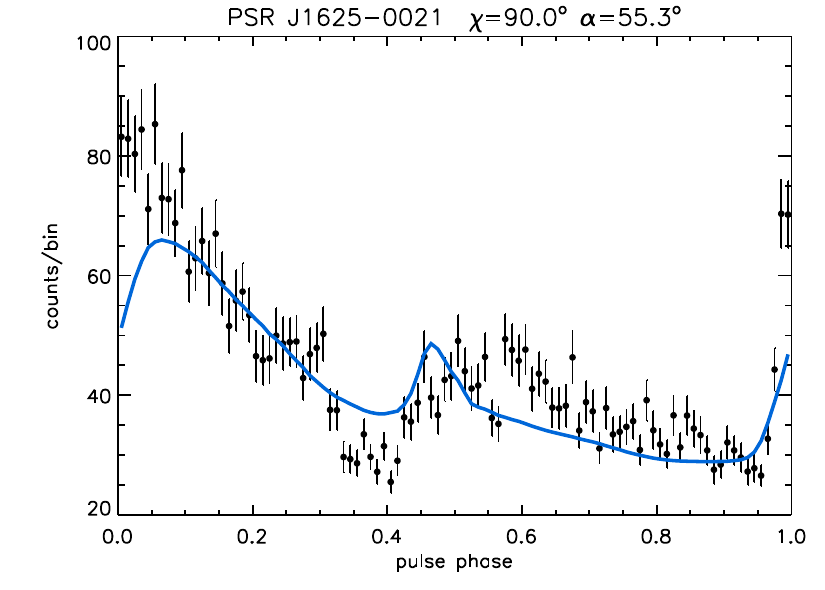}
\includegraphics[width=4.5cm]{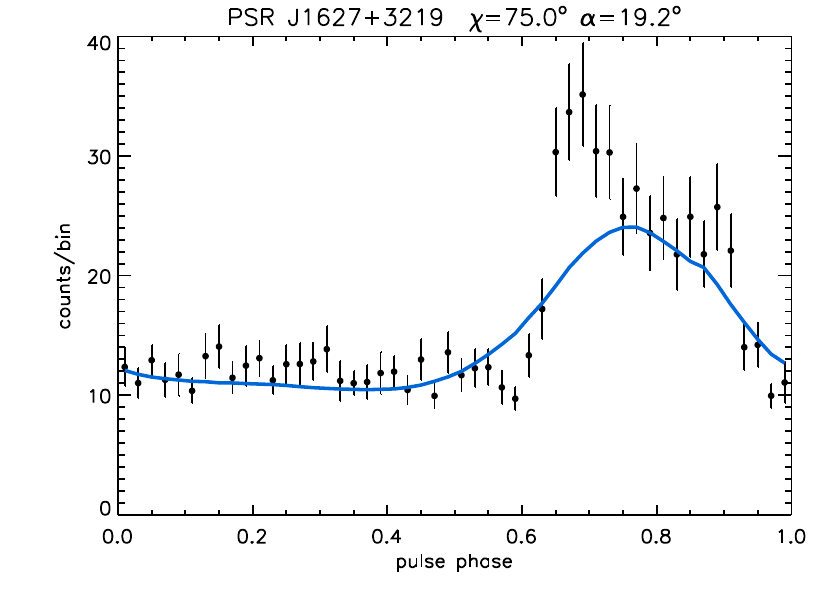}
\includegraphics[width=4.5cm]{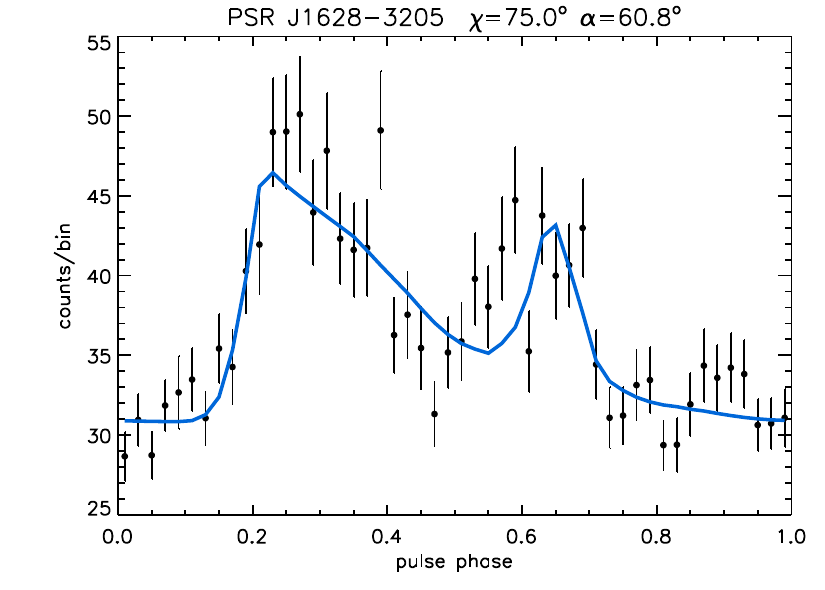}
\includegraphics[width=4.5cm]{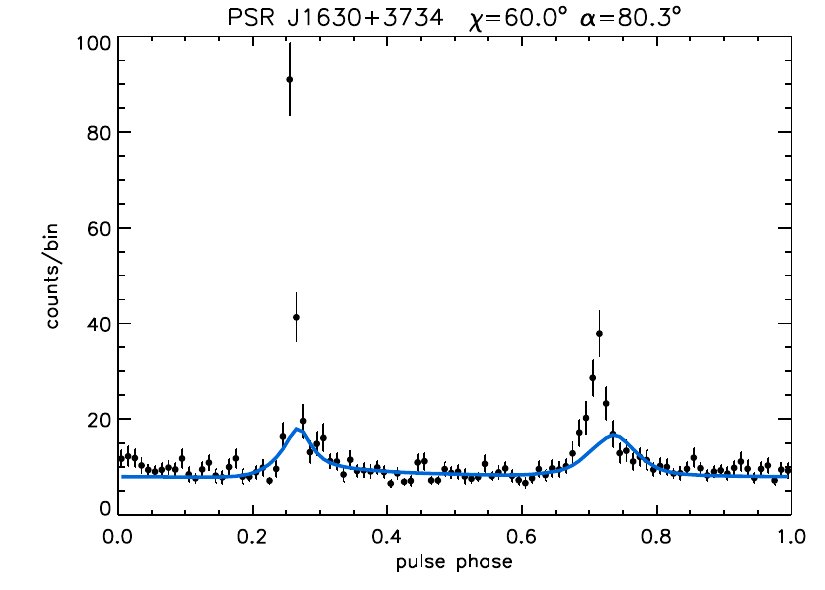}
\includegraphics[width=4.5cm]{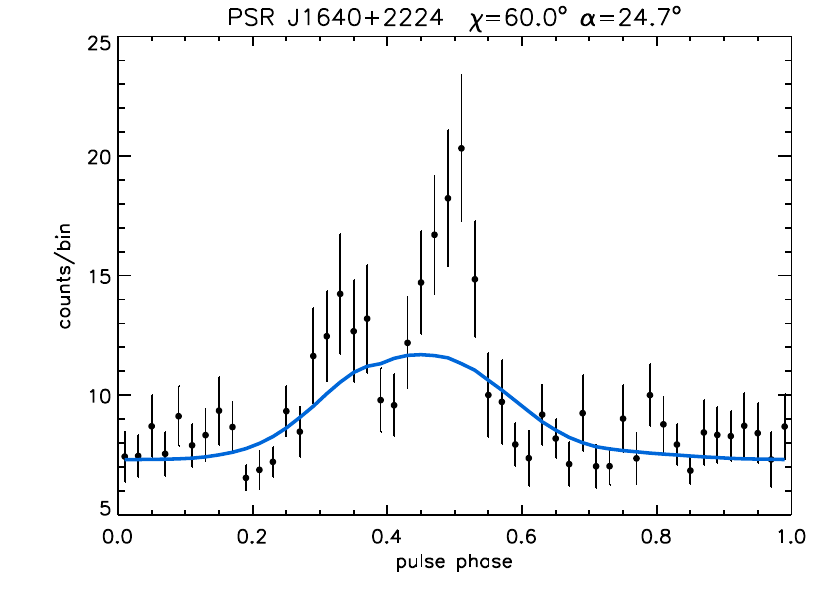}
\includegraphics[width=4.5cm]{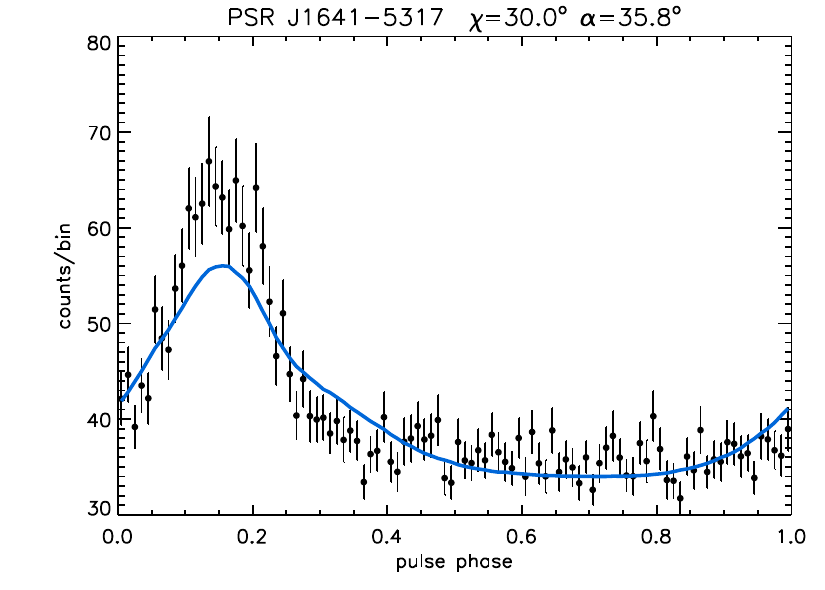}
\includegraphics[width=4.5cm]{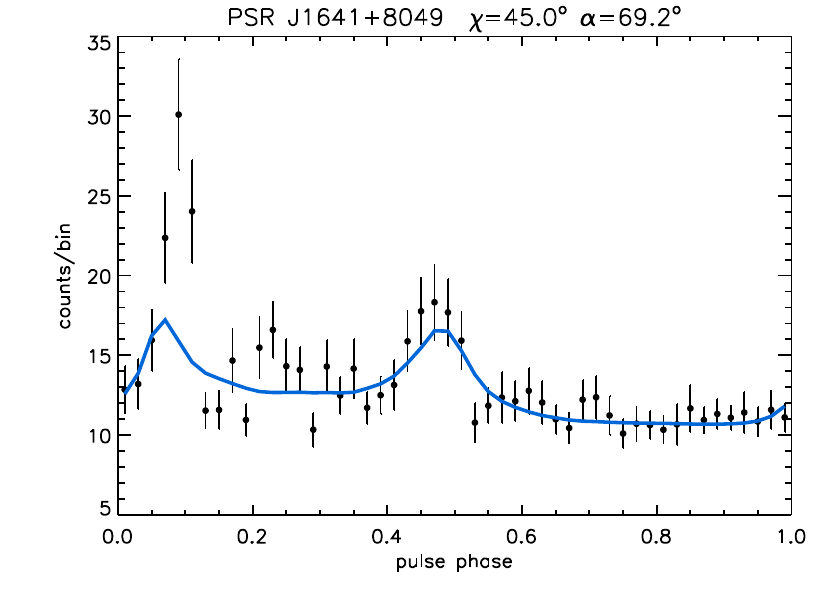}
\includegraphics[width=4.5cm]{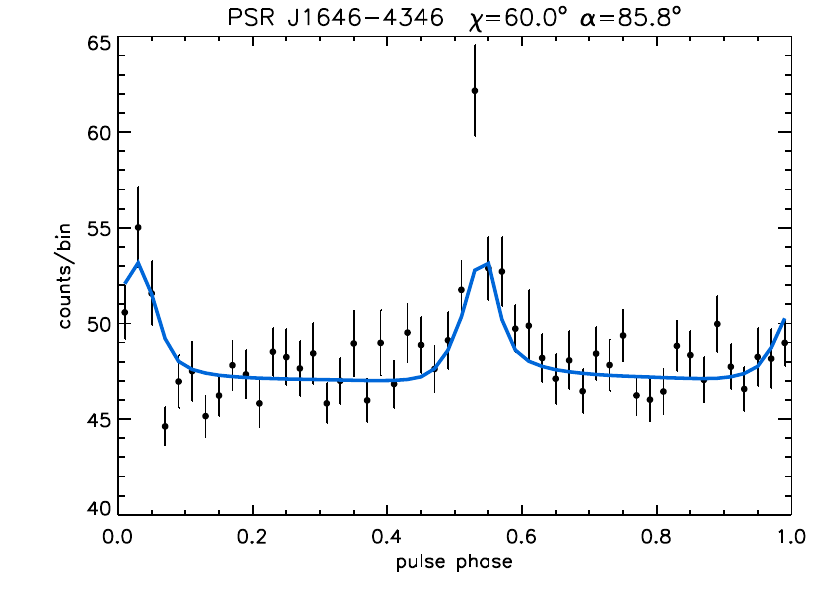}
\includegraphics[width=4.5cm]{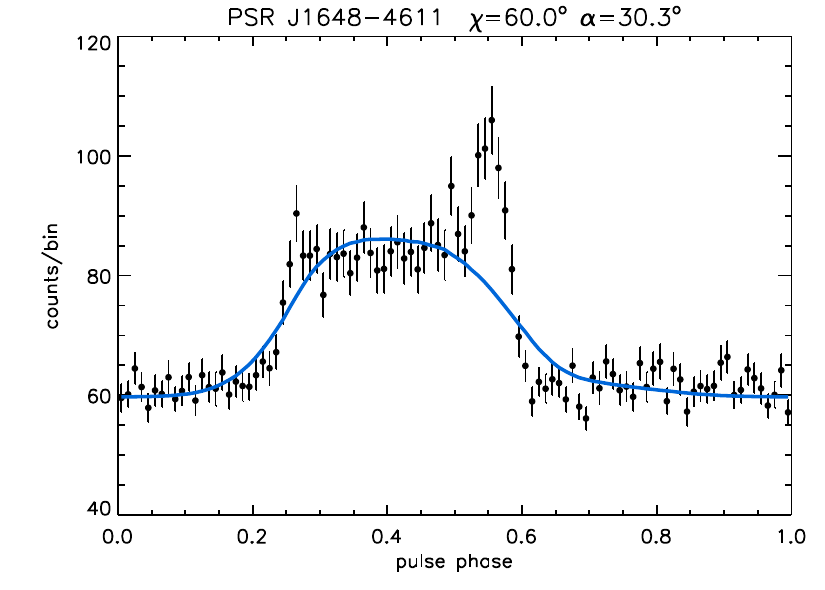}
\includegraphics[width=4.5cm]{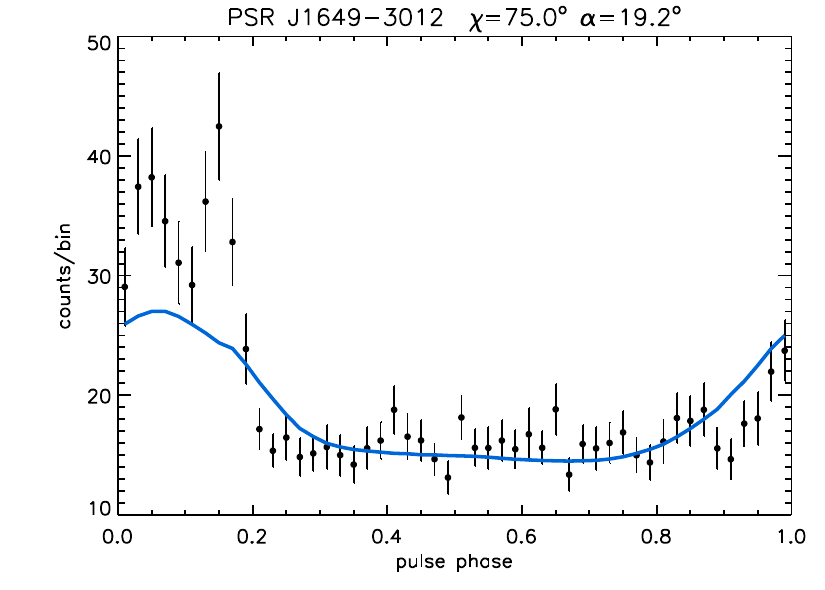}
\includegraphics[width=4.5cm]{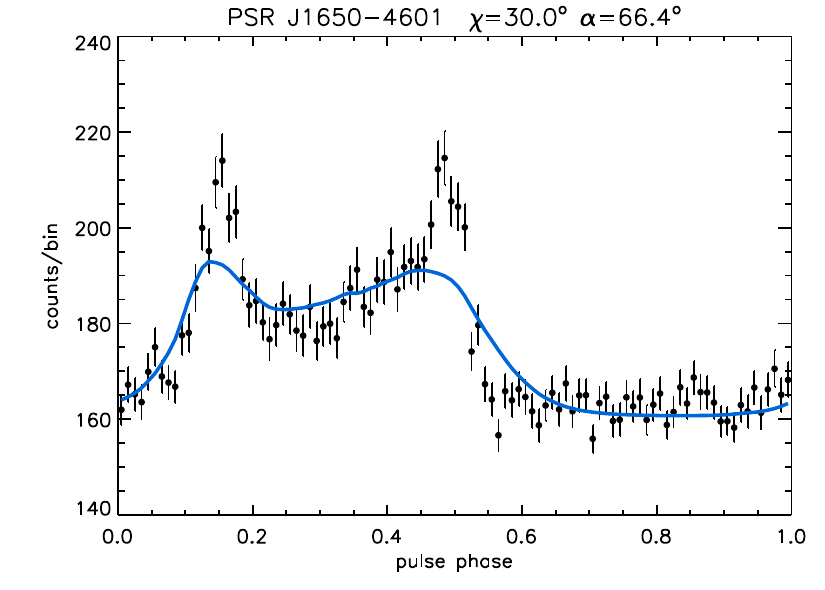}
\includegraphics[width=4.5cm]{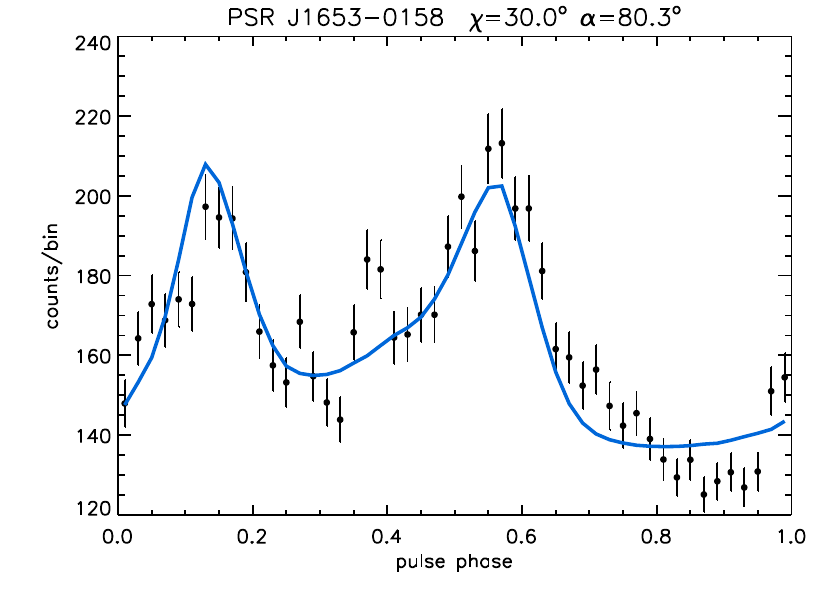}
\includegraphics[width=4.5cm]{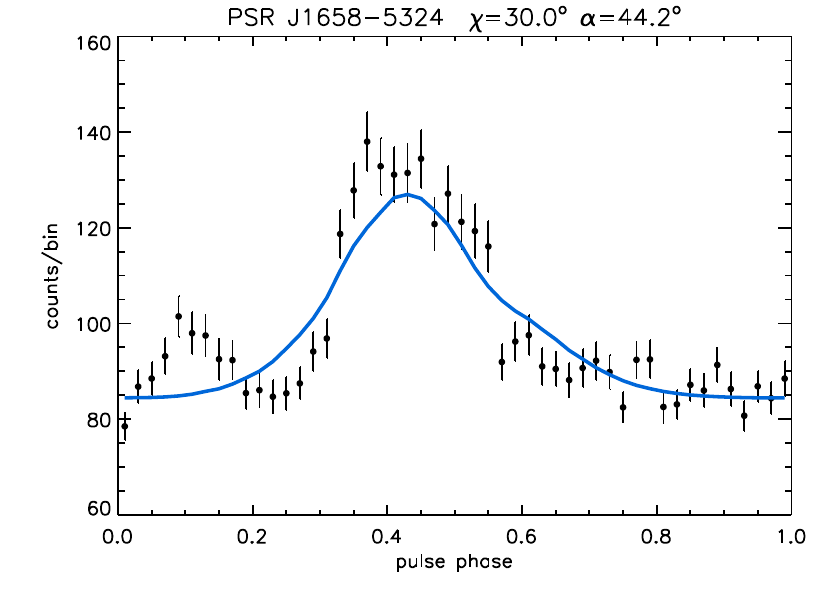}
\includegraphics[width=4.5cm]{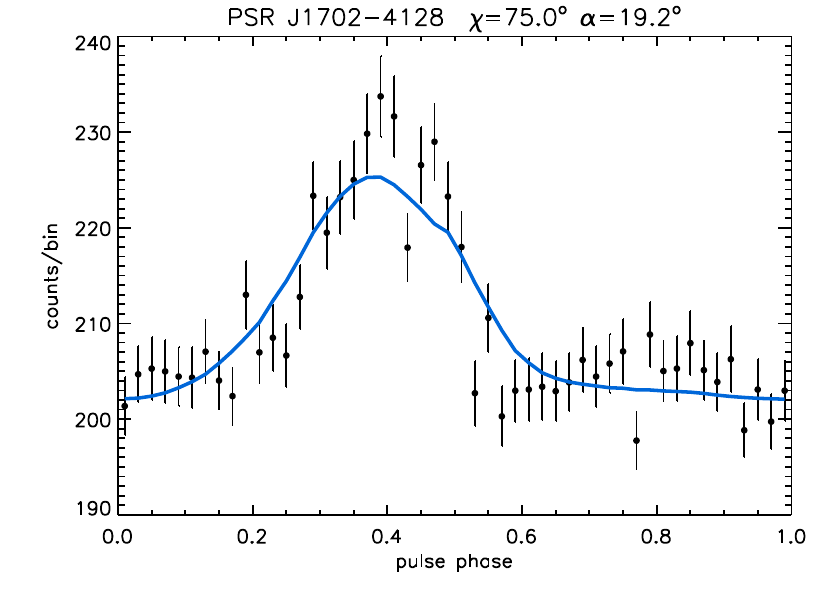}
\includegraphics[width=4.5cm]{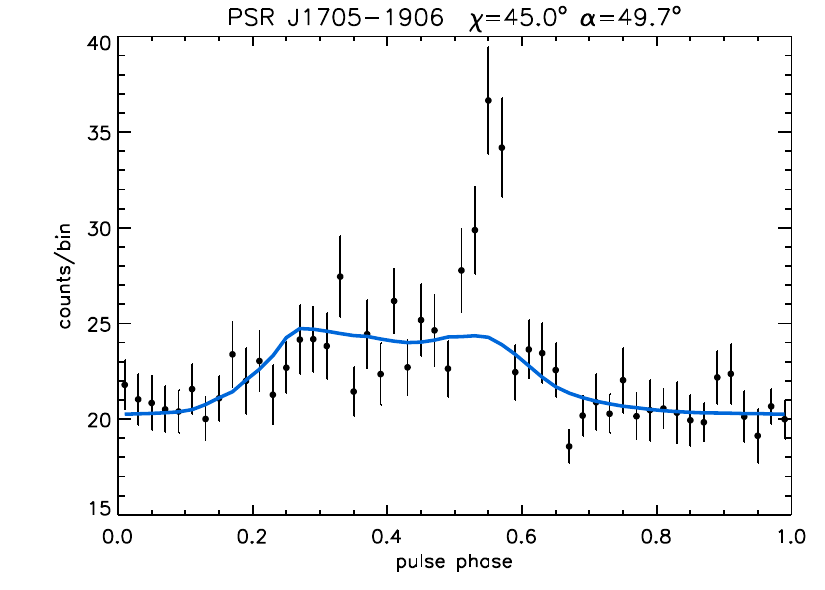}
\includegraphics[width=4.5cm]{J1709-4429.pdf}
\includegraphics[width=4.5cm]{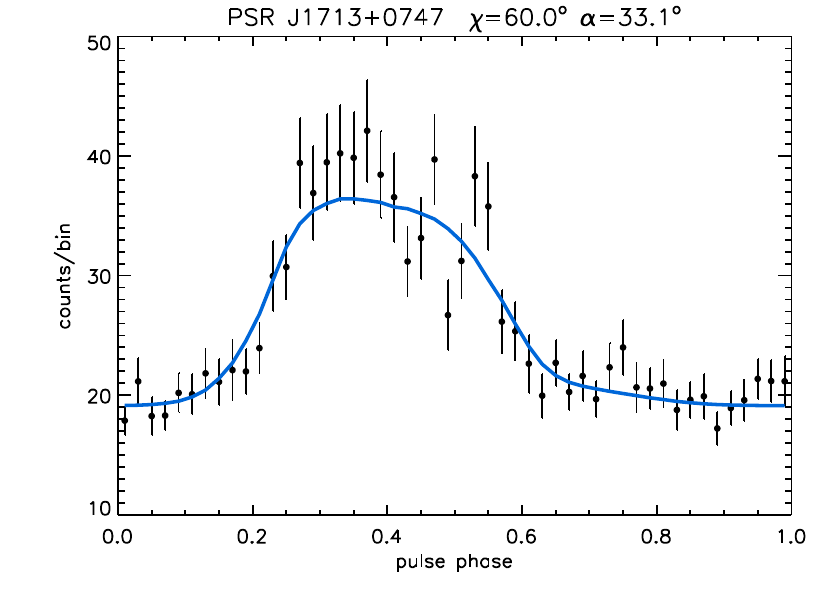}
\includegraphics[width=4.5cm]{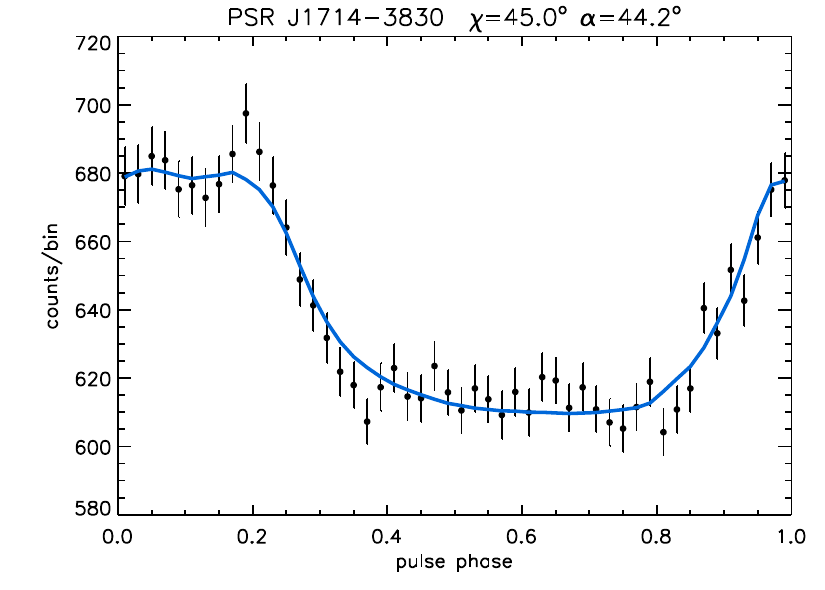}
\caption{continued.}
\end{figure*}

\begin{figure*}
\addtocounter{figure}{-1}
\centering
\includegraphics[width=4.5cm]{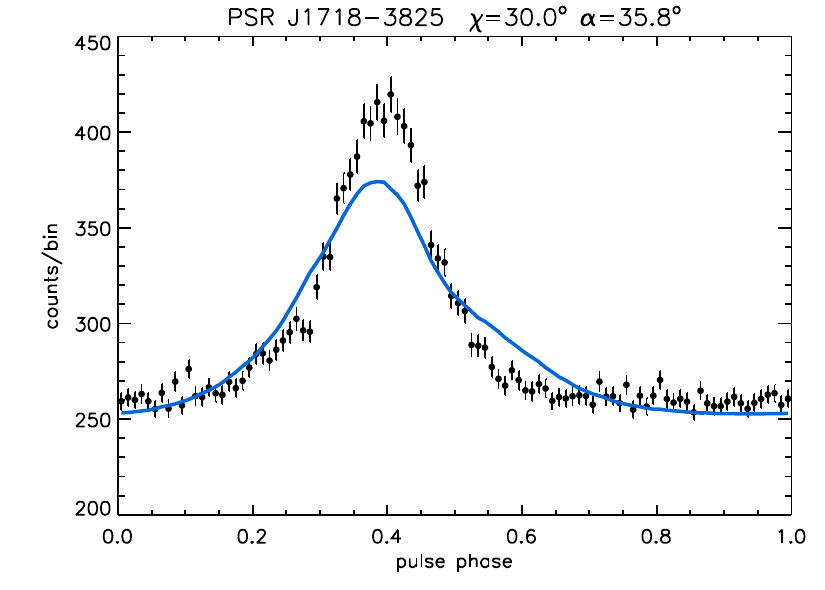}
\includegraphics[width=4.5cm]{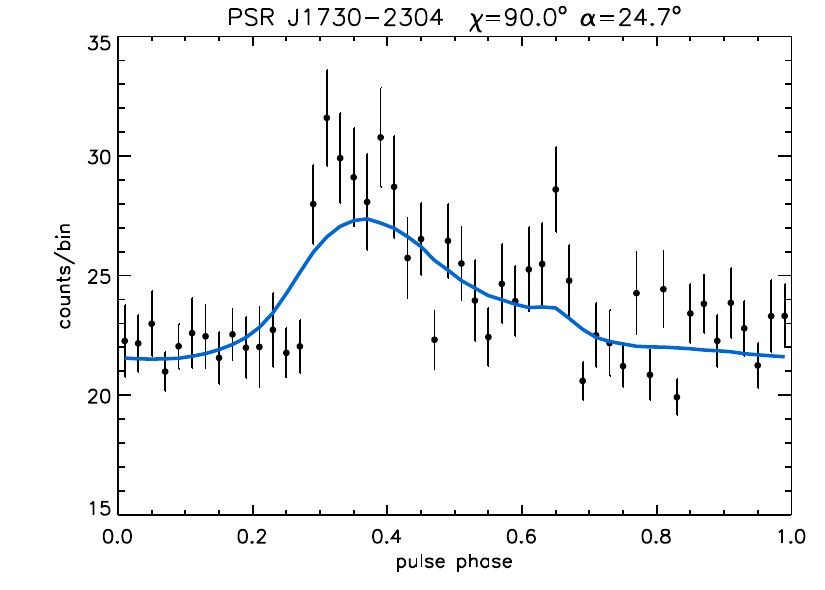}
\includegraphics[width=4.5cm]{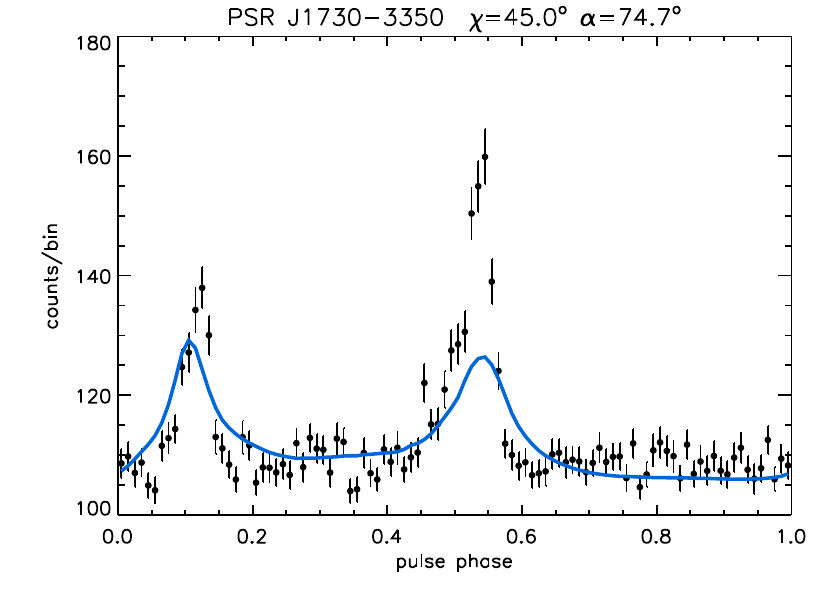}
\includegraphics[width=4.5cm]{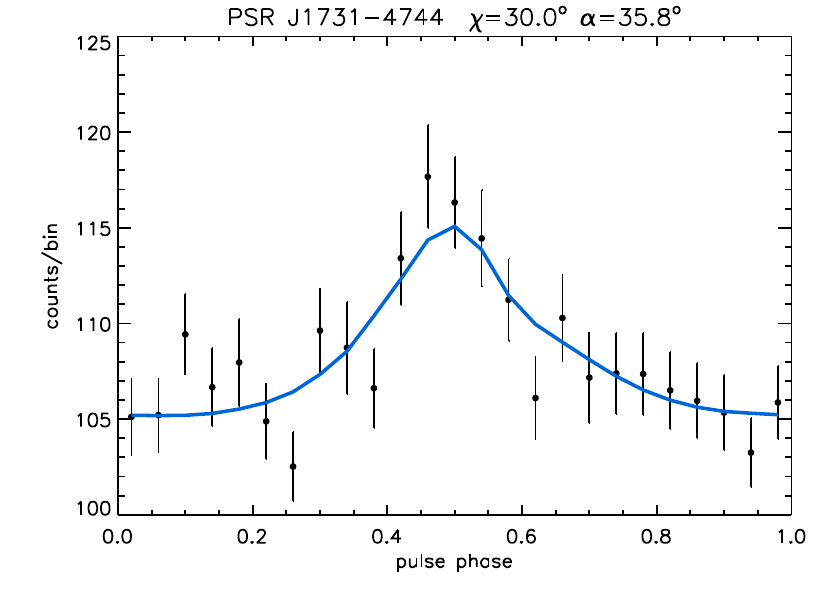}
\includegraphics[width=4.5cm]{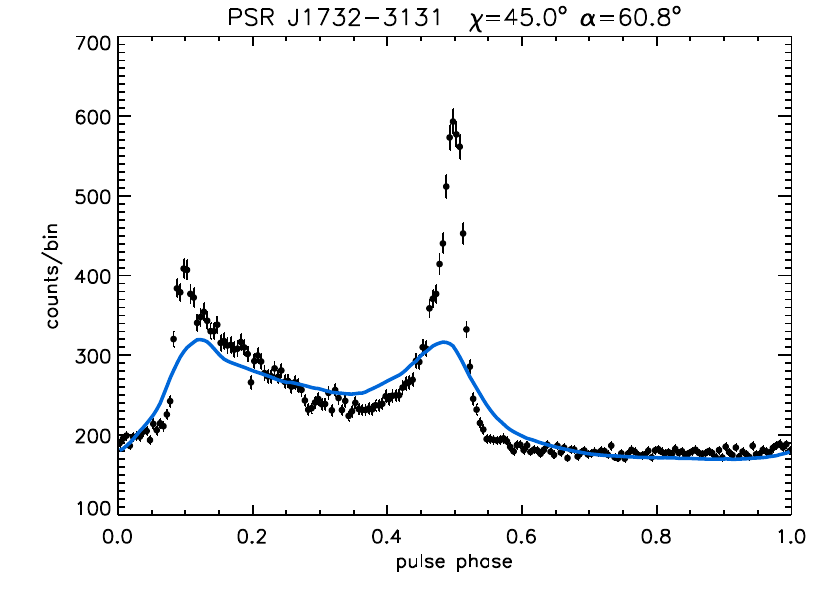}
\includegraphics[width=4.5cm]{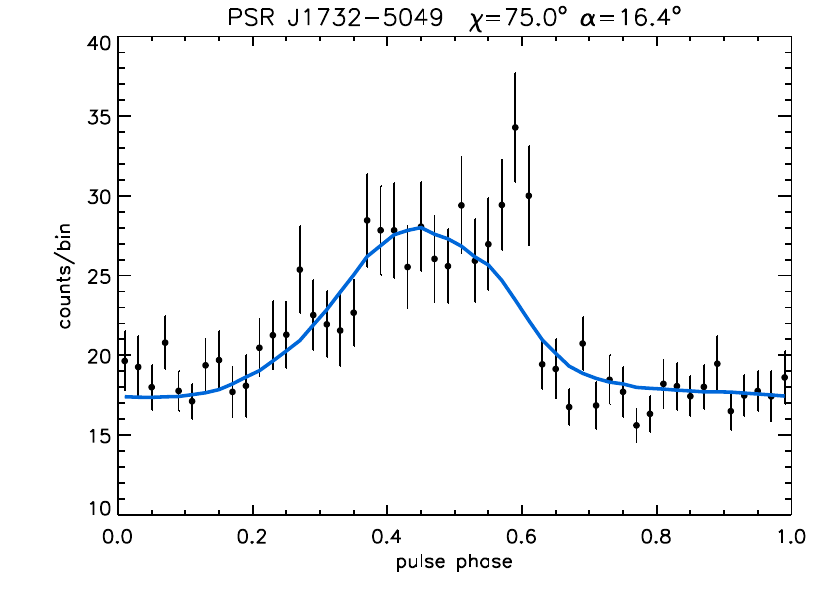}
\includegraphics[width=4.5cm]{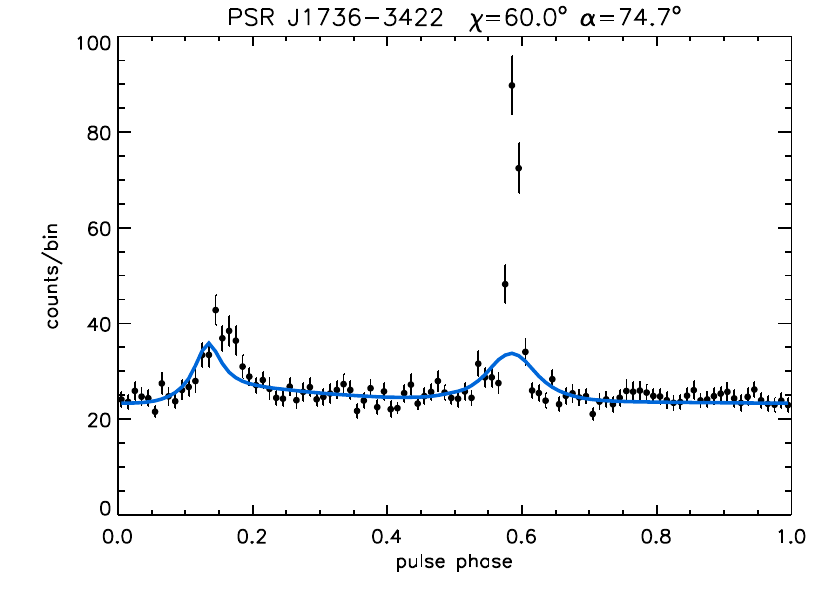}
\includegraphics[width=4.5cm]{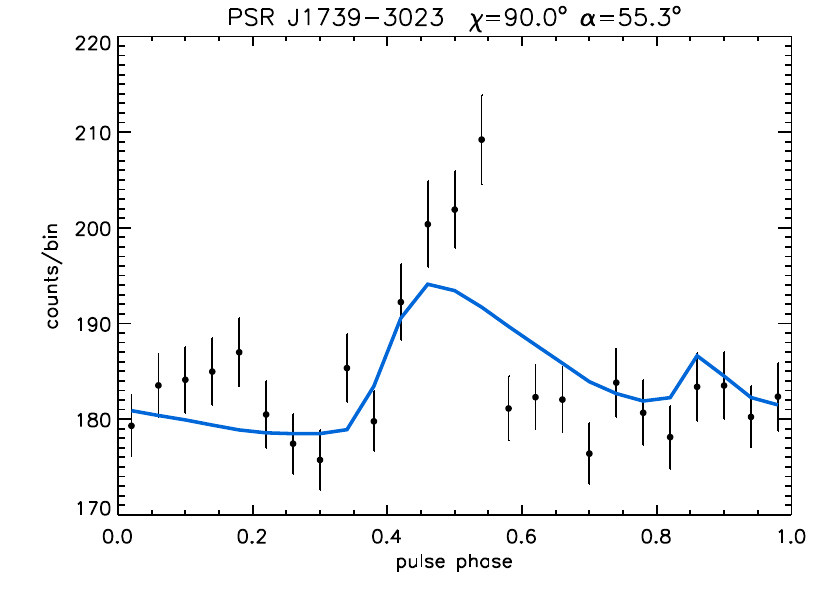}
\includegraphics[width=4.5cm]{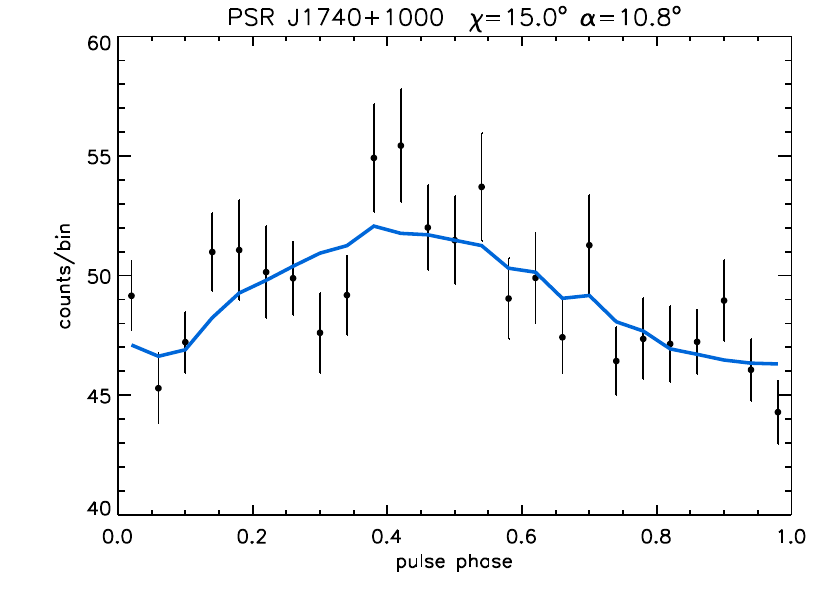}
\includegraphics[width=4.5cm]{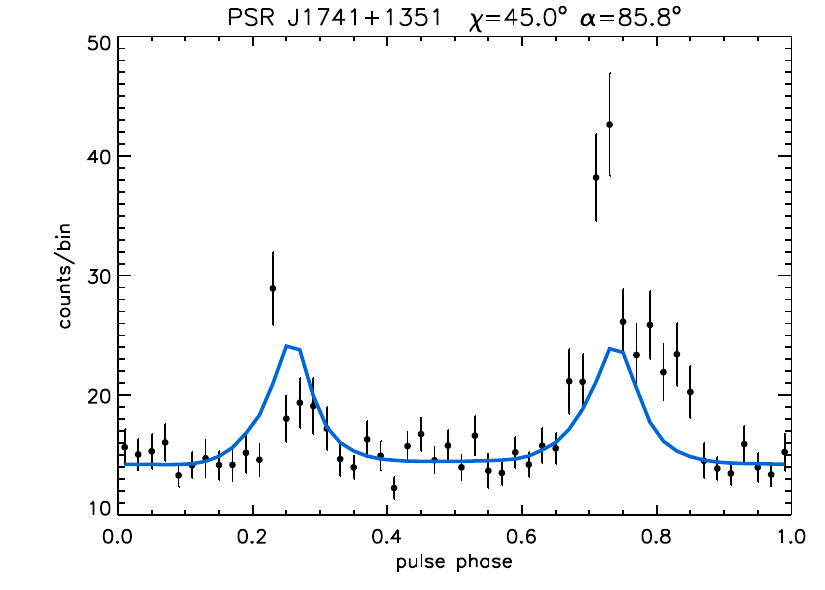}
\includegraphics[width=4.5cm]{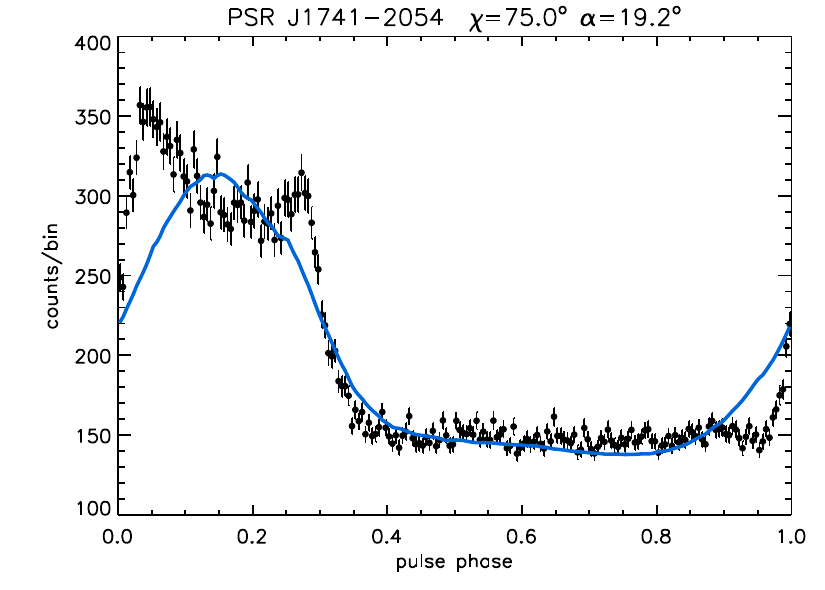}
\includegraphics[width=4.5cm]{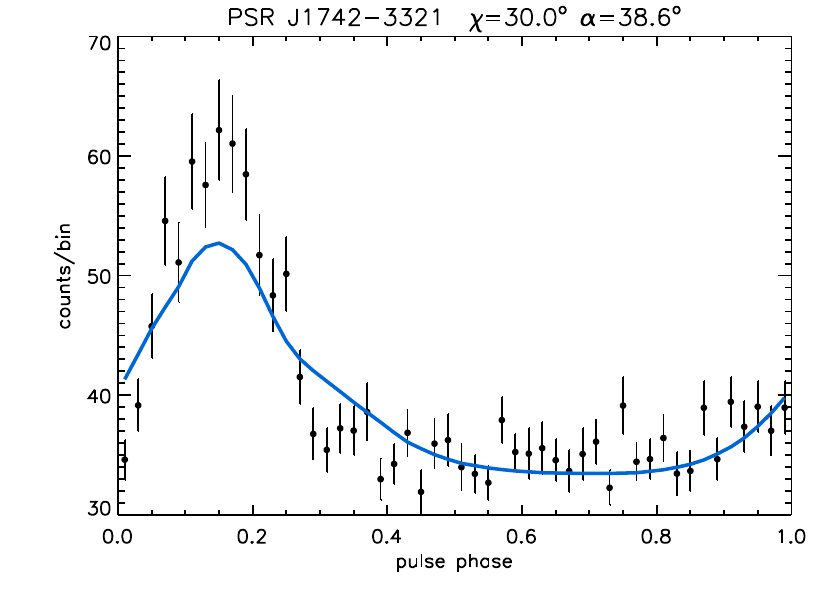}
\includegraphics[width=4.5cm]{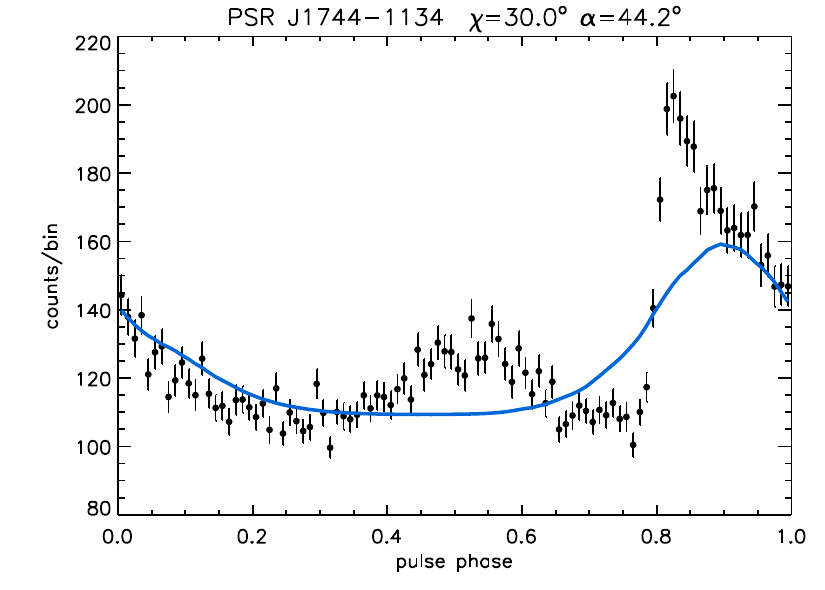}
\includegraphics[width=4.5cm]{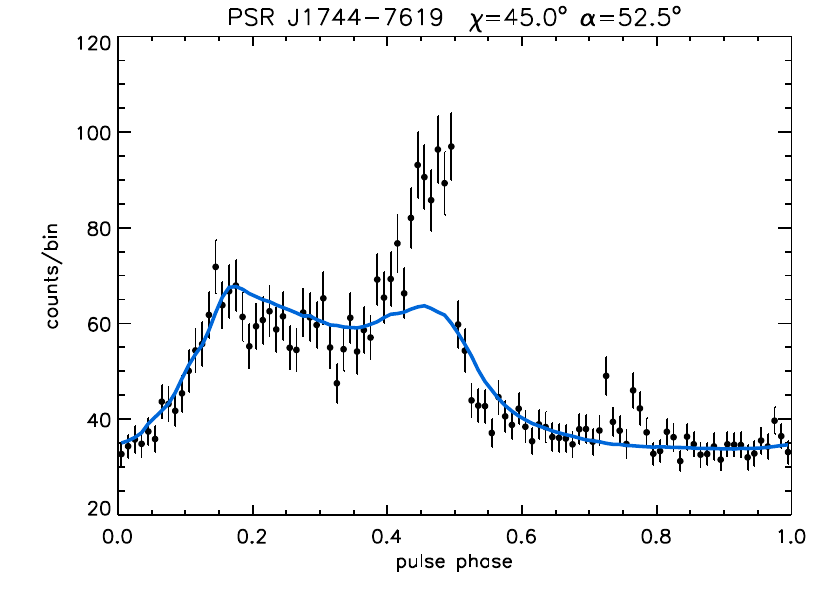}
\includegraphics[width=4.5cm]{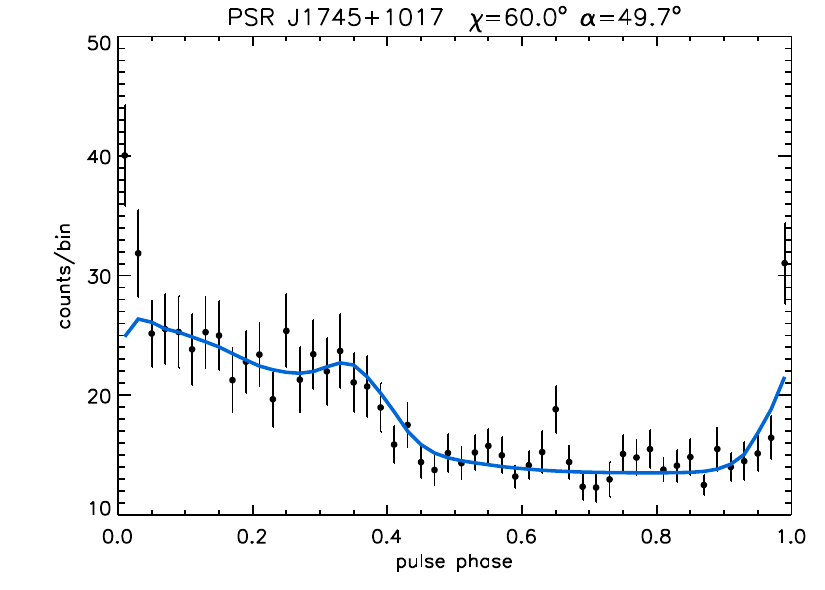}
\includegraphics[width=4.5cm]{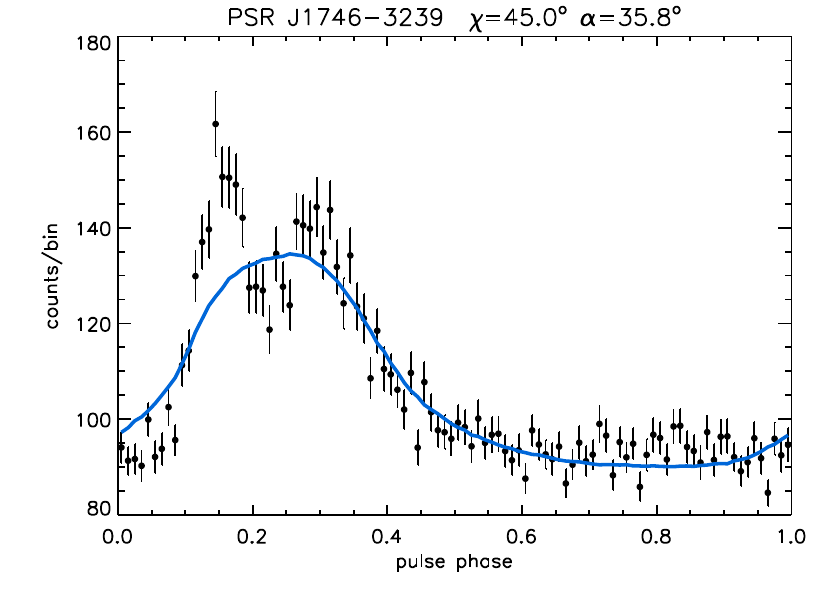}
\includegraphics[width=4.5cm]{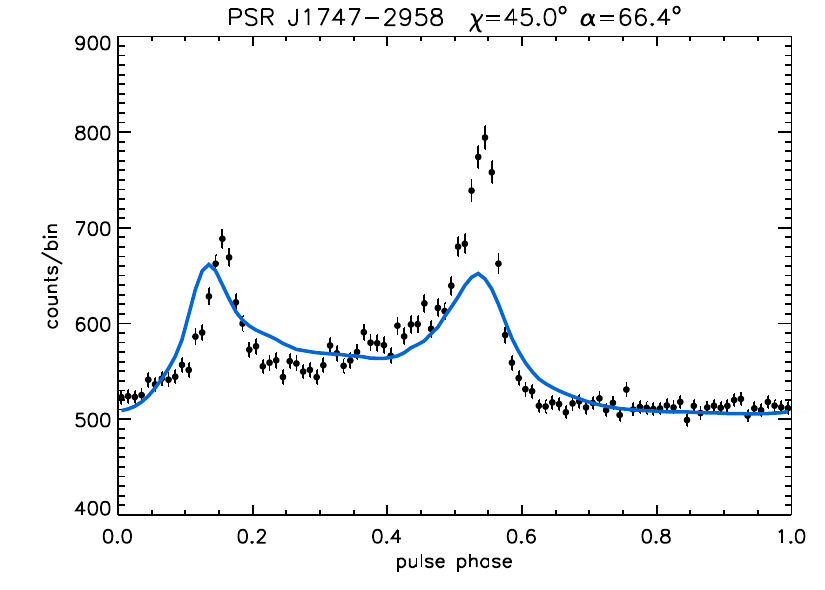}
\includegraphics[width=4.5cm]{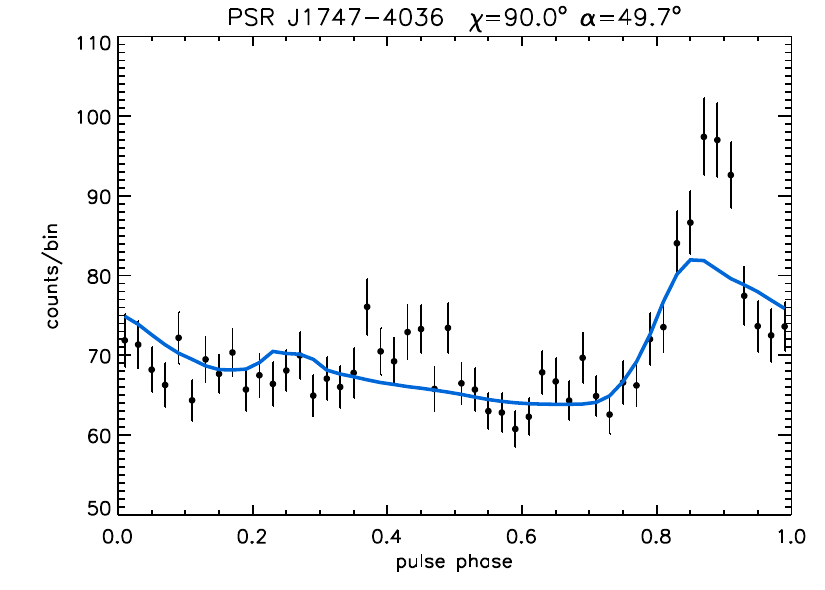}
\includegraphics[width=4.5cm]{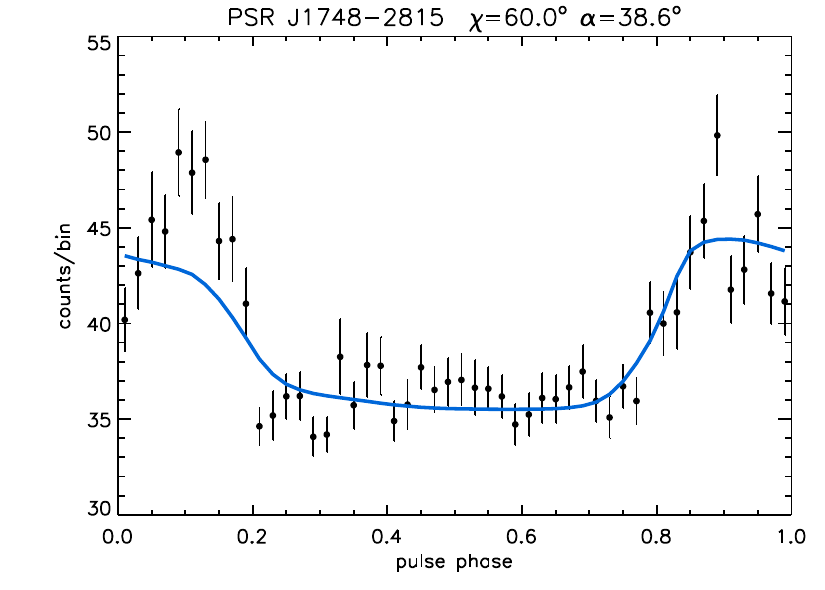}
\includegraphics[width=4.5cm]{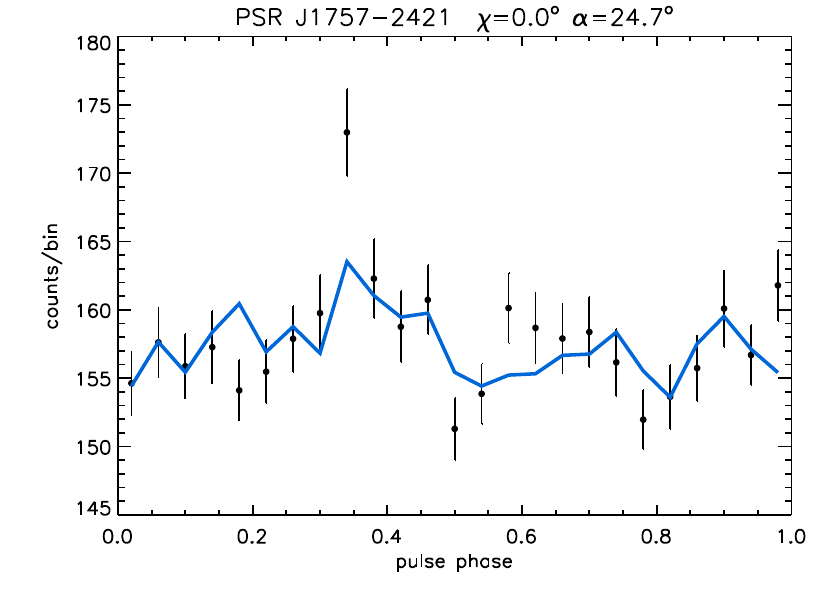}
\includegraphics[width=4.5cm]{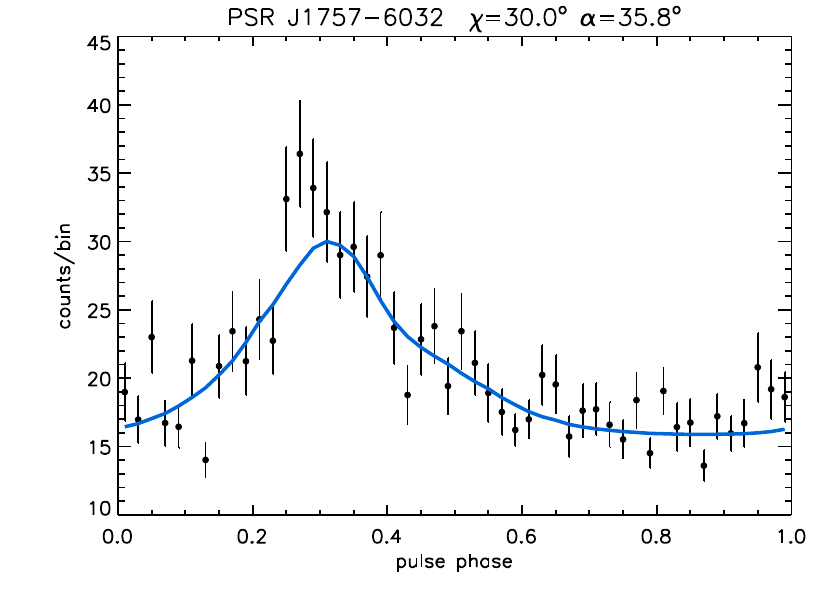}
\includegraphics[width=4.5cm]{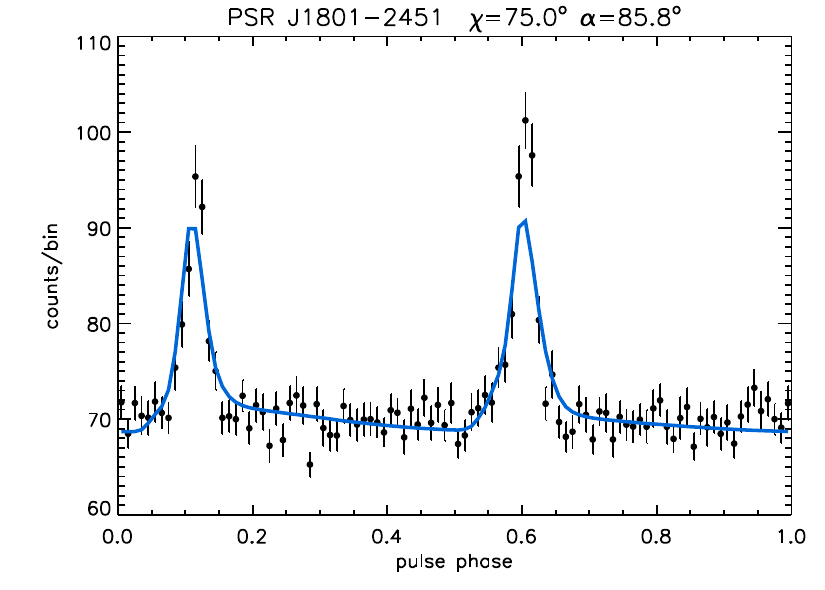}
\includegraphics[width=4.5cm]{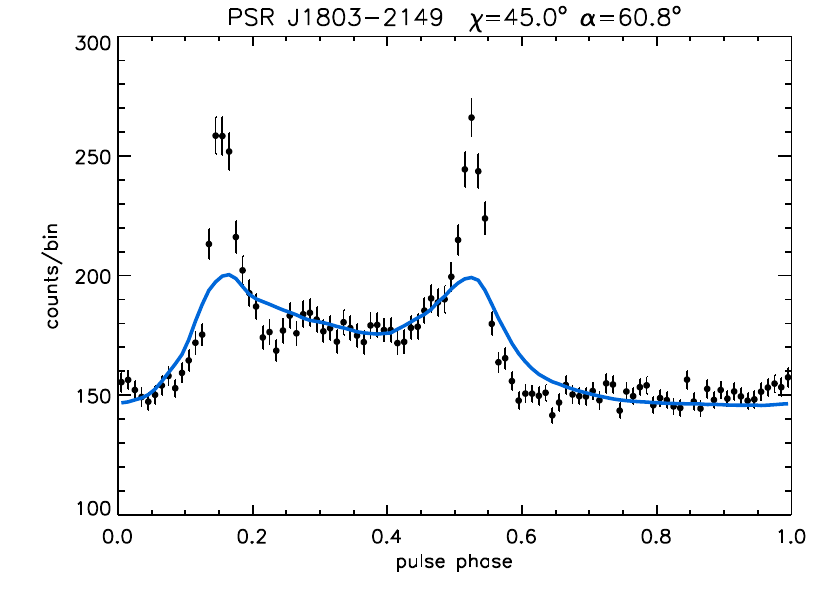}
\includegraphics[width=4.5cm]{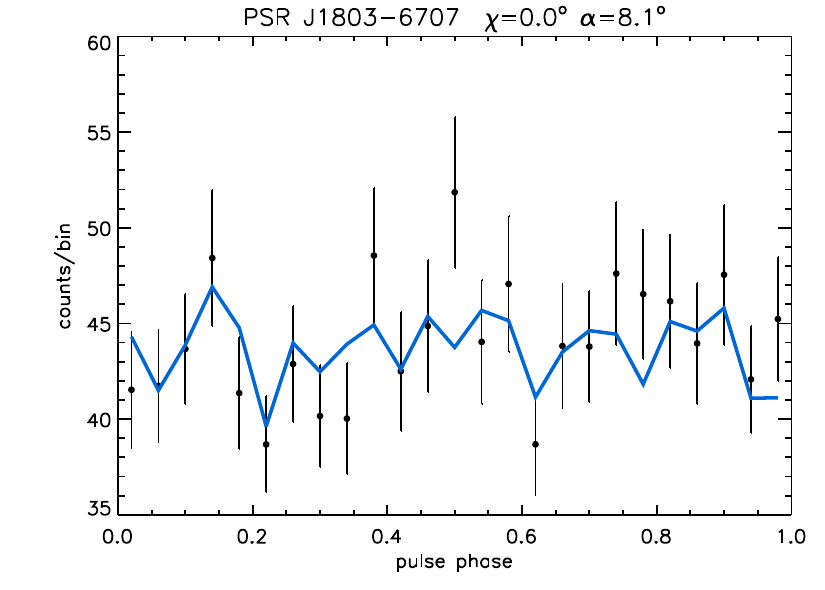}
\includegraphics[width=4.5cm]{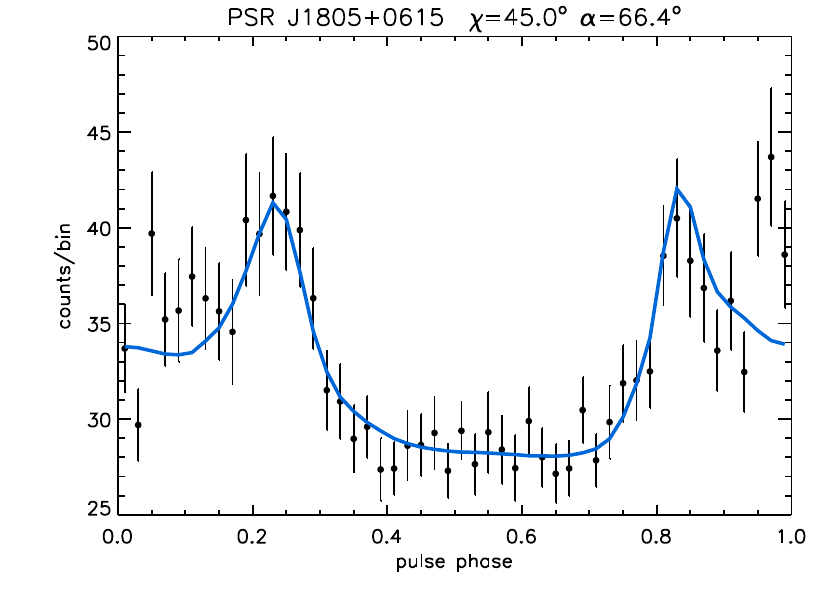}
\includegraphics[width=4.5cm]{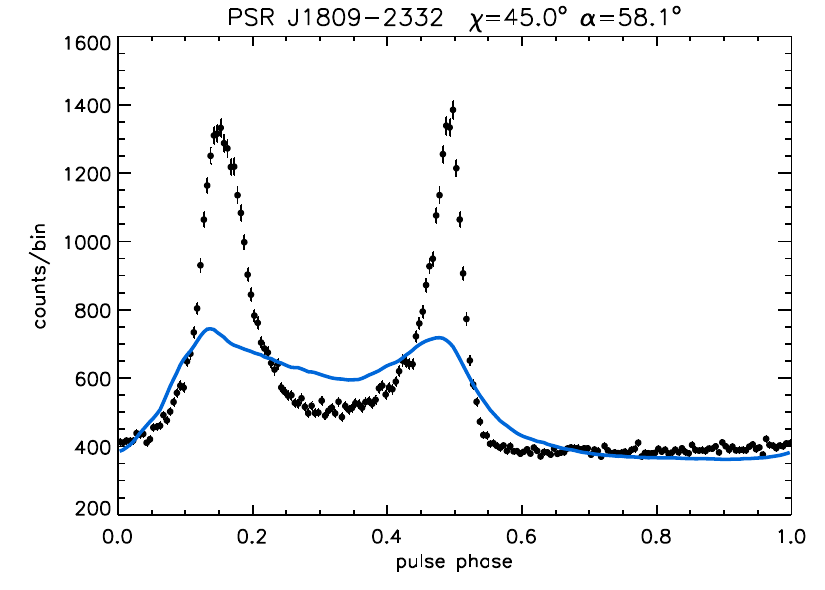}
\includegraphics[width=4.5cm]{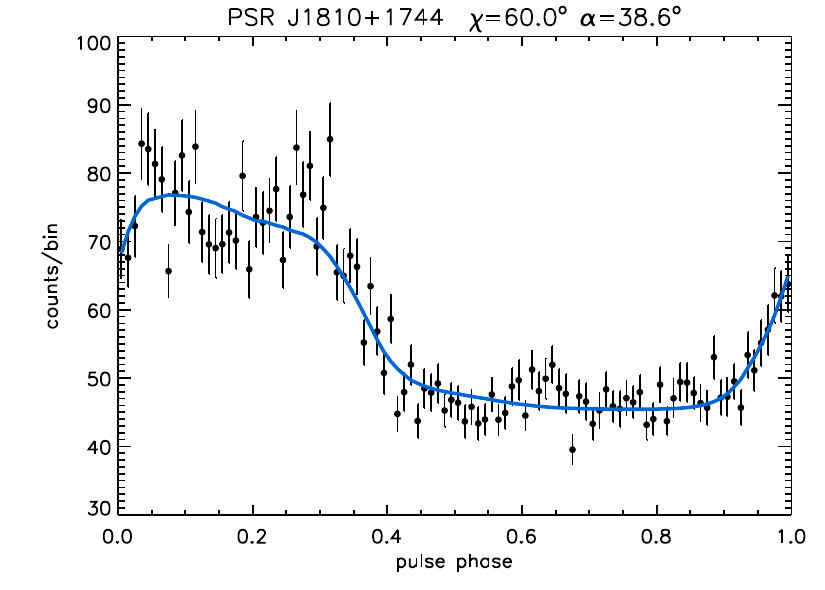}
\includegraphics[width=4.5cm]{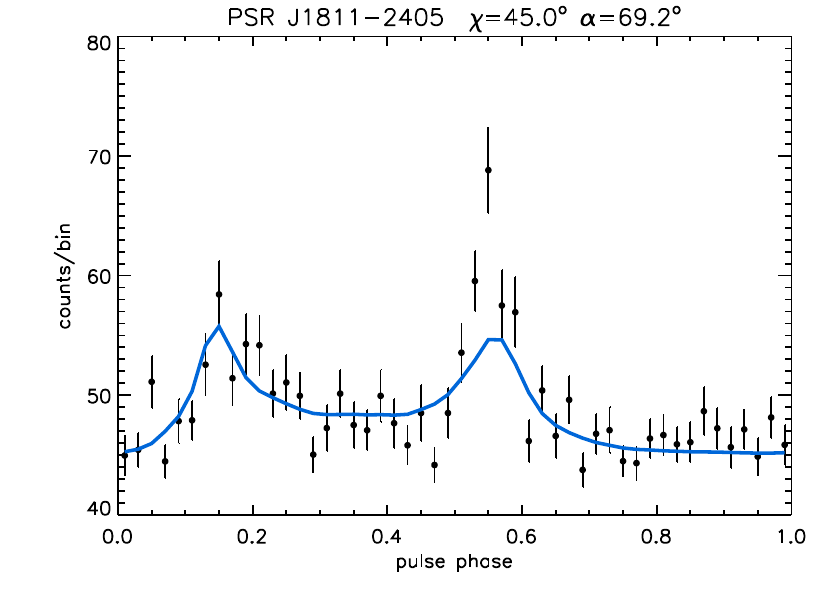}
\caption{continued.}
\end{figure*}

\begin{figure*}
\addtocounter{figure}{-1}
\centering
\includegraphics[width=4.5cm]{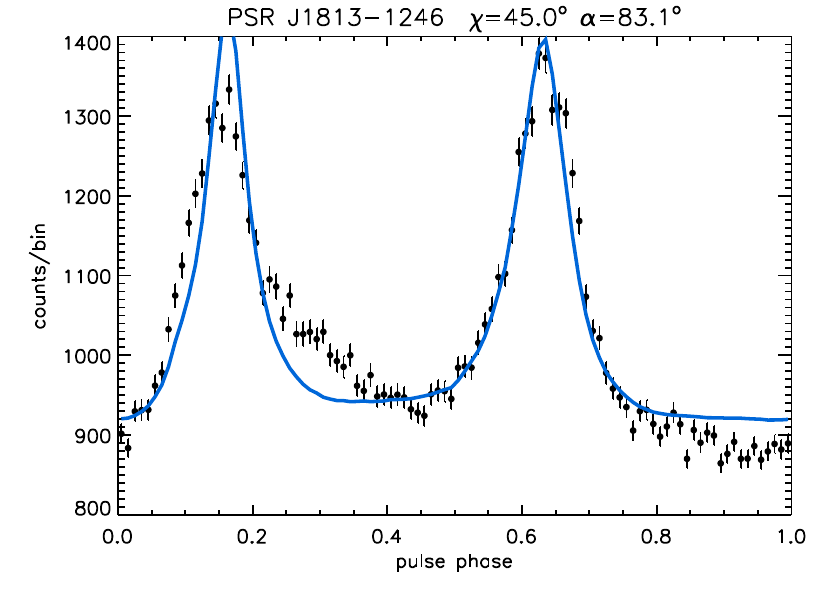}
\includegraphics[width=4.5cm]{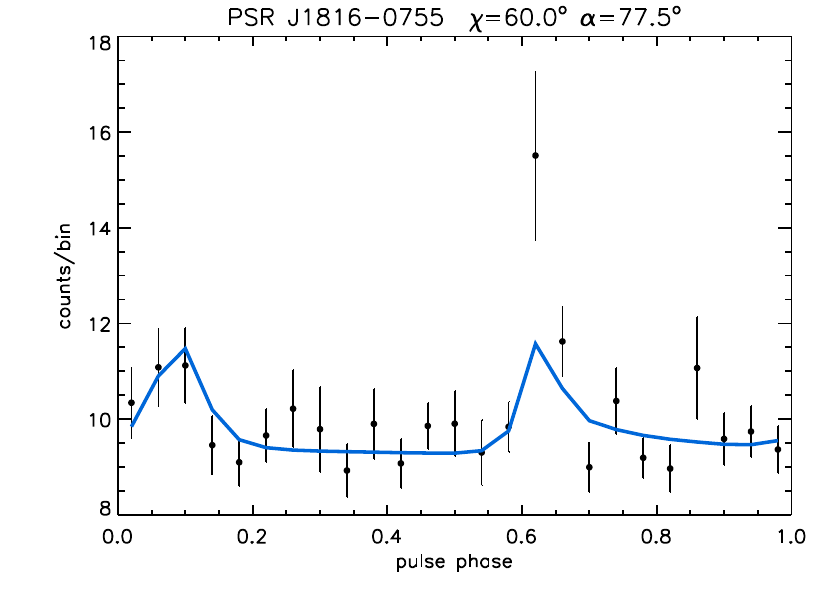}
\includegraphics[width=4.5cm]{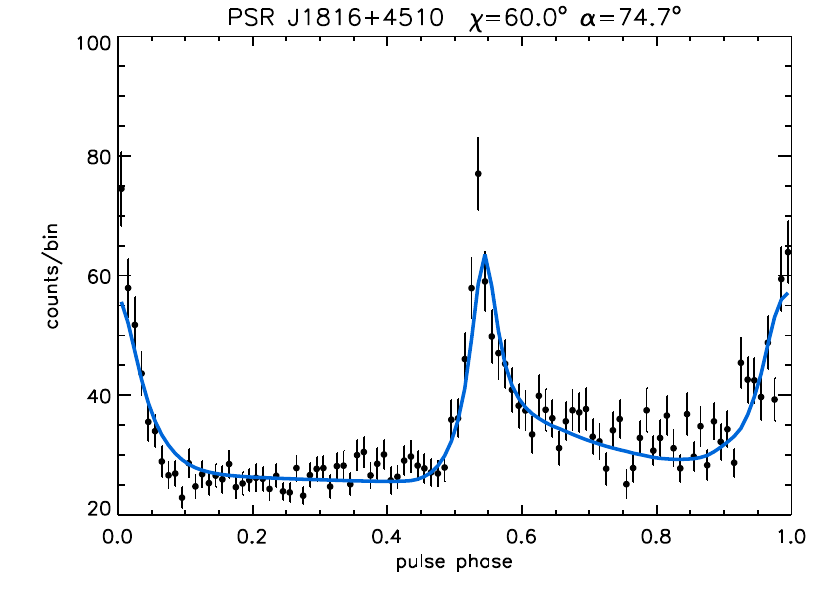}
\includegraphics[width=4.5cm]{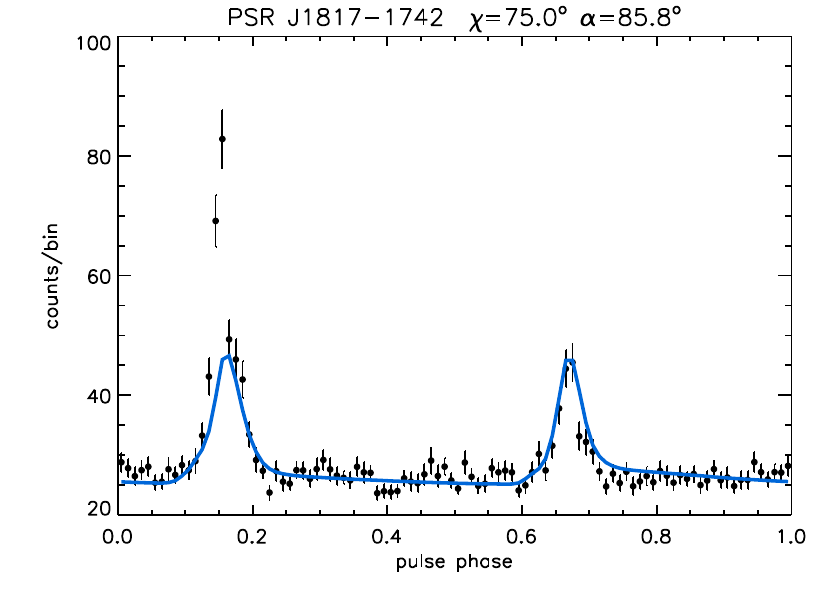}
\includegraphics[width=4.5cm]{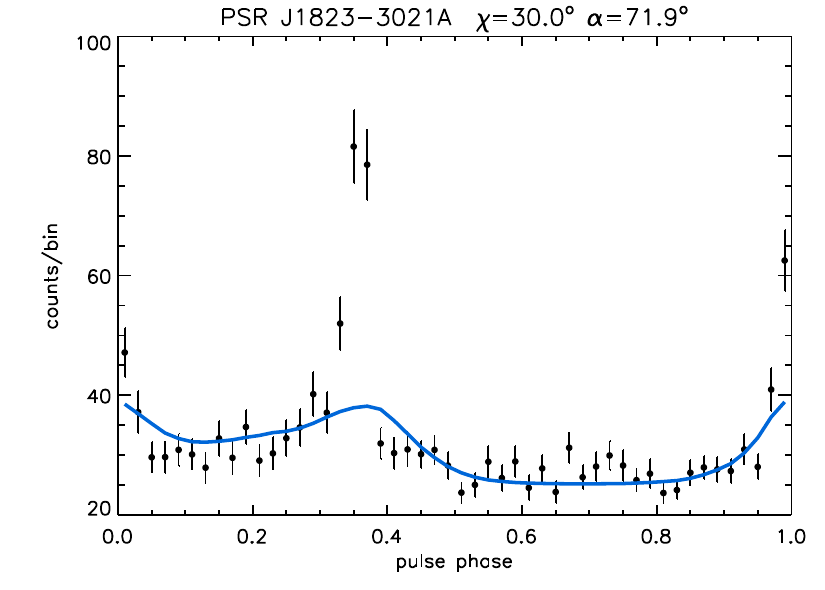}
\includegraphics[width=4.5cm]{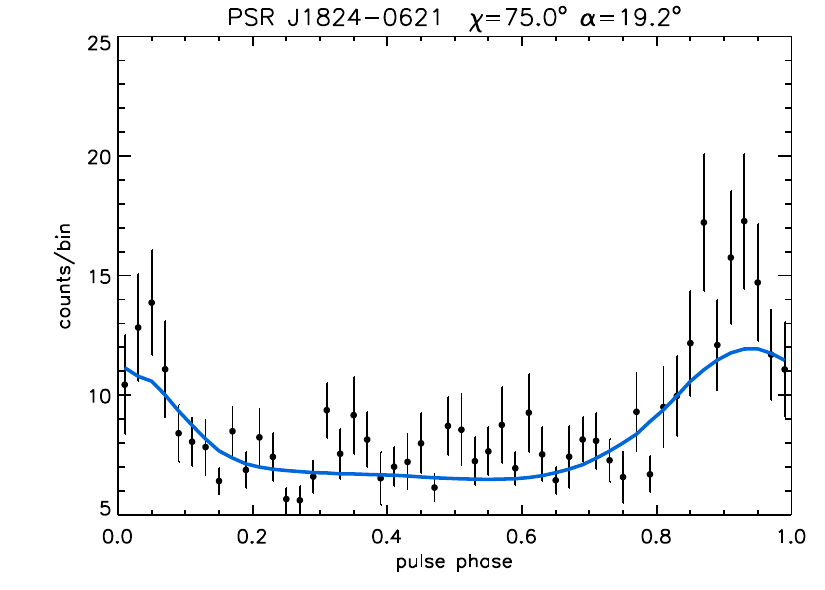}
\includegraphics[width=4.5cm]{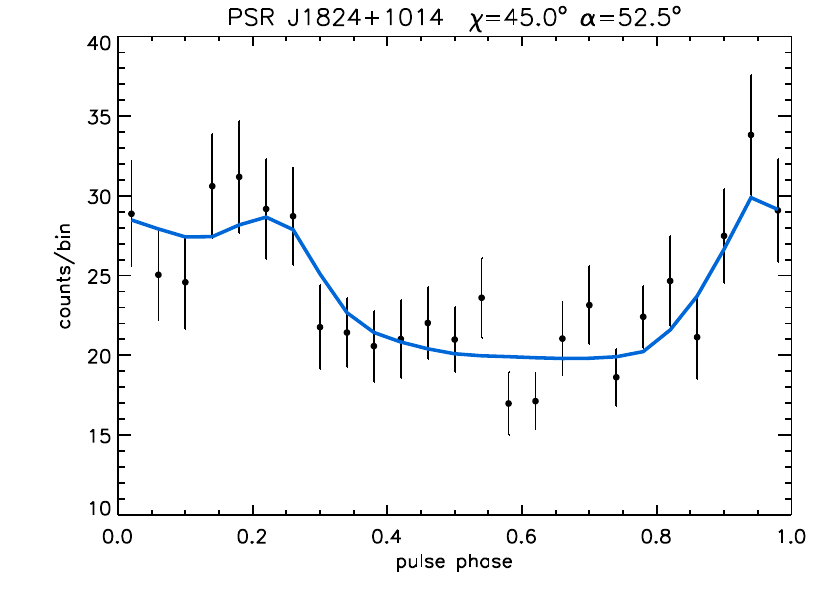}
\includegraphics[width=4.5cm]{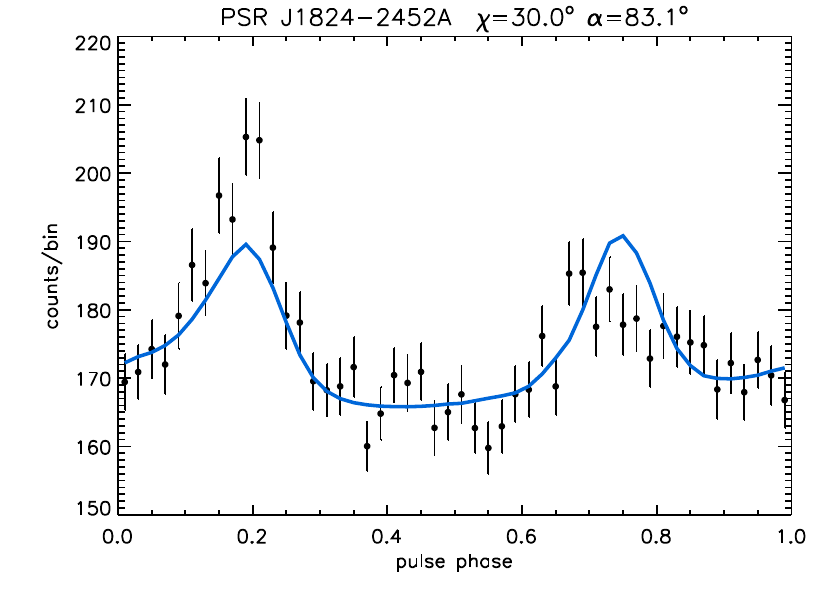}
\includegraphics[width=4.5cm]{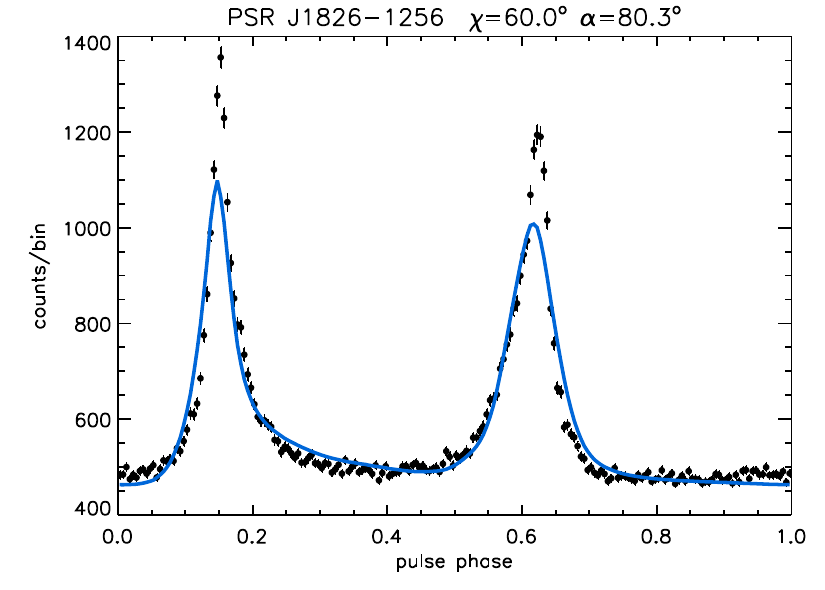}
\includegraphics[width=4.5cm]{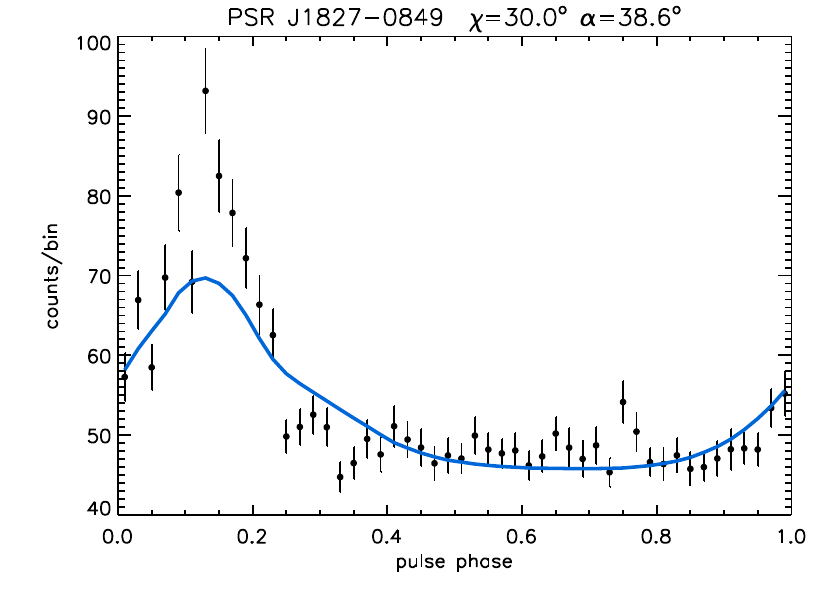}
\includegraphics[width=4.5cm]{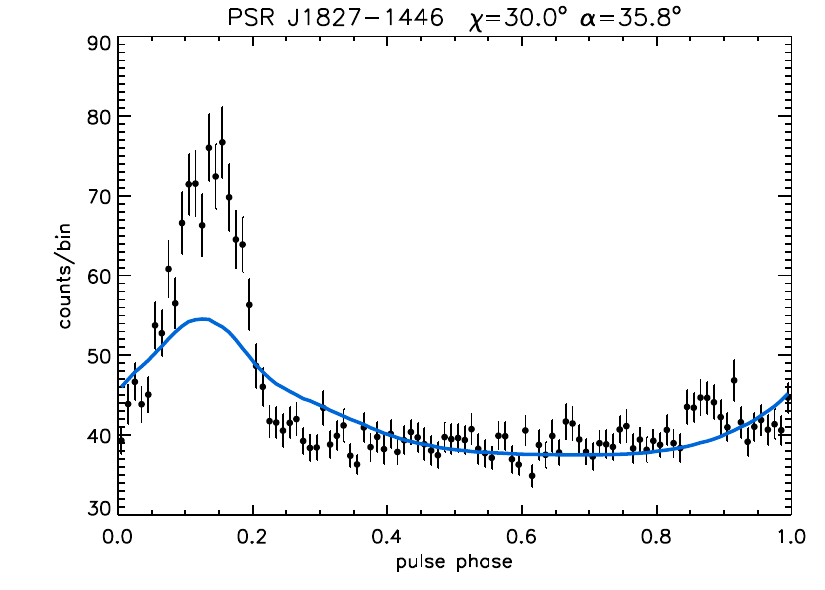}
\includegraphics[width=4.5cm]{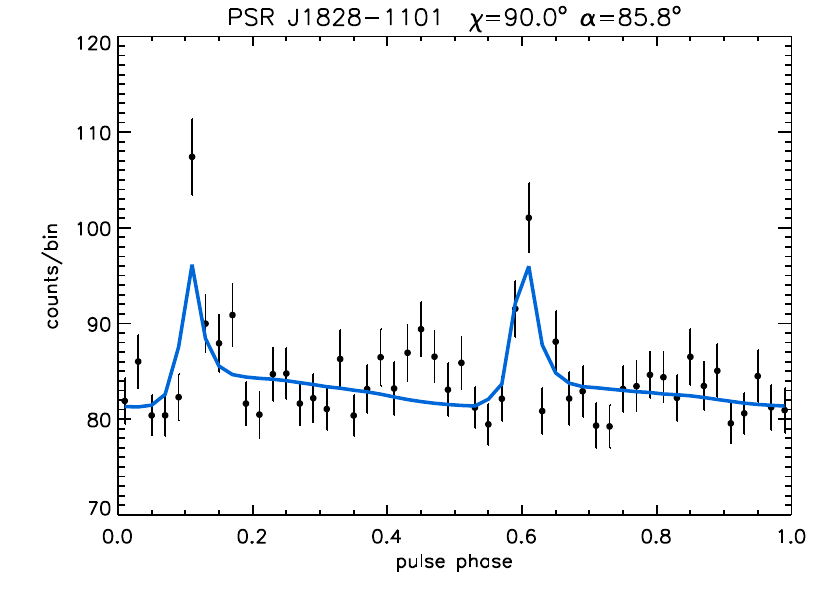}
\includegraphics[width=4.5cm]{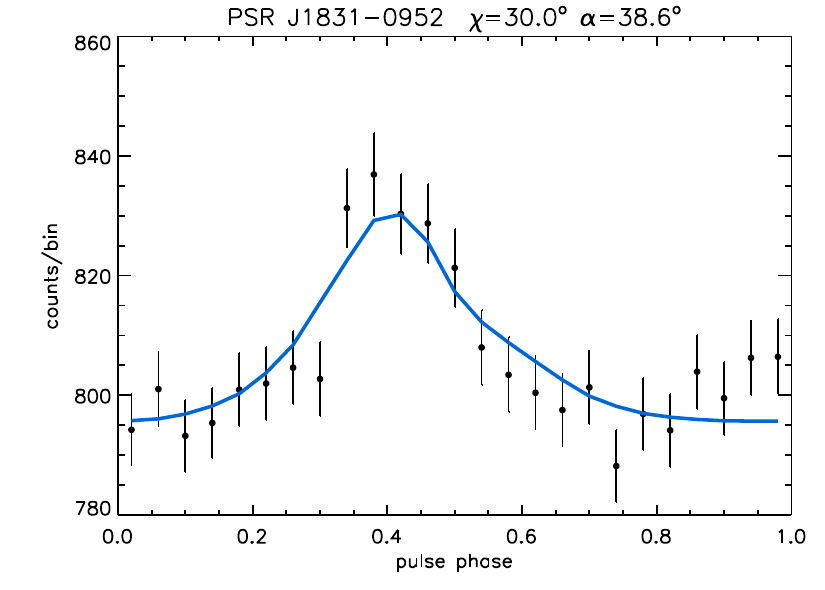}
\includegraphics[width=4.5cm]{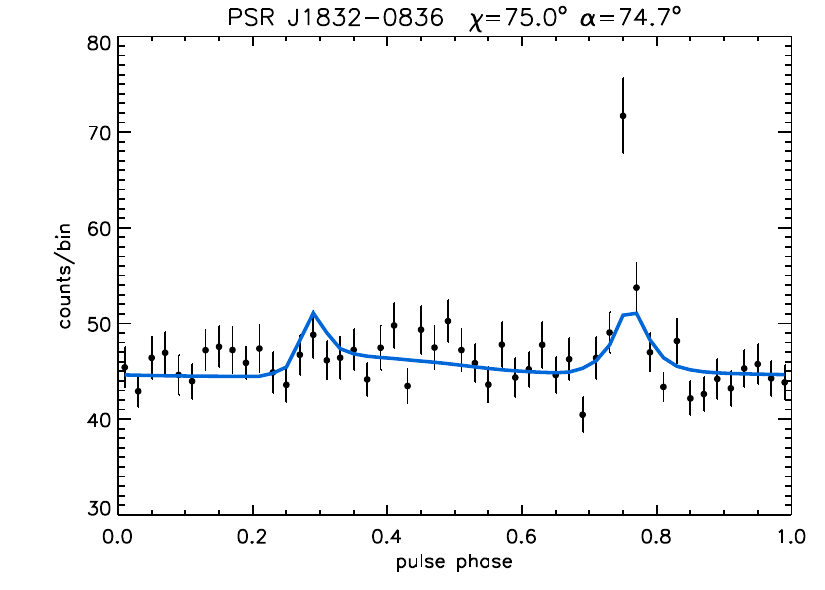}
\includegraphics[width=4.5cm]{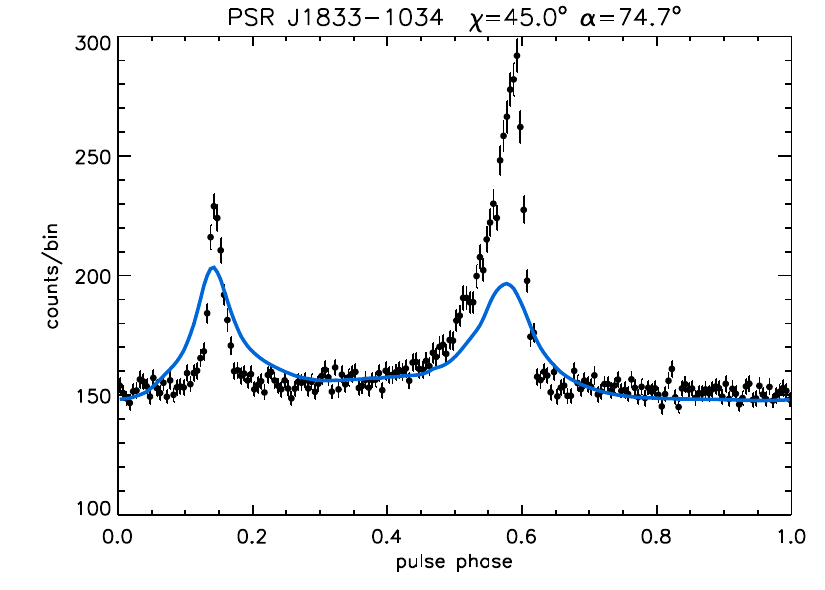}
\includegraphics[width=4.5cm]{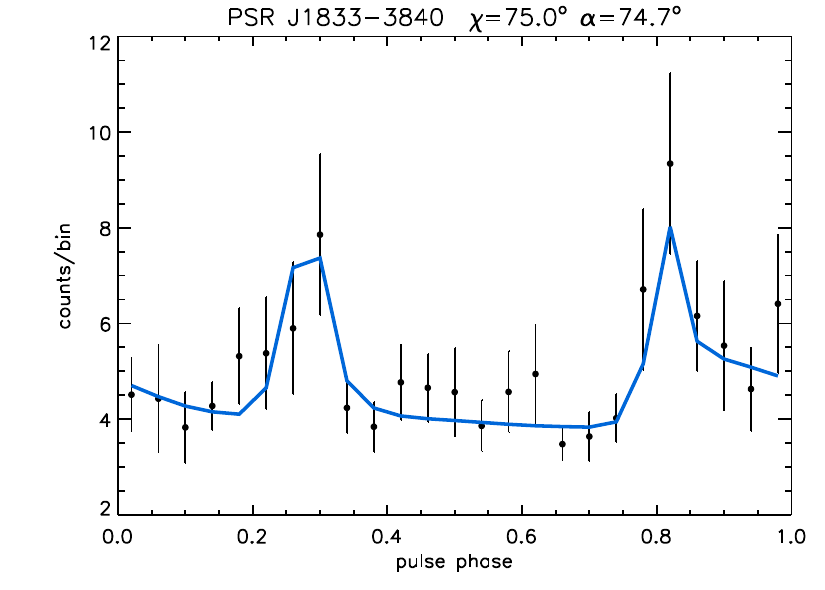}
\includegraphics[width=4.5cm]{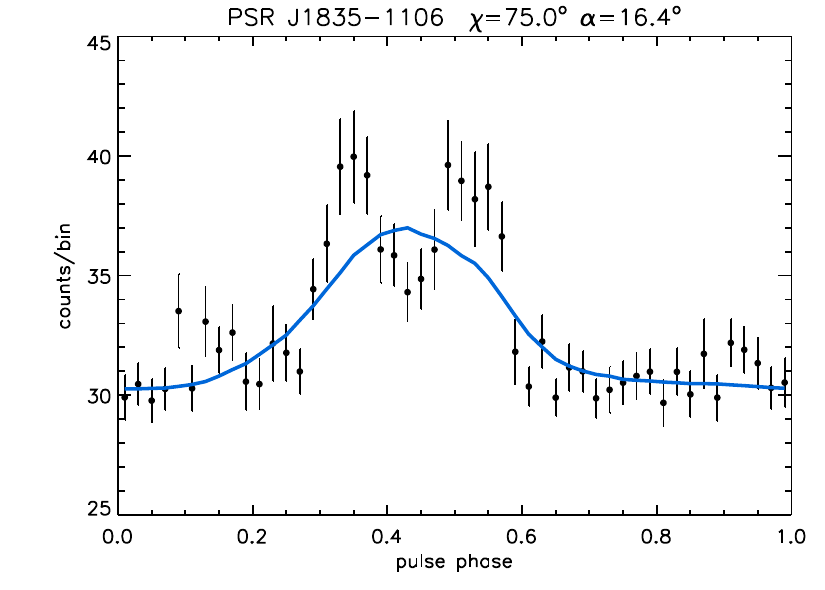}
\includegraphics[width=4.5cm]{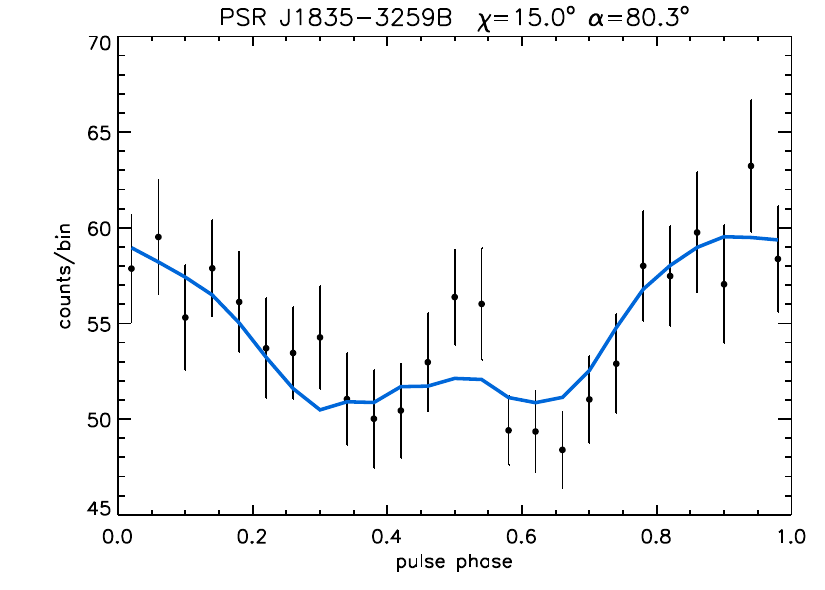}
\includegraphics[width=4.5cm]{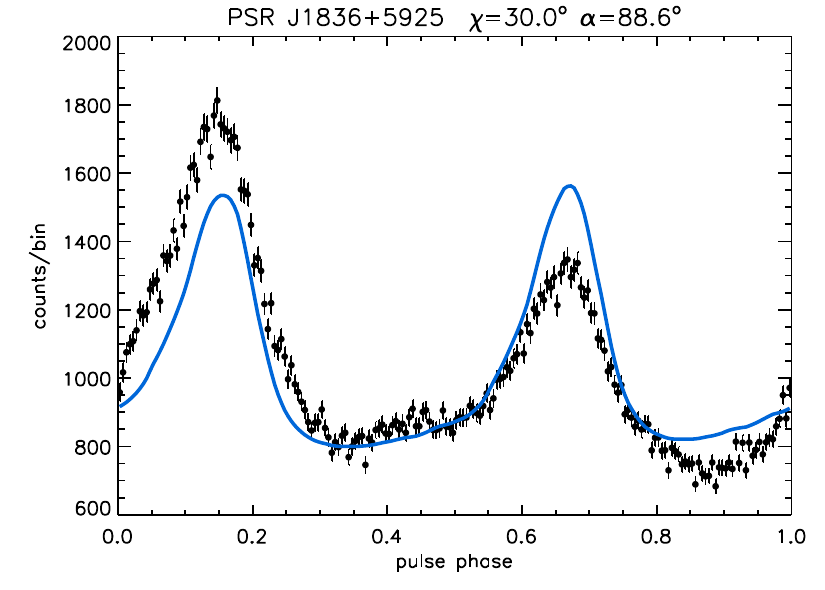}
\includegraphics[width=4.5cm]{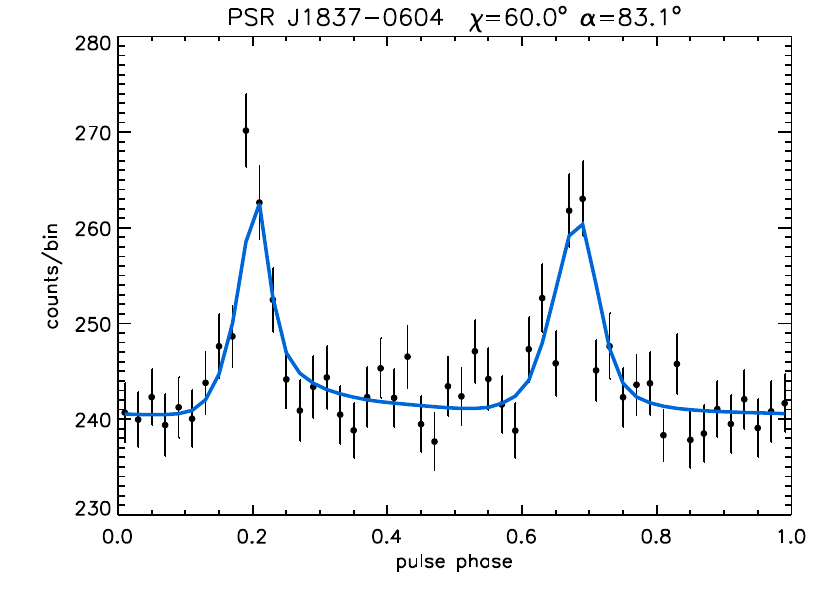}
\includegraphics[width=4.5cm]{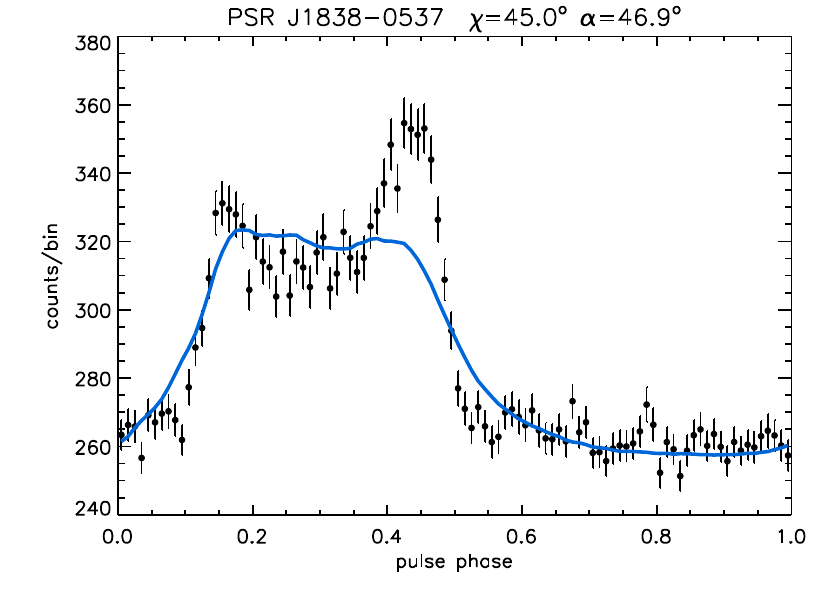}
\includegraphics[width=4.5cm]{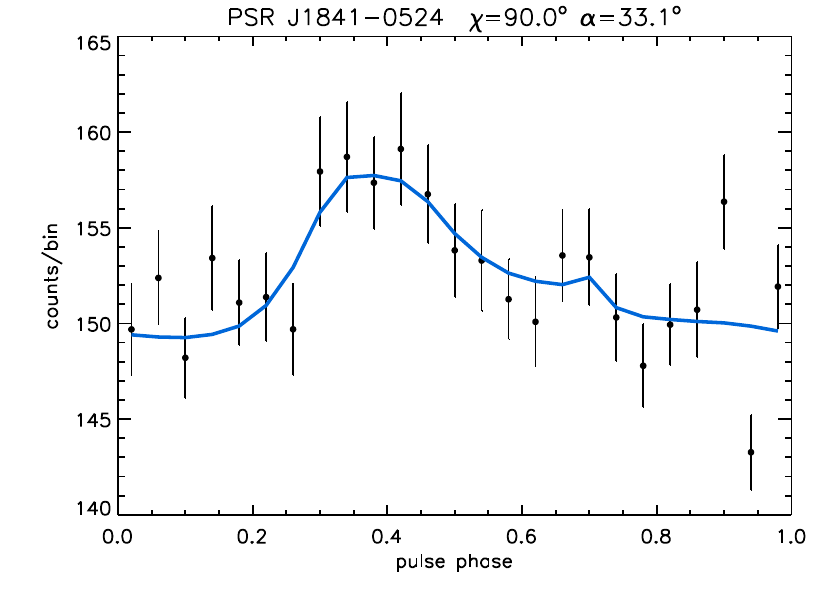}
\includegraphics[width=4.5cm]{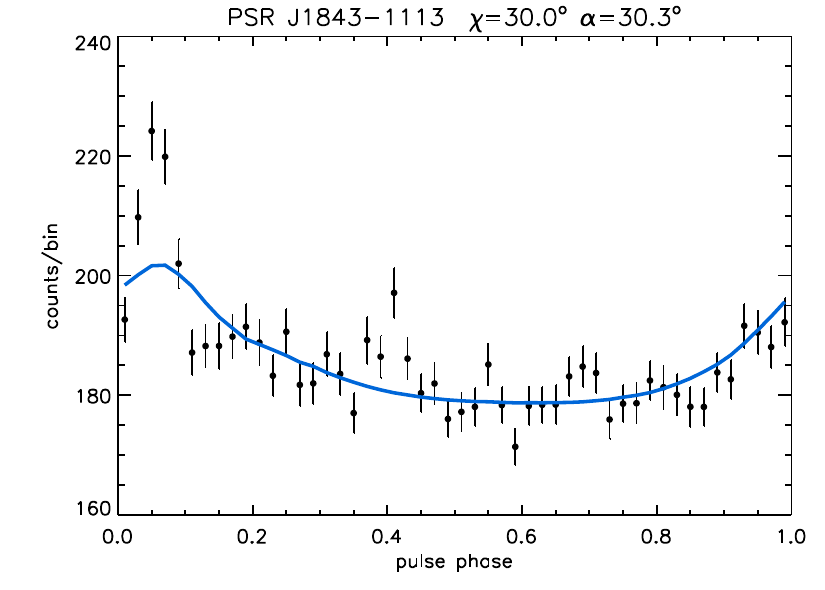}
\includegraphics[width=4.5cm]{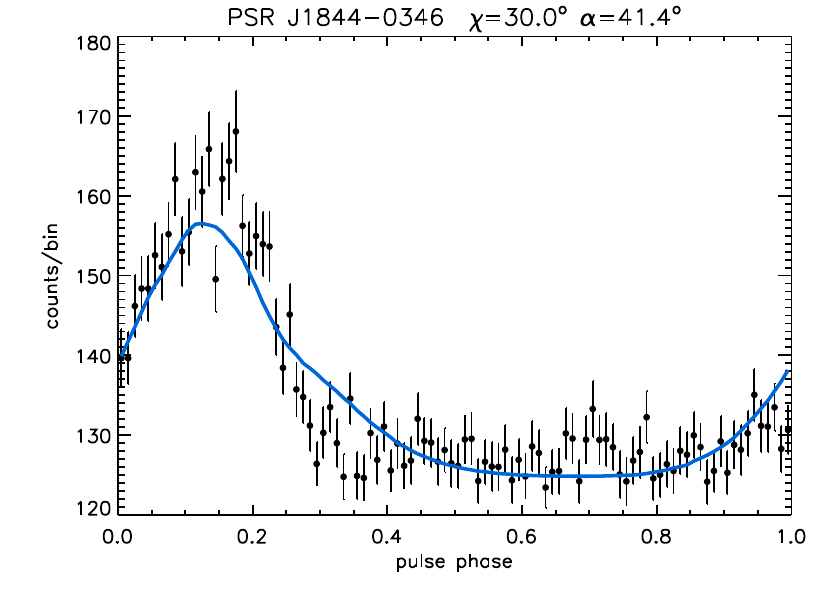}
\includegraphics[width=4.5cm]{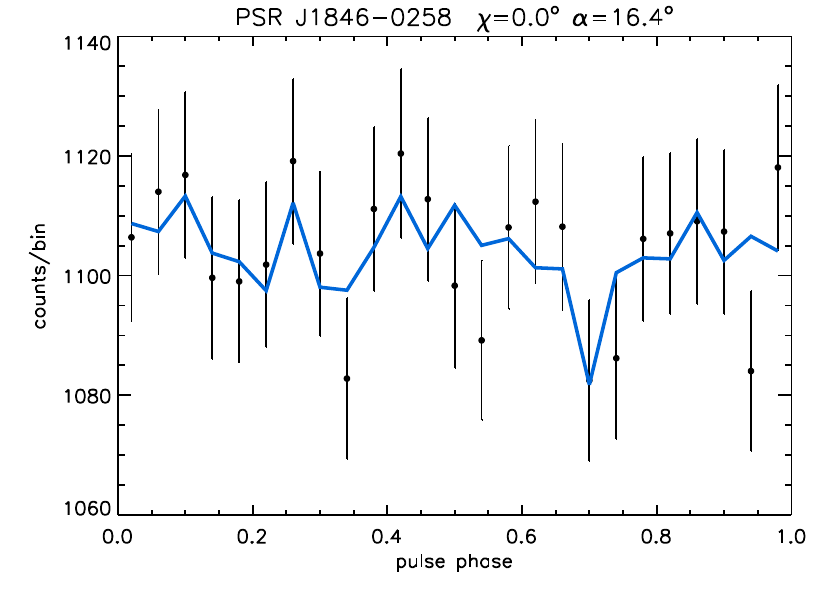}
\includegraphics[width=4.5cm]{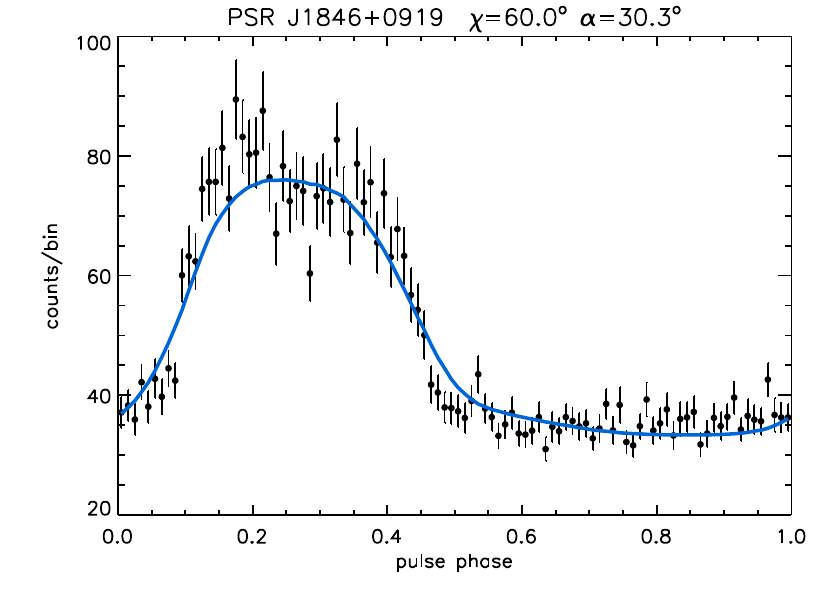}
\includegraphics[width=4.5cm]{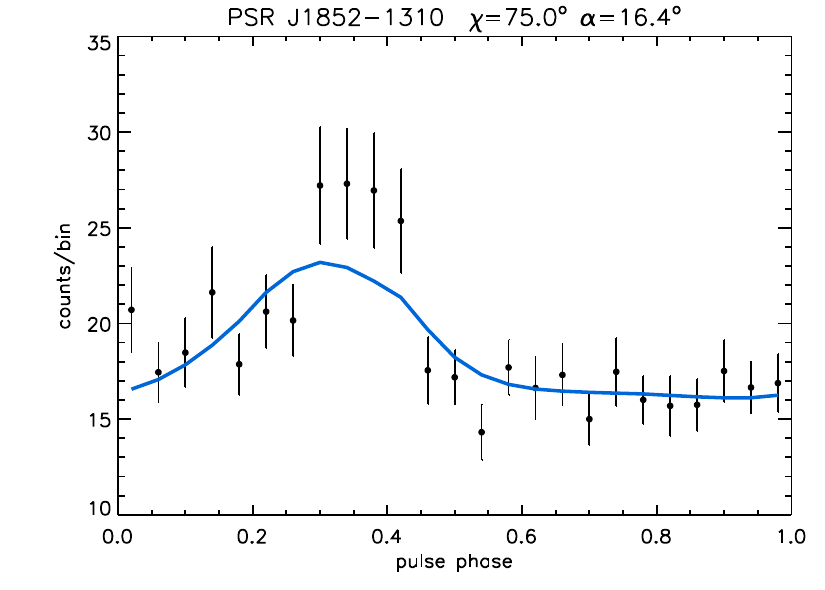}
\includegraphics[width=4.5cm]{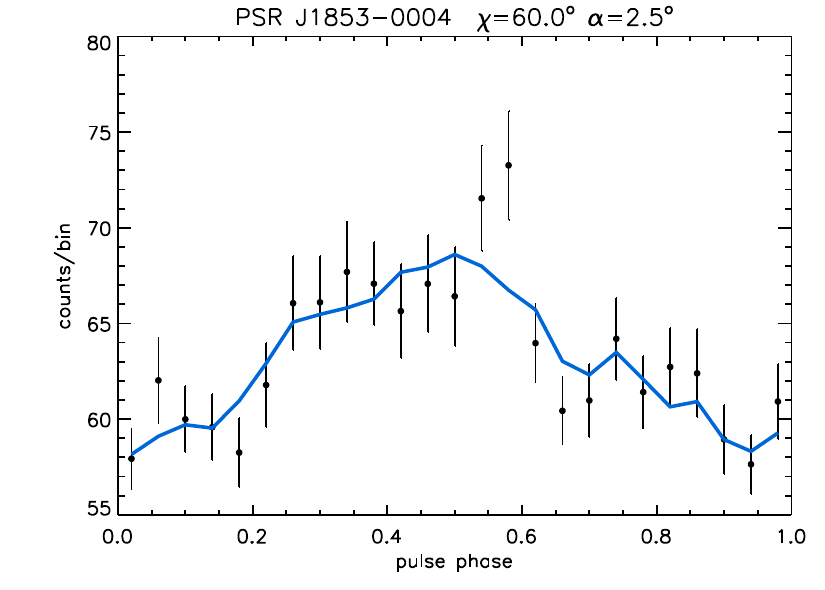}
\caption{continued.}
\end{figure*}

\begin{figure*}
\addtocounter{figure}{-1}
\centering
\includegraphics[width=4.5cm]{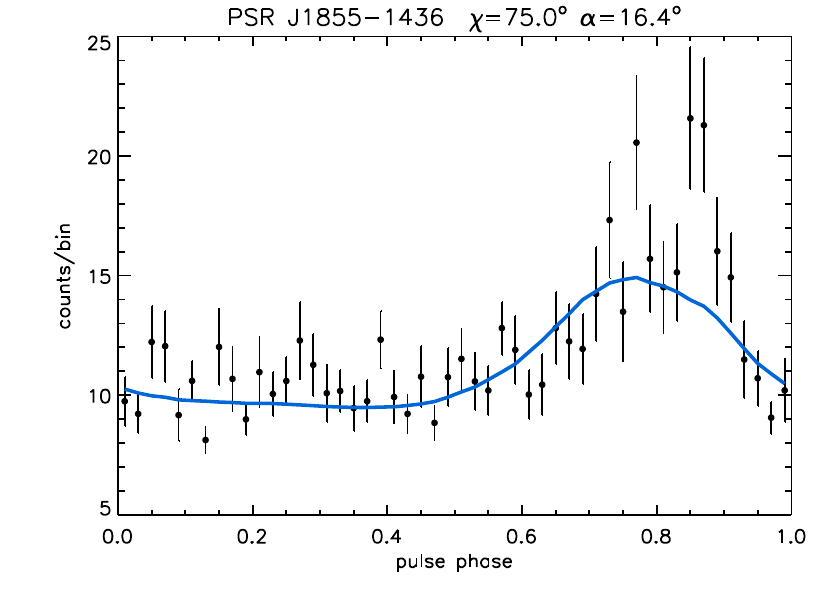}
\includegraphics[width=4.5cm]{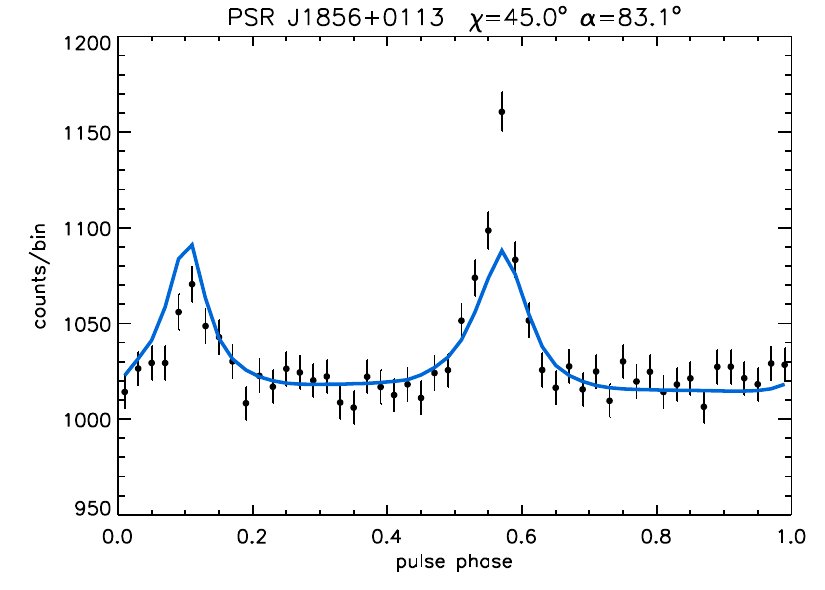}
\includegraphics[width=4.5cm]{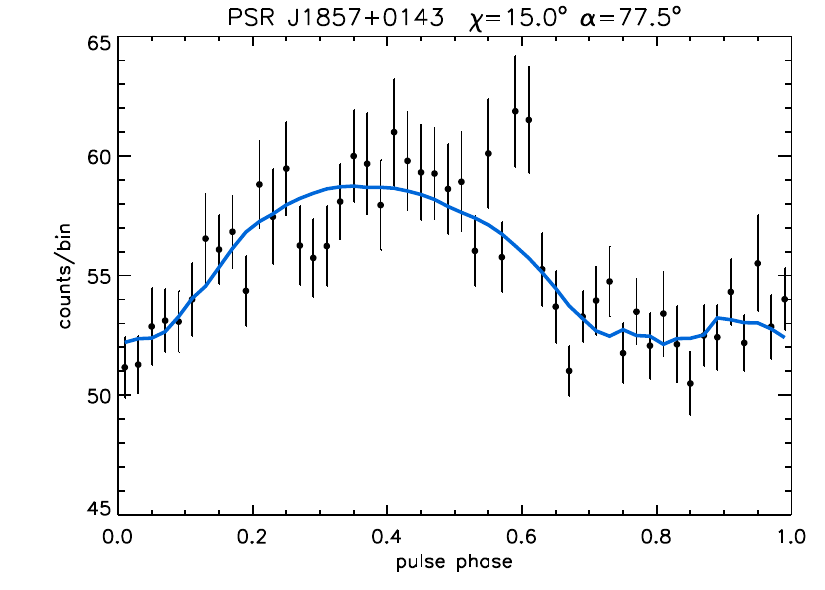}
\includegraphics[width=4.5cm]{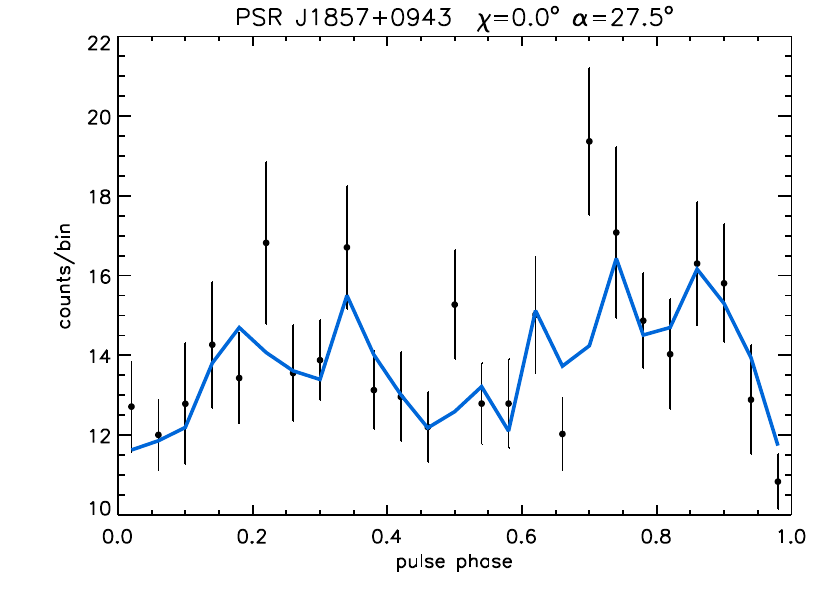}
\includegraphics[width=4.5cm]{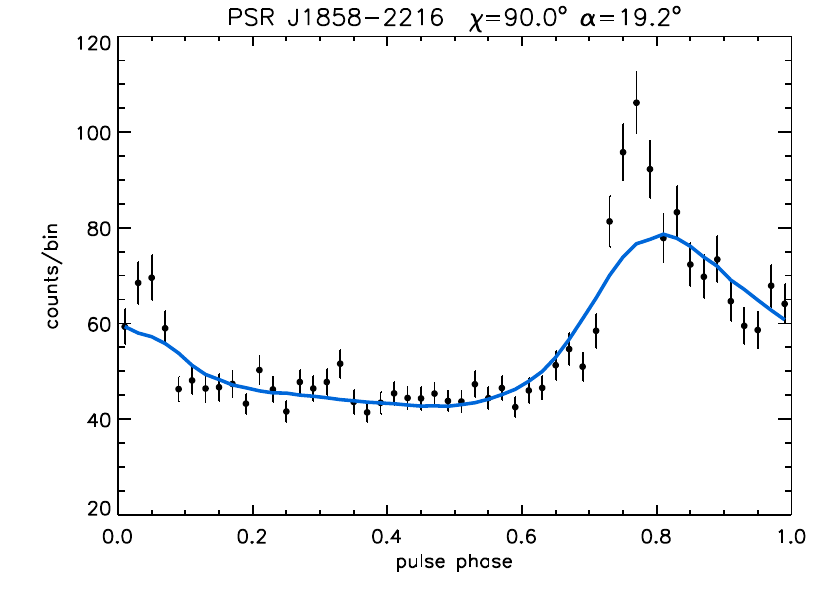}
\includegraphics[width=4.5cm]{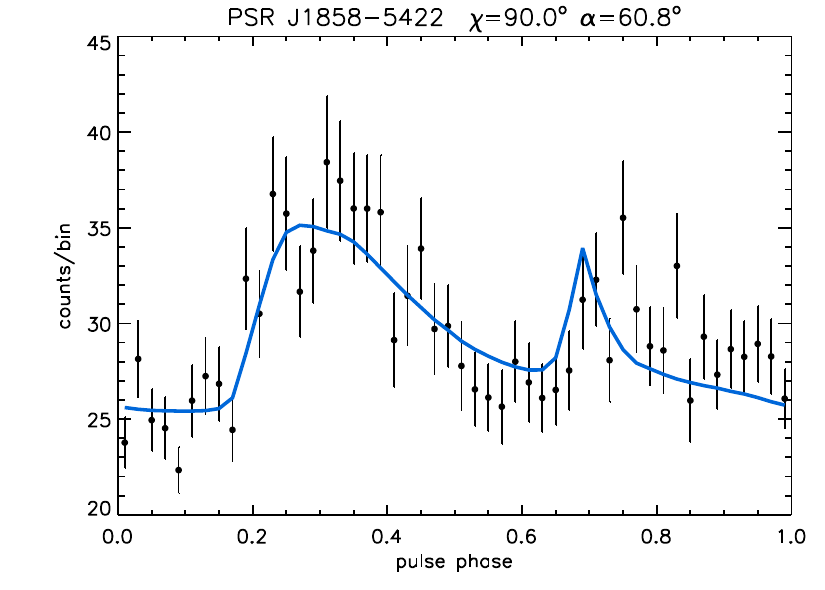}
\includegraphics[width=4.5cm]{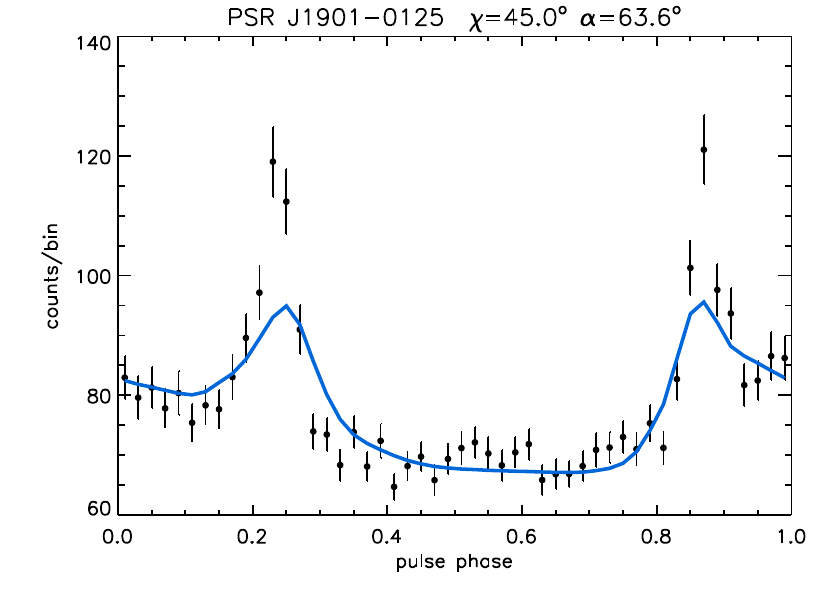}
\includegraphics[width=4.5cm]{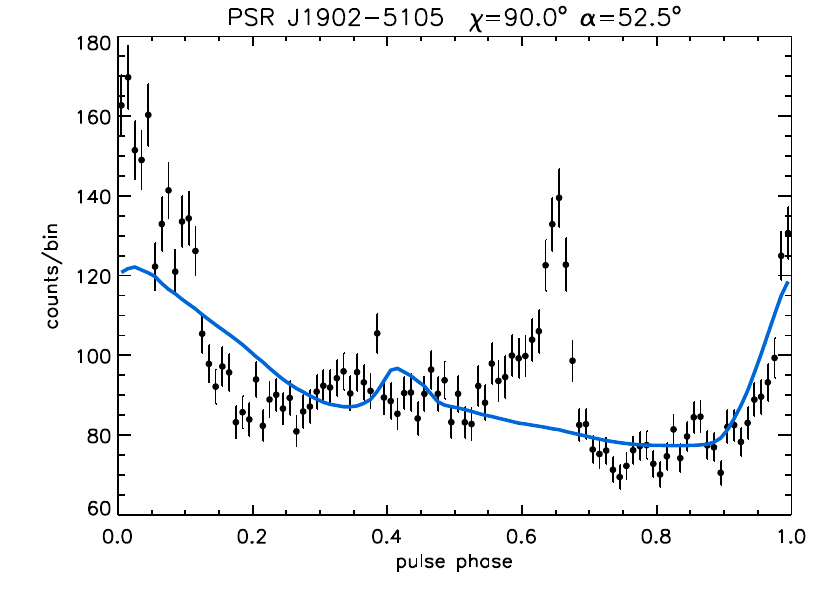}
\includegraphics[width=4.5cm]{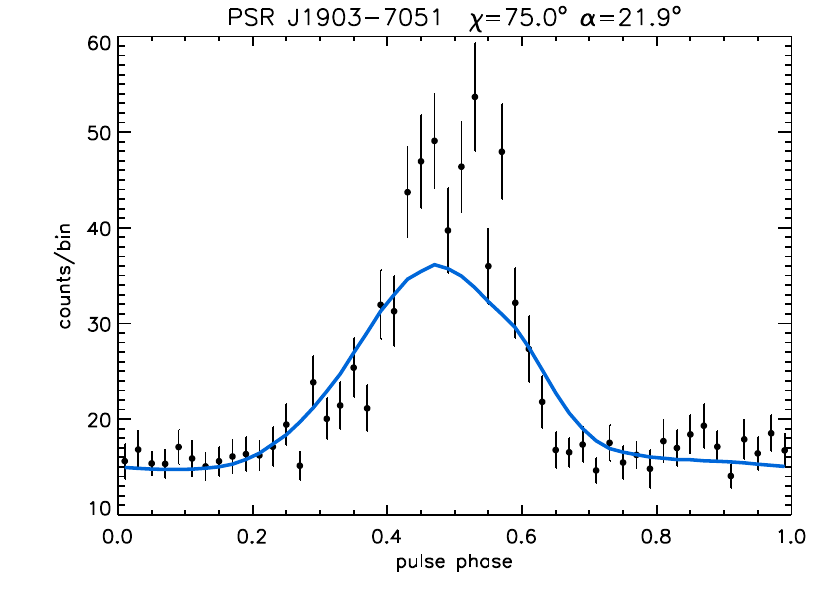}
\includegraphics[width=4.5cm]{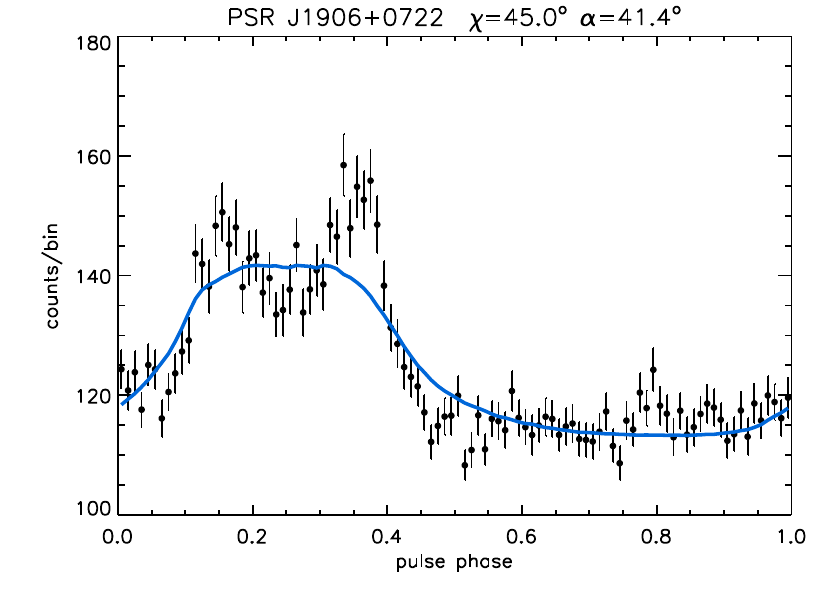}
\includegraphics[width=4.5cm]{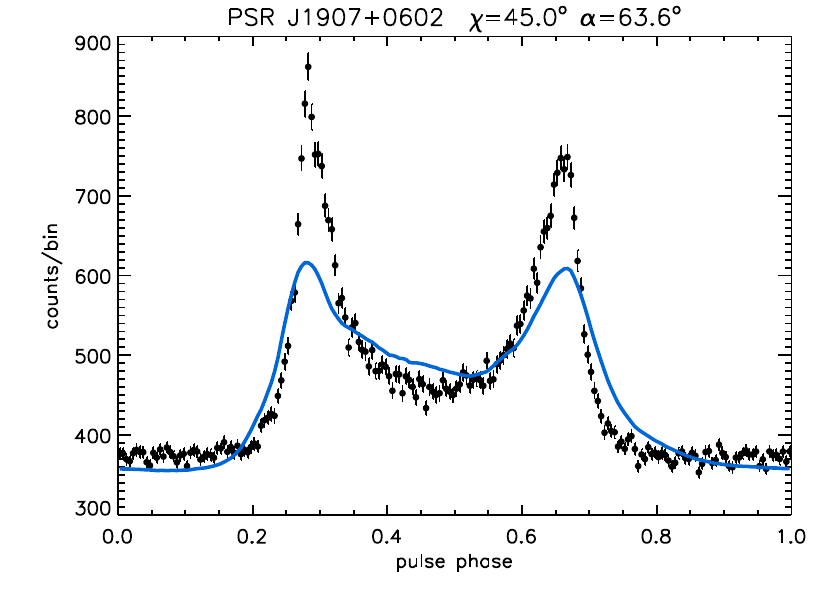}
\includegraphics[width=4.5cm]{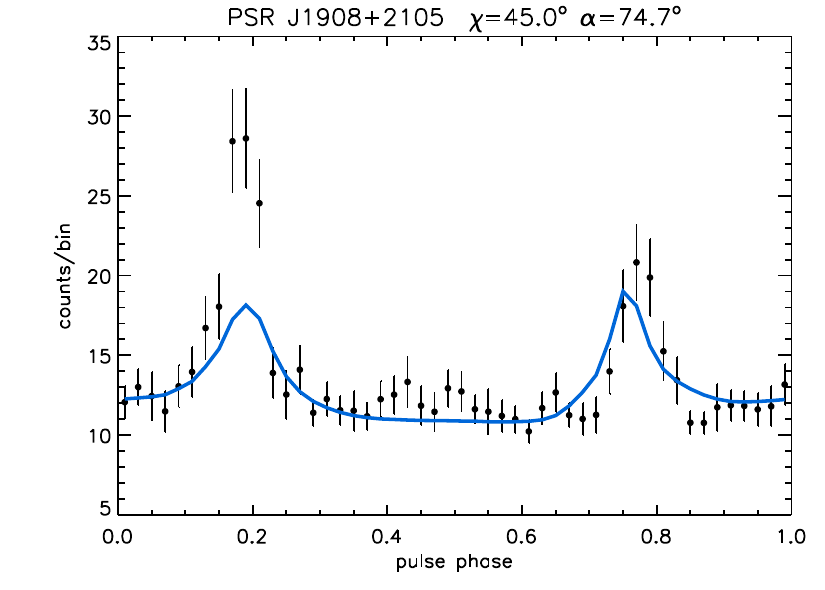}
\includegraphics[width=4.5cm]{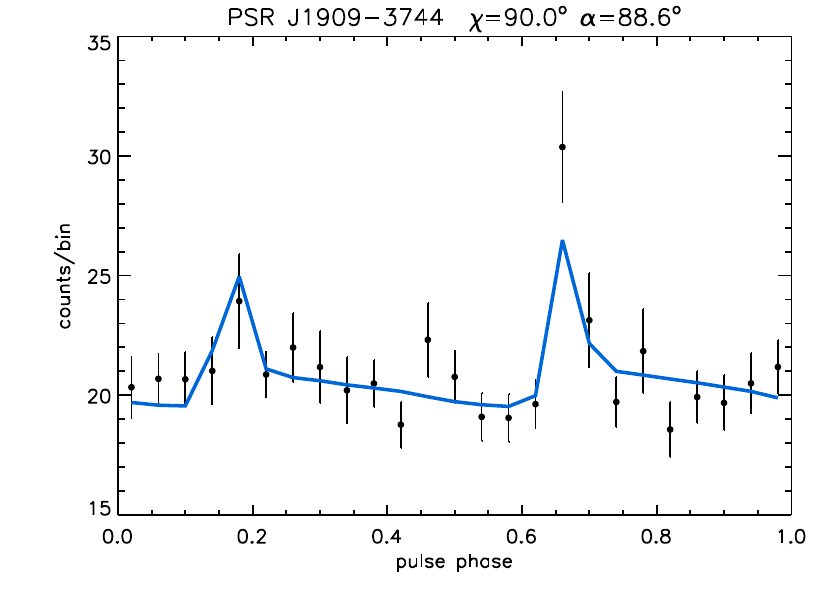}
\includegraphics[width=4.5cm]{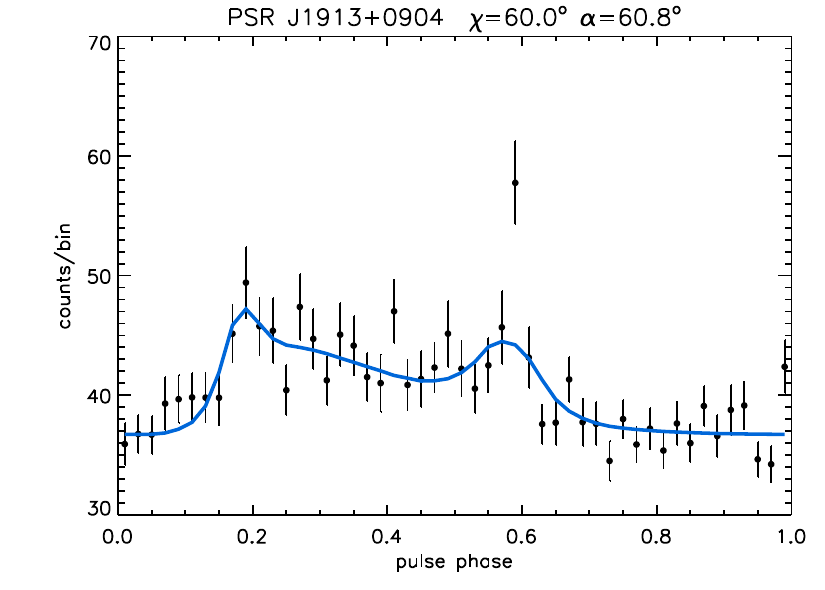}
\includegraphics[width=4.5cm]{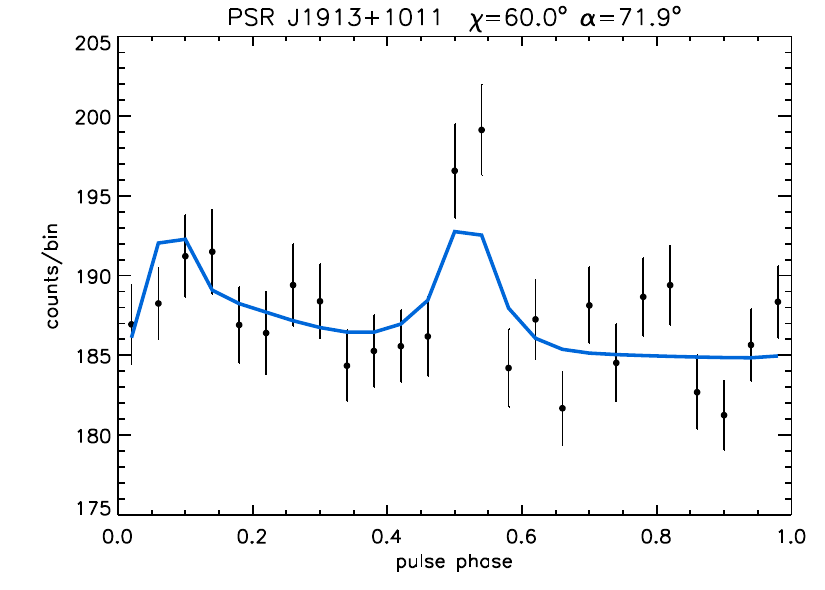}
\includegraphics[width=4.5cm]{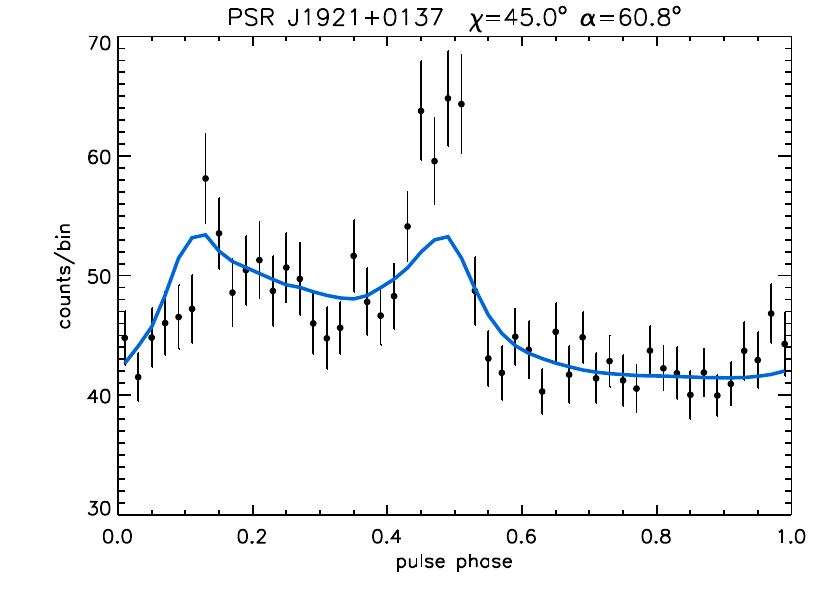}
\includegraphics[width=4.5cm]{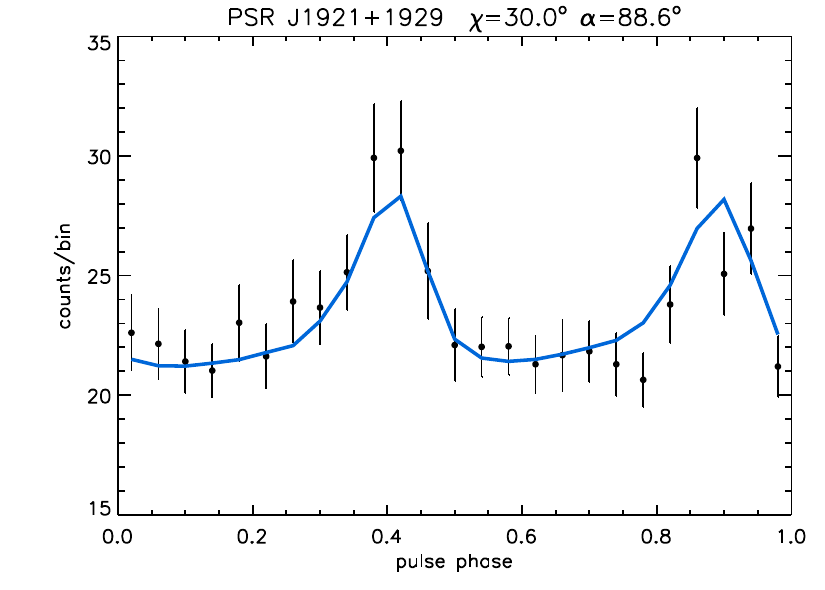}
\includegraphics[width=4.5cm]{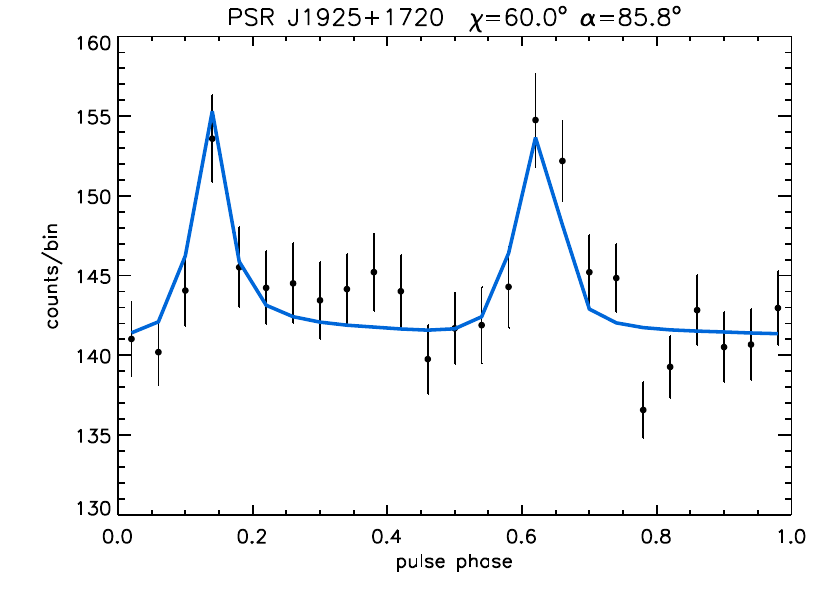}
\includegraphics[width=4.5cm]{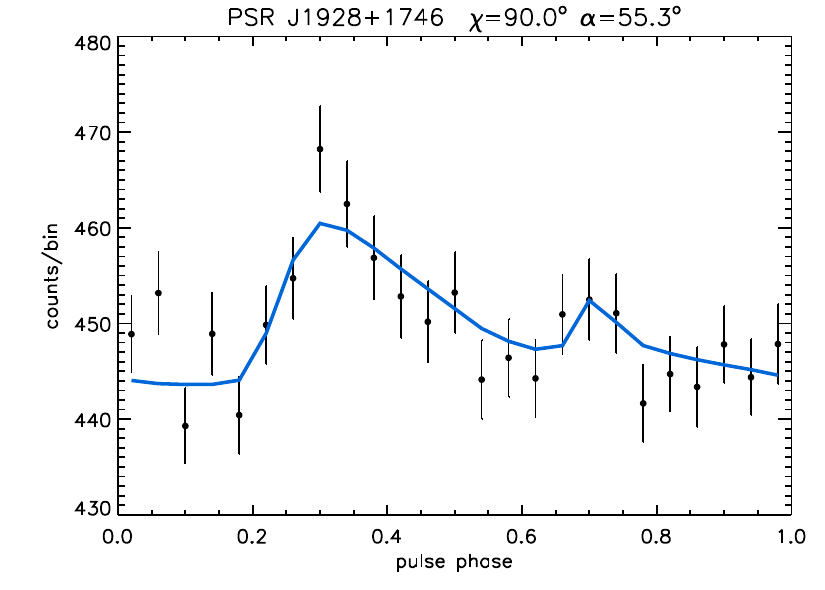}
\includegraphics[width=4.5cm]{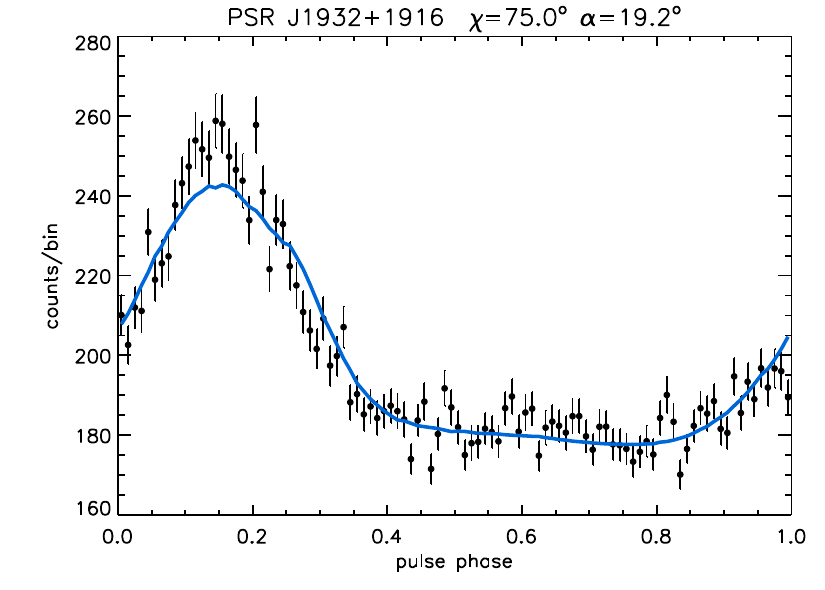}
\includegraphics[width=4.5cm]{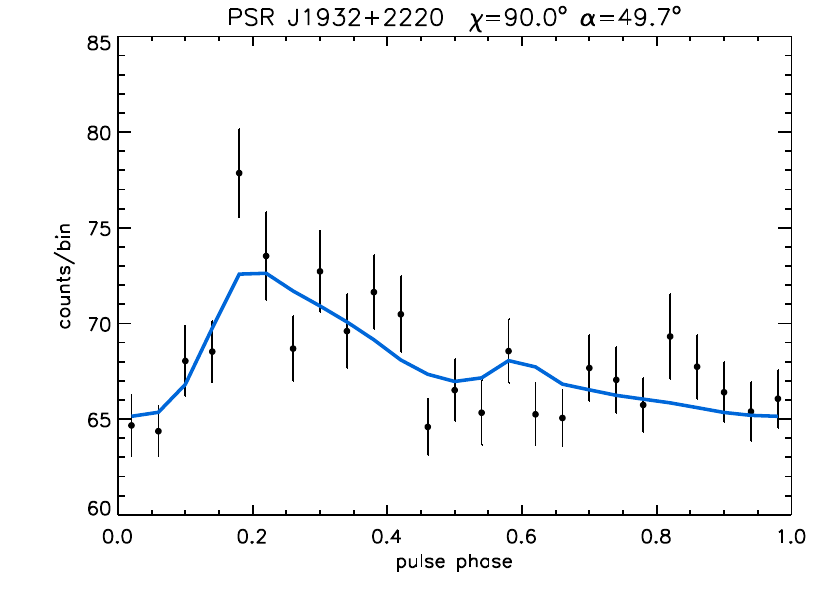}
\includegraphics[width=4.5cm]{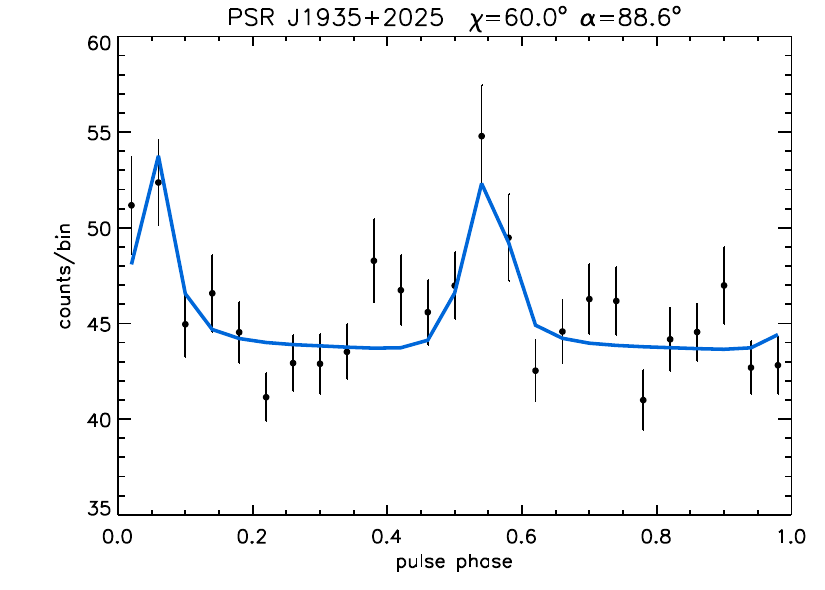}
\includegraphics[width=4.5cm]{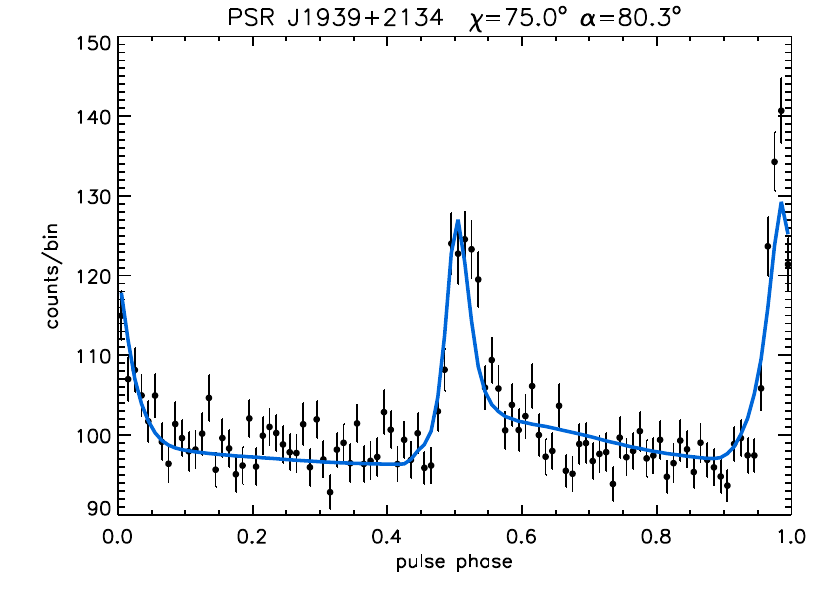}
\includegraphics[width=4.5cm]{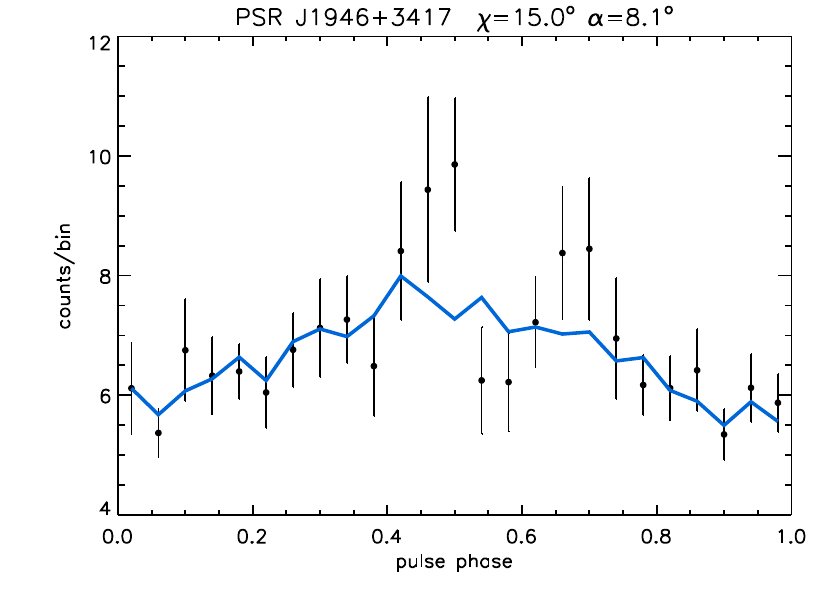}
\includegraphics[width=4.5cm]{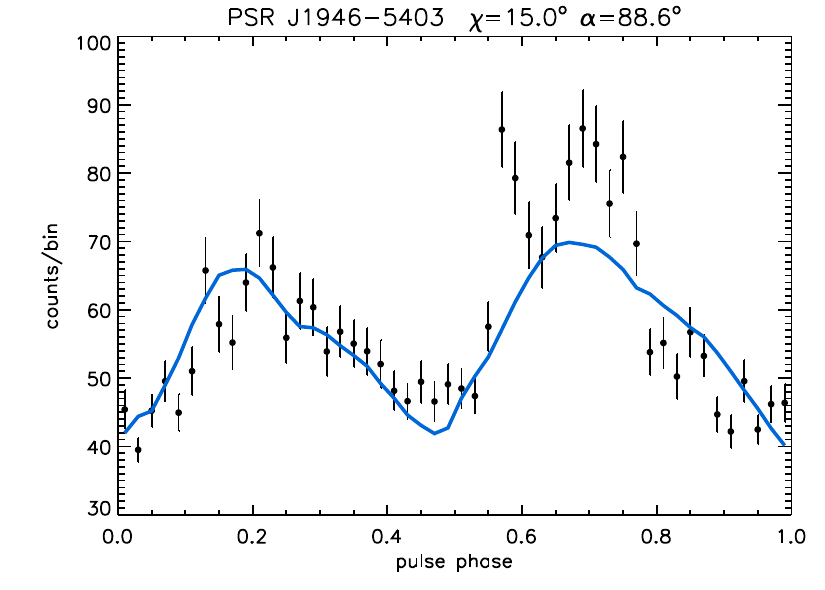}
\includegraphics[width=4.5cm]{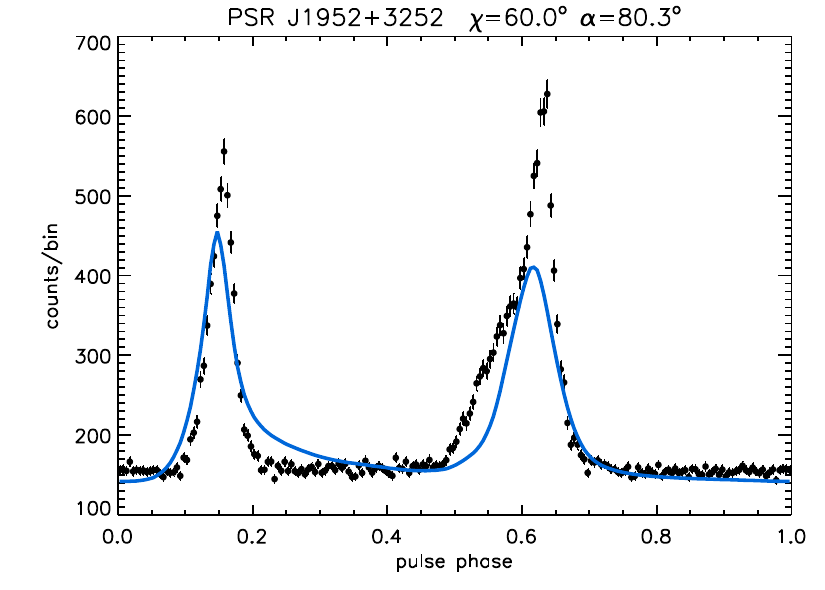}
\includegraphics[width=4.5cm]{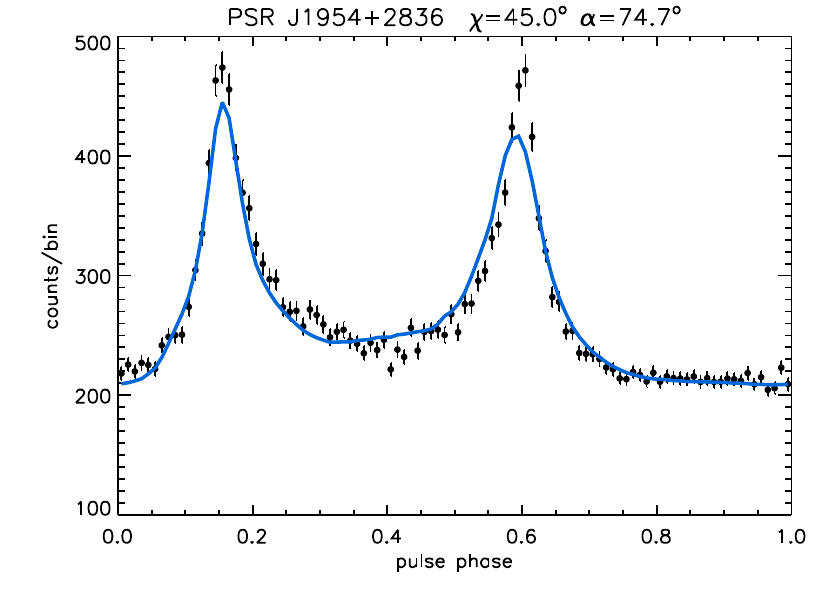}
\includegraphics[width=4.5cm]{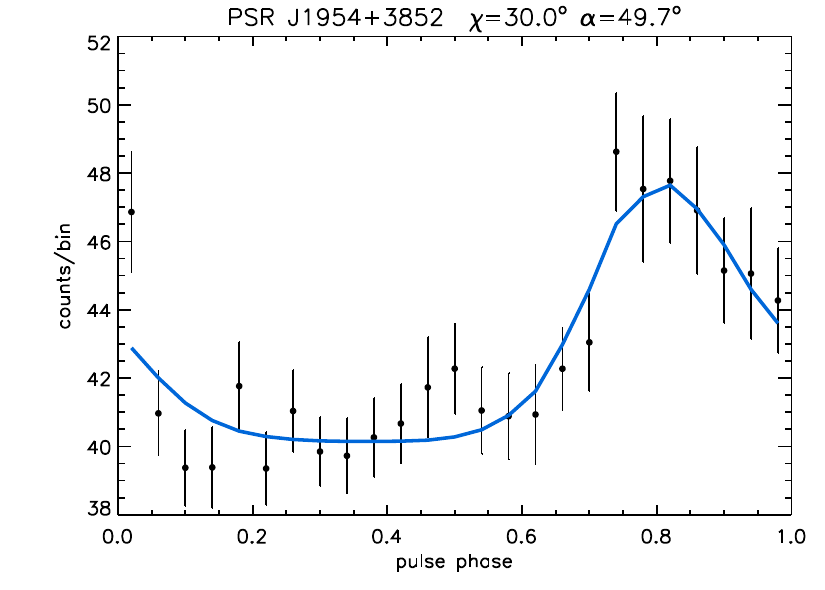}
\caption{continued.}
\end{figure*}

\begin{figure*}
\addtocounter{figure}{-1}
\centering
\includegraphics[width=4.5cm]{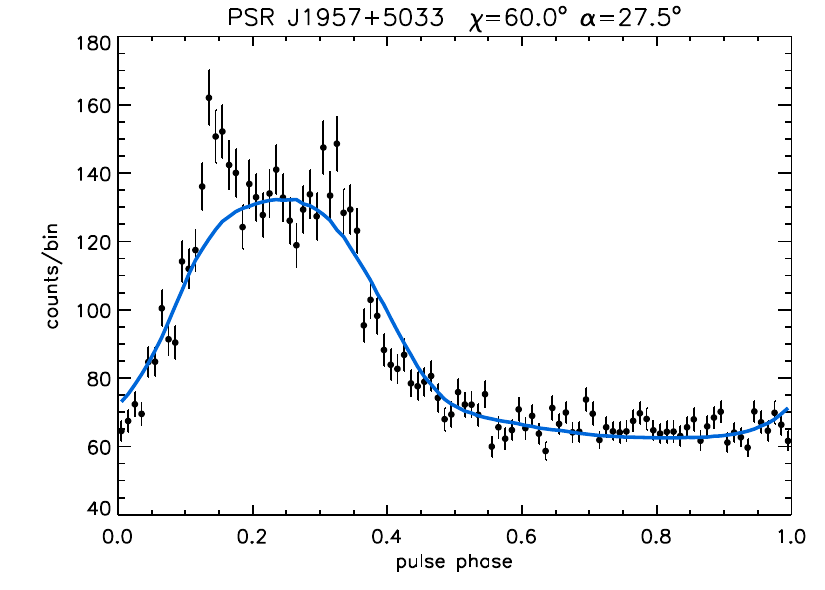}
\includegraphics[width=4.5cm]{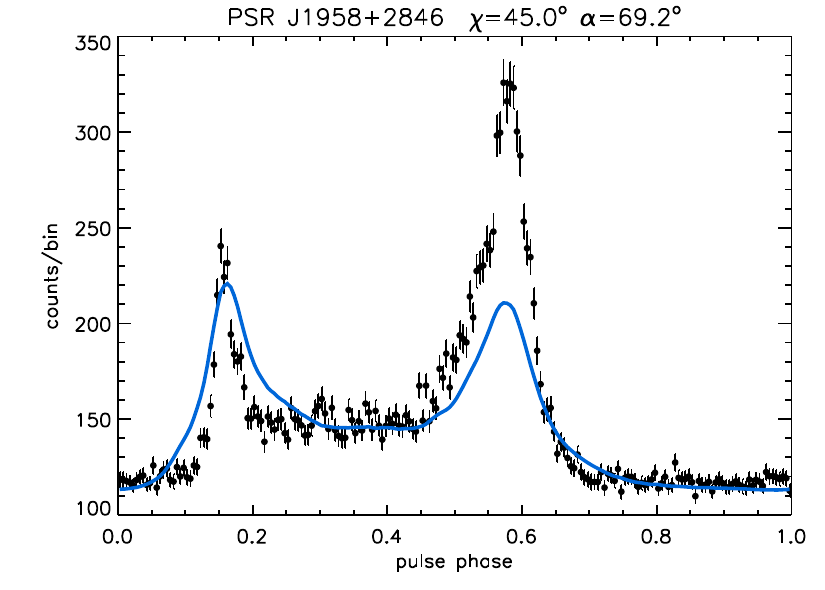}
\includegraphics[width=4.5cm]{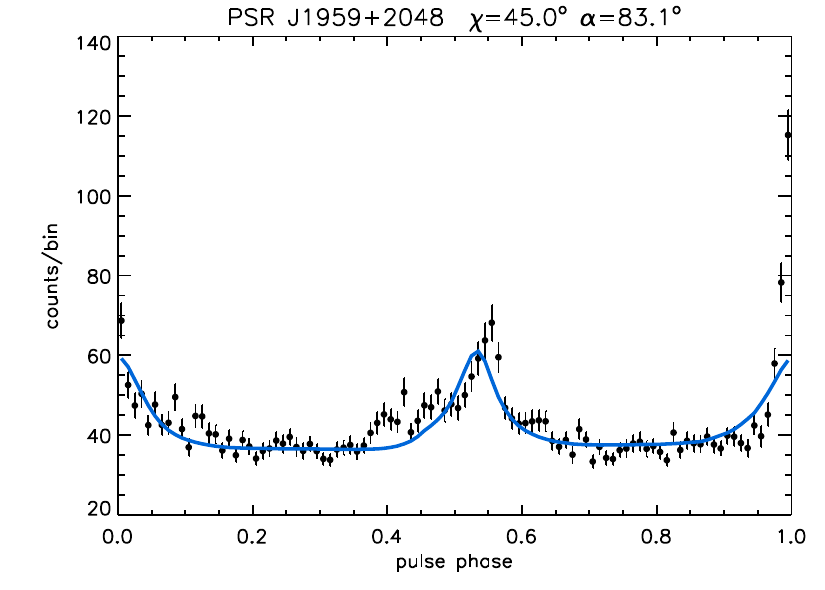}
\includegraphics[width=4.5cm]{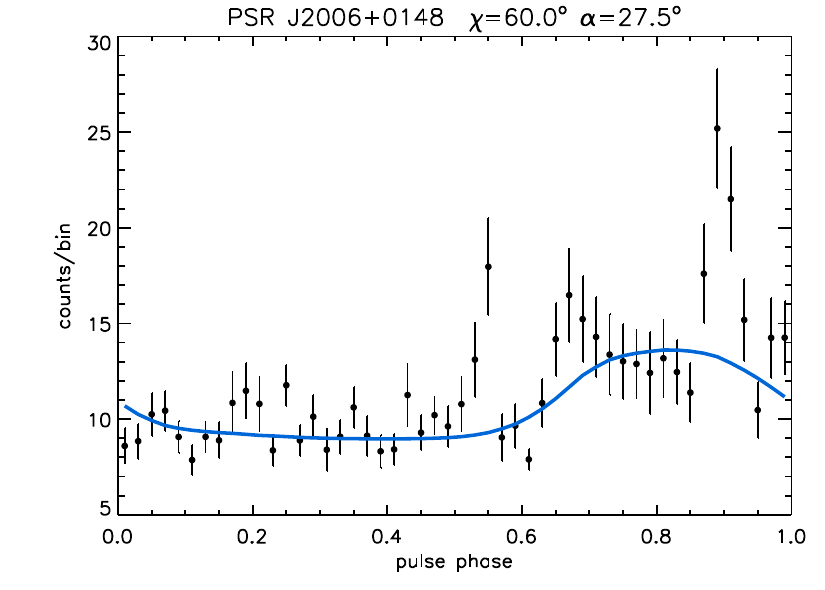}
\includegraphics[width=4.5cm]{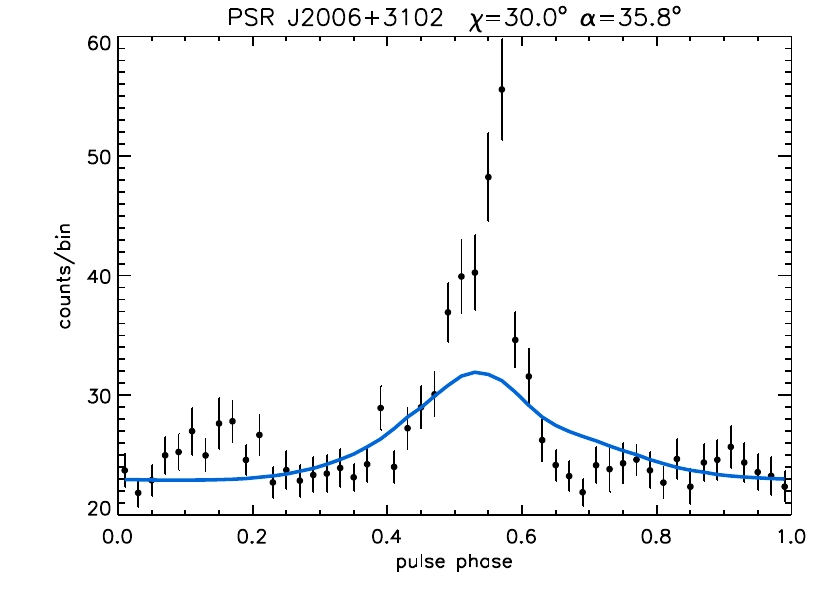}
\includegraphics[width=4.5cm]{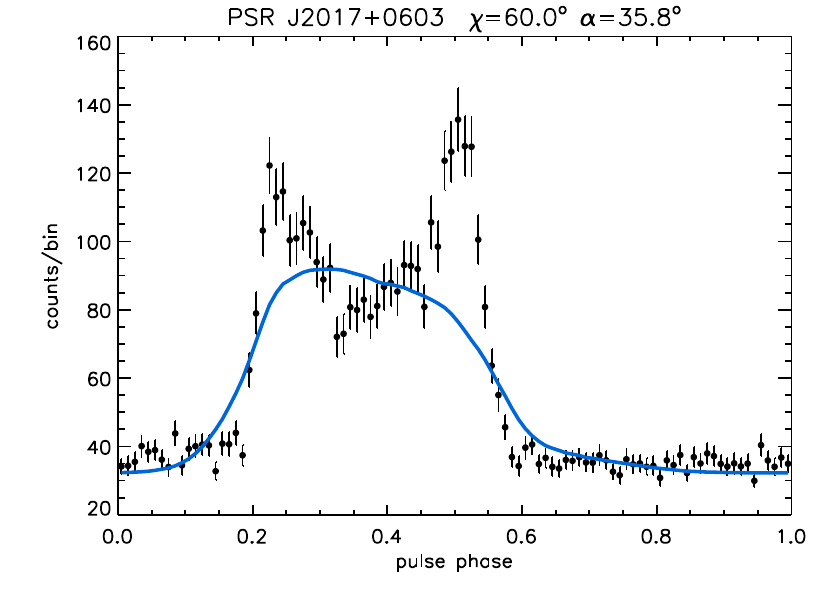}
\includegraphics[width=4.5cm]{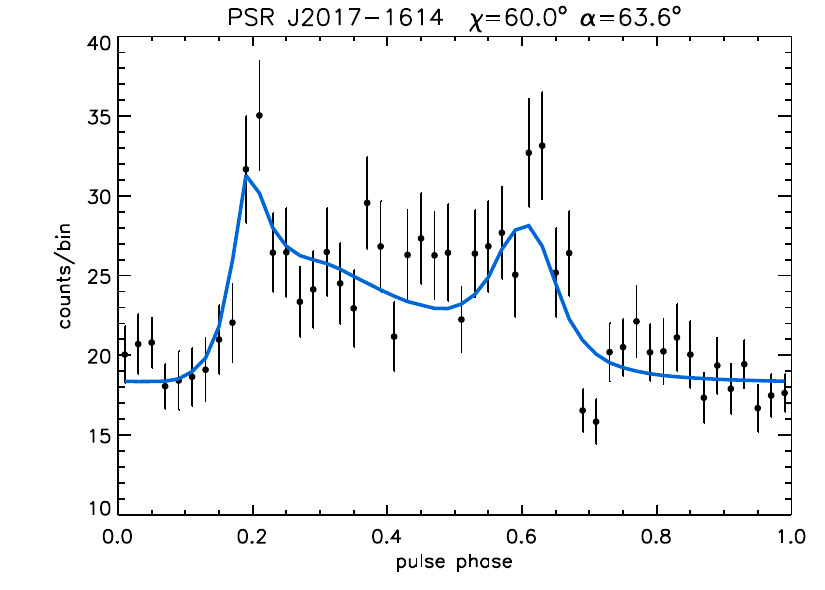}
\includegraphics[width=4.5cm]{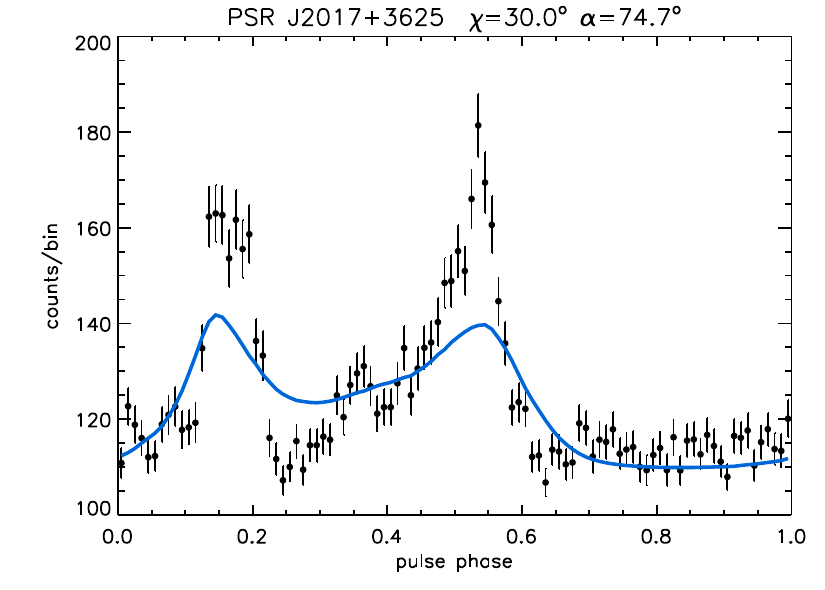}
\includegraphics[width=4.5cm]{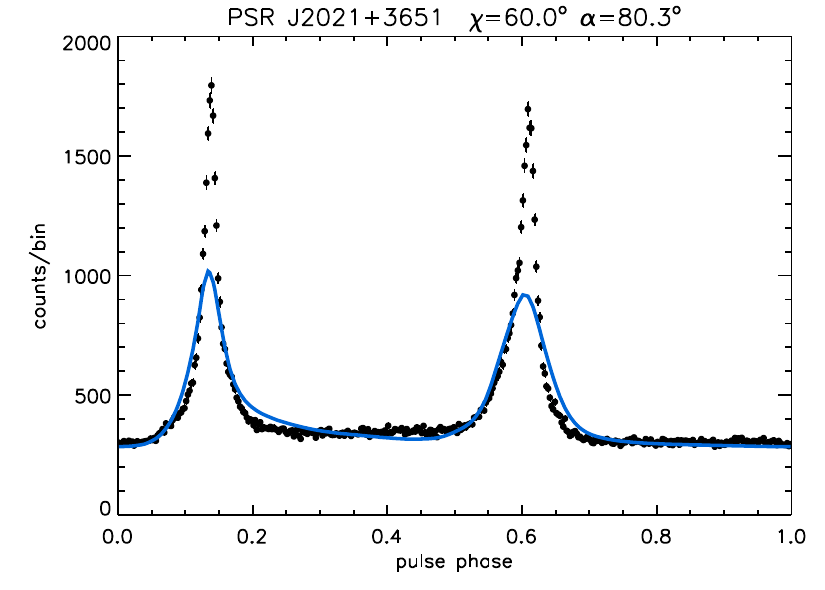}
\includegraphics[width=4.5cm]{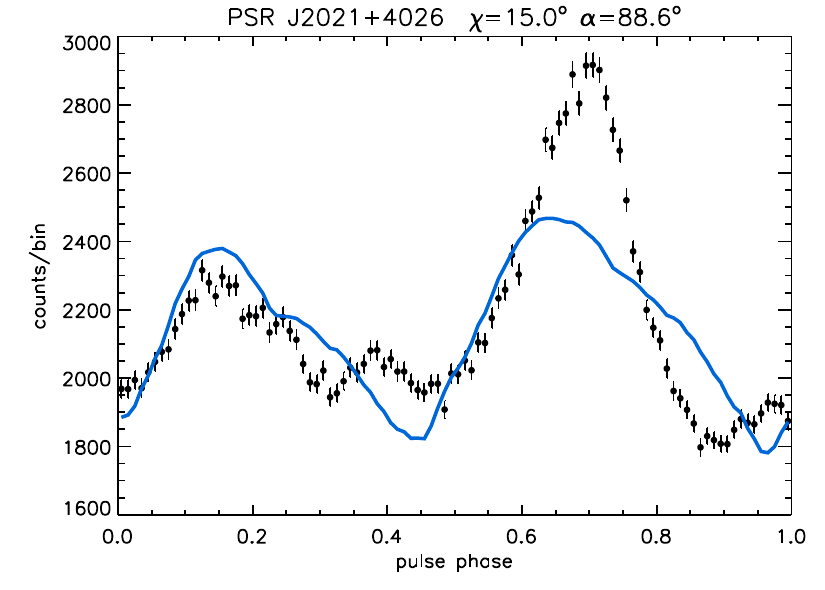}
\includegraphics[width=4.5cm]{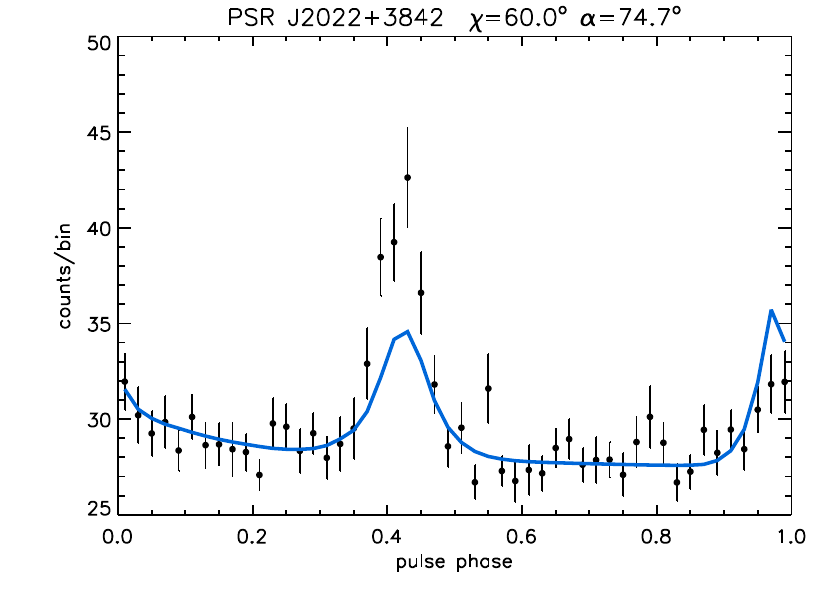}
\includegraphics[width=4.5cm]{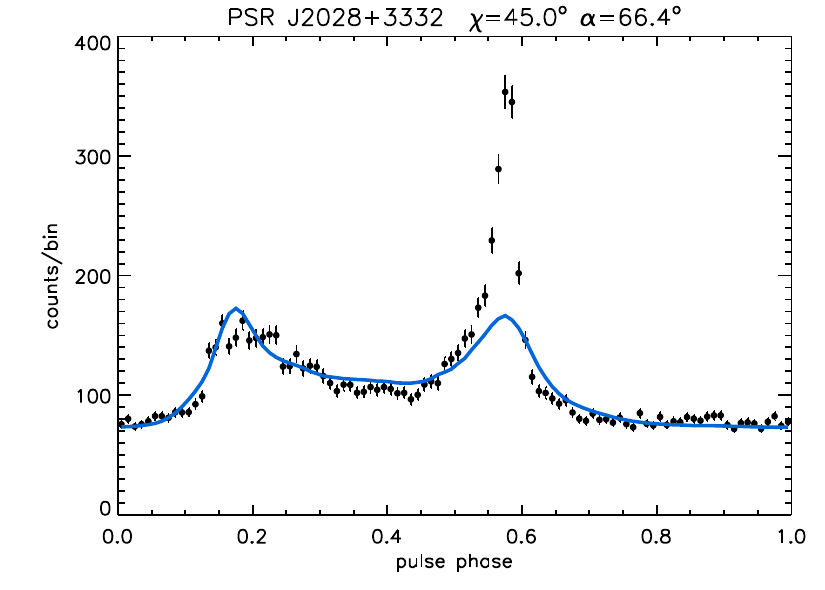}
\includegraphics[width=4.5cm]{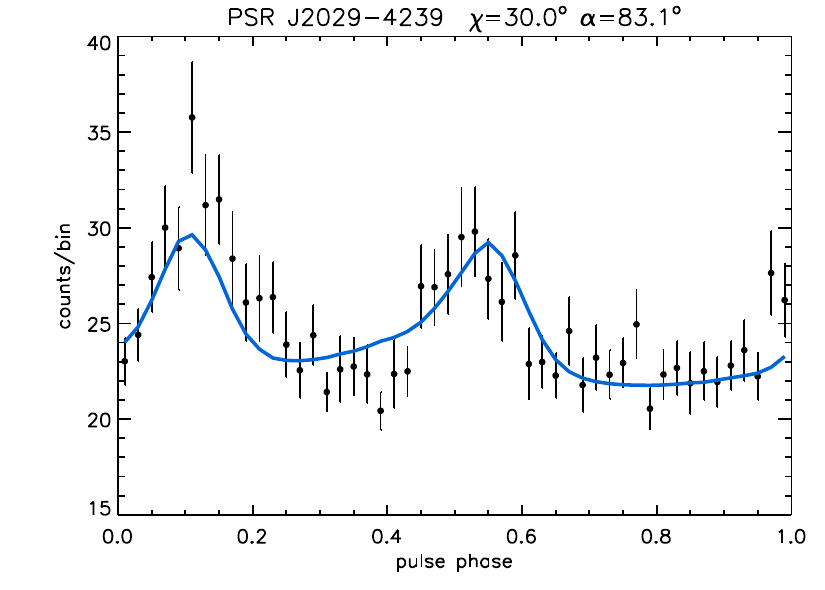}
\includegraphics[width=4.5cm]{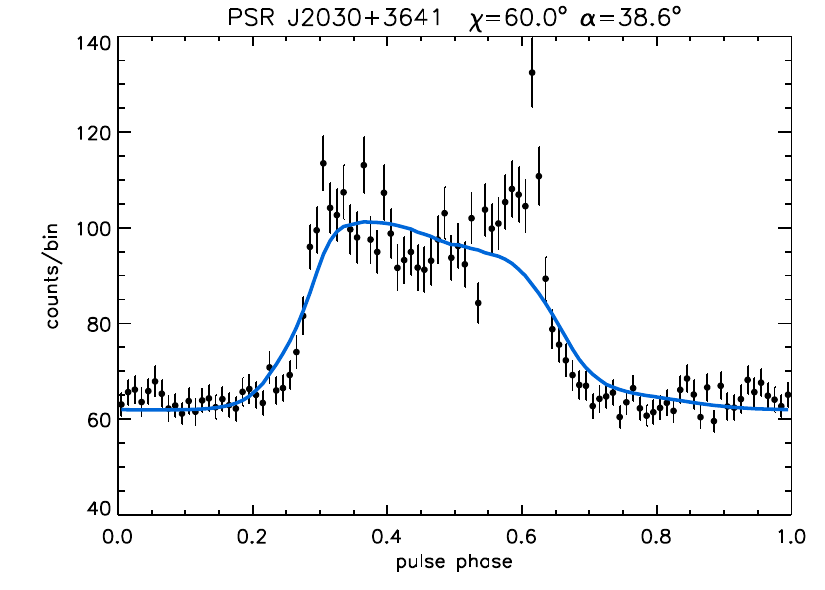}
\includegraphics[width=4.5cm]{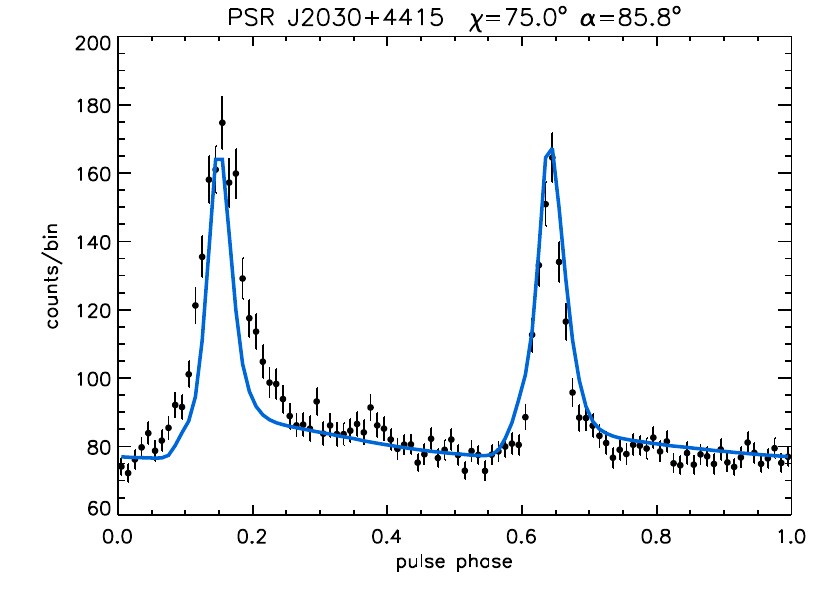}
\includegraphics[width=4.5cm]{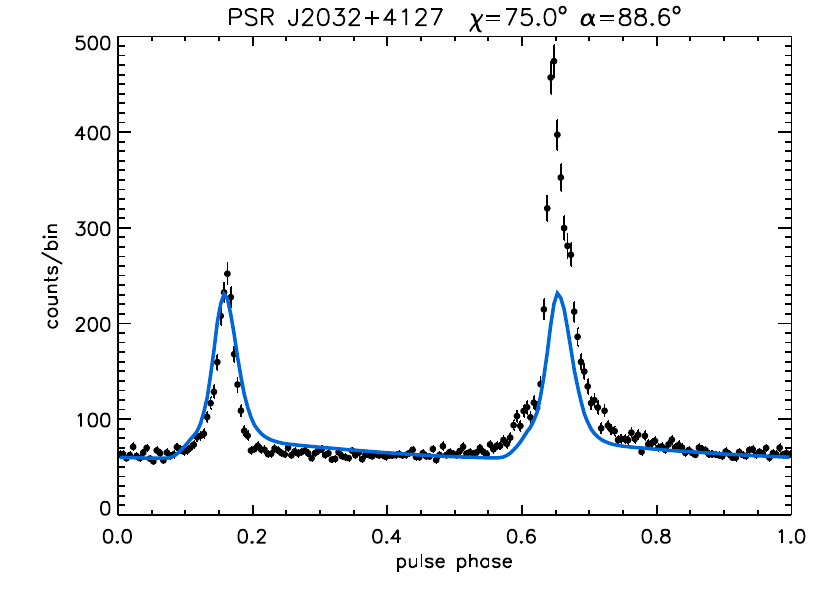}
\includegraphics[width=4.5cm]{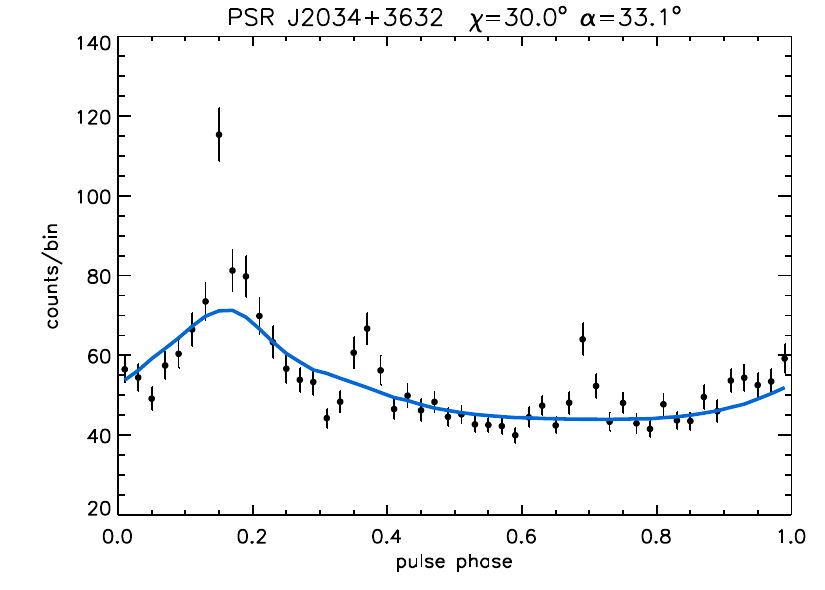}
\includegraphics[width=4.5cm]{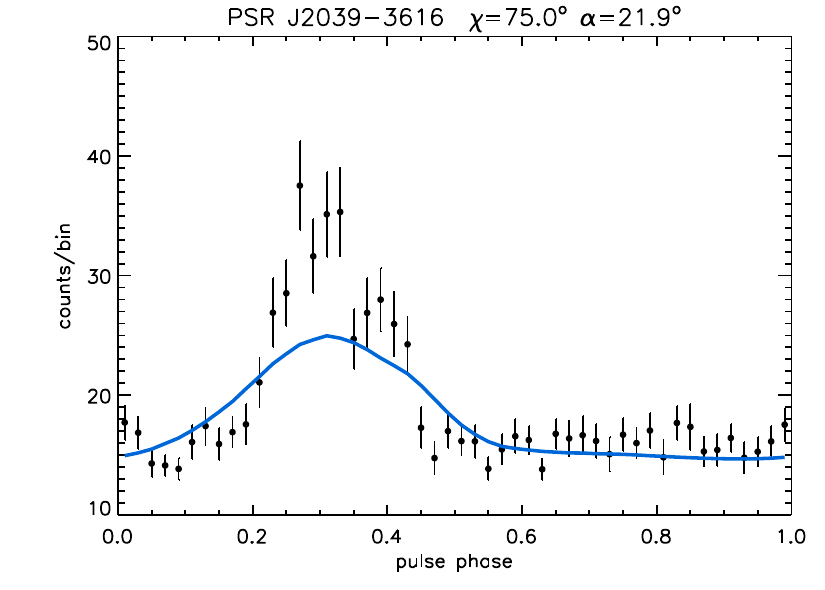}
\includegraphics[width=4.5cm]{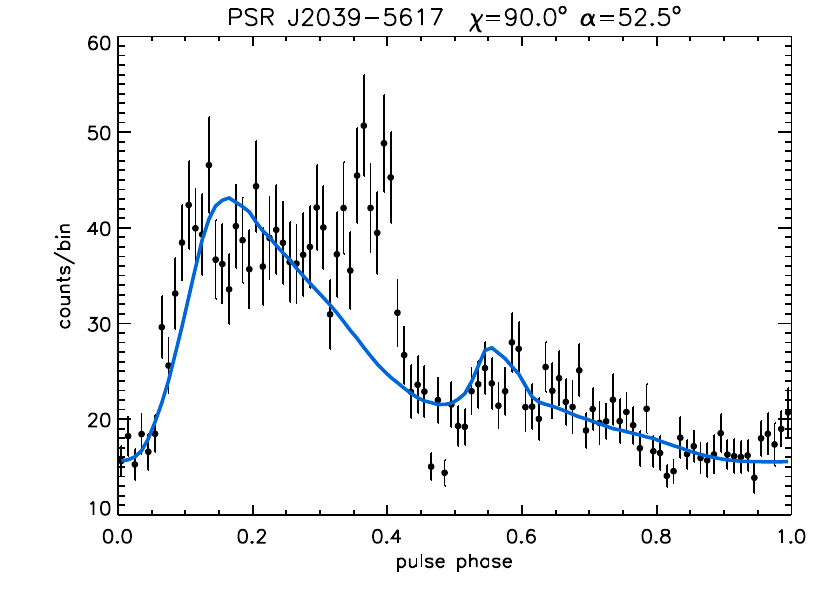}
\includegraphics[width=4.5cm]{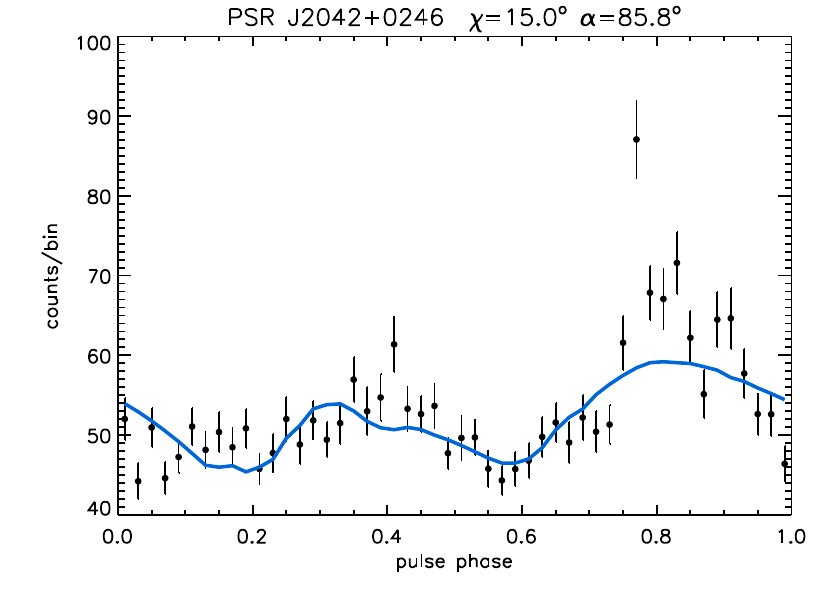}
\includegraphics[width=4.5cm]{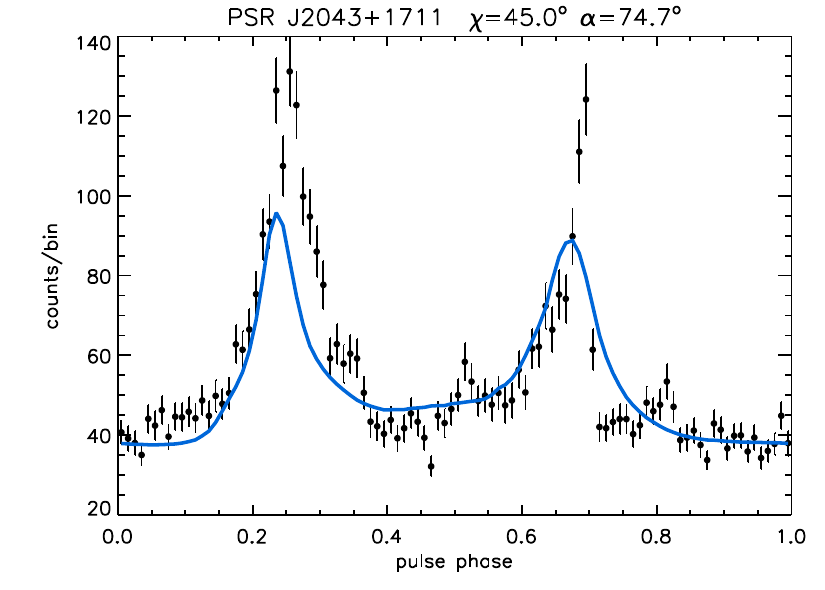}
\includegraphics[width=4.5cm]{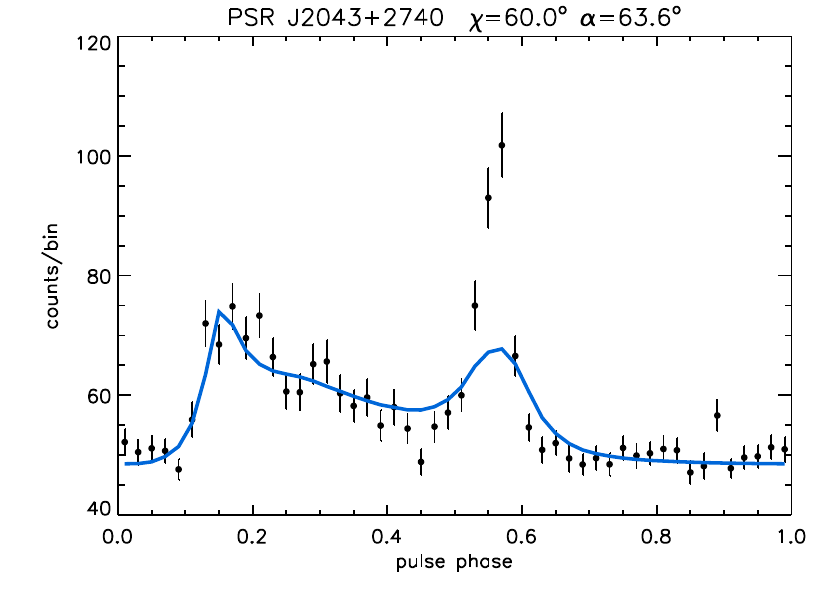}
\includegraphics[width=4.5cm]{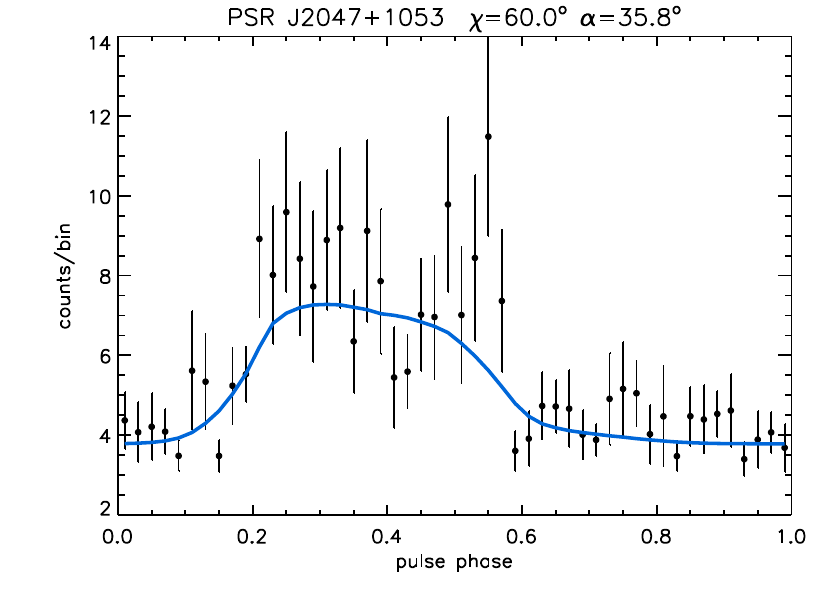}
\includegraphics[width=4.5cm]{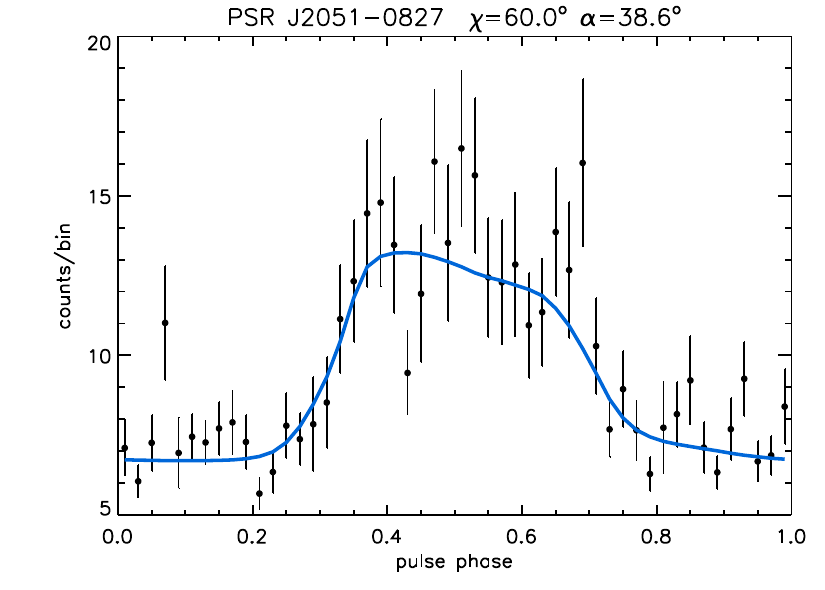}
\includegraphics[width=4.5cm]{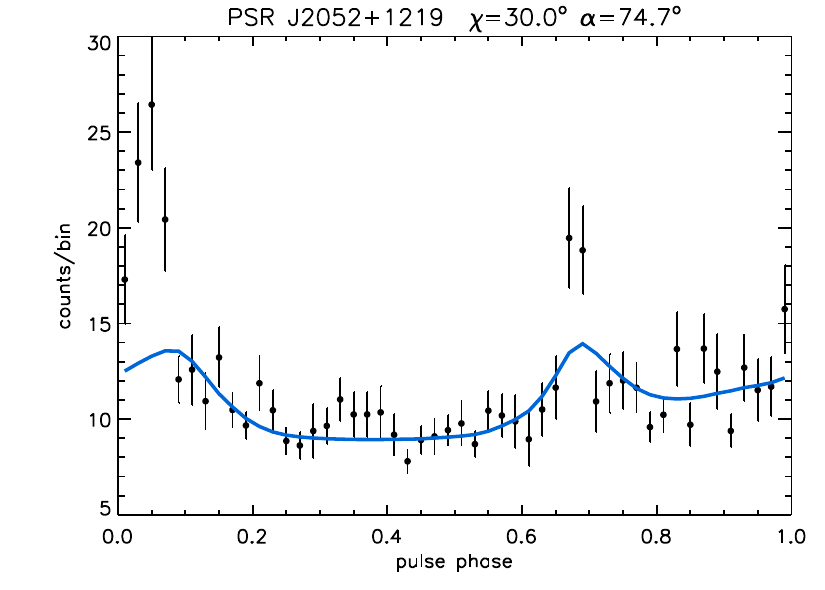}
\includegraphics[width=4.5cm]{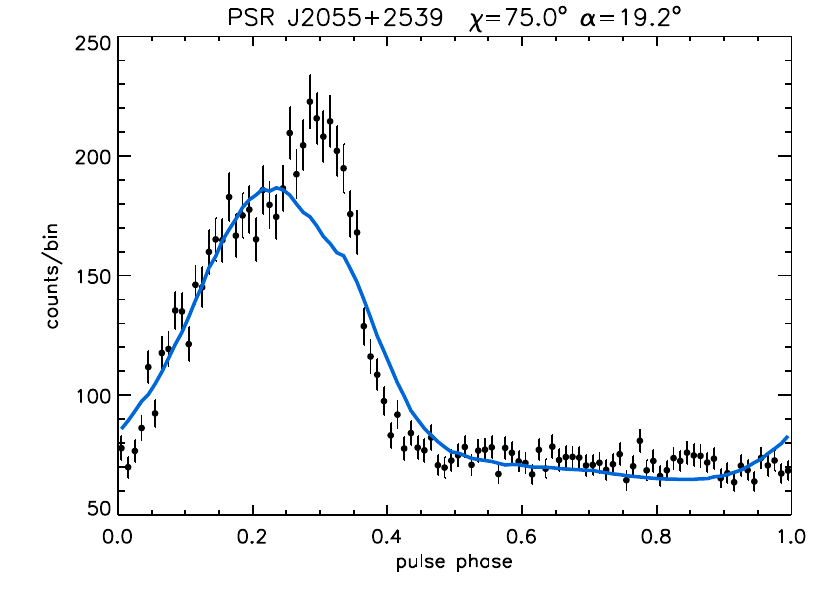}
\includegraphics[width=4.5cm]{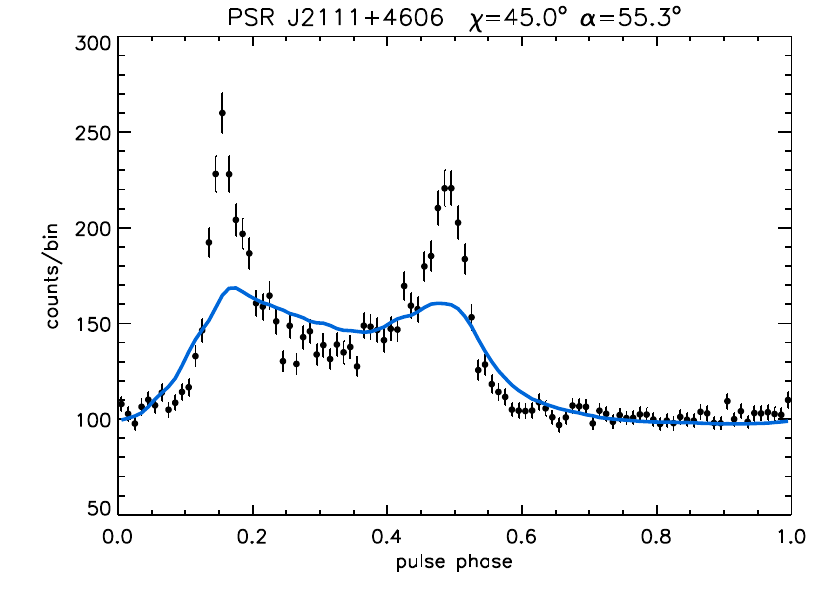}
\includegraphics[width=4.5cm]{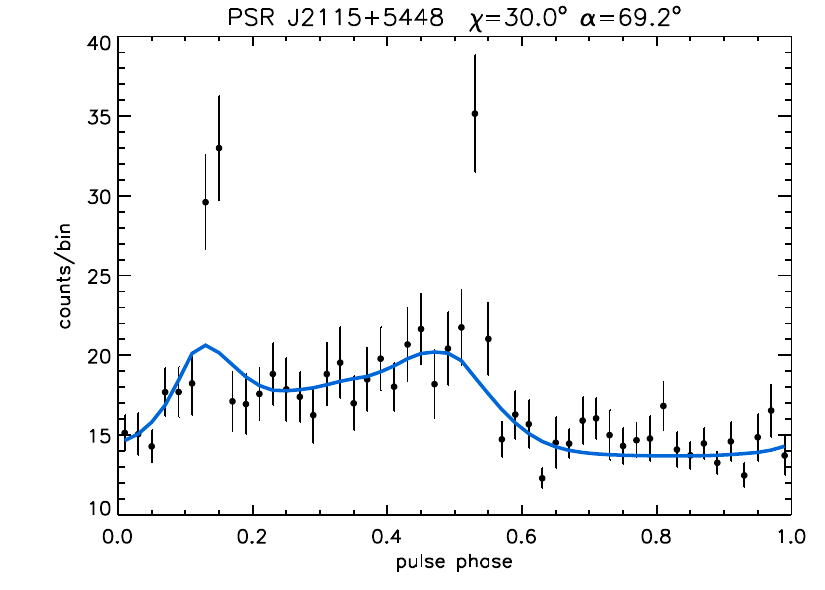}
\caption{continued.}
\end{figure*}

\begin{figure*}
\addtocounter{figure}{-1}
\centering
\includegraphics[width=4.5cm]{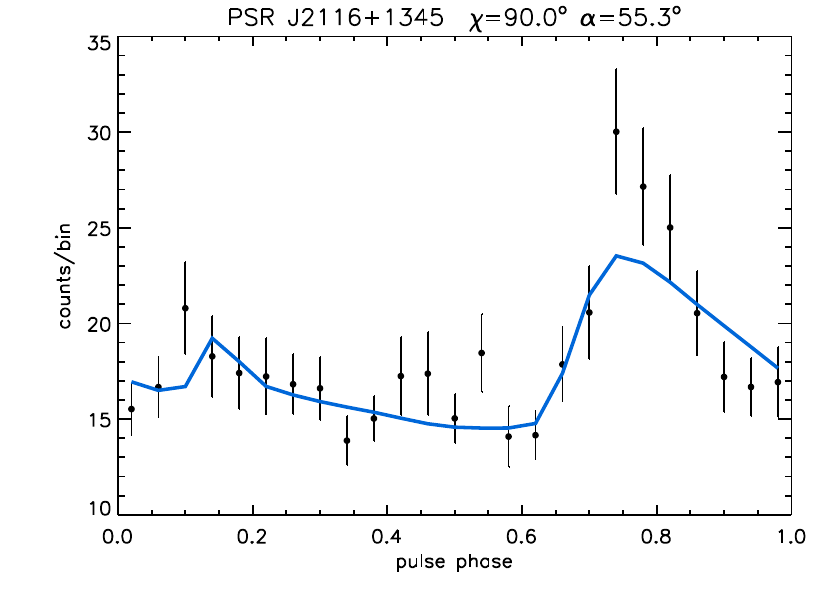}
\includegraphics[width=4.5cm]{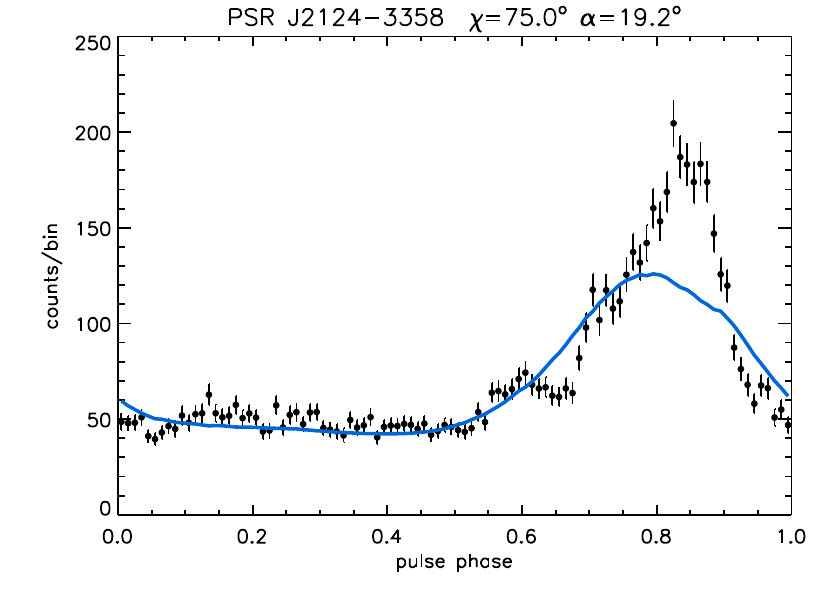}
\includegraphics[width=4.5cm]{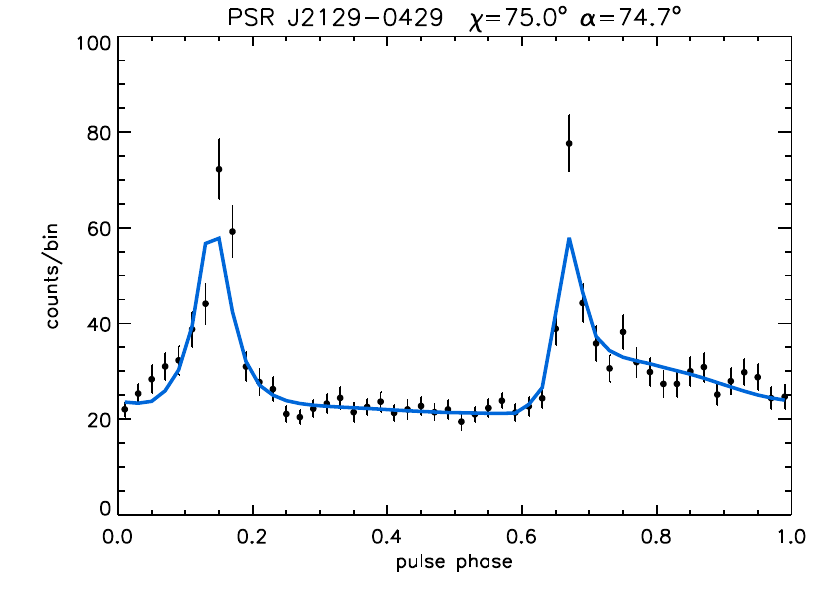}
\includegraphics[width=4.5cm]{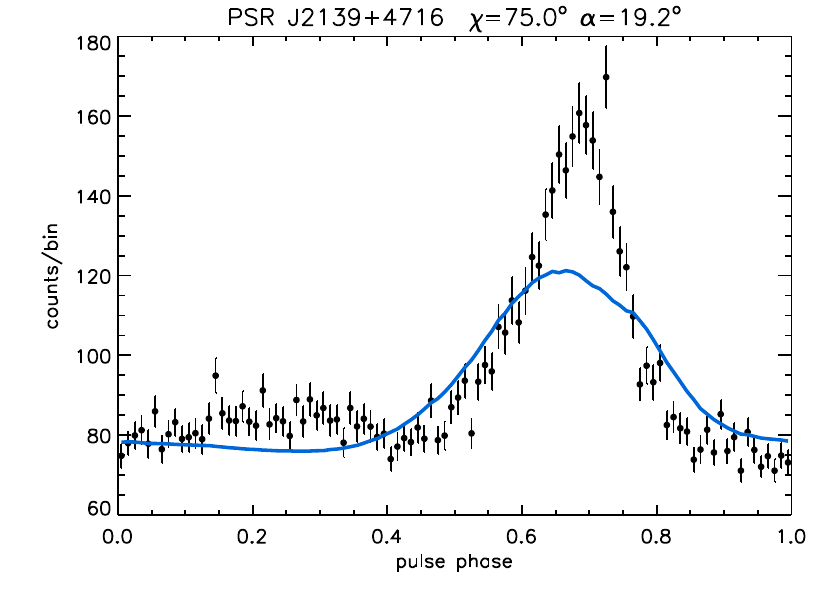}
\includegraphics[width=4.5cm]{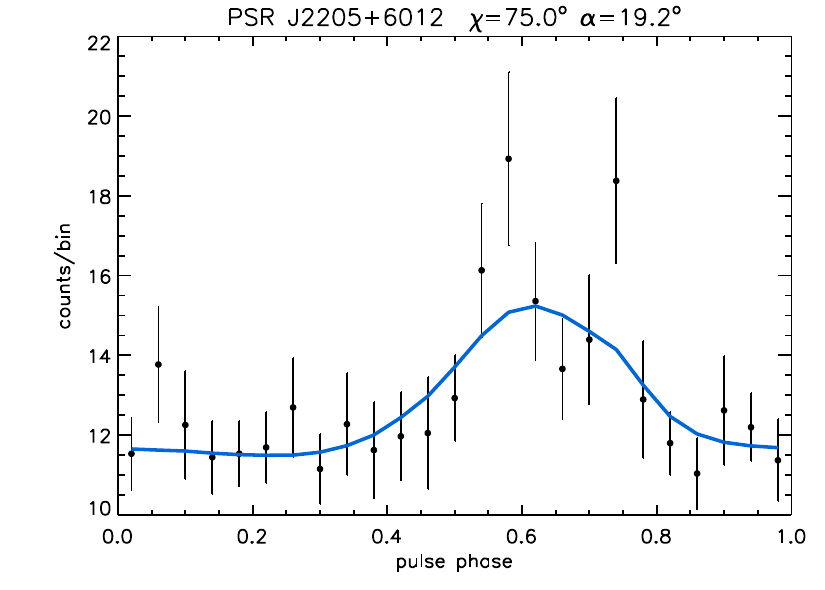}
\includegraphics[width=4.5cm]{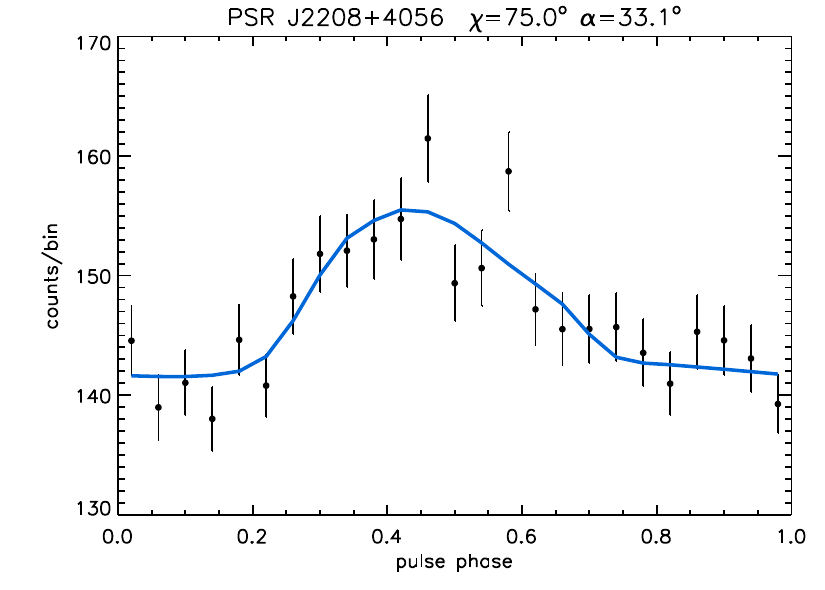}
\includegraphics[width=4.5cm]{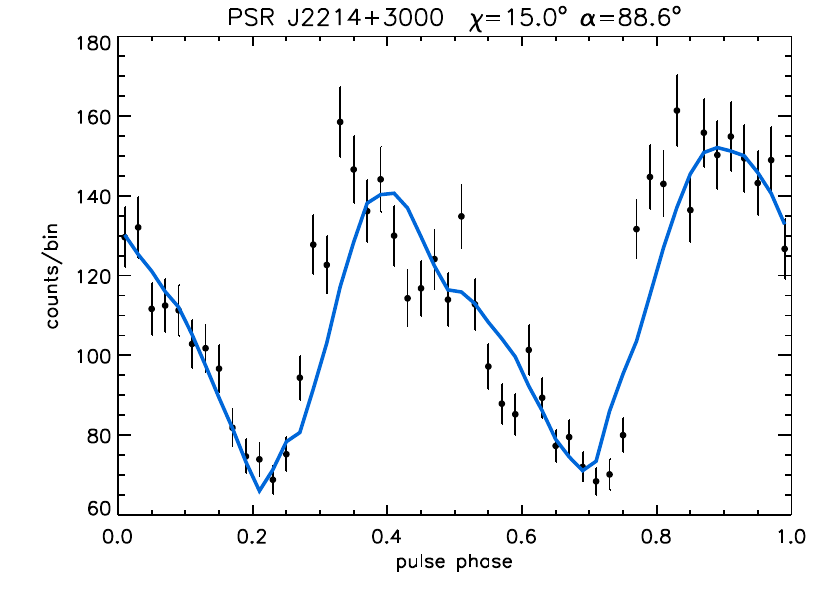}
\includegraphics[width=4.5cm]{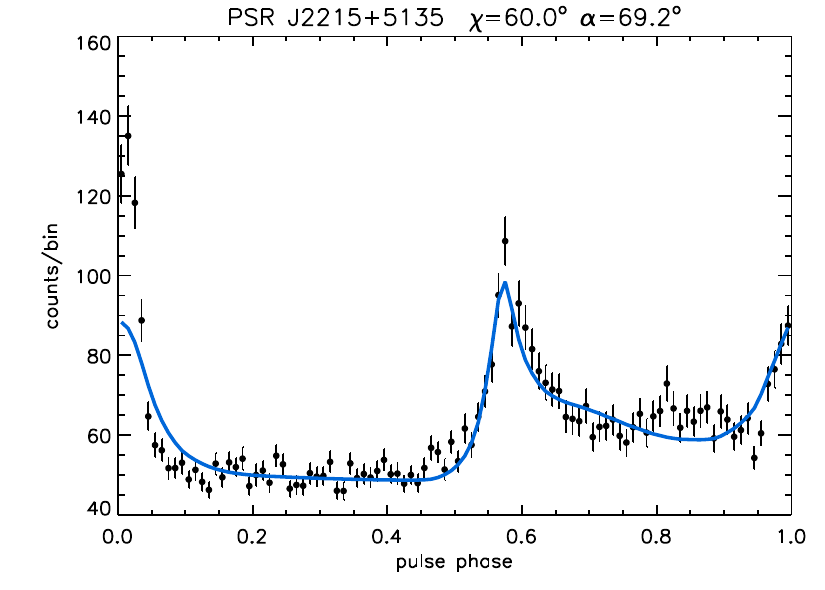}
\includegraphics[width=4.5cm]{J2229+6114.pdf}
\includegraphics[width=4.5cm]{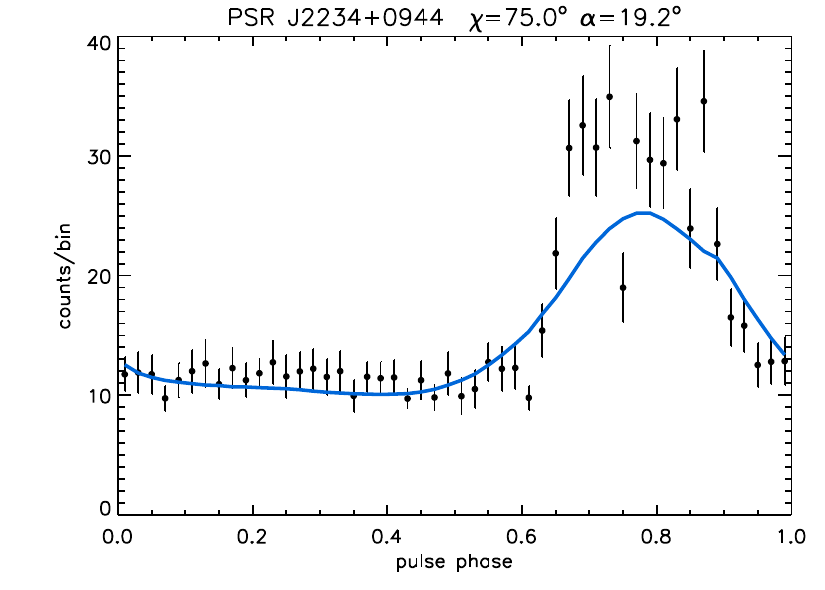}
\includegraphics[width=4.5cm]{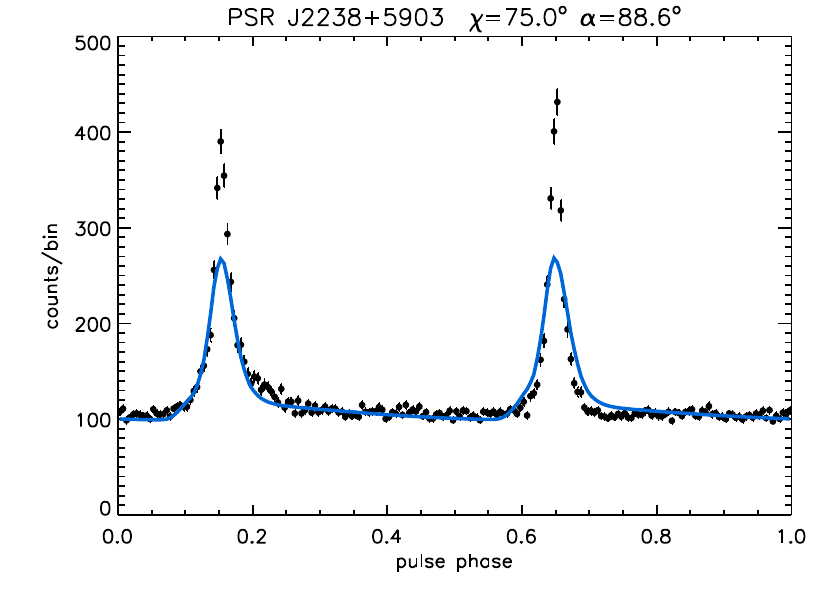}
\includegraphics[width=4.5cm]{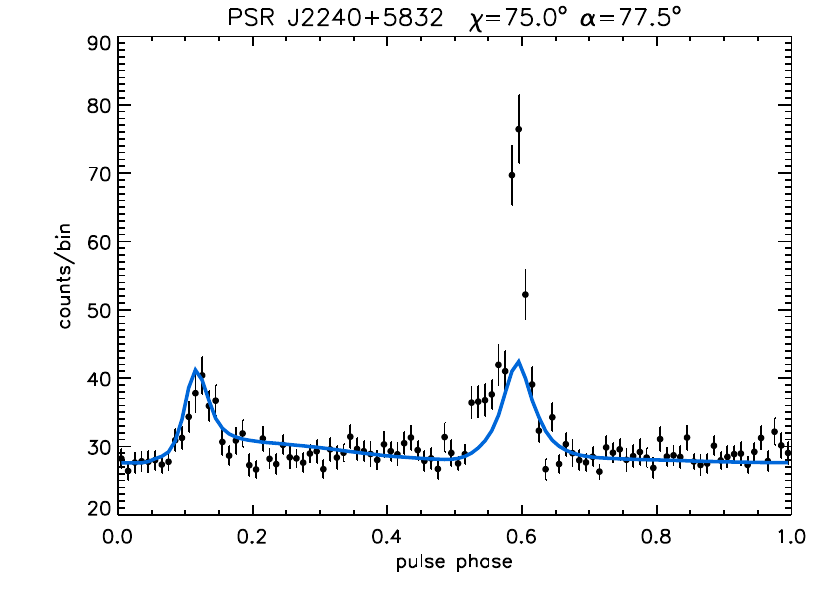}
\includegraphics[width=4.5cm]{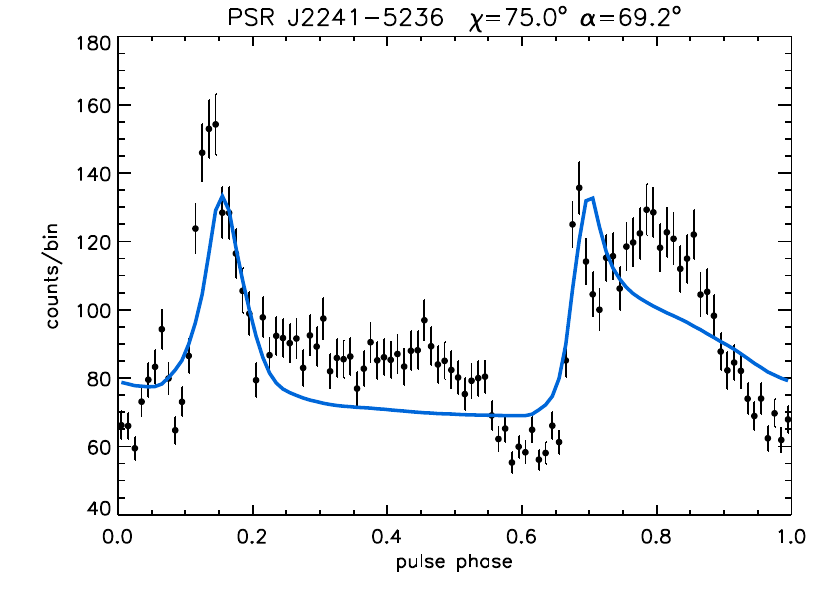}
\includegraphics[width=4.5cm]{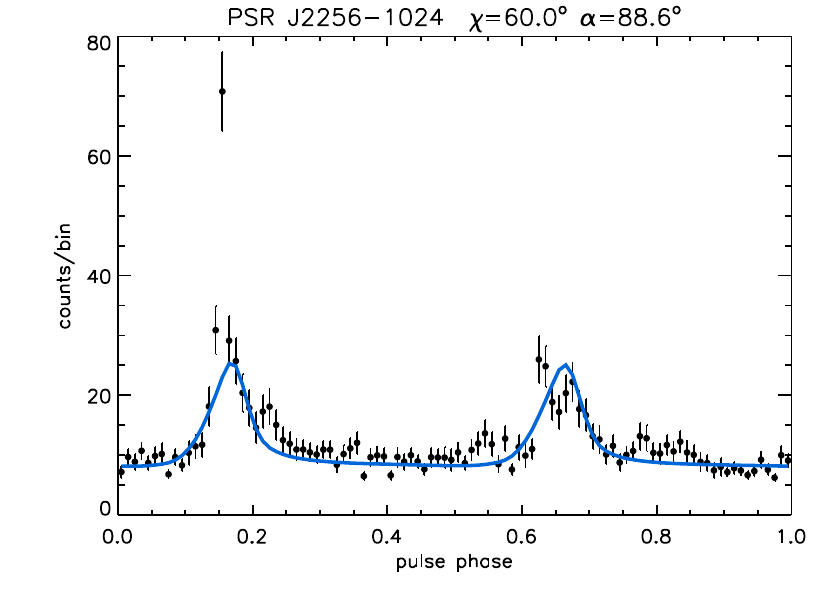}
\includegraphics[width=4.5cm]{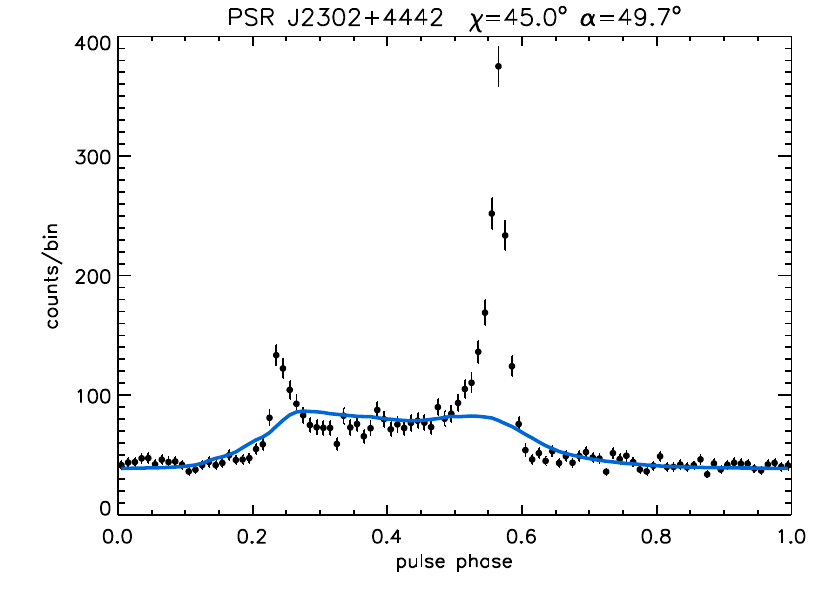}
\includegraphics[width=4.5cm]{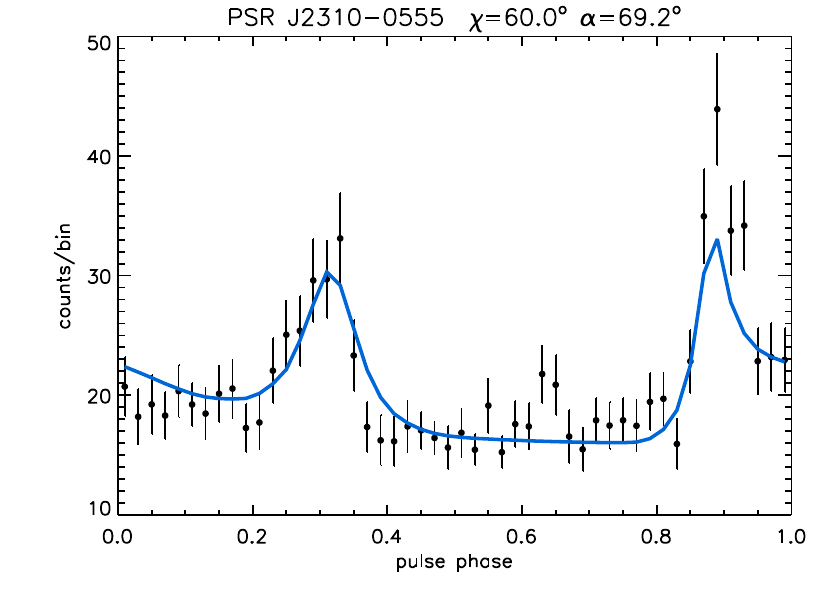}
\includegraphics[width=4.5cm]{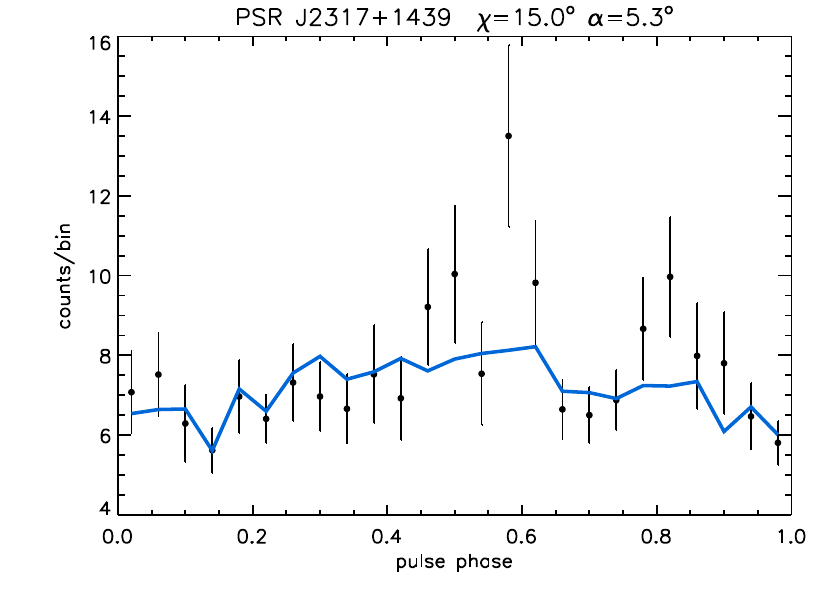}
\includegraphics[width=4.5cm]{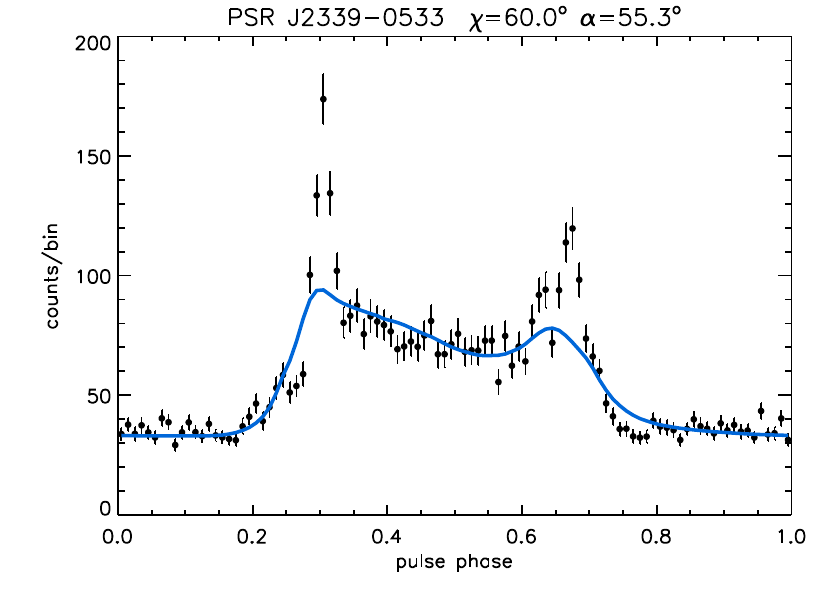}
\caption{end.}
\end{figure*}

\end{appendix}

\end{document}